\documentclass[11pt,a4paper]{article}

\pdfoutput=1

\usepackage{jheppub}
\usepackage{multicol,bbm}
\usepackage{bm}
\usepackage{multirow}
\usepackage{xcolor}
\usepackage{mathrsfs}
\usepackage[T1]{fontenc}
\usepackage{lmodern}
\usepackage{tocloft}
\usepackage[normalem]{ulem}

\newcommand{\nn}{\nonumber} 
\newcommand{\bn}{{\bar n}}

\newcommand{\be}{\begin{equation}}
\newcommand{\ee}{\end{equation}}
\newcommand{\bea}{\begin{eqnarray}}
\newcommand{\eea}{\end{eqnarray}}

\newcommand{\SCETa}{\mbox{${\rm SCET}_{\rm I}$ }}

\newcommand{\event}{\texttt{EVENT2}}

\newcommand{\vect}[1]{\mathbf{#1}}
\newcommand{\abs}[1]{\left\lvert #1\right\rvert}

\newcommand{\Lqcd}{\Lambda_{\text{QCD}}}

\newcommand{\nkll}{N$^k$LL\ }

\newcommand{\as}{\alpha_s}
\newcommand{\minus}{\!-\!}
\newcommand{\plus}{\!+\!}
\newcommand{\Gcusp}{\Gamma_{\rm cusp}}

\newcommand{\MSbar}{\overline{\text{MS}}}

\newcommand{\cO}{\mathcal{O}}

\newcommand{\cG}{\mathcal{G}}

\newcommand{\HypF}{{}_2F_{1}}

\newcommand{\wt}[1]{\widetilde{#1}}

\newcommand{\eq}[1]{Eq.~\eqref{eq:#1}}
\newcommand{\eqs}[2]{Eqs.~\eqref{eq:#1} and \eqref{eq:#2}}

\renewcommand{\sec}[1]{Sec.~\ref{sec:#1}}
\newcommand{\ssec}[1]{Sec.~\ref{ssec:#1}}
\newcommand{\appx}[1]{App.~\ref{app:#1}}
\newcommand{\fig}[1]{Fig.~\ref{fig:#1}}
\newcommand{\figs}[2]{Figs.~\ref{fig:#1} and \ref{fig:#2}}
\newcommand{\tab}[1]{Table~\ref{tab:#1}}

\newcommand{\red}[1]{{\color{red}#1}}
\newcommand{\blue}[1]{{\color{blue}#1}}
\newcommand{\green}[1]{{\color{teal}#1}}

\newcommand{\LP}{\mathscr{L}}
\newcommand{\iLP}{\mathscr{L}^{-1}}

\allowdisplaybreaks[1]

\title{$e^{+}e^{-}$ angularity distributions at NNLL$'$ accuracy}

 \author[a]{Guido Bell,}
 \author[b]{Andrew Hornig,}
 \author[b]{Christopher Lee,}
 \author[c]{Jim Talbert} 

\affiliation[a]{Theoretische Physik 1, Naturwissenschaftlich-Technische Fakult\"at, Universit\"at Siegen, \\Walter-Flex-Strasse 3, 57068 Siegen, Germany}
\affiliation[b]{Theoretical Division, Group T-2, MS B283, Los Alamos National Laboratory, P.O. Box 1663, \\Los Alamos, NM  87545, USA}
\affiliation[c]{Theory Group, Deutsches Elektronen-Synchrotron (DESY), Notkestra{\ss}e 85, 22607 Hamburg, Germany}

\emailAdd{bell@physik.uni-siegen.de}
\emailAdd{andrew.hornig@gmail.com}
\emailAdd{clee@lanl.gov}
\emailAdd{james.talbert@desy.de}

\abstract{We present predictions for the $e^{+}e^{-}$ event shape {\it angularities} 
at NNLL$^{\prime}$ resummed and $\mathcal{O}(\as^{2})$ matched accuracy and compare them
to LEP data at center-of-mass energies $Q=91.2$~GeV and $Q=197$~GeV. We perform the resummation 
within the framework of Soft-Collinear Effective Theory, 
and make use of recent results for the two-loop angularity soft function.  
We determine the remaining NNLL$^{\prime}$ and $\cO(\as^2)$ ingredients from a fit to the 
\event\ generator, and implement a shape function with a renormalon-free gap parameter 
to model non-perturbative effects. Using values of the strong coupling 
$\as(m_Z)$ and the universal non-perturbative shift parameter 
$\Omega_1$ that are consistent with those obtained in previous 
fits to the thrust and $C$-parameter distributions, we find excellent agreement between our 
predictions and the LEP data for all angularities with $a\in[-1,0.5]$. This  provides a robust 
test of the predictions of QCD, factorization, and the universal scaling of the non-perturbative 
shift across different angularities. 
Promisingly, our results indicate that current degeneracies in the 
$\{\as(m_Z),\Omega_1\}$ parameter space could be alleviated upon fitting these parameters to experimental data for the angularity distributions.
}

\keywords{Event shapes, QCD, Resummation, Soft-Collinear Effective Theory}

\allowdisplaybreaks[1]

\begin{document}

{\flushright LA-UR-18-24071 \\[-9ex]}
{\flushright DESY 18-083 \\[-9ex]}
{\flushright SI-HEP-2018-19 \\[-9ex]}
\maketitle

%%%%%%%%%%%%%%%%%%%%%%%%%%%%%%%%%%%%%%%%%%%%%%%%%%%%%%%%%%%%%%%
%%%%%%%%%%%%%%%%%%%%%%%%%%%%%%%%%%%%%%%%%%%
\newpage
\section{Introduction}
\label{sec:ANGINTRO}

Event-shape variables are classic QCD observables that characterize the geometric
properties of a hadronic final-state distribution in high-energy collisions \cite{Dasgupta:2003iq}. They have
been measured to high accuracy at LEP and other $e^+ e^-$ colliders in the past, and are commonly used 
for determinations of the strong coupling constant $\alpha_s$.
Whereas past $\alpha_s$ extractions from event shapes were often limited by
perturbative uncertainties, various theoretical developments triggered a renewed
interest in event-shape data in the past decade.

The theoretical description of event-shape distributions combines elements from QCD in
different regimes. For generic values of the event shape, the distributions can be
described in fixed-order perturbation theory where, currently, $\mathcal{O}(\alpha_s^3)$
corrections are known for a large class of
variables~\cite{GehrmannDeRidder:2007hr,Weinzierl:2009ms,Ridder:2014wza,DelDuca:2016ily}. 
In phase-space regions in which the QCD radiation is confined to jet-like configurations,
the perturbative expansion develops large logarithmic corrections
that need to be controlled to all orders. The resummation of these Sudakov logarithms
was pioneered in~\cite{Catani:1992ua} using the coherent branching formalism, and 
has since been reformulated and pushed to new levels of accuracy
with methods from Soft-Collinear Effective Theory
(SCET)~\cite{Bauer:2000ew,Bauer:2000yr,Bauer:2001yt,Beneke:2002ph}. To date, resummations for a variety of
event-shape variables have been worked out to impressively high accuracies, e.g.~to next-to-next-to-leading logarithmic 
(NNLL)~\cite{deFlorian:2004mp,Becher:2012qc,Banfi:2014sua} and even N$^3$LL 
order~\cite{Becher:2008cf,Chien:2010kc,Hoang:2014wka}.

The resummation of the logarithmic corrections tames the singular behaviour of the 
cross section near the kinematic endpoint and generates a characteristic peak
structure whose shape is affected by non-perturbative physics. For the tails of the 
distributions---the region commonly used for the $\alpha_s$ 
determinations---the non-perturbative effects typically result in a shift of the
perturbative distribution that is driven by a non-perturbative 
but universal parameter $\Omega_1$.
Current $\alpha_s$ extractions from event shapes (see 
e.g.~\cite{Davison:2008vx,Abbate:2010xh,Abbate:2012jh,Gehrmann:2012sc,Hoang:2015hka})
are therefore organized as two-dimensional $\{\alpha_{s}(m_{Z}),\Omega_1\}$ fits to the experimental data. 
The parameter $\Omega_1$ is universal in the sense that the leading non-perturbative shift to a large 
number of event shapes depends only on this single parameter, scaled by exactly calculable 
observable-dependent coefficients, as can be demonstrated using factorization and an operator 
product expansion~\cite{Lee:2006fn,Lee:2006nr,Becher:2013iya}. This universality does, however, 
require a careful consideration of hadron mass effects~\cite{Salam:2001bd,Mateu:2012nk}. It is 
therefore possible to reduce the correlation between $\alpha_{s}(m_{Z})$ and {$\Omega_1$ in the 
two-dimensional fits by including data from different event-shape variables.

The most precise $\alpha_s$ extractions from event shapes to date are based on the
 thrust~\cite{Abbate:2010xh,Abbate:2012jh} and $C$-parameter~\cite{Hoang:2015hka} variables. 
In these works the authors combine an N$^3$LL resummation of 
large logarithmic corrections
in the two-jet region together with fixed-order predictions up to $\cO(\as^3)$ 
accuracy
in the multi-jet region, and
they use a shape function that reproduces the aforementioned shift in the tail region
to model non-perturbative effects. While the values of $\alpha_{s}(m_{Z})$ the authors find
from these fits are very precise and consistent with each
other, they are noticeably below the world average~\cite{PhysRevD.98.030001} (see also \cite{Bethke:2011tr}).
As can be seen, e.g., in Fig.~4 of~\cite{Hoang:2015hka}, the low value of 
$\alpha_{s}(m_{Z})$ seems to be associated with the implementation of 
non-perturbative effects. It is therefore important to test the underlying assumptions
of the non-perturbative physics with other event-shape variables.

In the present work we analyze a class of event shapes known as 
\emph{angularities}, which are defined as~\cite{Berger:2003iw}
\be
\label{eq:tauadef}
\tau_a = \frac{1}{Q} \, \underset{i}{\sum} \;
|{\bf p}_{\perp}^{i}|\; e^{-|\eta_i|(1-a)}\,,
\ee
where $Q$ is the center-of-mass energy of the collision and the sum runs over all 
final-state particles $i$ with rapidity $\eta_{i}$ and transverse momentum 
${\bf p}_{\perp}^{i}$ with respect to the thrust axis.  The angularities depend on a
continuous parameter $a$, and they include thrust ($a=0$) and total jet broadening 
($a=1$) as special cases. Whereas infrared safety requires that $a<2$, we restrict
our attention to values of $a\leq0.5$ in this work, since soft recoil effects 
which complicate the resummation are known to become increasingly more important 
as $a\to1$~\cite{Dokshitzer:1998kz}. 
It is also possible to define $\tau_a$ in \eq{tauadef} with respect to an axis other than the thrust 
axis, such as the broadening axis or another soft-recoil-insensitive axis \cite{Larkoski:2014uqa}. We 
stick to the standard thrust-axis-based definition here, to coincide with the available data. 
See~\cite{Procura:2018zpn} for a recent calculation with an alternative axis.

The phenomenological effect of varying $a$ is to change the proportions of two-jet-like 
events and three-or-more-jet-like events that populate the peak region of the $\tau_a$ distributions 
(see \fig{2vs3jet}). The relevant collinear scale that enters the factorization of angularity distributions 
in the two-jet limit then varies accordingly with $a$, to properly reflect the transverse size 
of the jets that are dominating each region of the distributions.

\begin{figure}[t]
\vspace{-1em}
\begin{center}
\includegraphics[width=0.32\columnwidth]{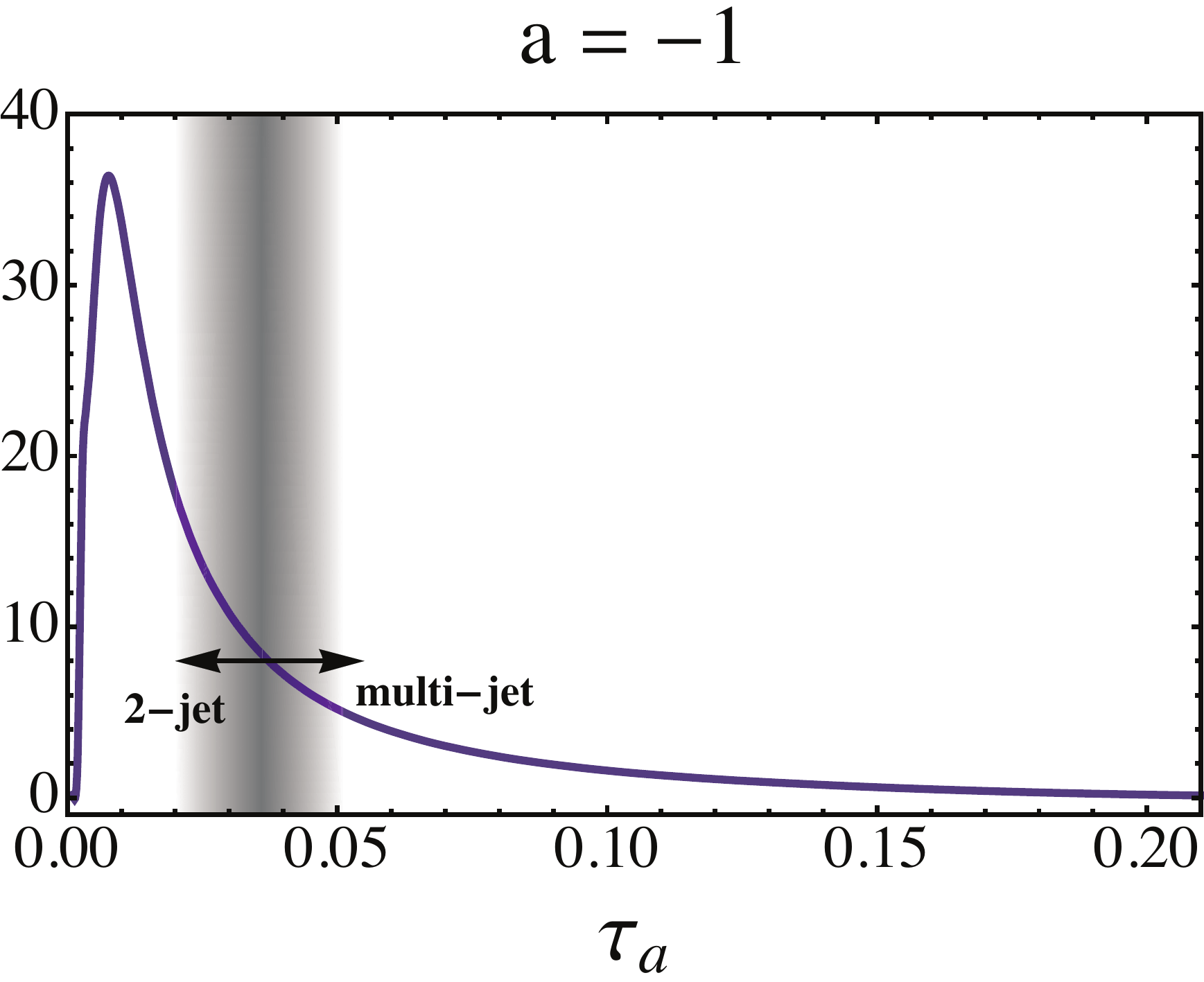} \
\includegraphics[width=0.32\columnwidth]{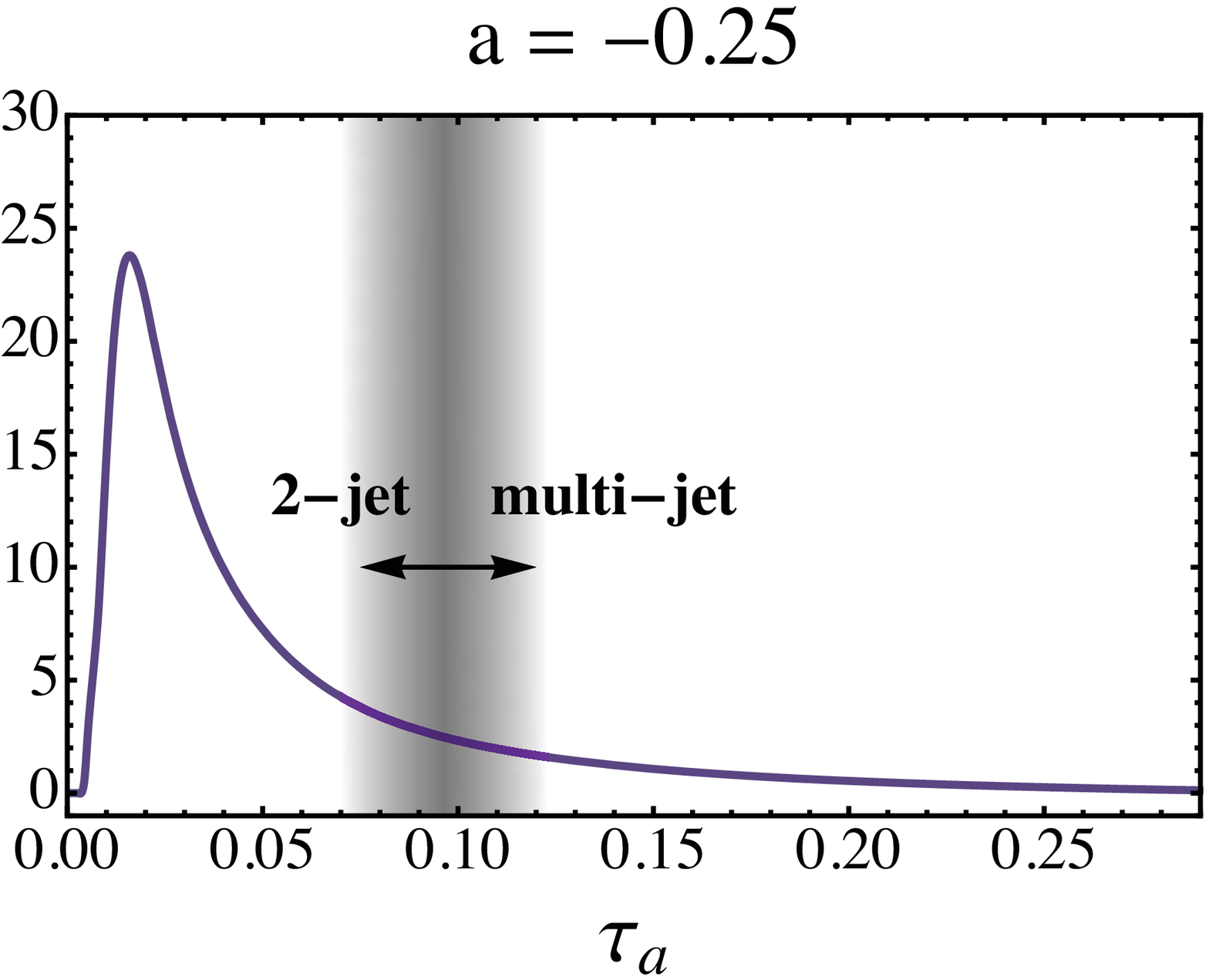} \
\includegraphics[width=0.32\columnwidth]{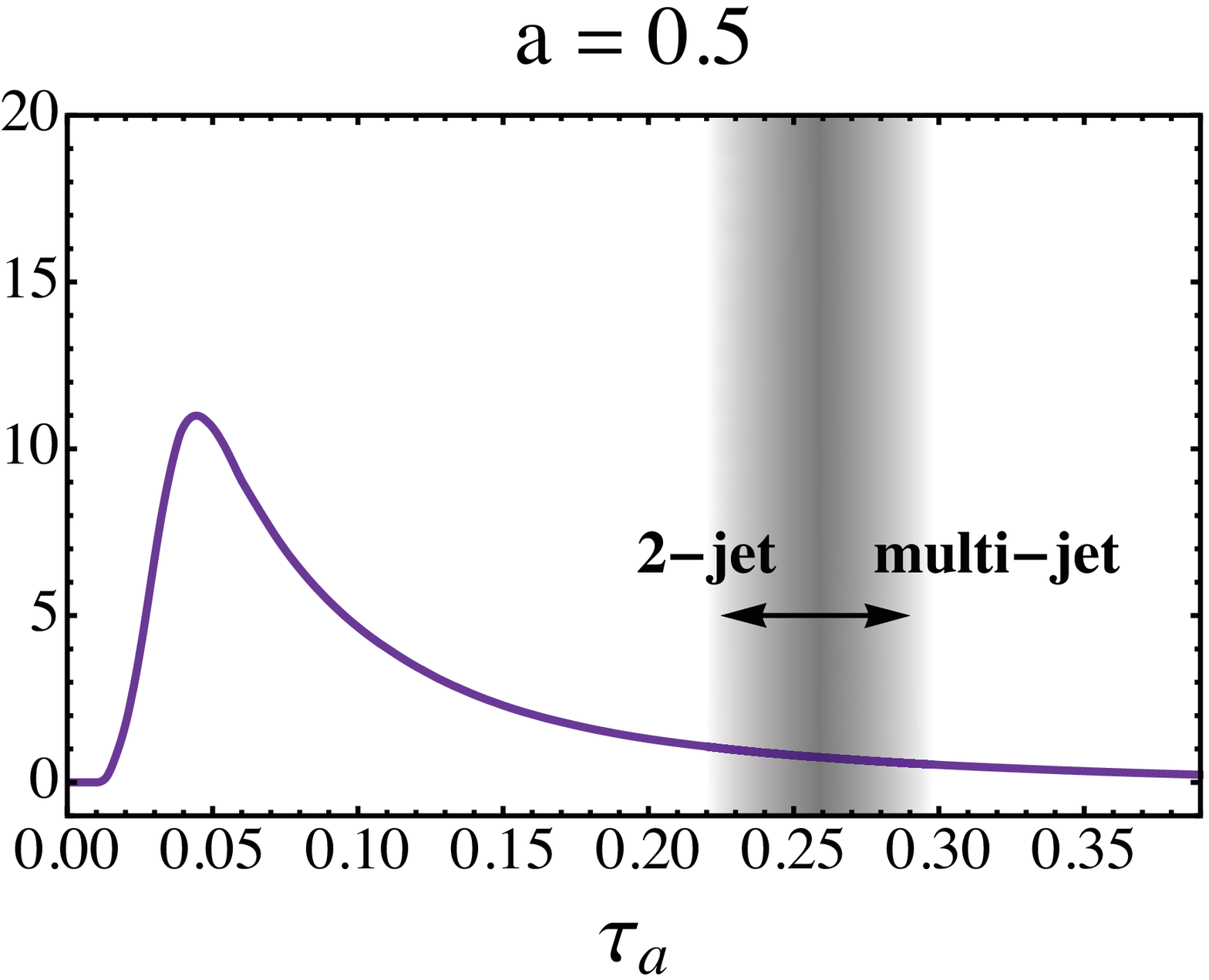}
\end{center}
\vspace{-2em}
\caption{Angularity distributions at NNLL$'+\cO(\as^2)$ accuracy, convolved with a renormalon-free 
non-perturbative shape function, whose calculation is the subject of this paper. We display the predictions 
for three values of $a$ (for now without uncertainties), illustrating roughly where two-jet and 
three-or-more-jet events lie in each $\tau_a$ spectrum. For this illustration, the boundary is drawn at the 
value of $\tau_a$ for a four-particle state that is grouped into pairs of jets with opening angle $30^\circ$. 
As $a$ becomes larger (smaller), the peak region is more (less) dominated by purely two-jet events.
}
\label{fig:2vs3jet}
\end{figure}

The resummation of Sudakov logarithms for the angularity distributions is based on the
factorization theorem~\cite{Berger:2003iw,Bauer:2008dt,Hornig:2009vb,Almeida:2014uva}
\be
\label{eq:factorization}
\frac{1}{\sigma_0}
\frac{d\sigma}{d\tau_a}(\tau_a) = 
H(Q^2,\mu) 
\int dt_{n}^a\, dt_{\bar{n}}^a\, dk_{s}\, 
J_n^a(t_{n}^a, \mu)\, J_\bn^a(t_{\bar{n}}^a, \mu)\, S^a(k_{s}, \mu)\; 
\delta\Bigl(\tau_a - \frac{t_{n}^a+t_{\bar{n}}^a}{Q^{2-a}}-\frac{k_{s}}{Q}\Bigr)\,,
\ee
which arises in the two-jet limit $\tau_a\to 0$. Here $H$ is a hard function that contains the virtual corrections to $e^+e^-\to q\bar q$ scattering at center-of-mass
energy $Q$ (normalised to the Born cross section $\sigma_0$); 
$J_{n,\bn}^a$ are quark jet functions that describe the collinear
emissions into the jet directions, 
and are functions of a variable $t_{n,\bn}^a$ of mass dimension $(2-a)$; 
and $S^a$ is a soft function that encodes the low-energetic cross talk between the two jets 
and depends on a variable $k_s$ of mass dimension $1$. In the language of
SCET, the angularities are of SCET$_{\rm I}$ type, since the virtuality of the soft
modes $\mu_S^2\sim Q^2\tau_a^2$ is smaller than that of the collinear modes
$\mu_J^2\sim Q^2\tau_a^{2/(2-a)}$ for $a<1$. In contrast, jet broadening with $a=1$
corresponds to a SCET$_{\rm II}$ observable, which requires somewhat different
techniques for the resummation of the logarithmic corrections~\cite{Becher:2011pf,Chiu:2012ir}.

For very small values of $\tau_a$, the soft scale becomes non-perturbative and the 
soft effects are often parametrized by a non-perturbative shape function that is 
convolved with the perturbative distribution~\cite{Korchemsky:1998ev,Korchemsky:1999kt}.
The dominant effect of the shape function in the tail region is a shift of the perturbative distribution:
\be
\label{eq:shift}
\frac{d\sigma}{d\tau_{a}} (\tau_{a}) \underset{\text{NP}}{\longrightarrow} 
\frac{d\sigma}{d\tau_{a}} \Big(\tau_{a}-c_{\tau_{a}} \frac{\Omega_1}{Q}\Big)\,.
\ee
This form of the shift, scaling as $1/Q$ and with a value of $c_{\tau_a}=2/(1-a)$ for the 
angularity-dependent coefficient, was argued for in~\cite{Berger:2003pk,Berger:2004xf} based 
on the infrared behavior of the resummed perturbative exponent \cite{Korchemsky:1999kt} and on 
dressed gluon exponentation \cite{Gardi:2001di},
 building upon earlier models of an effective infrared coupling leading to the proposed universal form of the shift in \eq{shift}  for numerous event shapes \cite{Dokshitzer:1995qm,Dokshitzer:1995zt,Dokshitzer:1998pt}.
This universal form was later proven to all orders in soft 
emissions by an operator analysis in~\cite{Lee:2006nr}. 
In the latter work, a definition of the non-perturbative parameter $\Omega_1$ was obtained as a 
vacuum matrix 
element of soft Wilson lines and a transverse energy-flow operator. More importantly, it turns out 
that $\Omega_1$ is the same 
non-perturbative parameter that enters the thrust analysis, related through the coefficient 
$c_{\tau_{a}}$. 
The growth of the shift with larger $a$  reflects the greater dominance of narrow 
two-jet events in the peaks of the $\tau_a$ distributions than for smaller $a$} (see \fig{2vs3jet}).
It is one of the purposes of this work to provide a state-of-the-art description of the angularity 
distributions in order to verify if the angularity-dependent shift is reflected in the experimental data. 

To this end we will significantly improve the theoretical description of the angularity 
distributions
 for which, up until this year, the NLL$'$+$\mathcal{O}(\alpha_s)$ calculation from \cite{Hornig:2009vb} represented the highest accuracy achieved.  However, a recent calculation of the two-loop angularity soft function~\cite{Bell:2015lsf,Bell:2018vaa,Bell:2018oqa} allows the distributions to immediately be extended to NNLL accuracy, and indeed NNLL predictions using these two-loop results appeared recently in~\cite{Procura:2018zpn} in the context of multi-differential observables.  NNLL predictions were also obtained using the \texttt{ARES} formalism in \cite{Banfi:2018mcq}.  In this paper, we will further extend these results to NNLL$' + \cO(\as^2)$ accuracy\footnote{We also presented preliminary results at this accuracy in~\cite{Bell:2017wvi}. The precise prescriptions for which parts of the resummed distribution are needed to achieve these quoted accuracies will be given in \tab{LogAcc} in \ssec{finalcs} below.} by utilizing the \texttt{EVENT2} generator to determine the missing singular ingredients (the constant in the two-loop angularity jet function) and the full non-singular part of the fixed-order $\cO(\as^2)$ prediction.
Furthermore, in order to treat non-perturbative effects, we convolve the improved perturbative 
distribution with a shape function, 
\be
\label{eq:ultimate}
\frac{1}{\sigma_0}\,\sigma(\tau_a) = \int dk\, \sigma_\text{PT}\Bigl(\tau_a - 
\frac{k}{Q};\mu_H,\mu_J,\mu_S,\mu_{\text{ns}}\Bigr) 
\Bigl[ e^{-2\delta_a(\mu_S,R)\frac{d}{dk}}f_\text{mod}
\bigl(k - 2\Delta_a(\mu_S,R)\bigr)
\Bigr]  \,,
\ee
where 
$\sigma_\text{PT}$ represents the perturbative prediction. In our notation, $\sigma$ and 
$\sigma_\text{PT}$ can represent either the differential or integrated spectra,
\be
\label{eq:diffint}
\sigma = \frac{d\sigma}{d\tau_a} \qquad\text{or}\qquad 
\sigma = \sigma_c \equiv \int^{\tau_a} d\tau_a' \,\frac{d\sigma}{d\tau_a'}\,.
\ee
The perturbative cross section $\sigma_\text{PT}$ depends on the scales $\mu_{H,J,S}$
associated with the factorization into hard, jet, and soft functions in the two-jet
region (see \eq{factorization}), and a scale $\mu_{\text{ns}}$ related to 
nonsingular terms in $\tau_a$ which are predicted in 
fixed-order perturbation theory, but which are not contained in the factorized part of 
$\sigma_\text{PT}$. That is, one can write the perturbative distribution as
\be
\label{eq:singandns}
\sigma_\text{PT}(\tau_a;\mu_H,\mu_J,\mu_S,\mu_{\text{ns}}) = \sigma_\text{sing}(\tau_a;\mu_H,
\mu_J,\mu_S) + \sigma_\text{ns}(\tau_a,\mu_{\text{ns}})\,.
\ee
The non-perturbative shape function $f_\text{mod}$ contains a \emph{gap parameter} 
$\Delta_a$ 
accounting for the minimum value of $\tau_a$ for a hadronic (as opposed to 
partonic) spectrum, while $R$ is the scale of subtraction terms $\delta_a(\mu_S,R)$ that 
remove renormalon ambiguities between $\Delta_a$ and the perturbative soft 
function~\cite{Hoang:2007vb}. Formal definitions of all these objects 
will be given below.

%------------------------------------------------
\begin{figure}[t]
\centering 
\includegraphics[scale=0.38]{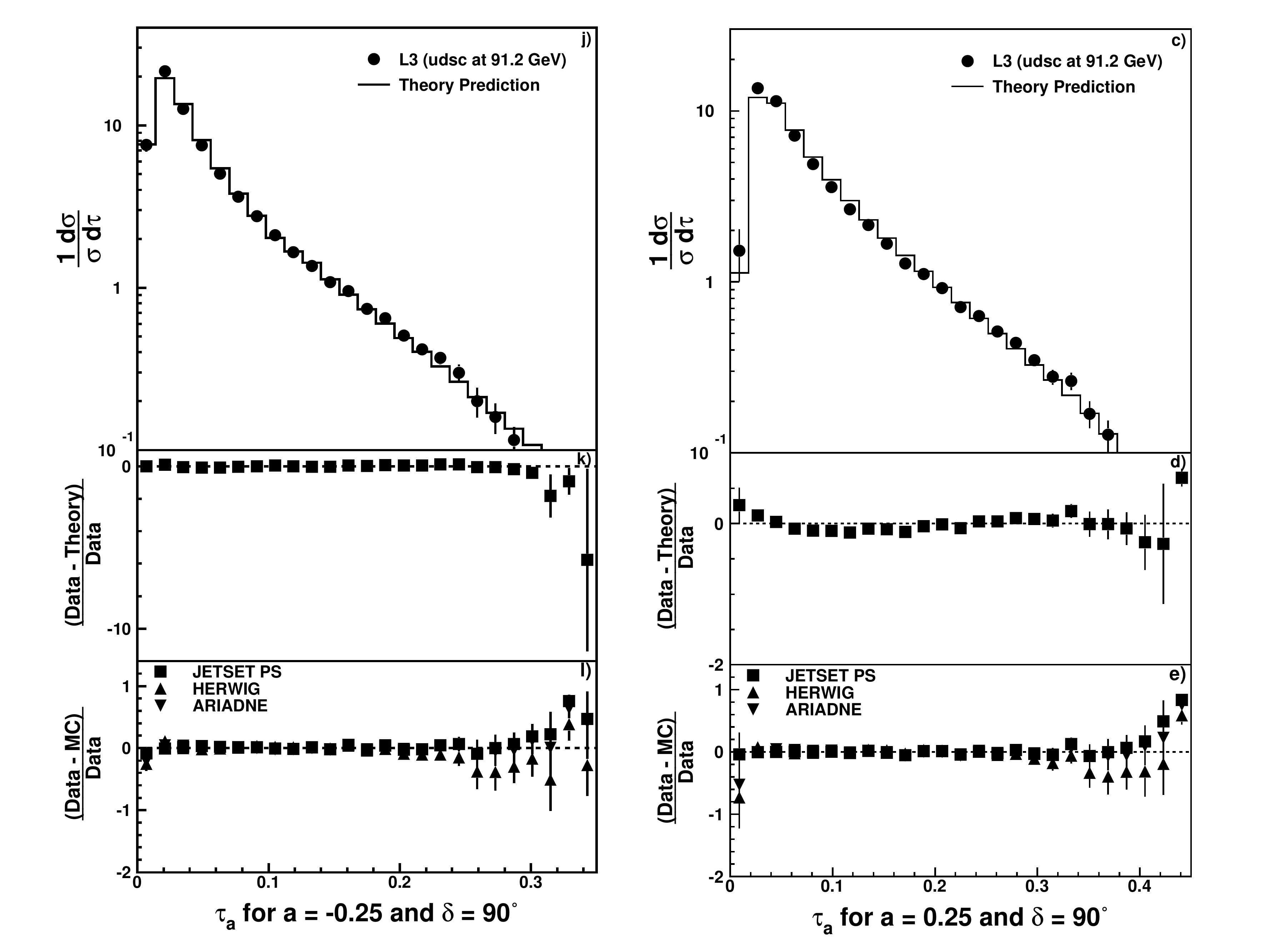}
\vspace{-1em}
\caption{\small Sample data at $Q=91.2$ GeV from the L3 collaboration \cite{Achard:2011zz} for 
two angularities:  $a = -0.25$ and $a =0.25$.  The theory predictions in the top panels are based 
on \cite{Berger:2003pk,Berger:2003zh,Berger:2004xf}, which include the prediction of the non-perturbative 
scaling rule in \eq{shift}.}
\label{fig:L3plots}
\end{figure}
%------------------------------------------------

Finally, we compare our predictions to experimental data from the
L3 Collaboration~\cite{Achard:2011zz}, which provides measurements of the angularity 
distributions for eight different values of 
$a \in \lbrace-1.0,\,-0.75,\,-0.5,\,-0.25,\,0.0,\,0.25,\,0.5,\,0.75\rbrace$ at 
center-of-mass energies $Q=91.2$~GeV and $Q = 197$~GeV (though we will stick to seven values in our analysis, 
leaving out $a=0.75$ due to uncontrolled $a\to 1$ corrections). The $Q=91.2$~GeV data for two 
values of $a$ is shown in \fig{L3plots} for illustration. The L3 analysis also 
includes a comparison of this data with different Monte-Carlo event generators and NLL theory 
predictions from \cite{Berger:2003pk,Berger:2003zh,Berger:2004xf}, which include the non-perturbative 
scaling rule.  However, with the higher-order perturbative contributions and more sophisticated treatment 
of non-perturbative effects now available, we think that the time is ripe for an updated comparison. In particular, 
our setup in \eq{ultimate} allows for a clear separation of perturbative and non-perturbative effects, 
which is not possible with Monte Carlo hadronization 
models that were tuned to LEP data and which entered many of the theory comparisons 
in \cite{Achard:2011zz}. We can therefore rigorously assess the impact of the 
non-perturbative corrections in our framework.

This paper is organized as follows: In Section \ref{sec:ANGCALC} we collect the
formulae required for the resummation of Sudakov corrections in the two-jet region, which
includes the new two-loop ingredients from the soft function calculation
in~\cite{Bell:2015lsf,Bell:2018vaa,Bell:2018oqa}. In order to achieve NNLL$'$ accuracy,
one in addition needs to obtain the corresponding two-loop jet function terms,
which we determine from a fit to the {\tt EVENT2} generator in Section 
\ref{sec:CONSTANTS}. In this section we also perform the matching of the 
resummed distribution to the fixed-order $\mathcal{O}(\as^2)$ prediction.
Then, in Section \ref{sec:NONPERT}, we discuss our implementation of non-perturbative 
effects and we present the final expressions of our analysis after renormalon subtraction. 
We further discuss our scale choices in Section \ref{sec:SCALES}, and compare our results 
to the L3 data in Section \ref{sec:ANGRESULTS}.  Finally, we conclude and give an outlook about a future $\as$ determination from a fit to the angularity distributions
in Section \ref{sec:CONCLUSIONS}. Some technical details of our analysis are discussed
in the Appendix.
\section{NNLL$^{\prime}$ resummation}
\label{sec:ANGCALC}

The formalism for factoring and resumming dijet event shapes within a SCET$_{\text{I}}$ 
factorization framework is well developed and documented in many places 
(see, e.g., \cite{Fleming:2007xt,Bauer:2008dt,Almeida:2014uva}) and will not be re-derived here. Below we 
will simply display the 
final results of these analyses and collect the required ingredients 
to achieve the NNLL$^{\prime}$ resummation we desire. The precise prescriptions for which 
parts of \eq{ultimate} are needed to which order in $\as$ will be given in 
\tab{LogAcc}  in \ssec{finalcs}. In particular, to reach NNLL$'$ accuracy, we  
need to know the heretofore unknown two-loop jet and soft anomalous dimensions 
$\gamma_{J,S}^1$ and finite terms of the two-loop jet and soft functions 
$c_{J,S}^2$ (in a notation we shall define below). These have recently been determined or 
can be obtained from results in \cite{Bell:2015lsf,Bell:2018vaa,Bell:2018oqa} and the \event\ simulations 
we report in this paper.  The rest of this section details what these ingredients are and how 
they enter the final cross sections that we use to predict the angularity distributions.  

%%%%%%%%%%%%%%%%%%%%%%%%%%%%%%%%%%%%%%%%%%%%%%%%%%%%%%%
\subsection{Resummed cross section}
\label{subsec:resum}

The analytic forms for the resummed differential or integrated cross sections in $\tau_{a}$, derived 
in standard references like \cite{Hornig:2009vb,Almeida:2014uva}, are given by
\begin{align}
\label{eq:differential}
  \frac{\sigma_\text{sing}(\tau_a)}{\sigma_0}&= e^{K(\mu,\mu_H,\mu_J,\mu_S)} \biggl(\frac{\mu_H}{Q}\biggr)^{\! \omega_H(\mu,\mu_H)}\!  \biggl(\frac{\mu_J^{2-a}}{Q^{2-a}\tau_a}\biggr)^{\! 2\omega_J(\mu,\mu_J)}  \biggl(\frac{\mu_S}{Q\tau_a}\biggr)^{\! \omega_S(\mu,\mu_S)} \! H(Q^2,\mu_H) \nn \\
&\quad\times \wt J\Bigl(\partial_\Omega \plus \ln\frac{\mu_J^{2-a}}{Q^{2-a}\tau_a},\mu_J\Bigr)^2 \;\wt S\Bigl(\partial_\Omega \plus \ln\frac{\mu_S}{Q\tau_a},\mu_S\Bigr)
\times
\begin{cases}
\frac{1}{\tau_a}\mathcal{F}(\Omega) & \sigma = \frac{d\sigma}{d\tau_a} \\
\mathcal{G}(\Omega) & \sigma = \sigma_c
\end{cases}\,,
\end{align}
where the two cases are for $\sigma$ being the differential or integrated distributions in \eq{diffint}, and 
with the two functions $\mathcal{F},\mathcal{G}$ given by
\be
\mathcal{F}(\Omega) = \frac{e^{\gamma_E\Omega}}{\Gamma(-\Omega)} \,, \quad 
\mathcal{G}(\Omega) = \frac{e^{\gamma_E\Omega}}{\Gamma(1-\Omega)}\,.
\ee
The Born cross-section
\be
\sigma_0 = \frac{4\pi\alpha_{\text{em}}^2 N_C}{3Q^2}\sum_f\biggl[ Q_f^2 - \frac{2Q^2 v_e v_f Q_f}{Q^2-m_Z^2} 
+ \frac{Q^4(v_e^2+a_e^2)(v_f^2+a_f^2)}{(Q^2 - m_Z^2)^2}\biggr]\,
\ee
contains a sum over massless quark flavours $f=\{u,d,s,c,b\}$ with $Q_{f}$ being the charge of the associated 
flavour in units of the electronic charge $e$, and $v_{f}$ and $a_{f}$ are the vector and axial charges of the 
flavour:
\begin{equation}
v_{f} = \frac{1}{2 \sin \theta_{W} \cos \theta_{W}}(T_{f}^{3} - 2 Q_{f} \sin^{2}\theta_{W})\,, \,\,\,\,\,\,\,\,\,\, a_{f} = \frac{1}{2 \sin \theta_{W} \cos \theta_{W}}T^{3}_{f}\,.
\end{equation}
The jet and soft functions $\wt J, \wt S$ appearing in \eq{differential} are the Laplace 
transforms of $J_{n,\bn}^a, S^a$ from \eq{factorization}, with their arguments written in terms of the 
logarithms on which they naturally depend
(we suppress their indices to simplify the notation).  The total evolution kernels $K,\Omega$ accounting for the running of 
the hard function $H$ and the jet and soft functions $\wt J, \wt S$ are given by
\be
\label{eq:KOmega}
\begin{split}
K(\mu,\mu_H,\mu_J,\mu_S) &= K_H(\mu,\mu_H) + 2K_J(\mu,\mu_J) + K_S(\mu,\mu_S)\,, \\
\Omega\equiv \Omega(\mu_J,\mu_S) &= 2\omega_J(\mu,\mu_J) + \omega_S(\mu,\mu_S)\,,
\end{split}
\ee
constructed out of the individual evolution kernels
\begin{equation}
\label{eq:KomegaF}
\begin{split}
K_F(\mu,\mu_F) &\equiv -j_F\kappa_F K_\Gamma(\mu,\mu_F) + K_{\gamma_F}(\mu,\mu_F)\,, \\
\omega_F(\mu,\mu_F) &\equiv -\kappa_F \,\eta_\Gamma(\mu,\mu_F)\,,
\end{split}
\end{equation}
which are determined from the anomalous dimensions of the functions $F = H,\wt J,\wt S$:
\begin{align}
\label{eq:Keta}
K_\Gamma(\mu,\mu_F) &\equiv \int_{\mu_F}^\mu \frac{d\mu'}{\mu'} \Gcusp[\as(\mu')]\ln\frac{\mu'}{\mu_F}\,, \\
\eta_\Gamma(\mu,\mu_F) &\equiv \int_{\mu_F}^{\mu}\frac{d\mu'}{\mu'} \Gcusp[\as(\mu')]\,, \qquad K_{\gamma_F}(\mu,\mu_F) \equiv \int_{\mu_F}^\mu \frac{d\mu'}{\mu'} \gamma_F[\as(\mu')] \, .\nn
\end{align}
The coefficients $j_F,\kappa_F$ in \eq{KomegaF} are given by
\begin{align}
\label{eq:jFkF}
j_H &=1 \,, & \kappa_H &= 4\,, \\
j_J &= 2-a \,, & \kappa_J &= -\frac{2}{1-a}\,, \nn \\
j_S &= 1\,,&  \kappa_S &= \frac{4}{1-a}\, , \nn
\end{align}
and RG invariance of the cross section \eq{differential} imposes two consistency relations on these 
anomalous dimension coefficients,
\begin{align} 
\label{eq:consistency}
\kappa_H + 2j_J\kappa_J + \kappa_S  &= 0\,, \\
2\kappa_J + \kappa_S &= 0\,. \nn
\end{align}
The appearance of partial derivatives $\partial_{\Omega}$ in the arguments of \eq{differential} 
uses the notation of \cite{Becher:2006mr,Becher:2006nr}, and is due to the implementation of the following 
identity for arbitrary powers of logarithms:
\begin{equation}
\label{eq:LOGIDENT}
\left[ \ln^{m}\left(\frac{x^{2}}{y^{2}}\right) \right] \left(\frac{x^{2}}{y^{2}} \right)^{n} = \partial_{n}^{(m)} \left(\frac{x^{2}}{y^{2}} \right)^{n}\,,
\end{equation}
such that functions originally dependent on a logarithm $L$ can be rewritten as 
$F(L, \mu) \rightarrow F(\partial_{n}, \mu)$.  The arguments are further shifted by logarithms of scale ratios 
in \eq{differential} because we have pulled the evolution kernels through the fixed-order $H$, $\wt{J}$, 
and $\wt{S}$ functions from the right.

Note that in \eq{KOmega} $\Omega$ is actually independent of the
factorization scale $\mu$ due to \eq{consistency}, whereas $K$ still depends on $\mu$, but this dependence 
cancels against the $\omega_F$-dependent factors on the first line of \eq{differential}. Of course, even 
the dependence on $\mu_{H,J,S}$ cancels in the all-order cross section, but a residual dependence will remain 
at any finite order of resummed accuracy. The $\mu$-dependence, on the other hand, should cancel exactly 
because of the consistency relations satisfied by the hard, jet, and soft anomalous dimensions.
With standard perturbative expansions of $K_\Gamma,\eta_\Gamma$ in \eq{Keta}, however, this property does 
not precisely hold in practice. We explain why and show how to restore exact, explicit $\mu$-independence 
in \appx{invariance}. Though the difference is formally subleading and numerically quite small, it is our 
aesthetic preference to work in this paper with the explicitly $\mu$-invariant form. The result from 
\appx{invariance} for the cross section is
\begin{align}
\label{eq:cumulant2}
\frac{\sigma_\text{sing}(\tau_a)}{\sigma_0} &= e^{\wt K(\mu_H,\mu_J,\mu_S;Q) + K_\gamma(\mu_H,\mu_J,\mu_S)}  \biggl(\frac{1}{\tau_a}\biggr)^{\Omega(\mu_J,\mu_S)} H(Q^2,\mu_H)  \\
&\quad\times \wt J\Bigl(\partial_\Omega + \ln\frac{\mu_J^{2-a}}{Q^{2-a}\tau_a},\mu_J\Bigr)^2 \;\wt S\Bigl(\partial_\Omega + \ln\frac{\mu_S}{Q\tau_a},\mu_S\Bigr)
\times
\begin{cases}
\frac{1}{\tau_a}\mathcal{F}(\Omega) & \sigma = \frac{d\sigma}{d\tau_a} \\
\mathcal{G}(\Omega) & \sigma = \sigma_c
\end{cases}\,, \nn 
\end{align}
where
\be
\wt K(\mu_H,\mu_J,\mu_S;Q) \equiv -\kappa_H \wt K_\Gamma(\mu,\mu_H;Q) - 2j_J\kappa_J \wt K_\Gamma(\mu,\mu_J;Q) 
- \kappa_S \wt K_\Gamma(\mu,\mu_S;Q)\,,
\ee
in terms of the modified cusp evolution kernel,
\be
\label{eq:Ktildecusp}
\wt K_\Gamma(\mu,\mu_F;Q) = \int_{\mu_F}^\mu \frac{d\mu'}{\mu'} \Gcusp[\as(\mu')] \ln\frac{\mu'}{Q}\,,
\ee
and where $K_\gamma$ is the sum of just the non-cusp evolution kernels in \eq{Keta},
\be
K_\gamma(\mu_H,\mu_J,\mu_S) \equiv K_{\gamma_H}(\mu,\mu_H) + 2K_{\gamma_J}(\mu,\mu_J) + K_{\gamma_S}(\mu,\mu_S)\,.
\ee
The reorganization of individual pieces of the evolution kernels that leads to \eq{cumulant2} restores 
explicit $\mu$-invariance to every piece of this formula.
In what follows we use \eq{cumulant2} to generate our theory predictions. (The numerical differences with \eq{differential} will be negligible.)

%%%%%%%%%%%%%%%%%%%%%%%%%%%%%%%%%%%%%%%%%%%%%%%%%%%%%%%
\subsection{Evolution kernels}

The evolution kernels in \eq{Keta} or \eq{Ktildecusp} can be computed explicitly in terms of the coefficients 
of the perturbative expansions of the anomalous dimensions given by
\be
\label{eq:gammaexpansion}
\Gcusp[\as] = \sum_{n=0}^\infty \Bigl(\frac{\as}{4\pi}\Bigr)^{n+1}\Gamma_n \,,\quad \gamma_F[\as]  = \sum_{n=0}^\infty \Bigl(\frac{\as}{4\pi}\Bigr)^{n+1}\gamma_F^n \,, 
\ee
and where the running of $\as$ is given by
\be
\label{eq:beta}
\mu\frac{d}{d\mu}\as(\mu) = \beta[\as(\mu)]\,,\quad \beta[\as] = -2\as \sum_{n=0}^\infty \Bigl(\frac{\as}{4\pi}\Bigr)^{n+1}\beta_n\,.
\ee
The values of the first few coefficients $\Gamma_n,\gamma_F^n,\beta_n$ are given in \appx{anomalous}.

The formulas for the evolution kernels $\wt K_\Gamma,\eta_\Gamma$ can be evaluated at any finite \nkll 
accuracy in closed form by first making the change of integration variables from $\mu$ to $\as$, using 
\eq{beta}:
\begin{align}
\label{eq:Ketaalpha}
\wt K_\Gamma(\mu,\mu_F;Q) &\equiv \int_{\as(\mu_F)}^{\as(\mu)} \frac{d\alpha}{\beta[\alpha]} \Gcusp[\alpha] \int_{\as(Q)}^\alpha\frac{d\alpha'}{\beta[\alpha']}\,, \\
\eta_\Gamma(\mu,\mu_F) &\equiv \int_{\as(\mu_F)}^{\as(\mu)}\frac{d\alpha}{\beta[\alpha]} \Gcusp[\alpha] \nn\,,
\end{align}
and similarly for $K_{\gamma_F}$. Up to N$^3$LL accuracy, the expansion of the kernel
$\eta_\Gamma$ becomes, for instance,
\begin{subequations}
\label{eq:etaclosedform}
\begin{align}
\eta_\Gamma^{\text{LL}}(\mu,\mu_F)  &=  -\frac{\Gamma_0}{2\beta_0} \ln r\,, \\
\eta_\Gamma^{\text{NLL}}(\mu,\mu_F)  &=  -\frac{\Gamma_0}{2\beta_0} \frac{\as(\mu_F)}{4\pi} \left( \frac{\Gamma_1}{\Gamma_0} - \frac{\beta_1}{\beta_0}\right) (r-1)\,, \\
\eta_\Gamma^{\text{NNLL}}(\mu,\mu_F)  &=  -\frac{\Gamma_0}{2\beta_0} \left(\frac{\as(\mu_F)}{4\pi}\right)^2 \left( B_2 + \frac{\Gamma_2}{\Gamma_0} - \frac{\Gamma_1\beta_1}{\Gamma_0\beta_0}\right)\frac{r^2-1}{2}\,,  \\
\eta_\Gamma^{\text{N$^3$LL}}(\mu,\mu_F) &= -\frac{\Gamma_0}{2\beta_0}  \left(\frac{\as(\mu_F)}{4\pi}\right)^3 \left( B_3 + \frac{\Gamma_1}{\Gamma_0}B_2 
- \frac{\Gamma_2\beta_1}{\Gamma_0\beta_0} + \frac{\Gamma_3}{\Gamma_0}\right)\frac{r^3-1}{3}\,,
\end{align}
\end{subequations}
where 
\be
\label{eq:rB}
r\equiv \frac{\as(\mu)}{\as(\mu_F)}\,, \quad B_2 \equiv \frac{\beta_1^2}{\beta_0^2} - \frac{\beta_2}{\beta_0}\,,\quad B_3 = -\frac{\beta_1^3}{\beta_0^3} + \frac{2\beta_1\beta_2}{\beta_0^2} - \frac{\beta_3}{\beta_0}\,
\ee
and $K_{\gamma_F}$ is directly obtained from \eq{etaclosedform} by using the replacement rules,
\be
\label{eq:kFclosedform}
K_{\gamma_F}^{\text{LL}} = 0 \,, \quad 
K_{\gamma_F}^{\text{NLL}} = \eta_{\gamma_F}^{\text{LL}}\,,\quad 
K_{{\gamma_F}}^{\text{NNLL}} = \eta_{\gamma_F}^{\text{NLL}}\,.\quad\dots
\ee
The non-cusp kernels only need to be evaluated to one order lower than the cusp kernels, since they do not 
multiply an additional logarithm of $\mu_F/Q_F$ in the anomalous dimension or in \eq{differential}. Even 
though an appropriate choice of $\mu_F$ will keep these logarithms small, cancellation of the $\mu$-dependence 
of the $K_\Gamma$ kernels requires $K_\Gamma$ and $\eta$ to be kept to the same accuracy, see 
\cite{Almeida:2014uva} or \appx{invariance}.

Meanwhile, the expansion of $K_\Gamma$ in \eq{Keta} is quoted in \eq{Kclosedform}.  As mentioned above, we 
observe that, when using these standard expansion formulae for $K_\Gamma$, the sum of evolution factors on 
the first line of \eq{differential} does not exhibit exact independence of the factorization scale $\mu$ at 
any \emph{truncated} order of resummed logarithmic accuracy, but instead has a residual (subleading) numerical 
dependence on it. This is due to a reliance on the relation used to obtain \eq{Kclosedform},
\be
\label{eq:dmu}
\ln\frac{\mu}{\mu_F} = \int_{\as(\mu_F)}^{\as(\mu)}\frac{d\alpha}{\beta[\alpha]} \,,
\ee
which is not exactly true if $\beta[\alpha]$ is truncated to finite accuracy. It is true at one-loop order 
and to infinite order, but not at any other fixed order. The corrections are, of course, subleading compared to 
the order to which \eq{dmu} is evaluated --- see \appx{invariance} for details. The expansion for 
$\wt K_\Gamma$ in \eq{cumulant2}, on the other hand, \emph{does} exhibit exact numerical independence on 
$\mu$ at any order. Its expansion up to N$^3$LL accuracy is given by
\begin{subequations}
\label{eq:Ktildeclosedform}
\begin{align}
\wt K_\Gamma^{\text{LL}}(\mu,\mu_F;Q) &=  \frac{\Gamma_0}{4\beta_0^2} \frac{4\pi}{\alpha_s(\mu_F)} \biggl\{  r_Q \ln r + \frac{1}{r} - 1  \biggr\}, \\
\wt K_\Gamma^{\text{NLL}}(\mu,\mu_F;Q)  &=  \frac{\Gamma_0}{4\beta_0^2} \biggl\{ \left(\frac{\Gamma_1}{\Gamma_0} \minus \frac{\beta_1}{\beta_0}\right) [ r_Q( r \minus 1)\minus \ln r] \minus \frac{\beta_1}{2\beta_0} (\ln^2 r \plus 2\ln r_Q\ln r)  \biggr\},\\
\!\wt K_\Gamma^{\text{NNLL}}(\mu,\mu_F;Q)  &=  \frac{\Gamma_0}{4\beta_0^2} \frac{\as(\mu_F)}{4\pi} \biggl\{  \left(\frac{\Gamma_2}{\Gamma_0} \minus \frac{\beta_1\Gamma_1}{\beta_0\Gamma_0}\right) \biggl[\frac{(1\minus r)^2}{2}  \plus  \frac{r^2 \minus 1}{2}(r_Q \minus 1)\biggr] \\
& \quad
+ B_2\bigg( r_Q\frac{r^2 \minus 1}{2}  - \frac{\ln r}{r_Q}\bigg) + \bigg(\frac{\beta_1 \Gamma_1}{\beta_0\Gamma_0} \minus  \frac{\beta_1^2}{\beta_0^2}\bigg) 
[ (r  \minus 1) (1 \minus  \ln r_Q)\minus r\ln r ] \biggr\}, \nn \\
\wt K_\Gamma^{\text{N$^3$LL}}(\mu,\mu_F;Q) &= \frac{\Gamma_0}{4\beta_0^2} \bigg(\frac{\as(\mu_F)}{4\pi}\bigg)^2  \biggl\{ \bigg(\frac{\Gamma_3}{\Gamma_0} -\frac{\Gamma_2\beta_1}{\Gamma_0\beta_0}  + \frac{B_2 \Gamma_1}{\Gamma_0} +B_3  \bigg)\frac{r^3-1}{3}r_Q  \\
&\quad   - \frac{\beta_1}{2\beta_0} \bigg( \frac{\Gamma_2}{\Gamma_0} - \frac{\Gamma_1\beta_1  }{\Gamma_0\beta_0} + B_2\bigg) [ r^2 \ln r + (r^2-1)\ln r_Q]  
- \frac{B_3 \ln r}{2r_Q^2}\nn \\
&\quad + \biggl( \frac{\beta_3}{\beta_0} - \frac{\beta_1\beta_2}{\beta_0^2} - \frac{2\Gamma_3}{\Gamma_0} + \frac{3\Gamma_2\beta_1}{\Gamma_0\beta_0} - \frac{\Gamma_1\beta_1^2}{\Gamma_0\beta_0^2}\biggr) \frac{r^2-1}{4} + B_2\bigg(\frac{\Gamma_1}{\Gamma_0}\minus\frac{\beta_1}{\beta_0}\bigg)\frac{1\minus r}{r_Q} \biggr\},\nn
\end{align}
\end{subequations}
where $r = \as(\mu)/\as(\mu_F)$ as in \eq{rB}, and 
\be
r_Q \equiv \frac{\as(\mu_F)}{\as(Q)}\,.
\ee
We use these expressions to evaluate \eq{cumulant2} at a given resummed accuracy.

%%%%%%%%%%%%%%%%%%%%%%%%%%%%%%%%%%%%%%%%%%%%%%%%%%%%%%%
\subsection{Fixed-order hard, jet, and soft functions}
\label{sec:hardjetsoft}

The hard, jet, and soft functions $H,\wt J,\wt S$ in \eqs{differential}{cumulant2} can be expanded to 
fixed order, as their corresponding logarithms are small near their natural 
scales $\mu_{H,J,S}$. We can obtain the generic form of their fixed-order expansions by making use of the 
solutions of the renormalization group equations (RGEs) that they satisfy,
\be
\label{eq:RGE}
\mu\frac{d}{d\mu} F(\mu) = \gamma_F(\mu) F(\mu)\,,
\ee
where $F=H,\wt J, \wt S$, and
\be
\gamma_F(\mu) = -\kappa_F\,\Gcusp[\as(\mu)]\ln\frac{\mu^{j_F}}{Q_F^{j_F}} + \gamma_F[\as(\mu)]\,,
\ee
where $Q_H=Q$, and $Q_F = Q/(e^{\gamma_E}\nu_a)^{1/j_F}$ for $\wt J,\wt S$.
Here $\nu_a$ is the Laplace-conjugate variable to $\tau_a$.
From the solution of \eq{RGE}, which is in general
\be
F(\mu) = F(\mu_F)\;e^{K_F(\mu,\mu_F)}\;
\Bigl(\frac{\mu_F}{Q_F}\Bigr)^{j_F \omega_F(\mu,\mu_F)}\,,
\ee
we can derive the form of each function $F$ order-by-order in their perturbative expansions,
\be
\label{eq:Fexpansion}
 F(L_F,\mu_F)  = \sum_{n=0}^\infty \left(\frac{\as(\mu_F)}{4\pi}\right)^n  F_n(L_F)\,,
\ee
 and to order $\as^2$ the coefficients $F_n$ are given by
 \begin{subequations}
\label{eq:Ffixedorder}
\begin{align}
F_0(L_F)  &= 1\,, \\
F_1(L_F) &= \frac{\Gamma_F^0}{j_F^2} L_F^2 + \frac{\gamma_F^0}{j_F} L_F + c_F^1\,, \\
F_2(L_F) &= \frac{1}{2j_F^4} (\Gamma_F^0)^2 L_F^4 + \frac{\Gamma_F^0}{j_F^3} \left( \gamma_F^0 + \frac{2}{3}\beta_0\right) L_F^3   \\
&\quad + \frac{1}{j_F^2}\left(\Gamma_F^1 + \frac{1}{2}(\gamma_F^0)^2 + \gamma_F^0\beta_0 + c_F^1 \Gamma_F^0 \right)L_F^2 + \frac{1}{j_F}(\gamma_F^1 + c_F^1\gamma_F^0 + 2c_F^1 \beta_0) L_F + c_F^2 \,. \nn
\end{align}
\end{subequations}
The corresponding logarithms are $L_H\equiv \ln(\mu_H/Q)$ for $F=H$ and 
$L_F = \ln(\mu_F^{j_F} e^{\gamma_E} \nu_a/Q^{j_F})$ for $F=\wt J,\wt S$. In \eqs{differential}{cumulant2}, 
each 
factor $L_F$ gets replaced by the differential operator shown in the argument of $\wt J,\wt S$ for the jet 
and soft functions.  
The quantities $\Gamma_F^n,\gamma_F^n$ are the coefficients of the perturbative expansions of the anomalous 
dimensions given by \eq{gammaexpansion}, and the individual cusp parts  $\Gamma_F$ of the anomalous dimensions 
are defined by
\be
\label{eq:cuspproportion}
\Gamma_F \equiv -\frac{j_F \kappa_F}{2}\Gcusp\,,
\ee
where $j_F,\kappa_F$ were given in \eq{jFkF}.
In evaluating the $\partial_\Omega$ derivatives acting on the functions $\mathcal{F},\mathcal{G}$ in 
\eqs{differential}{cumulant2} up to two-loop order, we need to know the first four derivatives. 
For $\mathcal{G}$ in the integrated distributions, this yields
\begin{align}
\label{eq:DG}
\partial_\Omega \cG(\Omega) &= H(-\Omega)\,\cG(\Omega)\,,  \\
\partial_\Omega^2\cG(\Omega) &= \bigl[ H(-\Omega)^2 - \psi^{(1)}(1-\Omega)\bigr] 
\,\cG(\Omega)\,, \nn \\
\partial_\Omega^3\cG(\Omega) &= \bigl[ H(-\Omega)^3  -3 H(-\Omega) \psi^{(1)}(1-\Omega) + \psi^{(2)}(1-\Omega)\bigr] \,\cG(\Omega)\,, \nn \\
\partial_\Omega^4\cG(\Omega) &= \bigl[ H(-\Omega)^4  -6 H(-\Omega)^2 \psi^{(1)}(1-\Omega) +4 H(-\Omega) \psi^{(2)}(1-\Omega)  \nn \\
&\qquad + 3\psi^{(1)}(1-\Omega)^2 - \psi^{(3)}(1-\Omega)\bigr] \,\cG(\Omega) \,, \nn 
\end{align}
where $H(-\Omega)\equiv  \gamma_E + \psi(1-\Omega)$ is the harmonic number function, $\psi(x)$ is the 
digamma function, and $\psi^{(n)}(x)$ is its $n$th derivative. For $\mathcal{F}$ in the differential 
distributions, the derivatives can be obtained using
$\mathcal{F}(\Omega) = -\Omega \,\mathcal{G}(\Omega)$, yielding
\begin{align}
\label{eq:DF}
\partial_\Omega\mathcal{F}(\Omega) &\equiv  - \Omega\,\partial_\Omega \mathcal{G}(\Omega)- \mathcal{G}(\Omega)\,, \\
\partial_\Omega^2\mathcal{F}(\Omega) &\equiv  - \Omega\,\partial_\Omega^2 \mathcal{G}(\Omega)- 2\partial_\Omega\mathcal{G}(\Omega)\,, \nn \\
\partial_\Omega^3\mathcal{F}(\Omega) &\equiv  - \Omega\,\partial_\Omega^3 \mathcal{G}(\Omega)
- 3\partial_\Omega^2\mathcal{G}(\Omega)\,, \nn \\
\partial_\Omega^3\mathcal{F}(\Omega) &\equiv  - \Omega\,\partial_\Omega^4 \mathcal{G}(\Omega)
- 4\partial_\Omega^3\mathcal{G}(\Omega)\,. \nn 
\end{align}
The forms of \eqs{DG}{DF} are convenient as they contain in closed, compact form the results of convolving 
fixed-order logarithmic plus-distributions in the momentum-space jet and soft functions with the 
generalized plus-distributions in the momentum-space evolution kernels (see, 
e.g., \cite{Ligeti:2008ac,Hornig:2009vb,Abbate:2010xh}).

%%%%%%%%%%%%%%%%%%%%%%%%%%%%%%%%%%%%%%%%%%%%%%%%%%%%%%%
\subsection{NNLL$^{\prime}$ ingredients}
\label{sec:nnllprime}

In order to achieve NNLL$^{\prime}$ accuracy, one needs to combine the NNLL evolution kernels
with the $\mathcal{O}(\as^2)$ fixed-order expansions of the hard, jet, and soft 
functions (the required ingredients are listed in 
\tab{LogAcc} in \ssec{finalcs}). Whereas the cusp and the non-cusp anomalous dimension for the hard 
function are currently
known to three-loop order, the non-cusp anomalous dimensions for the jet and soft 
function were previously only known to one-loop order for generic values of 
$a$~\cite{Hornig:2009vb}.\footnote{The $a=0$ jet and soft anomalous dimension are also
known to three-loop order and are given in \appx{anomalous}.} Recently, some of us developed a generic
framework for the computation of two-loop soft anomalous dimensions~\cite{Bell:2018vaa},
which we will use to determine $\gamma_S^1$. The two-loop jet 
anomalous dimension can then be determined from the consistency relation 
$\gamma_H + 2\gamma_J + \gamma_S = 0$.

According to~\cite{Bell:2018vaa}, the two-loop soft anomalous dimension can be written
in the form 
\be
\label{eq:gammaS1a}
\gamma_S^1(a) = \frac{2}{1-a} \,\Bigl[\gamma_1^{CA}(a)\, C_F C_A 
+ \gamma_1^{nf}(a)\, C_F T_F n_f\Bigr]\,,
\ee
where we have made the $a$-dependence of the anomalous dimension explicit, and
\begin{align}
\label{eq:gammaS1CANF}
\gamma_1^{CA}(a) &=  -\frac{808}{27} + \frac{11\pi^2}{9} + 28\zeta_3 - \Delta\gamma_1^{CA}(a)\,, \\
\gamma_1^{nf}(a) &=  \frac{224}{27} - \frac{4\pi^2}{9}  - \Delta\gamma_1^{nf}(a)  \,, \nn
\end{align}
are expressed in terms of the thrust anomalous dimension and $a$-dependent contributions expressed as integrals:
\begin{align}
\label{eq:DeltagammaS}
\Delta\gamma_1^{CA}(a) \!&=\! \int_0^1 \!\! dx\int_0^1 \!\! dy\frac{32 x^2(1\plus xy\plus y^2)\bigl[ x(1\plus y^2) + (x\plus y)(1\plus xy)\bigr]}{y(1-x^2)(x+y)^2(1+xy)^2}
\ln\Bigl[\frac{(x^a+ xy) (x+ x^a y)}{x^a(1+ xy)(x+y)} \Bigr], \nn \\
\Delta\gamma_1^{nf}(a) \!&=\!  \int_0^1 \!\! dx\int_0^1 \!\! dy\frac{64 x^2(1 + y^2)}{(1-x^2)(x+y)^2(1+xy)^2}
\ln\Bigl[\frac{(x^a+ xy) (x+ x^a y)}{x^a(1+ xy)(x+y)} \Bigr],
\end{align}
which vanish for $a=0$. The integral representations can easily be evaluated numerically to high accuracy 
for any value of $a$, and the relevant values for our work are given in \tab{gamma1}.

%----------------------------------------------------------------------------
\begin{table}[t]
\centering
\begin{tabular}{|c||c|c|c|c|c|c|c|}
\hline 
$a$ & $-1.0$ & $-0.75$ & $-0.5$ & $-0.25$ & 0.0 & 0.25 & 0.5 
\\ \hline \hline
$\gamma_1^{CA}$ & $1.0417$ & $5.8649$ & $9.8976$ & $13.190$ & $15.795$ & $17.761$ & $19.132$ 
\\ \hline
$\gamma_1^{nf}$ & $-0.9571$ & $0.5284$ &  $1.8440$ & $2.9751$ & $3.9098$ & $4.6398$ & $5.1613$
\\ \hline
\end{tabular}
\caption{Coefficients of the two-loop soft anomalous dimension as defined in
\eq{gammaS1a}.}
\label{tab:gamma1}
\end{table}
%-----------------------------------------------------------------------------

The constants in \eq{Ffixedorder} are known for the hard function at one \cite{Manohar:2003vb,Bauer:2003di}, 
two \cite{Idilbi:2006dg,Becher:2006mr}, and three loops \cite{Abbate:2010xh,Lee:2010cga}. At NNLL$^\prime$ 
accuracy, we will need these constants to two-loop order,
and they are given by
\begin{align}
\label{eq:cH}
c_H^1 &= C_F\biggl(\frac{7\pi^2}{3} - 16\biggr)\,, \\
c_H^2 &=  C_F^2 \biggl( \frac{511}{4} - \frac{83}{3}\pi^2 + \frac{67}{30}\pi^4 - 60\zeta_3\biggr) + C_F C_A \biggl( -\frac{51157}{324} + \frac{1061}{54}\pi^2 - \frac{8}{45}\pi^4 + \frac{626}{9}\zeta_3\biggr) \nn \\
&\quad + C_F T_F n_f \biggl( \frac{4085}{81} - \frac{182}{27}\pi^2 + \frac{8}{9}\zeta_3\biggr) \nn\,.
\end{align}
These results can be obtained from the two-loop quark form 
factor \cite{Gehrmann:2005pd,Matsuura:1987wt,Matsuura:1988sm,Moch:2005id}, also quoted 
in \cite{Becher:2006mr,Idilbi:2006dg}. From the latter results, we replace the logarithm 
$L=\ln(Q^2/\mu^2)$ present there for deep-inelastic scattering with $L=\ln(-Q^2-i\epsilon)/\mu^2$  
for $e^+e^-$ annihilation, keeping the resulting extra $\pi^2$ and $\pi^4$ terms in the hard function. 
The result in \eq{cH} agrees with the expressions in \cite{Becher:2008cf}. The expansion of the entire 
two-loop hard function given by \eq{Ffixedorder} with the values in \eq{cH} for the $c_H^{1,2}$ also 
agrees with \cite{Becher:2006mr,Idilbi:2006dg} with the replacement $L\to\ln(-Q^2-i\epsilon)/\mu^2$, and 
with the expressions in \cite{Becher:2008cf}.

The soft function constants are known to one \cite{Fleming:2007xt} and two 
loops \cite{Kelley:2011ng,Monni:2011gb} for $a=0$, but they were, until recently, only known to one-loop 
order for generic values of $a$ \cite{Hornig:2009vb}. The two-loop $a\neq 0$ soft function constants
have now been determined numerically in \cite{Bell:2015lsf,Bell:2018oqa}. 
In Laplace space, they are
\be
\label{eq:cS1}
c_{\tilde S}^1(a) = c_S^1(a) + \Gamma_S^0(a) \frac{\pi^2}{6} = -C_F\frac{\pi^2}{1-a} \,,
\ee
where \cite{Hornig:2009vb} 
\be
c_S^1(a) = C_F\frac{1}{1-a}\frac{\pi^2}{3}\,,
\ee
and $c_{\tilde S}^2(a)$ is given by \cite{Bell:2015lsf,Bell:2018oqa}
\be
\label{eq:cS2}
c_{\tilde S}^2(a) = c_2^{CA}(a) \,C_F C_A + c_2^{nf}(a) \,C_F T_F n_f + \frac{\pi^4}{2(1-a)^2}C_F^2\,.
\ee
Sample values of $c_{2}^{CA}$ and $c_{2}^{nf}$ for the values of $a$ we plot in this paper are given 
in \tab{c2soft}.

%----------------------------------------------------------------------------
\begin{table}[t]
\centering
\begin{tabular}{|c||c|c|c|c|c|c|c|}
\hline 
$a$ & $-1.0$ & $-0.75$ & $-0.5$ & $-0.25$ & 0.0 & 0.25 & 0.5 
\\ \hline \hline
$c_2^{CA}$ & $-22.430$ & $-29.170$ & $-36.398$ & $-44.962$ & $-56.499$ & $-74.717$ & $-110.55$ 
\\ \hline
$c_2^{nf}$ & 27.315 & 28.896 &  31.589 & 36.016 & 43.391 & 56.501 & 83.670 
\\ \hline
\end{tabular}
\caption{Coefficients of the two-loop soft function constant as defined in
\eq{cS2}.}
\label{tab:c2soft}
\end{table}
%-----------------------------------------------------------------------------

Finally, the jet function constants are known to one \cite{Bauer:2003pi,Becher:2009th}, 
two \cite{Becher:2006qw,Becher:2010pd}, 
and even three loops \cite{Bruser:2018rad,Banerjee:2018ozf} for $a=0$, but they were 
so far only known to one-loop order for generic values of $a$ \cite{Hornig:2009vb}. 
In Laplace space, they read
\begin{align}
\label{eq:cJ1}
c_{\tilde J}^1(a) &= c_J^1(a) +  \frac{\Gamma_J^0(a)}{(2-a)^2} \frac{\pi^2}{6}  \\
&= \frac{C_F}{2-a} \Bigl( 14 - 13a -\frac{\pi^2}{6}\,\frac{8-20a +9a^2}{1-a} - 4f(a)\Bigr) \nn
\,,
\end{align}
where $c_J^1$ is the momentum-space constant computed in \cite{Hornig:2009vb},
\be
c_J^1(a) = \frac{C_F}{2-a}\Bigl(14 - 13a - \frac{\pi^2}{6}\,\frac{12-20a +9a^2}{1-a} - 4f(a) \Bigr)\,,
\ee
and
\be
\label{eq:fint}
f(a) = \int_0^1\! dx \frac{2-2x+x^2}{x}\ln\bigl[ (1-x)^{1-a} + x^{1-a}\bigr]\,.
\ee
The two loop constant $c_{\tilde J}^2(a)$ is thus far unknown and will be derived in the next section. 
%%%%%%%%%%%%%%%%%%%%%%%%%%%%%%%%%%%%%%%%%%%%
\section{Numerical extraction of fixed-order contributions}
\label{sec:CONSTANTS}

In this section we determine the remaining fixed-order ingredients needed to predict the angularity 
distributions to NNLL$'$+$\cO(\as^2)$ accuracy: the constant $c_{\tilde J}^2$ in the two-loop jet function 
(as defined in \eq{Ffixedorder}) and the $\cO(\as^2)$ part of the 
nonsingular distribution in \eq{singandns} predicted by full QCD. We will determine these in \sec{cJ2event2} 
and \ref{sec:remainder}, respectively, using the \event\ generator
\cite{Catani:1996jh,Catani:1996vz}. In \sec{fixedorder} we first derive the analytic relation between 
$c_{\tilde J}^2$ and the singular part of the cross section that will allow us to perform this extraction. 

%%%%%%%%%%%%%%%%%%%%%%%%%%%%%%%%%%%%%%%%%%%%%%%%%%%%%%%
\subsection{Fixed-order expansion}
\label{sec:fixedorder}

To extract the unknown constant $c_{\tilde J}^2$ in the jet function,  we need to 
know the fixed-order expansion of the singular part of the cross section in \eq{differential} to 
$\cO(\as^2)$. We can do this by setting all the scales equal, $\mu=\mu_H=\mu_J=\mu_S$, and multiplying out 
the fixed-order expansions of the individual $H,\wt J,\wt S$ functions. In Laplace space,
\be
\label{eq:Laplacecs}
\frac{\wt\sigma_\text{sing}(\nu_a)}{\sigma_0}\equiv
 \int_{0}^{\infty} \!d\tau_a \,e^{-\tau_a\nu_a}\, \frac{1}{\sigma_0}\frac{d\sigma}{d\tau_a} 
=H(Q^2,\mu) \;\wt J(Q/\nu_a^{1/(2-a)},\mu)^2\; \wt S(Q/\nu_a,\mu)\,.
\ee
The expansions of each function are given in \eq{Ffixedorder}, with $\Gamma_F^n$ given 
by \eqs{cuspproportion}{jFkF}. The explicit expressions for the anomalous dimension coefficients
and the constants $c_F^{1,2}$ for $F=H,\wt J,\wt S$ can be found in \sec{nnllprime} and \appx{anomalous}.
The two-loop jet function constant $c_{\tilde J}^2$ is the only missing ingredient at
NNLL$^\prime$ accuracy, and its numerical extraction is the main goal of this and the next 
subsection.

Writing out the individual expansions for $H$, $\wt{J}$, and $\wt{S}$ explicitly and multiplying them 
together gives the Laplace-space cross section:
\begin{align}
\label{eq:Laplace2loop}
\frac{\wt\sigma_\text{sing}(\nu_a)}{\sigma_0}
= 1 &+ \frac{\as(Q)}{4\pi}\Bigl( - \frac{2\Gamma_0}{2-a} \ln^2\bar\nu_a + \frac{2\gamma_J^0}{2-a}\ln\bar\nu_a + c_H^1 + 2c_{\tilde J}^1 + c_{\tilde S}^1\Bigr) \\
&+ \Bigl(\frac{\as(Q)}{4\pi}\Bigr)^2 \biggl\{ \frac{2\Gamma_0^2}{(2\minus a)^2} \ln^4\bar\nu_a - \frac{4\Gamma_0\bigl[\gamma_J^0 + (1-\frac{a}{3})\beta_0\bigr]}{(2- a)^2}\ln^3\bar\nu_a \nn \\
&\qquad\qquad + \biggl[-\frac{2\Gamma_1 + 2\Gamma_0(c_H^1 + 2c_{\tilde J}^1 + c_{\tilde S}^1)}{2-a} +   \frac{2\gamma_J^0(\gamma_J^0+\beta_0)}{(2-a)^2}\biggr]\ln^2\bar\nu_a \nn \\
&\qquad\qquad + \Bigl[ \frac{2\gamma_J^1 + 2\gamma_J^0(c_H^1 + 2c_{\tilde J}^1 + c_{\tilde S}^1) + 4\beta_0 c_{\tilde J}^1}{2-a} + \gamma_S^1 + 2\beta_0 c_{\tilde S}^1\Bigr] \ln\bar\nu_a\nn \\
&\qquad\qquad + c_H^2 + 2c_{\tilde J}^2 + c_{\tilde S}^2 + (c_{\tilde J}^1)^2 + 2c_{\tilde J}^1 c_{\tilde S}^1 + c_H^1(2c_{\tilde J}^1 + c_{\tilde S}^1) \biggr\} \,,\nn
\end{align}
where we have used the consistency relation 
$\gamma_H + 2\gamma_J +\gamma_S = 0$ to eliminate $\gamma_H$, along with 
$\gamma_S^0(a)=0$. We have further expressed \eq{Laplace2loop} in terms of $\as(Q)$, using 
\eq{runningalpha} to one-loop order, and we introduced the notation 
$\bar\nu_a\equiv e^{\gamma_E}{\nu_a}$.

We can now use \eq{iLPs} to inverse Laplace transform \eq{Laplace2loop} back to momentum space.  
Specifically, we obtain the integrated distribution via
\be
\label{eq:Laplacecs}
\frac{\sigma_\text{c,sing}(\tau_a)}{\sigma_0}  \,=\,
\int_{\gamma-i\infty}^{\gamma+i\infty} \frac{d\nu_a}{2\pi i} \;e^{\tau_a\nu_a}\,
\frac{1}{\nu_a}\frac{\wt\sigma_\text{sing}(\nu_a)}{\sigma_0} 
\,,
\ee
where $\gamma$ is such that the integration path lies to the right of all the singularities of the integrand.
After performing the inverse Laplace transform and plugging in explicit values for the 
anomalous dimensions and fixed-order constants, we obtain 
\be
\label{eq:cumulant2loopexpanded}
\frac{\sigma_\text{c,sing}(\tau_a)}{\sigma_0} = 
\theta(\tau_a)\,\biggl\{1 + \frac{\as(Q)}{2\pi}\, \sigma_\text{c,sing}^{(1)}(\tau_a) + \Bigl(\frac{\as(Q)}{2\pi}\Bigr)^2 \sigma_\text{c,sing}^{(2)}(\tau_a) \biggr\}\,,
\ee
where
\be
\label{eq:sigmasing1}
\sigma_\text{c,sing}^{(1)}(\tau_a) = -\frac{C_F}{2-a}\biggl( 4 \ln^2\tau_a + 6\ln\tau_a + 2 + 5a - \frac{\pi^2}{3}(2+a) + 4f(a)\biggr)\,,
\ee
and
\begin{align}
\label{eq:sigmasing2}
\sigma_\text{c,sing}^{(2)}(\tau_a) &= \frac{C_F^2}{(2-a)^2} \biggl\{ 8\ln^4\tau_a + 24\ln^3\tau_a + \Bigl[ 26+20a - \frac{4\pi^2}{3}(6+a) + 16f(a)\Bigr]\ln^2\tau_a \\
&\qquad\qquad  + \Bigl[ 9 + \frac{63a}{2} - 4\pi^2(2+a) + 8\zeta_3(2+3a) + 24f(a) \Bigr]\ln\tau_a \biggr\} \nn \\
&\quad + \frac{4C_F\beta_0}{(2-a)^2}\Bigl(1-\frac{a}{3}\Bigr) \ln^3\tau_a + \biggl[ \frac{3C_F\beta_0}{(2-a)^2} - \frac{2C_F}{2-a}\frac{\Gamma_1}{\Gamma_0}\biggr]\ln^2\tau_a + \biggl[ \frac{\gamma_H^1\bigr\rvert_{\text{n.A.}}}{4(2\minus a)}  \nn \\
&\qquad \qquad  - \frac{1\minus a}{4(2\minus a)}\gamma_S^1(a) - \frac{C_F\beta_0}{(2\minus a)^2}\Bigl( 14\minus 13a \minus \frac{4\pi^2}{3}(1\minus a) \minus 4f(a)\Bigr)\biggr]\ln\tau_a \nn  + c^{(2)}\,, \nn
\end{align}
where we recall that the function $f(a)$ was defined in \eq{fint}. Explicit values for 
$\gamma_H^1\bigr\rvert_{\text{n.A.}}$ and $\gamma_S^1$ can then be inserted from \eqs{gammaHsplit}{gammaS1a}. 
The total two-loop constant $c^{(2)}$ is given in terms of the Laplace-space constants by 
\begin{align}
\label{eq:c2momentum}
c^{(2)} &= \frac{c_{\tilde J}^{2}}{2} + \frac{c_H^2 + c_{\tilde S}^{2} + (c_{\tilde J}^1)^2 + 2c_{\tilde J}^1 c_{\tilde S}^1 + c_H^1(2c_{\tilde J}^1 + c_{\tilde S}^1)}{4} + \frac{\pi^4}{120}\frac{\Gamma_0^2}{(2-a)^2} \\
&\quad  + \frac{2\zeta_3\Gamma_0}{(2-a)^2}\Bigl[ \gamma_J^0 + \Bigl(1 -\frac{a}{3}\Bigr)\beta_0\Bigr]  + \frac{\pi^2}{12}\frac{1}{2-a}\Bigl[ \Gamma_1 + \Gamma_0(c_H^1 + 2c_{\tilde J}^1 + c_{\tilde S}^1) - \frac{\gamma_J^0(\gamma_J^0 + \beta_0)}{2-a}\Bigr]\,. \nn
\end{align}
This formula immediately gives us $c_{\tilde J}^2$ as soon as we determine $c^{(2)}$ (which, we recall from 
\eq{sigmasing2}, is in \emph{momentum} space), whose extraction  from \event\ will be described in the next 
subsection. 

%%%%%%%%%%%%%%%%%%%%%%%%%%%%%%%%%%%%%%%%%%%%%%%%%%%%%%%
\subsection{Two-loop jet function constant}
\label{sec:cJ2event2}

The program \event\ \cite{Catani:1996jh,Catani:1996vz} gives numerical results for partonic QCD observables 
in $e^+e^-$ collisions to $\cO(\as^2)$. Using the method described by Hoang and 
Kluth \cite{Hoang:2008fs}, we can extract the singular constant $c^{(2)}$ in \eq{sigmasing2}, and thus the 
unknown jet function constant
$c_{\tilde J}^2$ via \eq{c2momentum}. For pedagogical purposes, we will give our own 
description of this method in the language of continuous distributions, which we find more intuitive to 
understand, rather than the language of discrete bins, which we encourage the reader to study 
in \cite{Hoang:2008fs}, as in practice one implements the discrete method.

The integrated (cumulative) angularity distribution in full QCD has a fixed-order 
expansion of the form:
\begin{align}
\label{eq:QCDexpansion}
\frac{\sigma_c(\tau_a)}{\sigma_0}
= \theta(\tau_a)\,\biggl\{& 1 + \frac{\as(Q)}{2\pi} \bigl[c_{12}\ln^2\tau_a + c_{11}\ln\tau_a + c_{10} + r_c^1(\tau_a)\bigr] \\
& + \Bigl(\frac{\as(Q)}{2\pi}\Bigr)^2\bigl[c_{24}\ln^4\tau_a + c_{23}\ln^3\tau_a + c_{22}\ln^2\tau_a + c_{21}\ln\tau_a + c_{20} + r_c^2(\tau_a) \bigr] \biggr\}, \nn 
\end{align}
to $\cO(\as^2)$. The $c_{nm}$ coefficients should agree with the SCET prediction in \eq{cumulant2loopexpanded} 
for the singular terms. The $r_c^{n}$ functions are the nonsingular remainders that vanish as $\tau_a\to 0$ 
and which are not predicted by the leading power expansion in SCET. Since SCET predicts the singular 
coefficients correctly, the difference of the QCD and SCET results is simply given by these remainders:
\be
\label{eq:difference}
\frac{\sigma_c(\tau_a)}{\sigma_0} -\frac{\sigma_\text{c,sing}(\tau_a)}{\sigma_0}=
r_c(\tau_a) 
= \theta(\tau_a)\,\biggl\{
\frac{\as(Q)}{2\pi} r_c^1(\tau_a) + \Bigl(\frac{\as(Q)}{2\pi}\Bigr)^2 r_c^2(\tau_a)
\biggr\},
\ee
which we will use in the next subsection to obtain the nonsingular remainder functions $r_c^n$ from the 
difference of the \event\ output and the SCET prediction. To do this, however, we must know all the $c_{nm}$ 
coefficients in \eq{QCDexpansion}, including the constants in $c_{20}\equiv c^{(2)}$ in \eq{c2momentum}. But 
we do not yet know $c_{\tilde J}^2$.

%%%%
\begin{figure}[t]
\begin{center}
\begin{tabular}{lc}
\quad \includegraphics[width=.4\columnwidth]{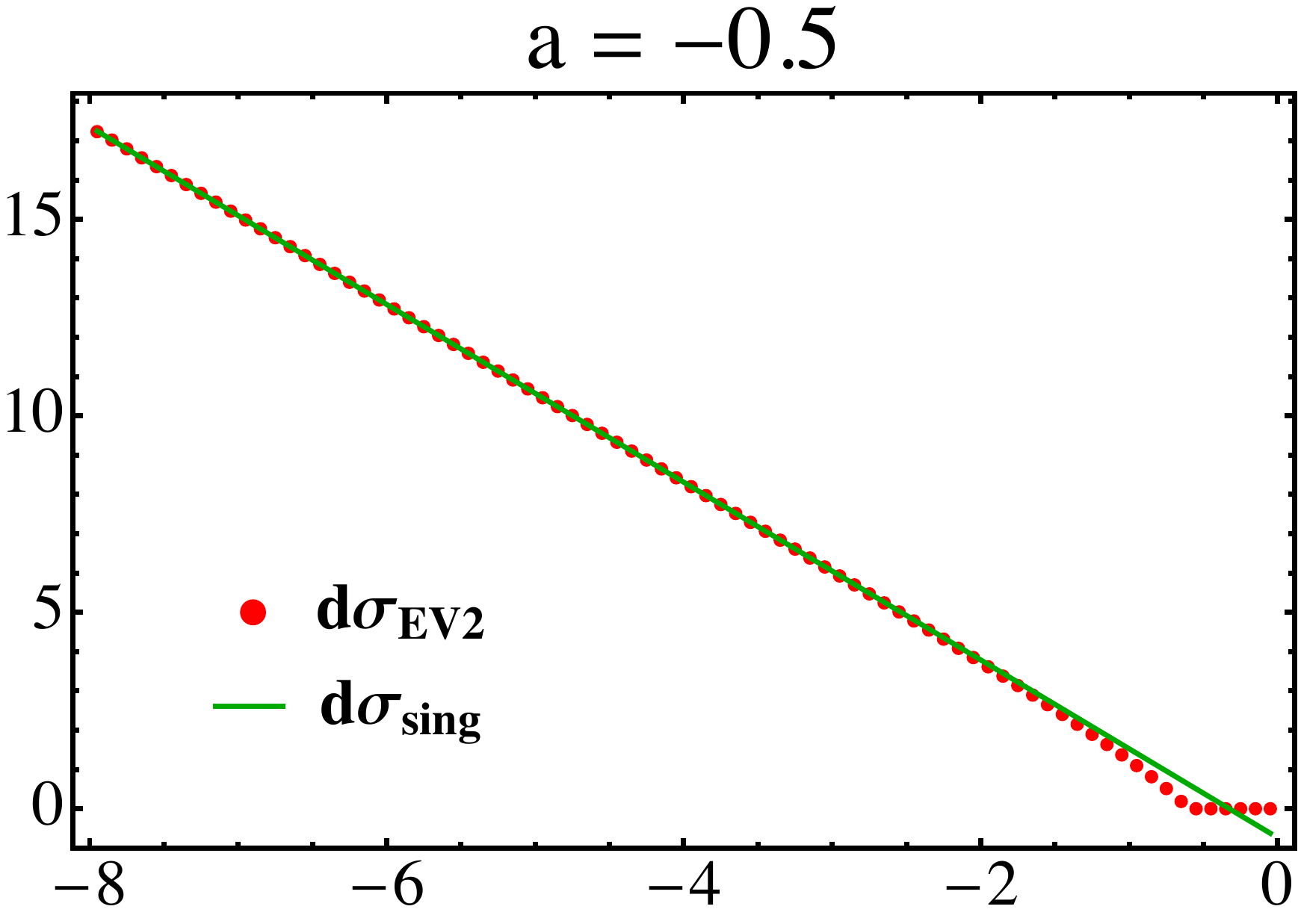} &
\includegraphics[width=.40\columnwidth]{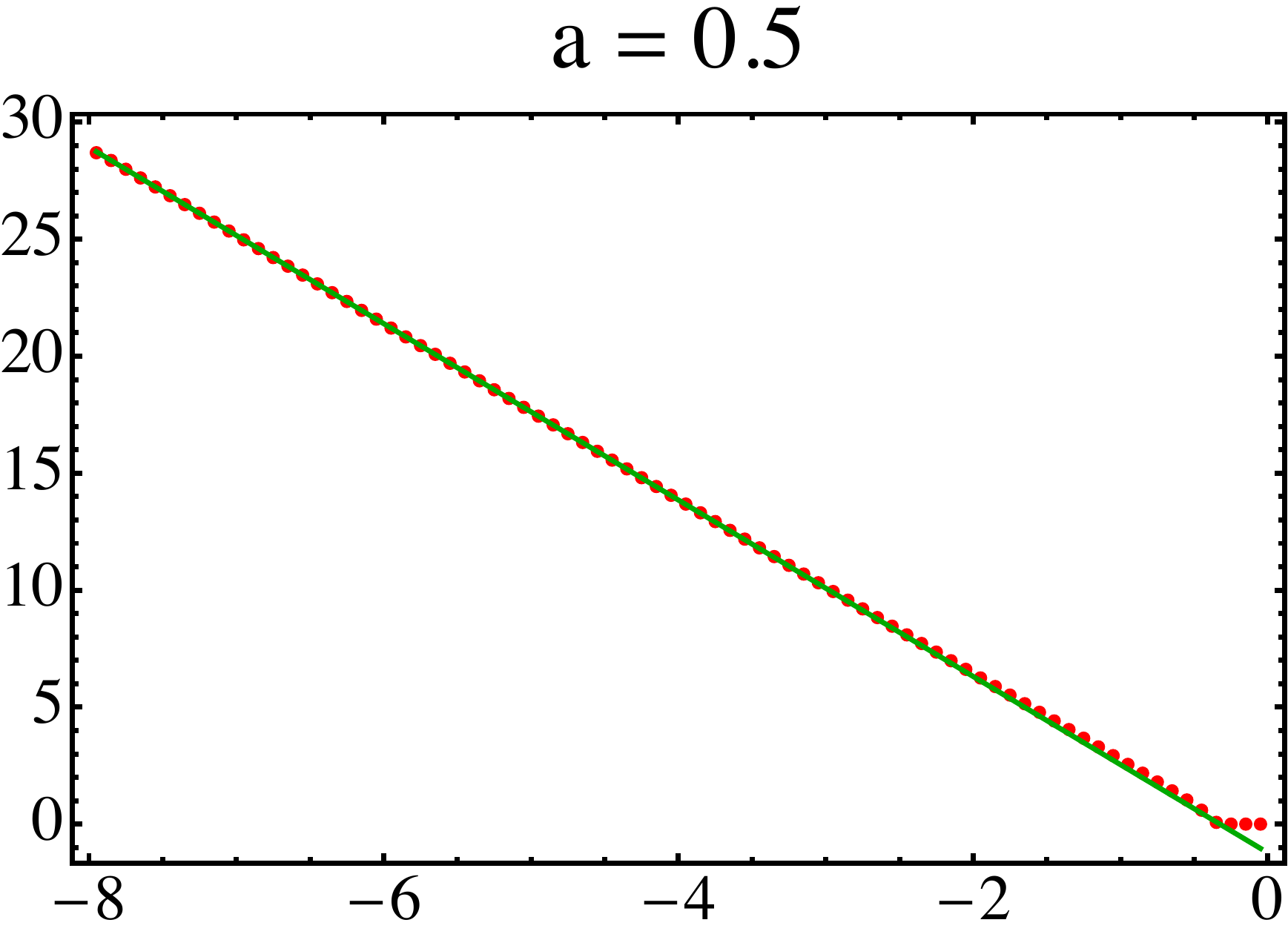} \\
\includegraphics[width=.42\columnwidth]{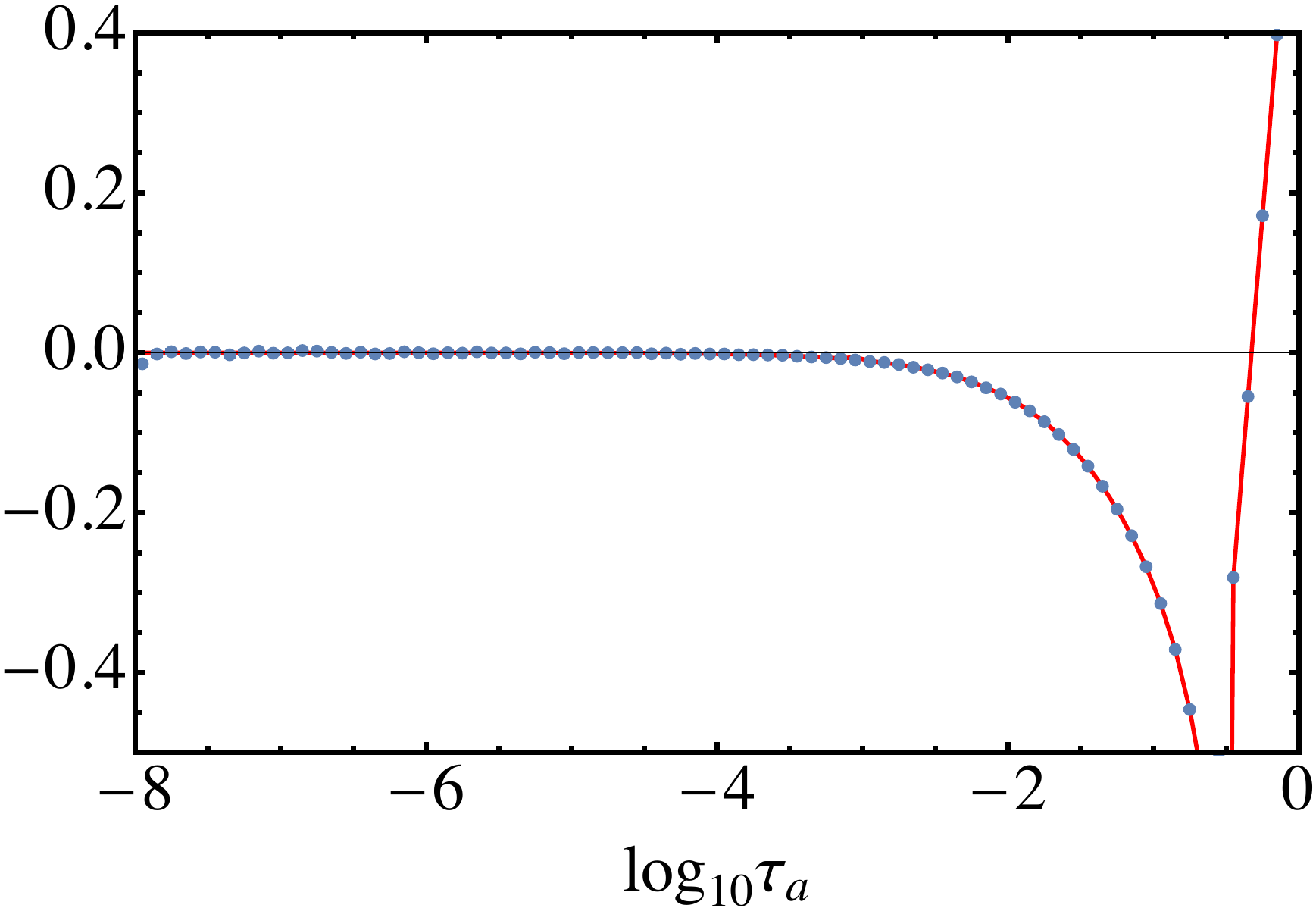} &
\includegraphics[width=.41\columnwidth]{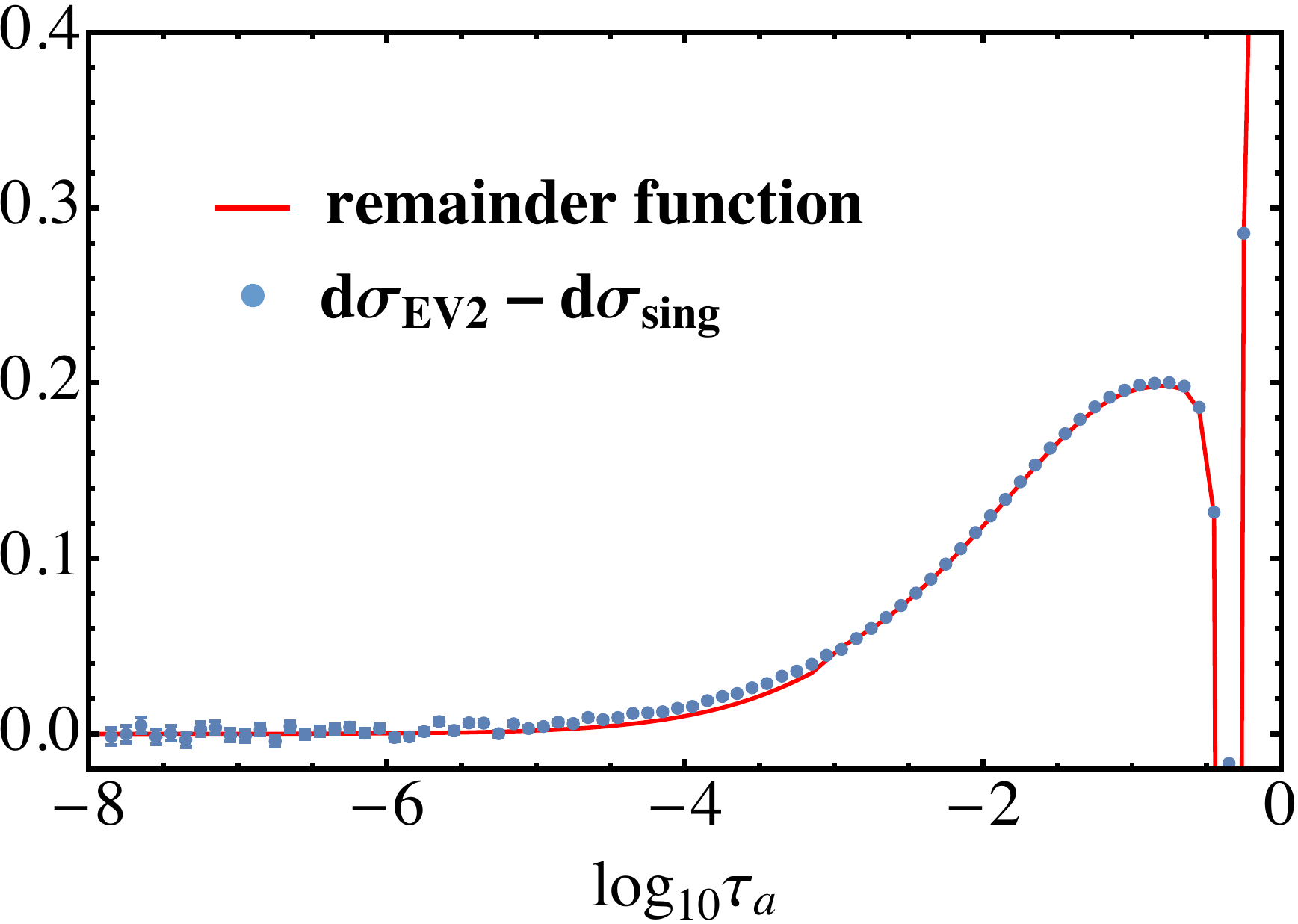}
\end{tabular}
\end{center}
\vspace{-2em}
\caption{\emph{Top}: Comparison of the \event\ prediction for two angularity distributions
at $\cO(\as)$ with the analytically-known singular terms that follow from \eq{sigmasing1}. \emph{Bottom}:
Comparison of their difference with the remainder function $r^1(\tau_a)$ from
\cite{Hornig:2009vb}.
}
\label{fig:alpha1singvsnonsing}
\end{figure}
%%%%

In the limit of zero bin size, \event\ is generating an approximation to the differential distribution, 
which takes the form:
\be
\label{eq:differentialdistribution}
\frac{1}{\sigma_0}\frac{d\sigma}{d\tau_a} = A\,\delta(\tau_a) + [B(\tau_a)]_+ + r(\tau_a)\,,
\ee
where $A$ is the constant coefficient of the delta function,  $B$ is a singular function, turned into an 
integrable plus-distribution, and $r  = dr_c/d\tau_a$ is nonsingular, that is, directly integrable as $\tau_a\to 0$. However, in practice we only obtain the distribution away from $\tau_a=0$ from 
\event, thus obtaining a histogram of the function:
\be
\frac{1}{\sigma_0}\frac{d\sigma}{d\tau_a} \biggr\rvert_{\tau_a>0} = B(\tau_a) + r(\tau_a)\,.
\ee
The delta function coefficient $A$, which gives the constants $c_{n0}$ in the integrated distribution 
\eq{QCDexpansion}, does not appear. The SCET prediction, on the other hand, gives just the singular parts:
\be
\label{eq:SCETprediction}
\frac{1}{\sigma_0}\frac{d\sigma_\text{sing}}{d\tau_a}  = A\,\delta(\tau_a) + [B(\tau_a)]_+\,,
\ee
so the difference between the \event\ and SCET predictions (away from $\tau_a=0$) gives
the nonsingular part:
\begin{align}
\label{eq:remainder}
\frac{1}{\sigma_0}\frac{d\sigma}{d\tau_a} - 
\frac{1}{\sigma_0}\frac{d\sigma_\text{sing}}{d\tau_a} \;\biggr\rvert_{\tau_a>0} 
= r(\tau_a) = \frac{\as(Q)}{2\pi} r^1(\tau_a) + \Bigl(\frac{\as(Q)}{2\pi}\Bigr)^2 r^2(\tau_a)
\,.
\end{align}
Integrating this difference over $0<\tau_a\leq 1$ then gives 
the expansion of the nonsingular part in terms of the $r_c^n$ functions in \eq{QCDexpansion}:
\be
\label{eq:remainderintegral}
\lim_{\tau_a\to 0} \,\int_{\tau_a}^1 d\tau_a'\,  r(\tau_a') = r_c(1)
= \frac{\as(Q)}{2\pi} r_c^1(1) +  \Bigl(\frac{\as(Q)}{2\pi}\Bigr)^2 r_c^2(1)  \,.
\ee
Now, the total hadronic cross section, normalized to $\sigma_0$, is simply the integrated 
distribution \eq{QCDexpansion} evaluated at $\tau_a=1$. As the plus-distributions integrate to zero over 
this region, we use  \eq{differentialdistribution} to obtain
\be
\label{eq:sigmatotsum}
\sigma_{\text{tot}} = A + r_c(1)\,.
\ee
The total cross section 
\be
\sigma_{\text{tot}}= 1 + \frac{\as(Q)}{2\pi} \,\sigma^{(1)}_{\text{tot}} + \Bigl(\frac{\as(Q)}{2\pi}\Bigr)^2 \sigma^{(2)}_{\text{tot}}
\ee
is known to the considered order from \cite{Chetyrkin:1979bj,Dine:1979qh,Celmaster:1979xr}:
\be
\label{eq:rn}
\begin{split}
\sigma^{(1)}_{\text{tot}}  &= \frac{3}{2}C_F \,, \\
\sigma^{(2)}_{\text{tot}}  &=  -\frac{3}{8} C_F^2 + \Bigl(\frac{123}{8} - 11\zeta_3\Bigr) C_F C_A + \Bigl( -\frac{11}{2} + 4\zeta_3\Bigr) C_F T_F n_f  \,.
\end{split}
\ee
\begin{figure}[t]
\vspace{-1em}
\begin{center}
\includegraphics[width=0.5\columnwidth]{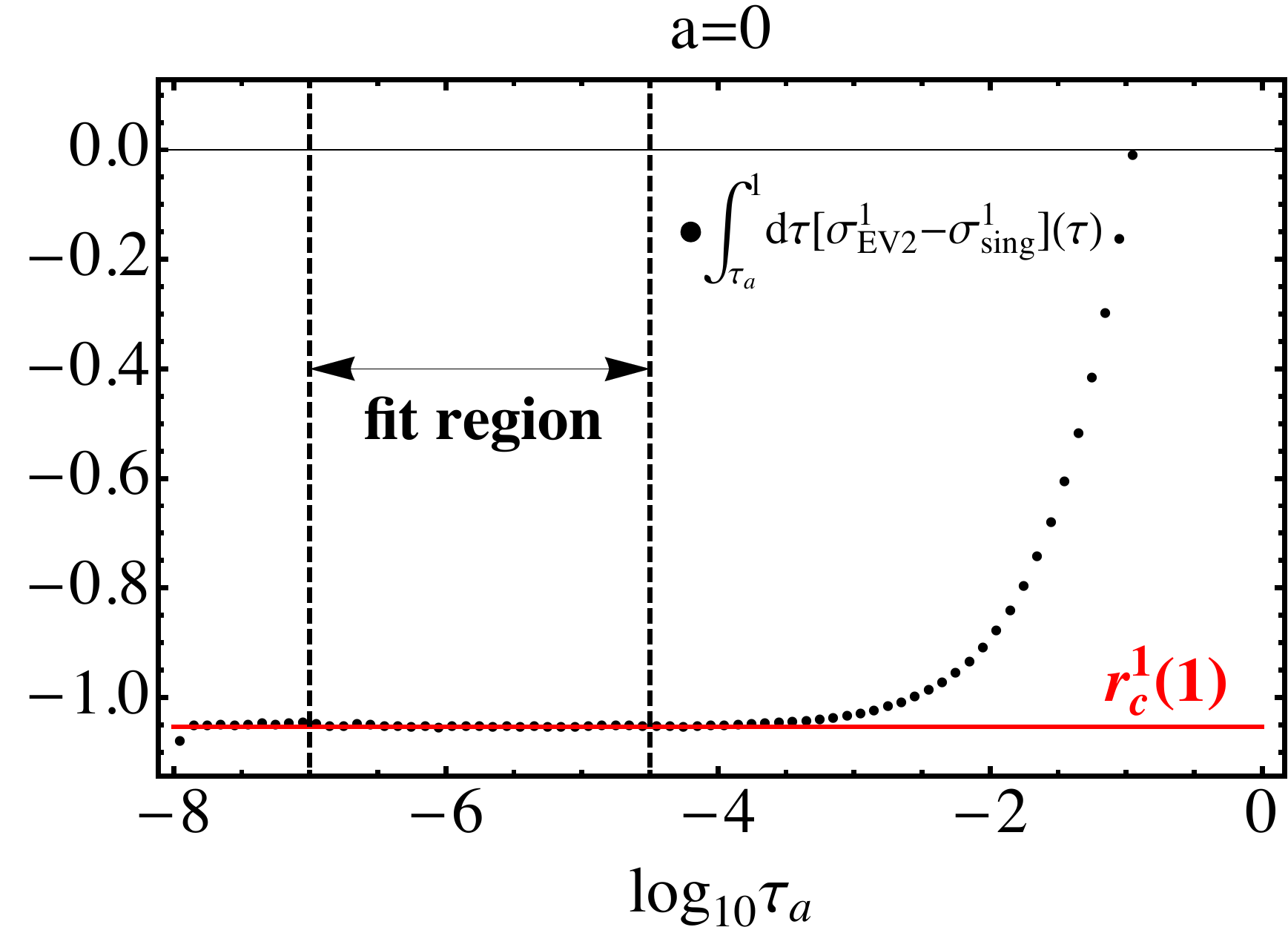}
\end{center}
\vspace{-2em}
\caption{Integral \eq{remainderintegral} of the $\cO(\as)$ remainder function from \event\ as a function of the 
lower integration limit $\tau_a$. As $\tau_a\to 0$, a suitable fitting region is indicated that gives the 
constant $r_c^1(1)$, for the case $a=0$.
}
\label{fig:alpha1Fita0Example}
\end{figure}
From the perturbative expansion of \eq{sigmatotsum}, we then obtain the relations
\be
\label{eq:totalQCD}
\begin{split}
\sigma^{(1)}_{\text{tot}}  &= c_{10} + r_c^1(1)\,, \\
\sigma^{(2)}_{\text{tot}}  &= c_{20} + r_c^2(1)\,.
\end{split}
\ee
The coefficient $c_{20}$ is precisely the constant $c^{(2)}$ in \eq{sigmasing2}, so we can use \eq{totalQCD} 
to determine $c^{(2)}$---and thus $c_{\tilde J}^2$ through \eq{c2momentum}---from the known 
$\sigma^{(2)}_{\text{tot}}$ in \eq{rn} and the \event\ results for $r_c^2(1)$.

Essentially, the method relies on the fact that the total cross section is the sum of the singular
constant $A$ and the integral over the nonsingular distribution $r_c(1)$. Summing the \event\ bins, with 
the singular terms subtracted off, between a small $\tau_a\sim 0$ and $\tau_a=1$ gives the latter. Its 
difference from the known total cross section then gives the singular constants $c_{n0}$. 

This is illustrated for the $\cO(\as)$ parts of the differential distributions in 
\figs{alpha1singvsnonsing}{alpha1Fita0Example}. The \event\ data for these plots is based on 
$1\cdot10^{11}$ events with a cutoff parameter of $10^{-12}$. In the upper plots of \fig{alpha1singvsnonsing} 
we show the output of \event\ for two values of the angularity parameter, 
$a=-0.5$ and $a=0.5$, compared to the known predictions for the singular terms. 
The difference is shown in the lower plots of \fig{alpha1singvsnonsing} together with the remainder 
function $r^1(\tau_a)$ from \cite{Hornig:2009vb}. One sees that the difference between the \event\ output 
and the singular terms vanishes as $\tau_a\to 0$ as expected.  In \fig{alpha1Fita0Example} we show the 
result of computing the integral \eq{remainderintegral} as a sum over \event\ bins, for $a=0$. The value 
of this integral as the lower limit $\tau_a\to 0$ gives the numerical result for the constant $r_c^1(1)$, 
which through \eq{totalQCD} gives an extraction of the singular constant $c_{10}$ in the fixed-order 
distribution. The $\cO(\as)$ terms are, of course, already known; we show this just to illustrate the 
logic of our procedure. 
\begin{figure}[t]
\centering
\begin{tabular}{cccc}
\rotatebox{90}{\qquad$a=-0.5$}&
\includegraphics[width=.3\columnwidth]{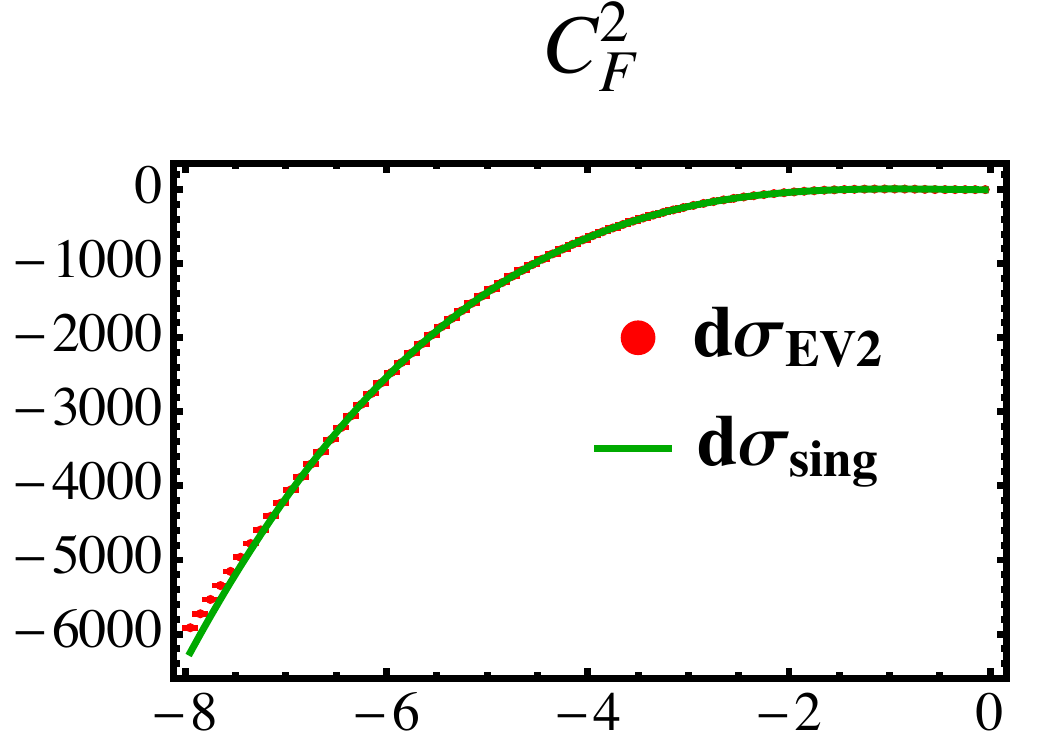}&
\includegraphics[width=.28\columnwidth]{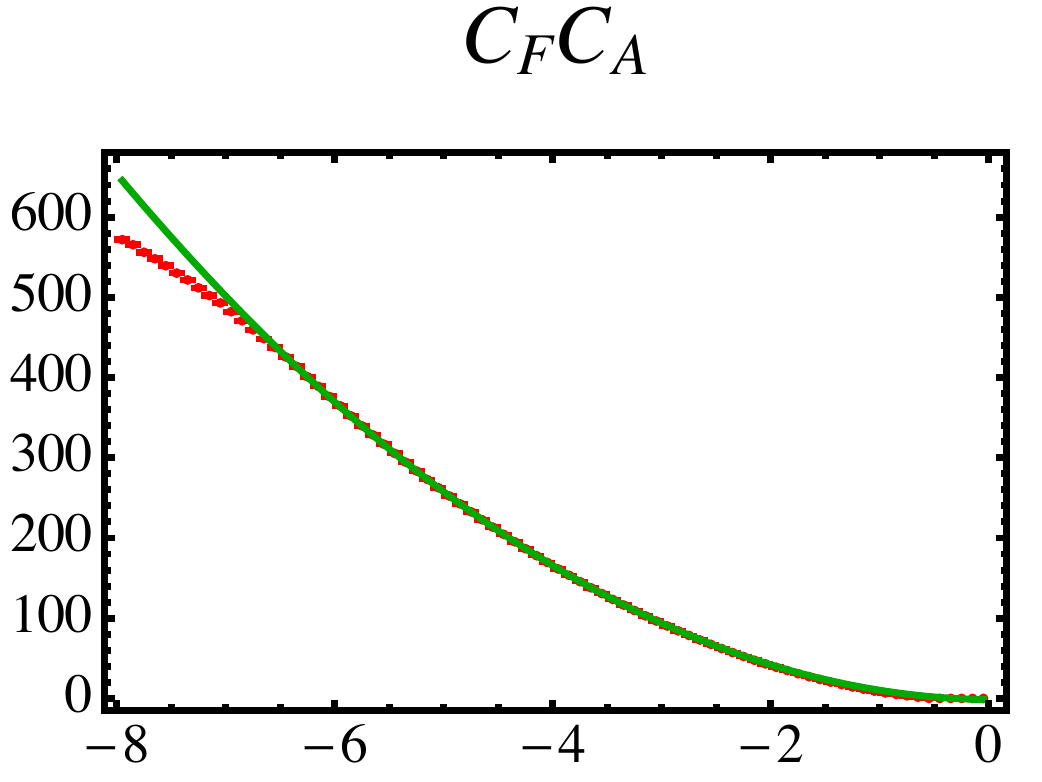}&
\includegraphics[width=.29\columnwidth]{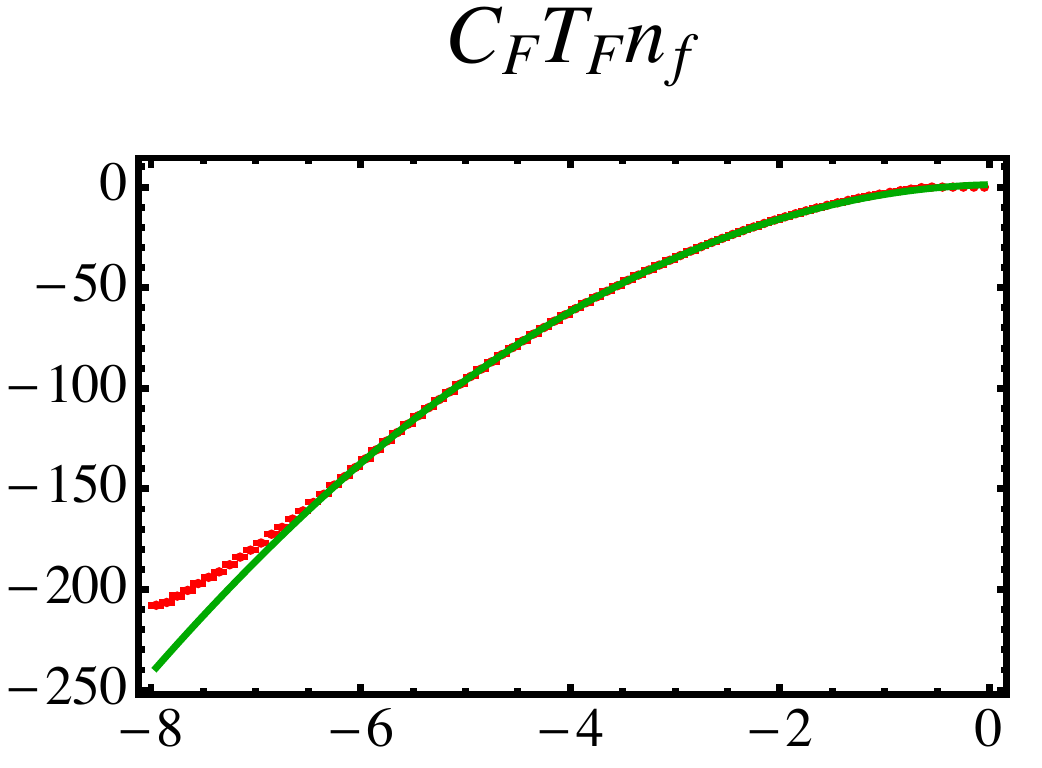} \\[-.5ex]
\rotatebox{90}{\qquad\quad$a=0.5$}&
\includegraphics[width=.3\columnwidth]{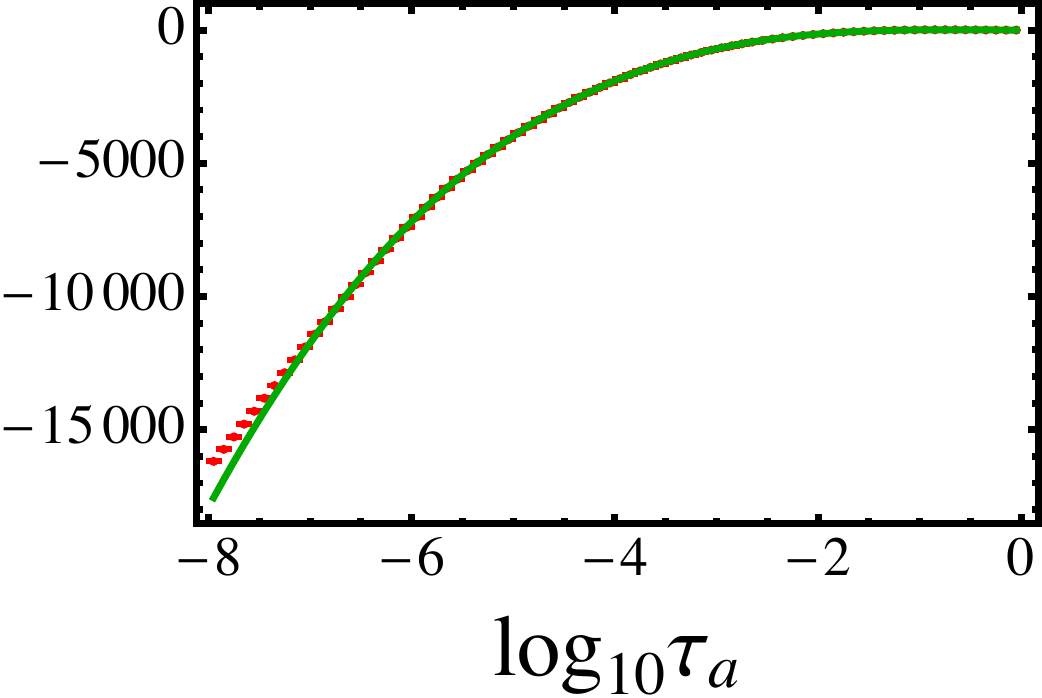}&
\includegraphics[width=.28\columnwidth]{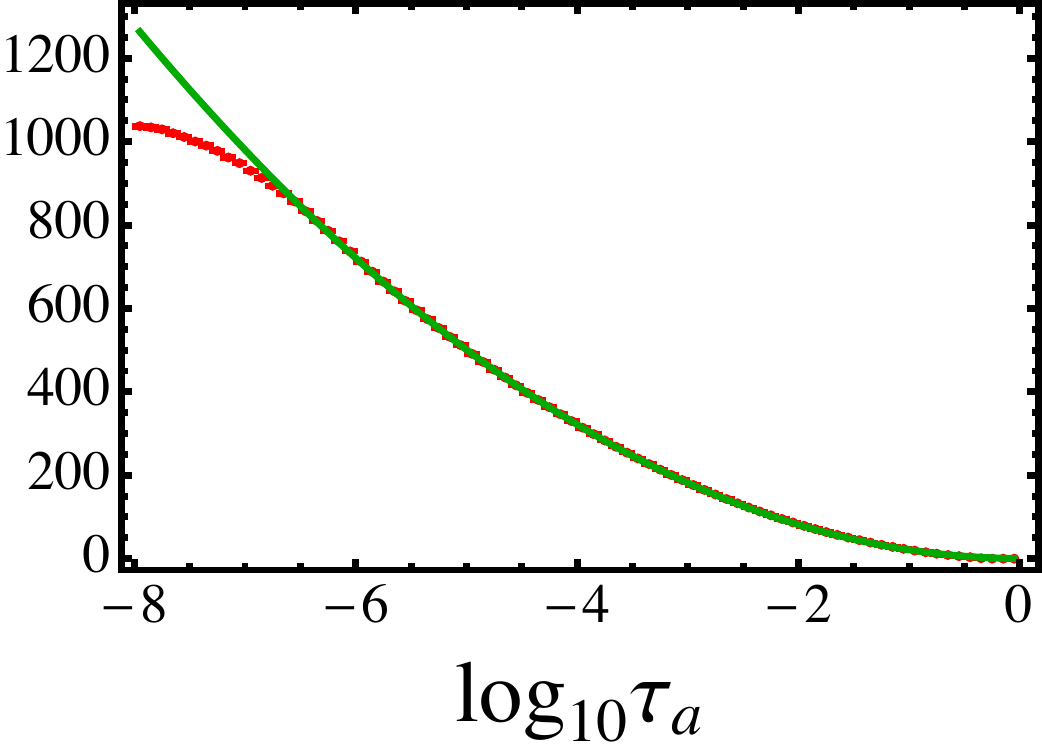}&
\includegraphics[width=.28\columnwidth]{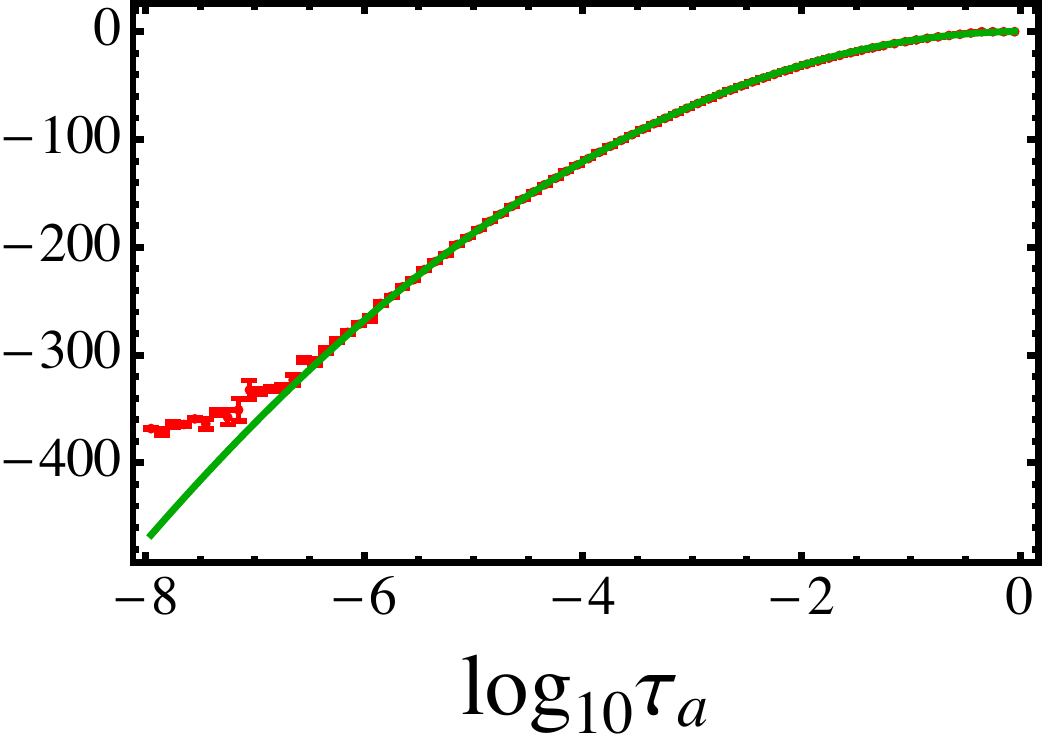}
\end{tabular}
\vspace{-1.5em}
\caption{
Comparison of the \event\ prediction for two angularity distributions
at $\cO(\as^2)$ with the analytically-known singular terms that follow from \eq{sigmasing2}.
The plots show the coefficients of the color structures $C_F^2$, $C_F C_A$, $C_F T_F n_f$. 
}
\label{fig:alpha2EV2}
\end{figure}
\begin{figure}[t]
\centering
\begin{tabular}{cccc}
\rotatebox{90}{\qquad$a=-1$} &
\includegraphics[width=.295\columnwidth]{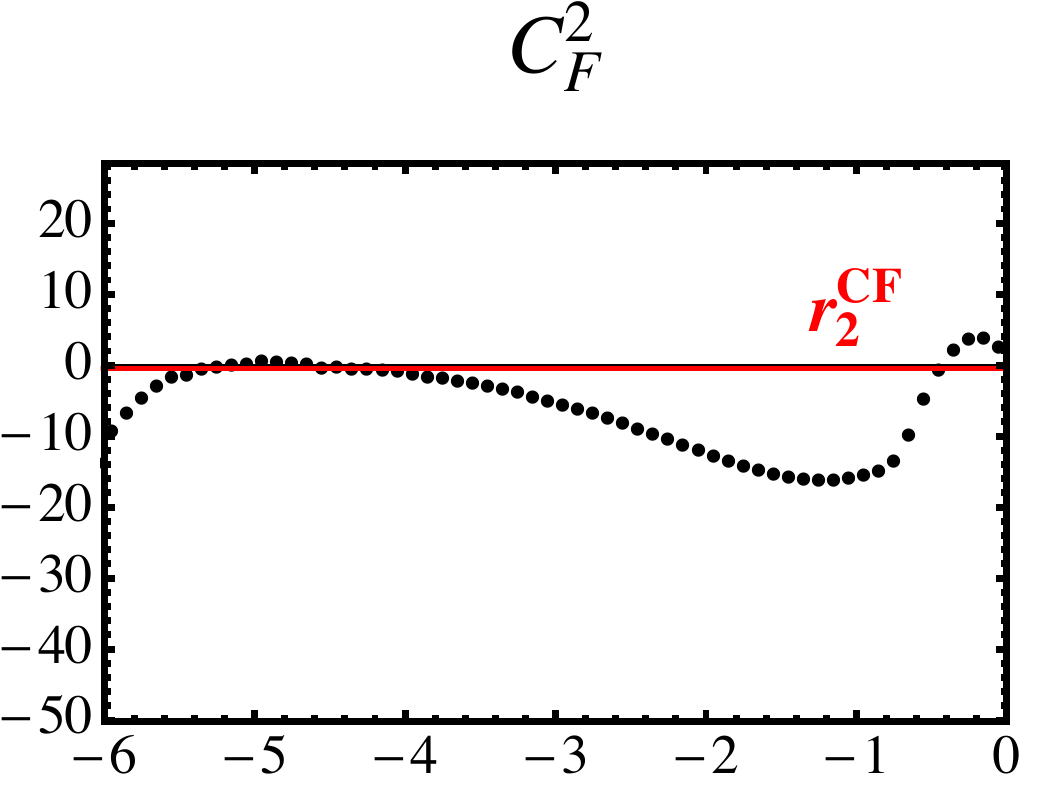}&
\includegraphics[width=.295\columnwidth]{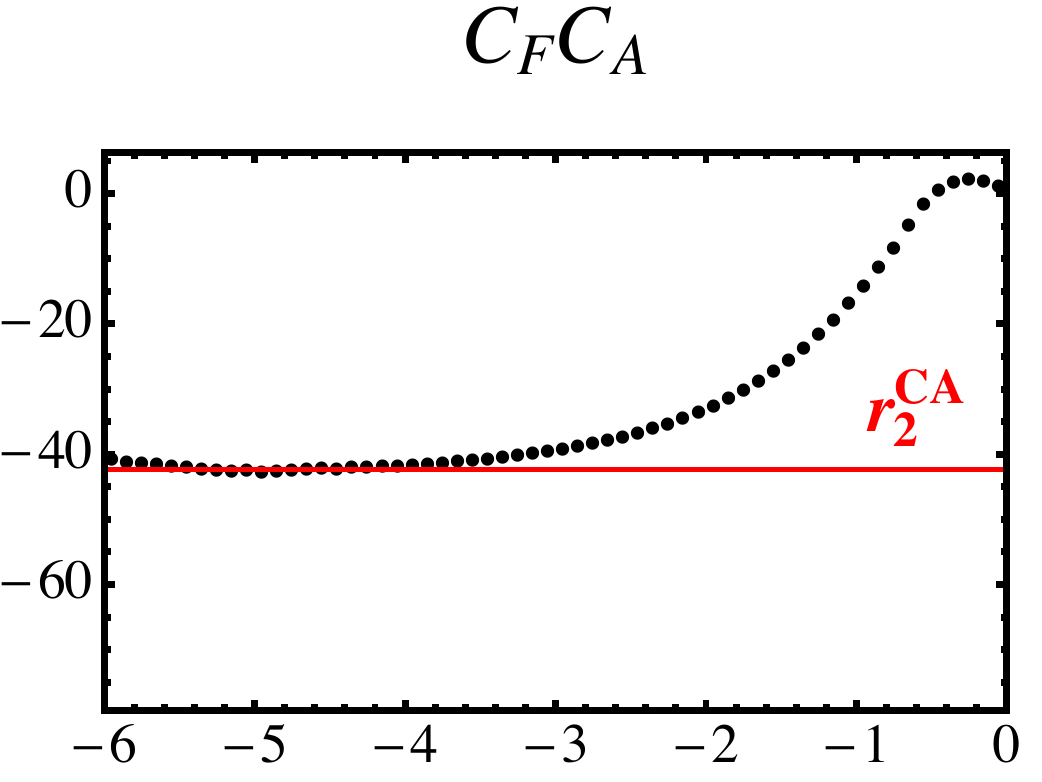}&
\includegraphics[width=.3\columnwidth]{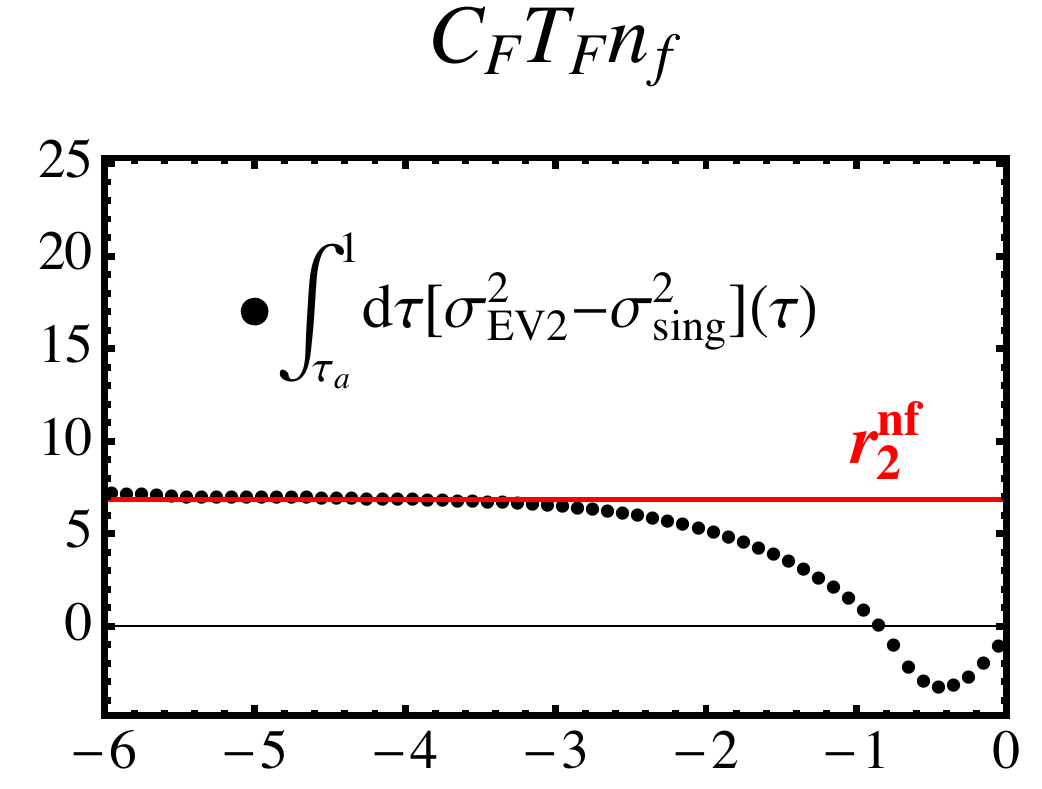} \\[-.5ex]
\rotatebox{90}{\quad$a=-0.75$}&
\includegraphics[width=.29\columnwidth]{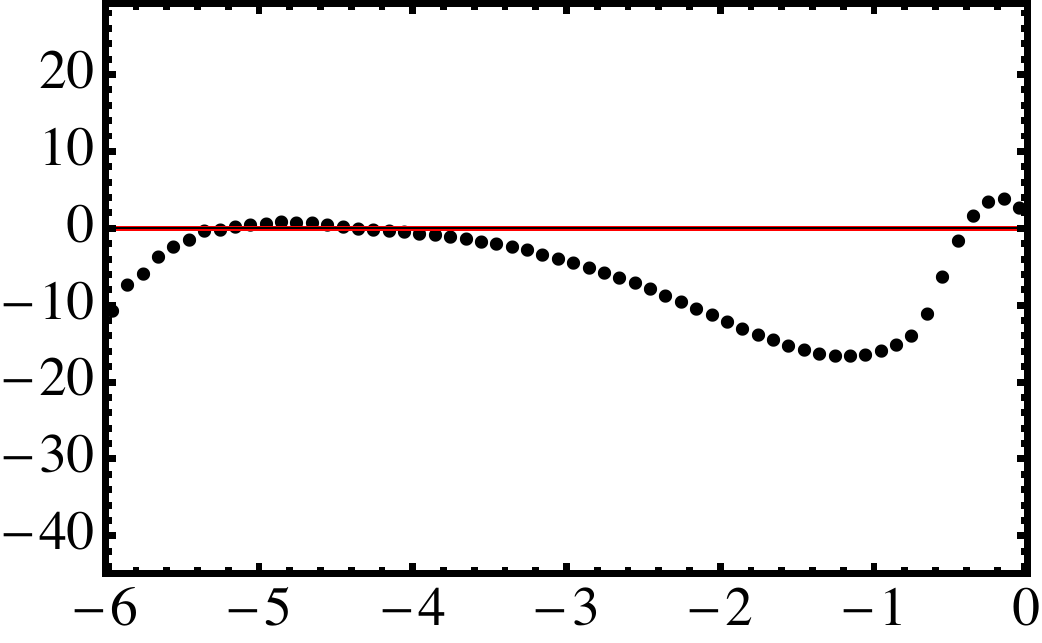}&
\includegraphics[width=.29\columnwidth]{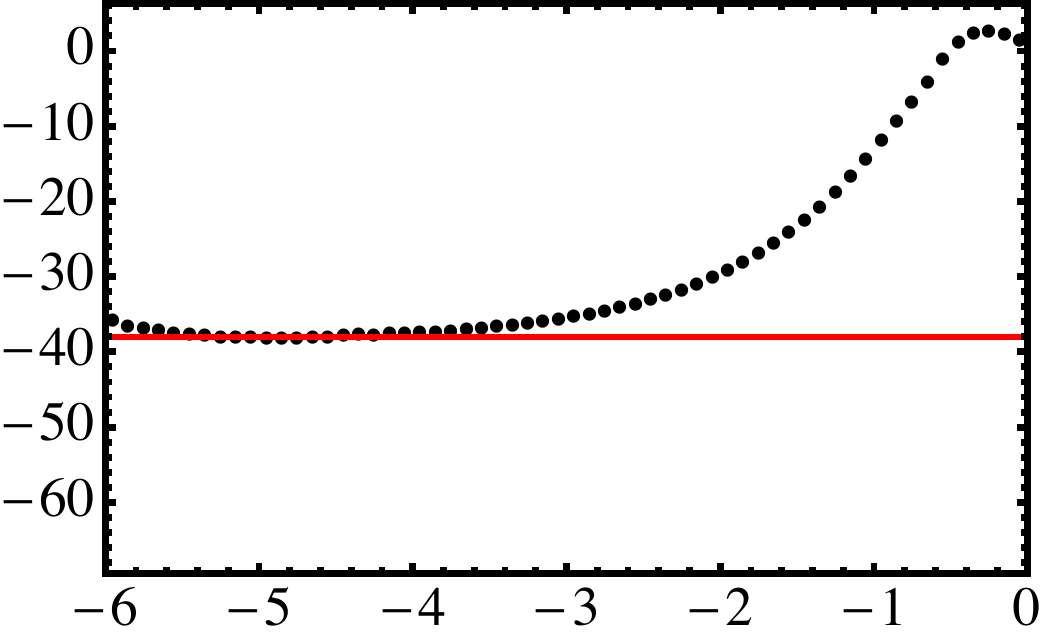}&
\includegraphics[width=.29\columnwidth]{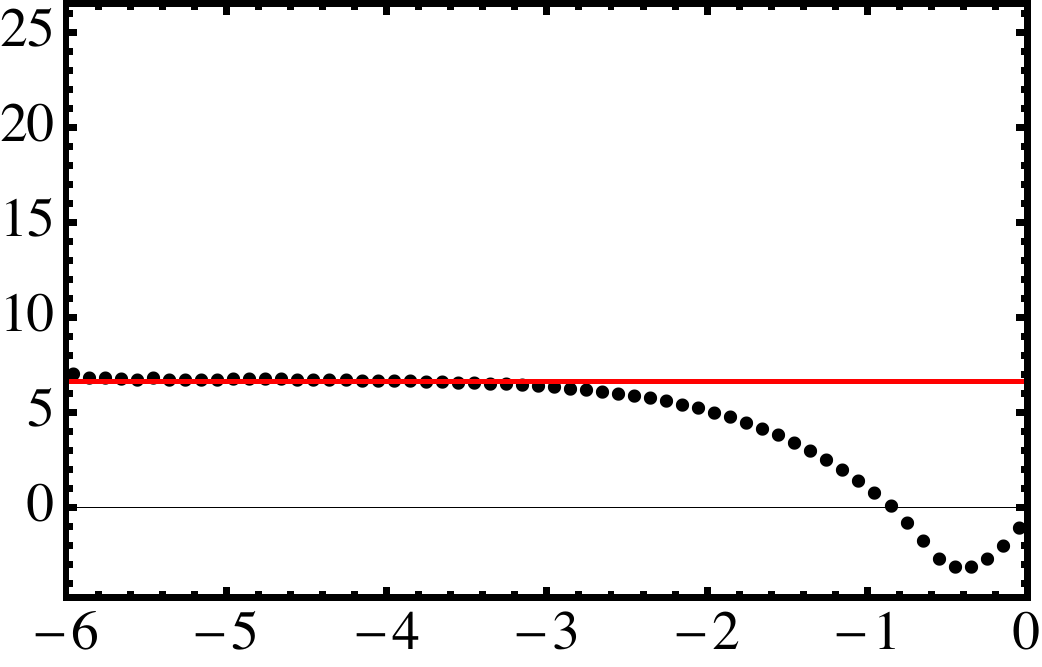} \\[-.5ex]
\rotatebox{90}{\qquad$a=-0.5$}&
\includegraphics[width=.29\columnwidth]{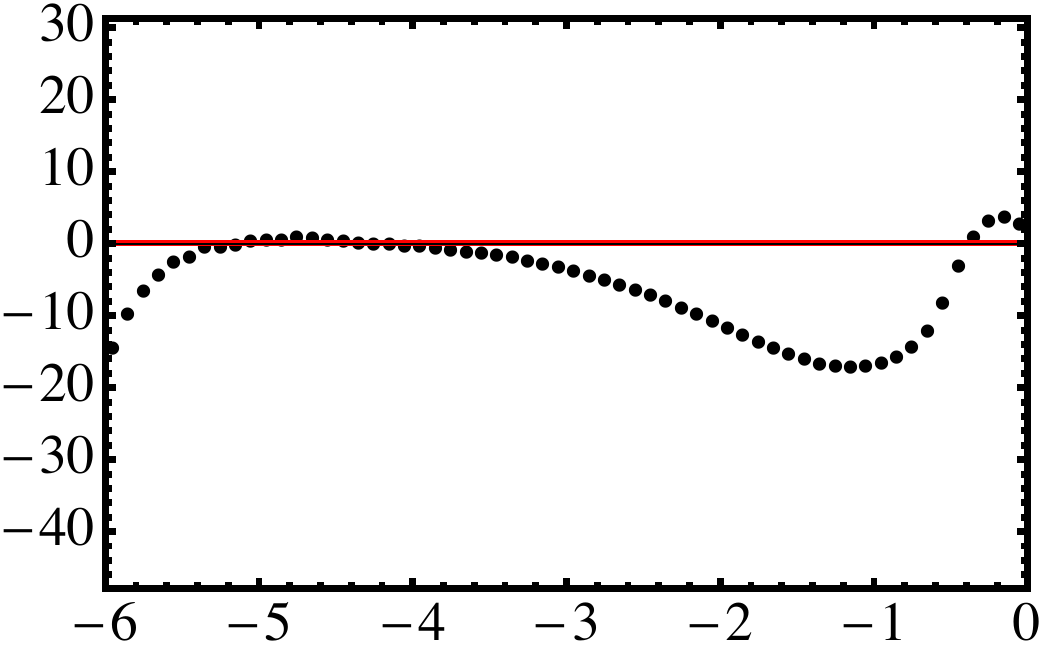}&
\includegraphics[width=.29\columnwidth]{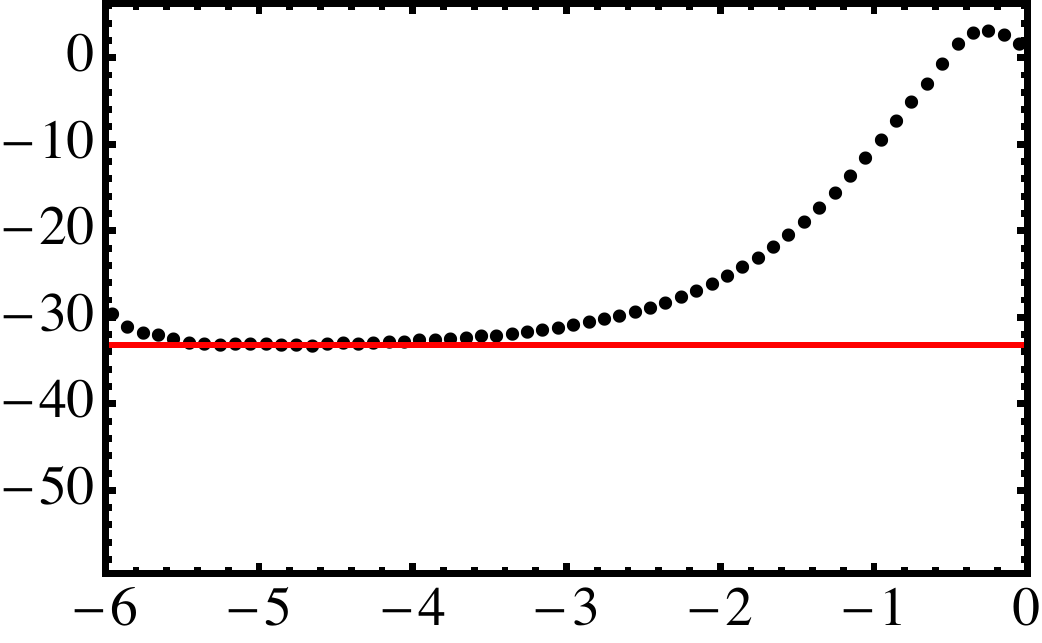}&
\includegraphics[width=.29\columnwidth]{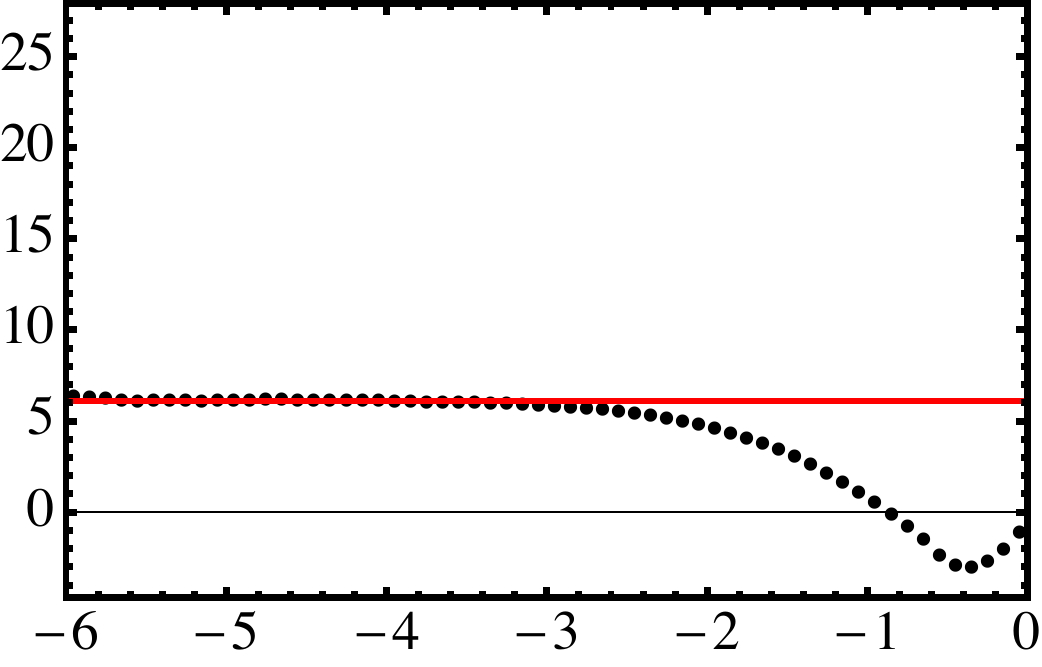} \\[-.5ex]
\rotatebox{90}{\quad$a=-0.25$}&
\includegraphics[width=.29\columnwidth]{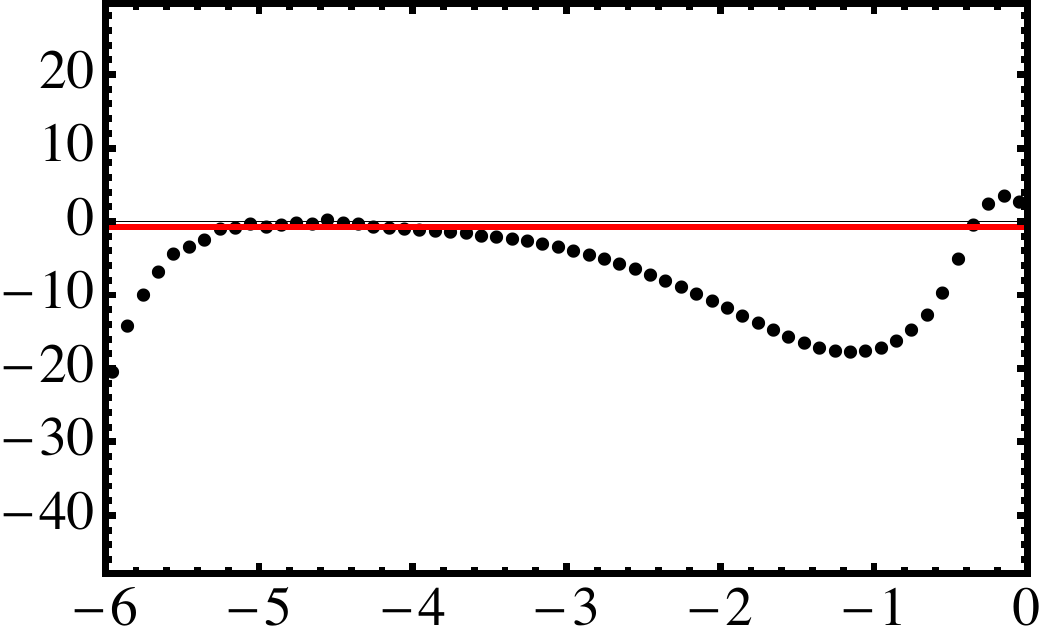}&
\includegraphics[width=.29\columnwidth]{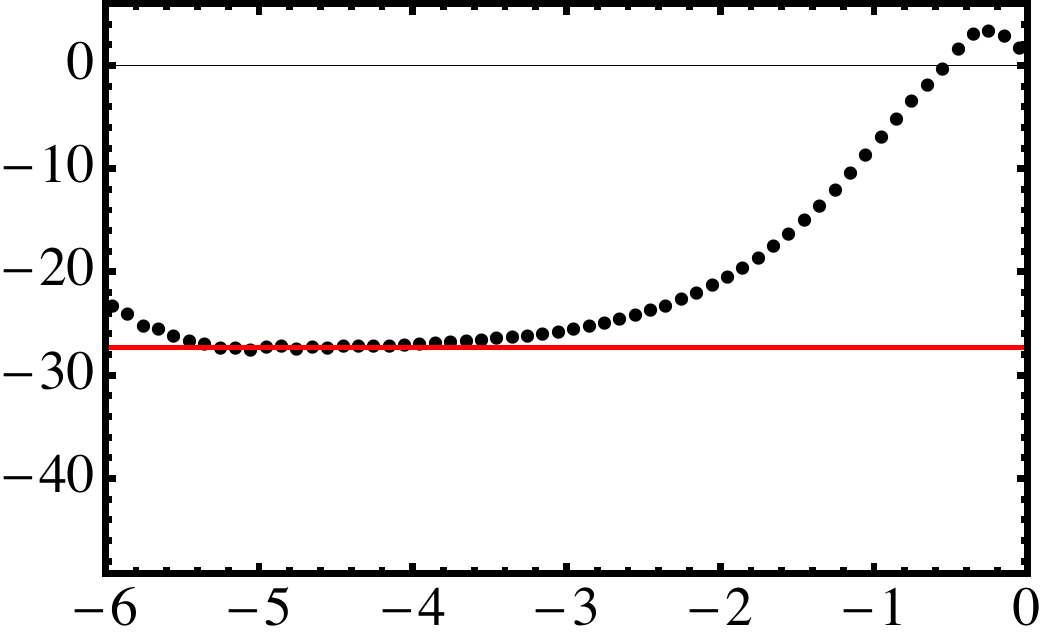}&
\includegraphics[width=.29\columnwidth]{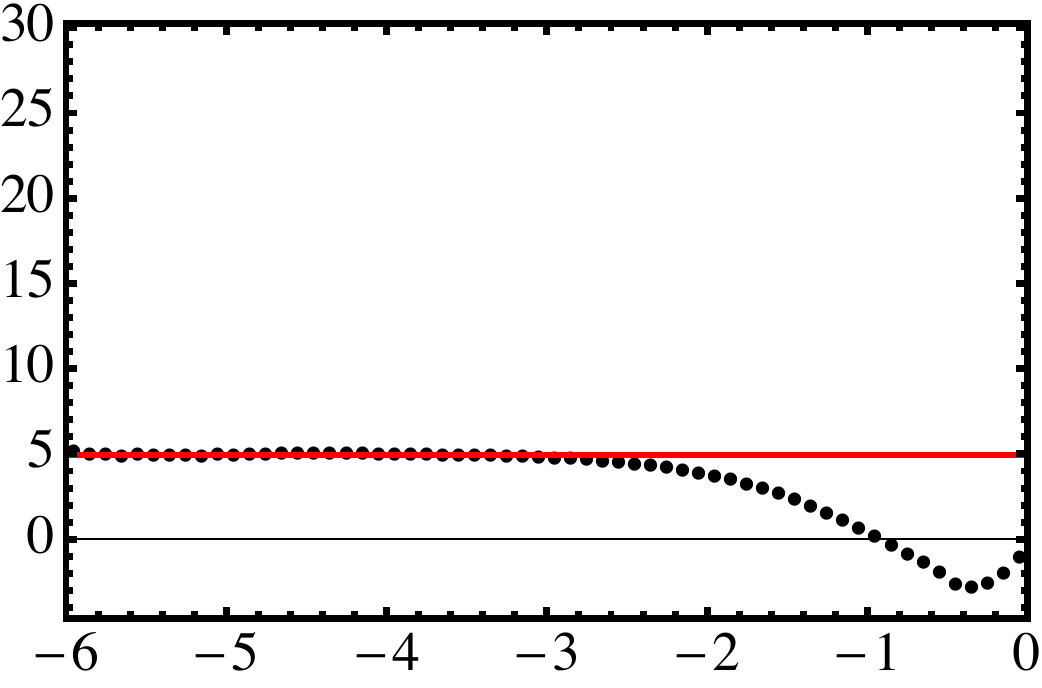} \\[-.5ex]
\rotatebox{90}{\qquad\quad$a=0$}&
\includegraphics[width=.29\columnwidth]{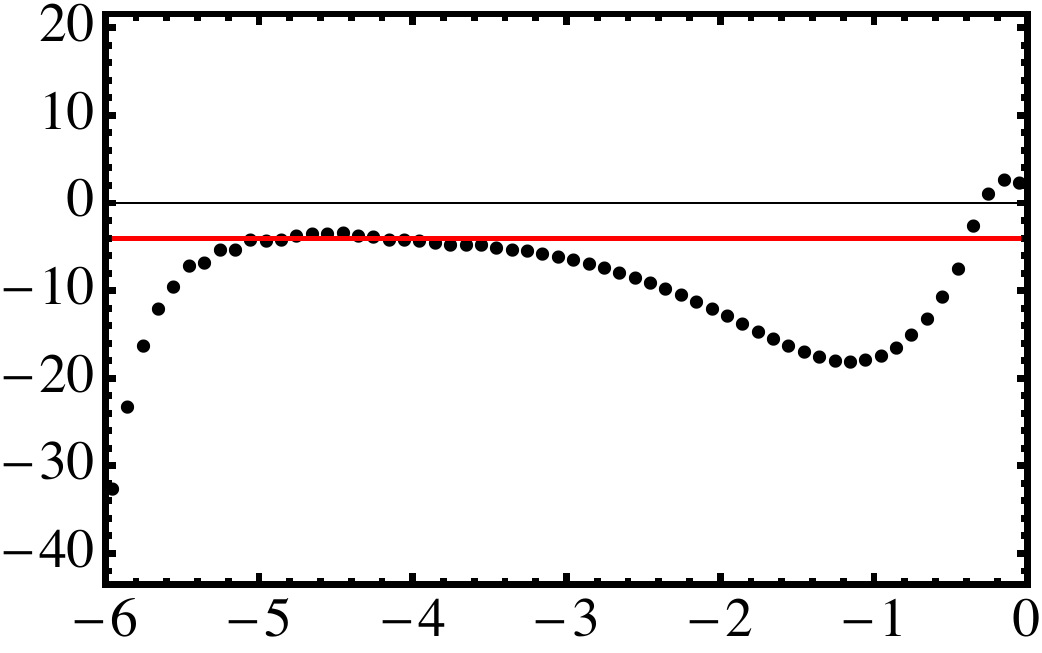}&
\includegraphics[width=.29\columnwidth]{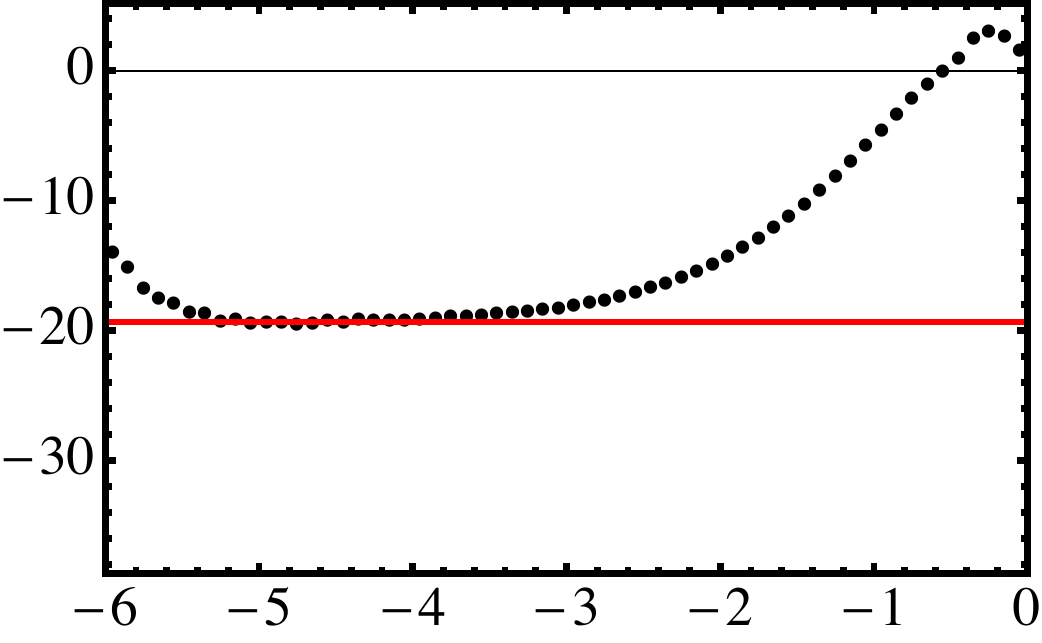}&
\includegraphics[width=.29\columnwidth]{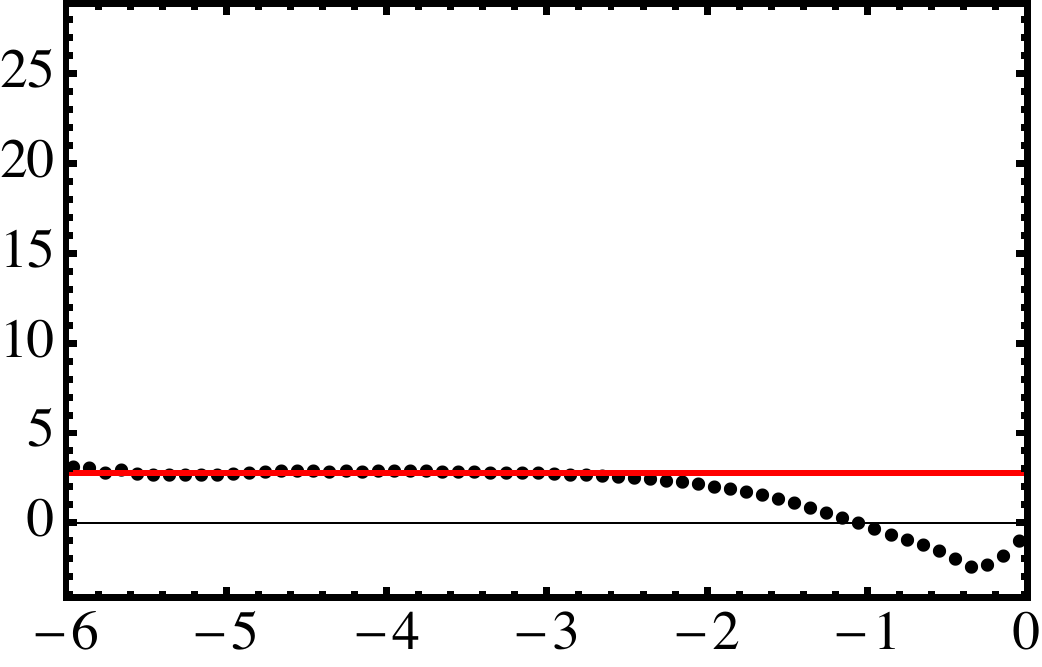} \\[-.5ex]
\rotatebox{90}{\qquad$a=0.25$}&
\includegraphics[width=.29\columnwidth]{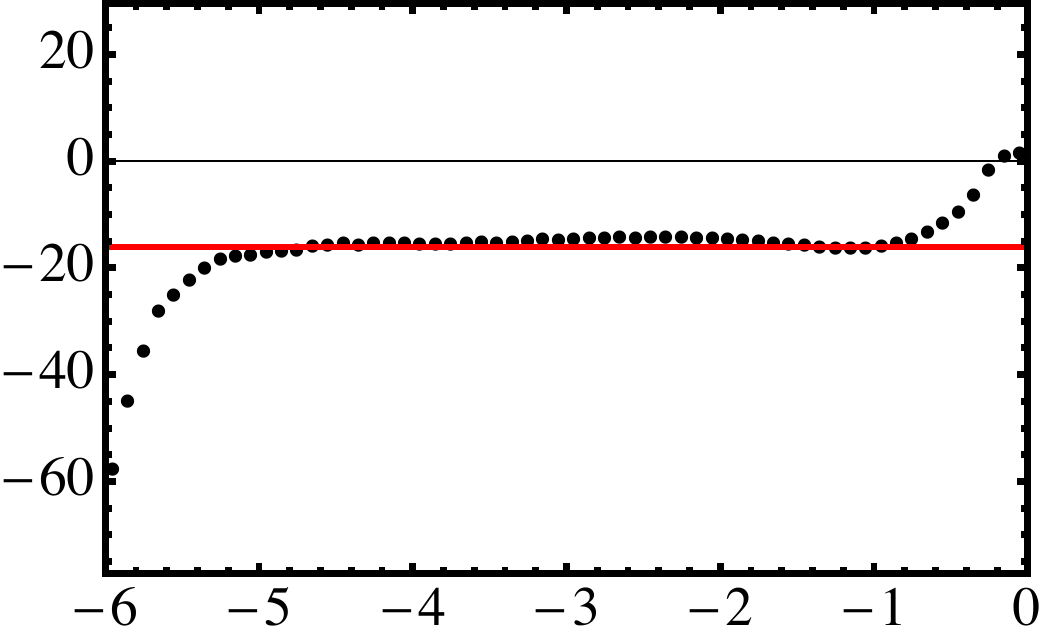}&
\includegraphics[width=.29\columnwidth]{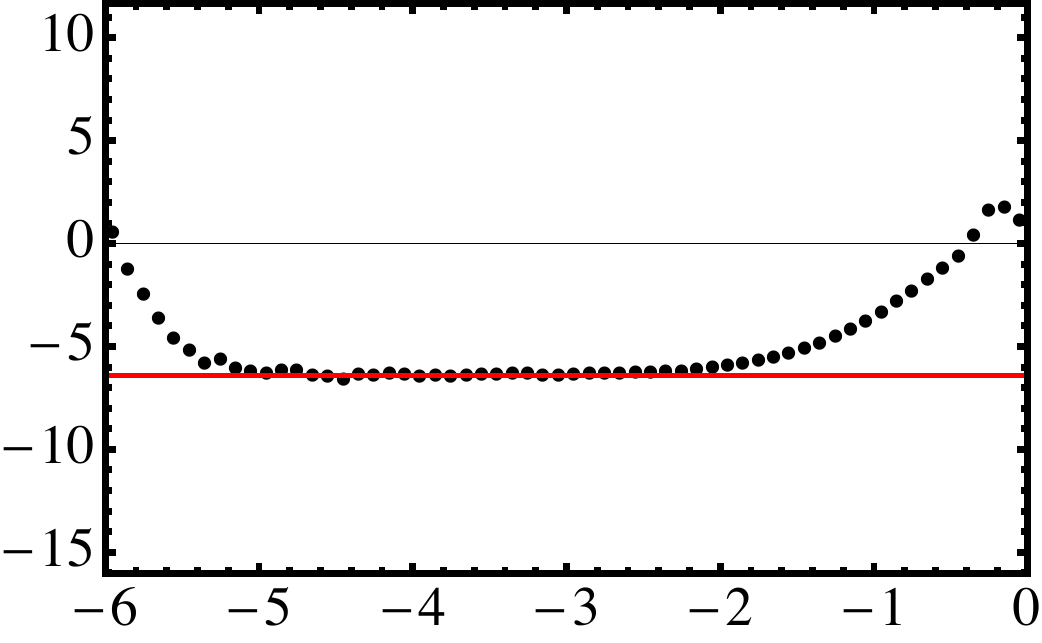}&
\includegraphics[width=.29\columnwidth]{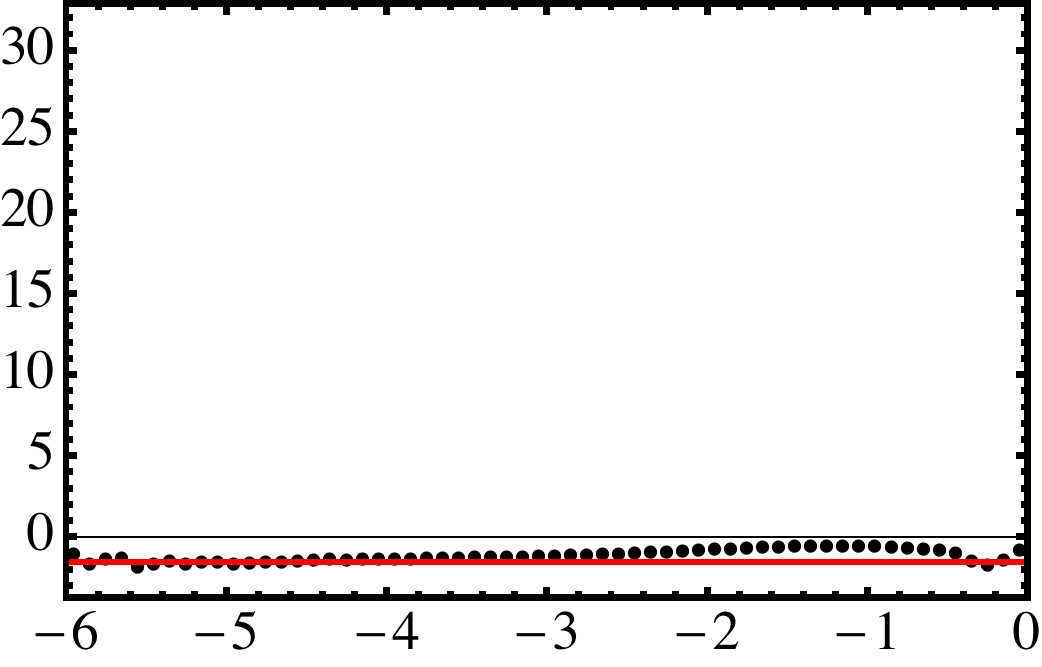} \\[-.5ex]
\rotatebox{90}{\qquad\quad$a=0.5$}&
\includegraphics[width=.29\columnwidth]{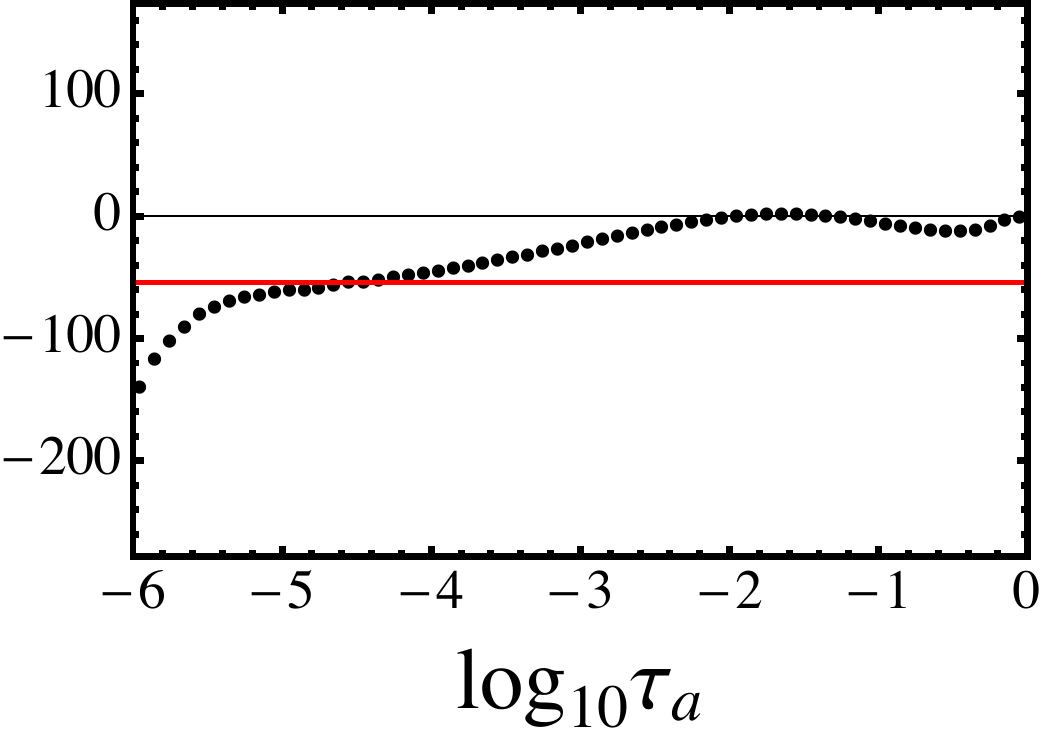}&
\includegraphics[width=.29\columnwidth]{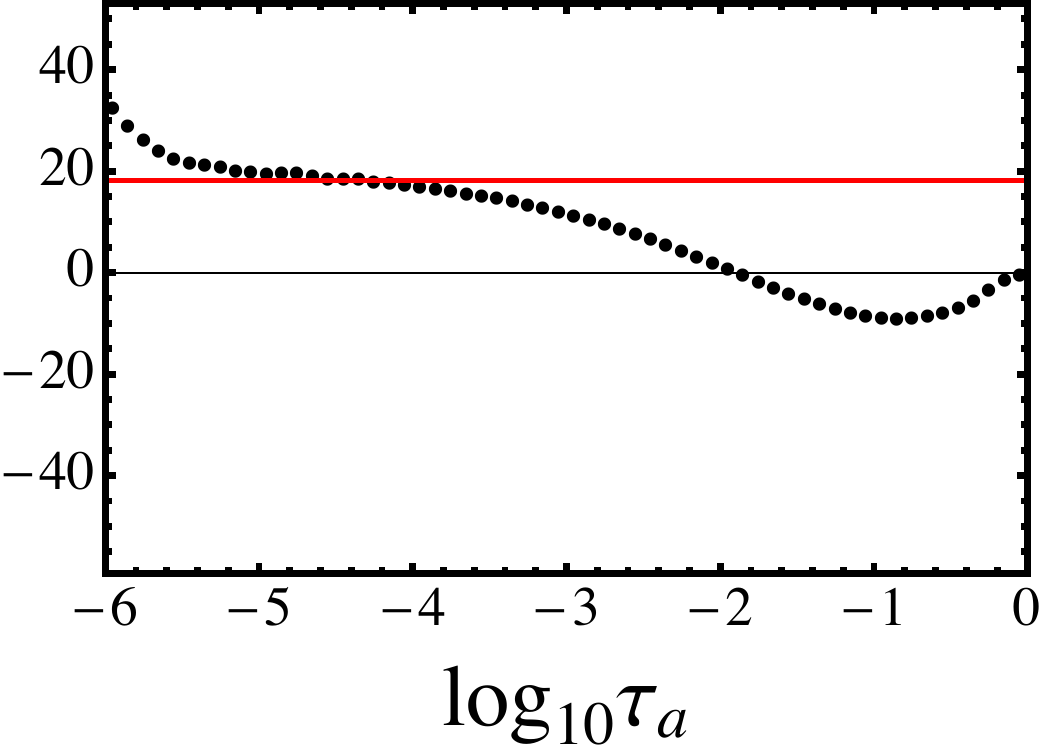}&
\includegraphics[width=.29\columnwidth]{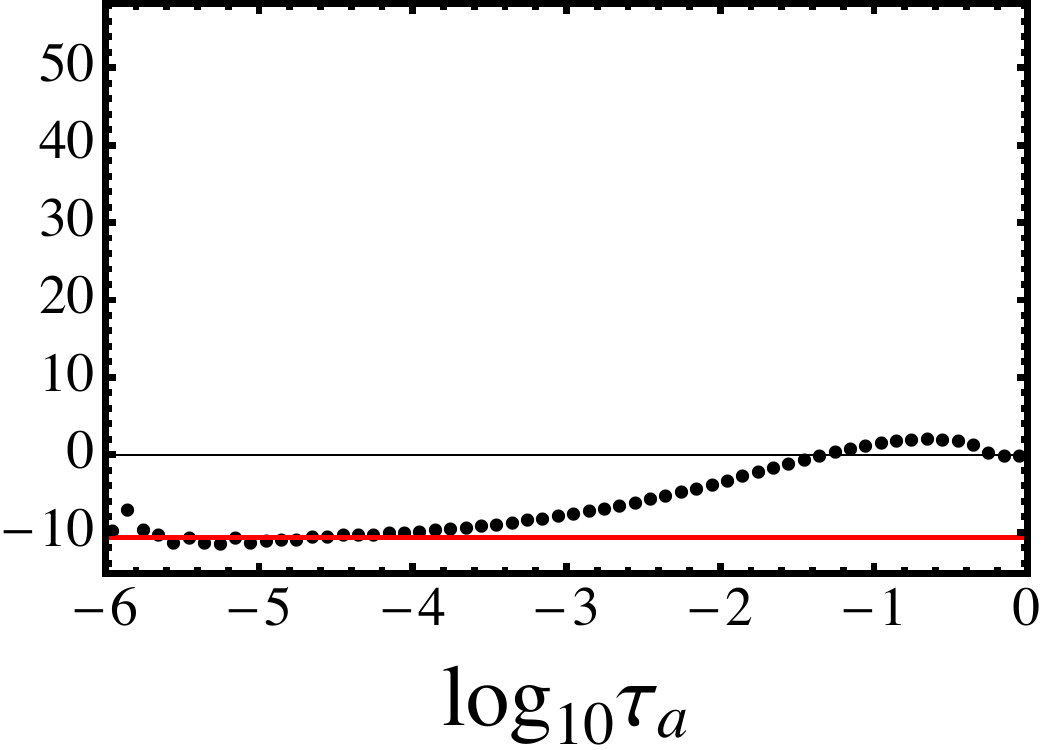}
\end{tabular}
\vspace{-1.5em}
\caption{Integral \eq{remainderintegral} of the $\cO(\as^2)$ remainder function from \event\ as a function of the lower integration limit $\tau_a$. Shown are the coefficients of the
$C_F^2$, $C_FC_A$, $C_FT_F n_f$ colour structures. Our best fit values for the constants 
$r_2^{CF}, r_2^{CA}, r_2^{nf}$ defined in \eq{rc2} are shown in red.
}
\vspace{-4em}
\label{fig:alpha2Fit}
\end{figure}

Our \event\ results for the $\cO(\as^2)$ parts of the differential distributions are displayed 
for $a=-0.5$ and $a=0.5$ in \fig{alpha2EV2}. The functions plotted are the coefficients of the color 
structures $C_F^2$, $C_F C_A$, $C_F T_F n_f$,
and the \event\ output is again shown in comparison to the singular terms predicted by 
\eq{cumulant2loopexpanded}. The agreement looks very good down to small values of $\tau_a$, until one gets so 
low that cutoff effects in \event\ play a role. For these plots we have again set the cutoff parameter 
to $10^{-12}$, and the results are based this time on 
$4\cdot 10^{10}$ events.\footnote{We also have $1\cdot 10^{11}$ events with a cutoff parameter set to $10^{-15}$, which are included 
in our results for the nonsingular remainder function, but which we do not use to extract the 
singular constants in the limit $\tau_a\to 0$ here. This is because the extracted values for the $C_F^2$ 
constant departs substantially, and the $C_F C_A$ constant marginally, from the known values at $a=0$ when 
such a small cutoff parameter is used. We thus stick to the $10^{-12}$ data, where the $a=0$ constants come 
out correct, for extracting the singular constants.
}
The difference between the \event\ output and the singular terms is the nonsingular remainder function $r^2(\tau_a)$, which we will show in \sec{remainder}. Integrating this difference according to \eq{remainderintegral} gives the numerical result for the integral $r_c^2(1)$. The relevant plots for computing these integrals as a sum over \event\ bins are shown in \fig{alpha2Fit}, and the values we extract for $r_c^2(1)$ are listed in \tab{d2}. The tabulated results are the coefficients of each color structure:
\be
\label{eq:rc2}
r_c^2(1) = r_2^{CF} C_F^2 + r_2^{CA} C_F C_A + r_2^{nf} C_F T_F n_f \,.
\ee
The errors we report on the extracted central values are determined by the values of the fitted parameters 
at which the $\chi^2$ per degree of freedom increases by 100\% relative to that at the best fit value over 
the fit range of 
$\log_{10}\!\tau_a \in (-5+4a/10, -4+4a/10)$ for $C_F^2$,
$\log_{10}\!\tau_a \in (-5, -4)$ for $C_F C_A$, and 
$\log_{10}\!\tau_a \in (-4.5, -3.5)$ for $C_F T_F n_f$, which we determined empirically by looking for 
stable plateau regions in \fig{alpha2Fit} from which to perform the fits. These plateaus are very stable 
for the $C_F C_A$ and $C_F T_F n_f$ color structures 
for most values of $a$,
but less so for many of the $C_F^2$ 
results --- stability issues with $C_F^2$ singular terms in \event\ have been encountered in 
prior analyses as well (e.g. \cite{Chien:2010kc,Chien:2015cka}). This is possibly due to an
undersampling of $C_F^2$ contributions in the infrared region.
Also from \fig{alpha2Fit} it can be seen that we did not achieve stable plateaus for $a=0.5$  (except arguably for $C_F T_F n_f$) before hitting the cutoff-affected region. However we went ahead and applied the same fit procedure as described above, and the result in \tab{cJ2} for $a=0.5$ correspondingly has a much larger uncertainty.
Despite these issues with $C_F^2$ and $a=0.5$ contributions, for our present analysis we were nevertheless able to obtain sufficiently reliable results  
for the final cross sections we present in \sec{ANGRESULTS}.
Namely, even doubling the uncertainties on $c_{\tilde J}^2$ shown in \tab{cJ2}, the total NNLL$'$ uncertainty bands on the cross sections in \sec{ANGRESULTS} are unchanged for all $a\leq 0.25$, and only change a few percent for $a=0.5$ in the resummation region, with the methods for uncertainty estimation described in \sec{scalevariations}. 
Of course, when other theoretical uncertainties are pushed sufficiently low to make the uncertainties on $r_2^{CF}$ and for $a=0.5$ more prominent (and before making a robust uncertainty estimate on any definitive extraction of $\as$), one should revisit the methods used to compute these numbers.

%______________
\begin{table}[t]
\centering
\small
\begin{tabular}{|c||c|c|c|c|c|c|c|}
\hline 
$a$ & $-1.0$ & $-0.75$ & $-0.5$ & $-0.25$ & 0.0 & 0.25 & 0.5 
\\ \hline \hline
$r_2^{CF}$ & 
$-0.16_{-0.28}^{+0.37} $ & 
$ 0.19_{-0.24}^{+0.34} $ & 
$ 0.24_{-0.18}^{+0.35} $ & 
$ -0.66_{-0.29}^{+0.36} $ & 
$ -4.03_{-0.27}^{+0.38} $ & 
$ -15.9_{-0.7}^{+0.4} $ & 
$ -49.9_{-8.4}^{+3.2}$\\
\hline
$r_2^{CA}$ & 
$-42.3_{-0.5}^{+0.2} $ & 
$ -38.0_{-0.5}^{+0.2} $ & 
$ -33.2_{-0.3}^{+0.1}$ & 
$ -27.3_{-0.2}^{+0.1} $ & 
$ -19.3_{-0.2}^{+0.1} $ & 
$ -6.42_{-0.11}^{+0.20} $ & 
$ 18.1_{-0.5}^{+1.5}$\\
 \hline
$r_2^{nf}$ & 
$6.76_{-0.03}^{+0.08} $ & 
$ 6.57_{-0.03}^{+0.08} $ & 
$ 6.03_{-0.03}^{+0.07} $ & 
$ 4.92_{-0.02}^{+0.06} $ & 
$ 2.78_{-0.02}^{+0.03} $ & 
$ -1.42_{-0.06}^{+0.02} $ & 
$  -9.92_{-0.87}^{+0.23} $ \\
 \hline
\end{tabular}
\caption{
Fit values for the coefficients of the integral $r_c^2(1)$ of the nonsingular QCD distribution 
as defined in \eq{rc2}. The central values and their uncertainties have been extracted from the plots in \fig{alpha2Fit} as described in the text.
}
\label{tab:d2}
\end{table}

%______________
\begin{table}[t]
\vspace{1em}
\begin{center}
\begin{tabular}{|c||c|c|c|c|c|c|c|}
\hline 
$a$ & $-1.0$ & $-0.75$ & $-0.5$ & $-0.25$ & 0.0 & 0.25 & 0.5 
\\ \hline \hline
$c_{\tilde J}^2$ & 
$66.0_{-3.4}^{+5.2}$ & 
$42.3_{-3.3}^{+5.1}$ & 
$17.3_{-2.5}^{+3.2}$ & 
$-9.34_{-2.48}^{+2.76}$ & 
$-36.3_{-2.4}^{+2.7}$ & 
$-57.6_{-3.2}^{+3.8}$ & 
$-79.8_{-24.9}^{+39.7}$ \\
 \hline
\end{tabular}
\end{center}
\vspace{-1em}
\caption{Extracted values of the two-loop jet function constants $c_{\tilde J}^2$, determined by \eq{c2momentum} and the value for $c^{(2)}$ implied by \eq{c2fromr2} and the numerical results for $r_c^{2}(1)$ in \tab{d2}.}
\label{tab:cJ2}
\end{table}
%______________

The constant part of the singular $\cO(\as^2)$ cross section is finally given by \eq{totalQCD},
\be
\label{eq:c2fromr2}
c^{(2)} \equiv c_{20} = \sigma^{(2)}_{\text{tot}} - r_c^2(1)\,,
\ee
which we can plug into \eq{c2momentum} to obtain the so-far unknown two-loop jet function
constant $c_{\tilde J}^2$. 
Our results are shown in \tab{cJ2}, where we have set 
$n_f=5$ and we have added the uncertainties of the
individual $r_c^2(1)$ coefficients in \tab{d2} linearly. In our phenomenological
analysis below, we will vary $c_{\tilde J}^2$ over the uncertainties shown in \tab{cJ2}, and we will 
account for its contribution to the overall uncertainties presented in  
\sec{ANGRESULTS}. 
This particular contribution to the total uncertainty turns out, however, to be almost negligible. Our 
fit result for $a=0$ agrees well with the analytically known result from \cite{Becher:2006qw}, 
$c_{\tilde J}^2=-36.3$.

%%%%%%%%%%%%%%%%%%%%%%%%%%%%%%%%%%%%%%%%%%%%%%%%%%%%%%%
\subsection{Remainder functions}
\label{sec:remainder}

%----------------------------------------------------------------------------
\begin{figure}[t]
\centering
\includegraphics[width=.46\columnwidth]{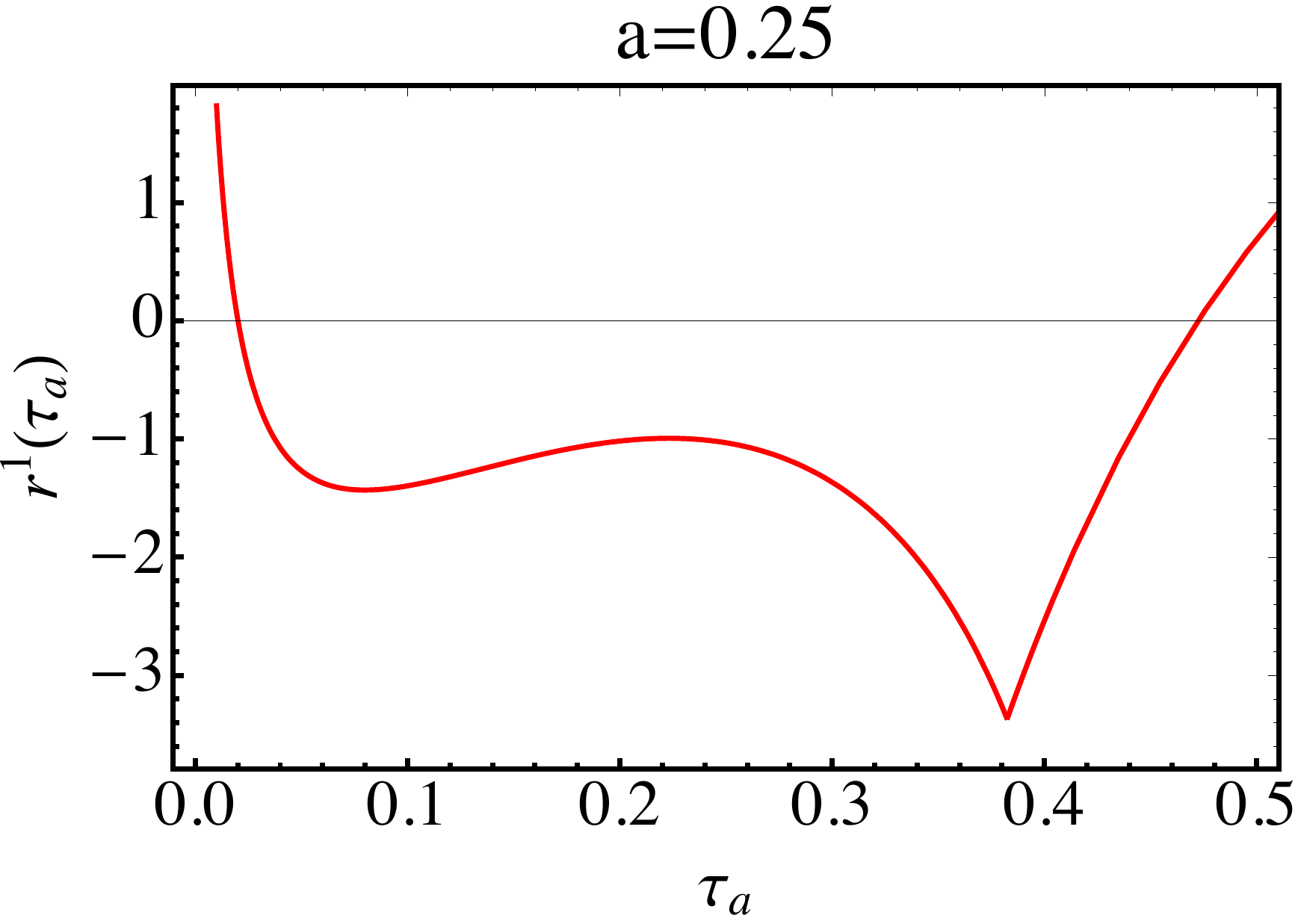} \quad
\includegraphics[width=.48\columnwidth]{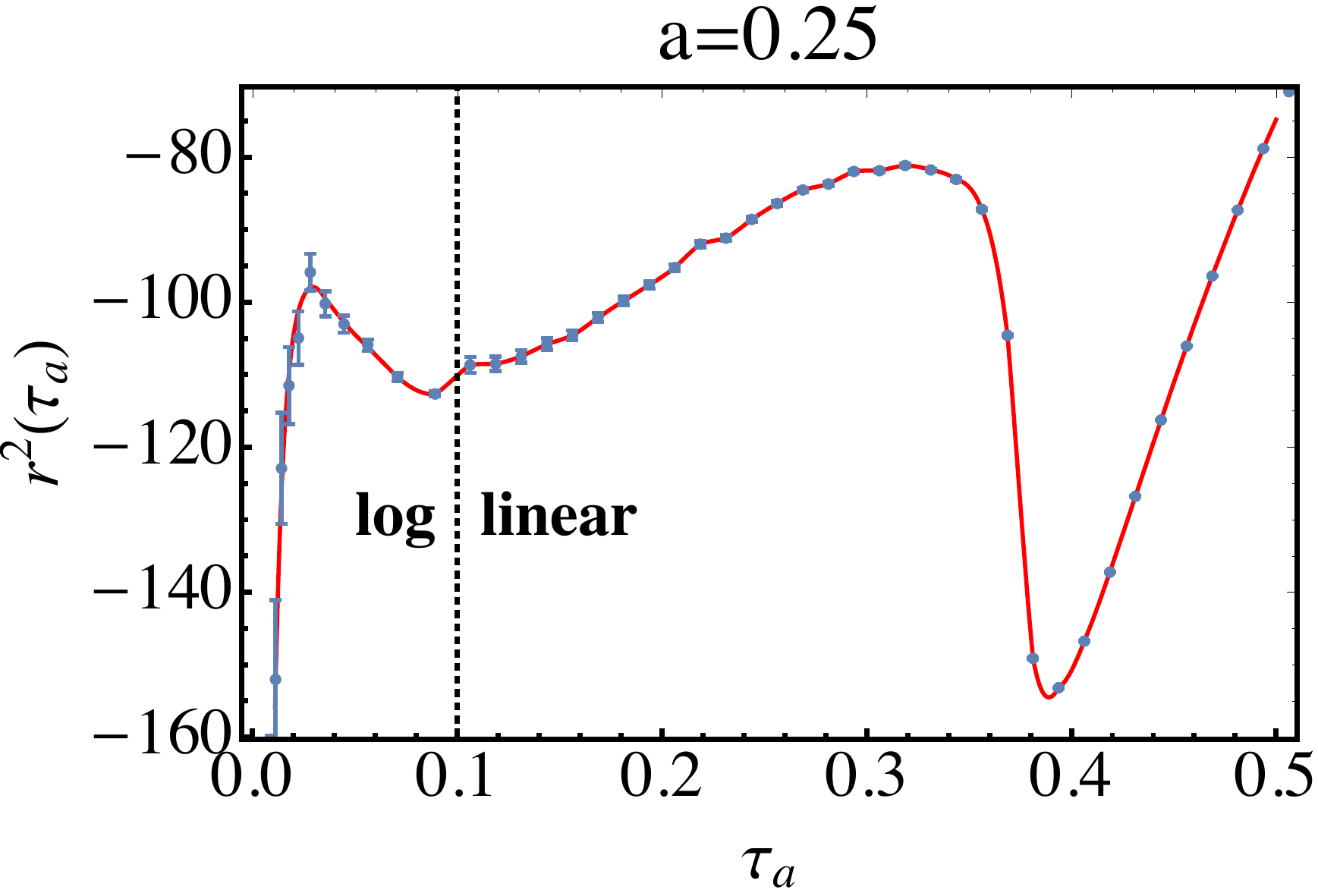}
\vspace{-1em}
\caption{
\emph{Left:} $\cO(\as)$ remainder function for $a=0.25$, which is known from \cite{Hornig:2009vb}.
\emph{Right:} The corresponding $\cO(\as^2)$ remainder function from \event\ (data points)
with a suitable interpolation as described in the text.
}
\label{fig:r2}
\end{figure}
%-----------------------------------------------------------------------------

We next move on to the integrable remainder function $r^2(\tau_a)$ in \eq{remainder}, 
and consequently its nonsingular integrated version $r_c^2(\tau_a)$ in \eq{QCDexpansion}. A prior issue to sort out 
in this context
is the kinematic endpoint of the $\tau_a$ distribution in full QCD, which the singular 
predictions from SCET do not ``see'' since they assume $\tau_a\ll \tau_a^\text{max}\sim 1$.

The full QCD distribution $d\sigma/d\tau_a$ in \eq{differentialdistribution} vanishes at 
$\mathcal{O}(\as^k)$ above the maximum kinematic value of $\tau_a$ for $(k+2)$ particles in the final state, 
which we will call $\tau_a^{k,\text{max}}$. The SCET distribution 
$d\sigma_\text{sing}/d\tau_a$, however, continues above $\tau_a>\tau_a^{k,\text{max}}$
at this order. The remainder distribution $r(\tau_a)$ in 
\eq{remainder} is therefore just the negative of the SCET distribution above 
$\tau_a^{k,\text{max}}$:
\begin{equation}
\label{eq:REMAINDERfunc}
r^k(\tau_a) = 
\theta(\tau_a^{k,\text{max}} - \tau_a) \,r^k(\tau_a) - 
\theta(\tau_a - \tau_a^{k,\text{max}})\,B^k(\tau_a)\,,
\end{equation}
where $B^k$ is the $\mathcal{O}(\as^k)$ coefficient of the singular distribution $B(\tau_a)$ in 
\eq{differentialdistribution} (in units of $(\as(Q)/(2\pi))^k$), and we have slightly abused notation in 
using the same symbol $r^k$ for the remainder function for $\tau_a<\tau_a^{k,\text{max}}$. 

At $\mathcal{O}(\as)$ there are up to three particles in the final state. The maximum value of $\tau_a$ occurs 
for the symmetric ``Mercedes-star'' configuration with three particles in a plane, with equal energy 
$E_{1,2,3} = Q/3$ and an angle $\theta_{ij} = 2\pi/3$ between any two of them (for the values of $a$ we consider, but see \cite{Hornig:2009vb} for exceptions). The thrust axis may then be 
taken to be along any of the three particles, and one easily derives 
$\tau_a^{1,\text{max}} = (1/3)^{1-a/2}$.

At $\mathcal{O}(\as^2)$ there are up to four particles in the final state, and it is similarly 
straightforward to compute the angularity $\tau_a$ for the symmetric four-particle configuration, 
which we derive in \appx{tetra}. Assuming that the symmetric configuration again determines 
the maximum for the values of $a$ relevant for our analysis, we obtain
\be
\tau_a^{2,\text{max}} 
= \Bigl(1 - \frac{\sqrt{3}}{3}\Bigr)^{1-a} \Bigl(\frac{2}{3}\Bigr)^{a/2}\,.
\ee

\begin{table}[t]
\begin{center}
\begin{tabular}{|c||c|c|c|c|c|c|c|}
\hline 
$a$ & $-1$ & $-0.75$ & $-0.5$ & $-0.25$ & 0 & 0.25 & 0.5 \\ \hline \hline
$a_1$ & $6212$ & $5401$ & $4328$ & $3526$ & $1475$ & $-209$ & $22756$  \\ \hline
$a_2$ & 1377 & 1210 &  986 & 803 & 343 & 50.8 & 6262  \\ \hline
$a_3$ & 94.4 & 82.5 &  67.3 & 54.2 & 23.5 & 14.0 & 512  \\ \hline
$b_2$ & $-1680$ & $-1139$ &  $-518$ & $-491$ & $-25.4$ & 793 & 4440  \\ \hline
$b_3$ & $-2455$ & $-2098$ &  $-1619$ & $-1317$ & $-465$ & 423 & $-7037$  \\ \hline\hline
$a_0$ & 5634 & 4694 &  3516 & 2884 & 1083 & $-598$ & 14658  \\ \hline
\end{tabular}
\end{center}
\caption{Results for the fit coefficients in \eq{logfit} obtained from the \event\ data displayed in \fig{r2all}. 
The uncertainties on these fits are illustrated in the plots on the right panel of \fig{r2all}.
}
\label{tab:fit}
\end{table}

%----------------------------------------------------------------------------
\begin{figure}[t]
\centering
\begin{tabular}{crrl}
\hspace{-1em}\rotatebox{90}{\qquad$a=-1$} \hspace{-1em} &
\includegraphics[width=.285\columnwidth]{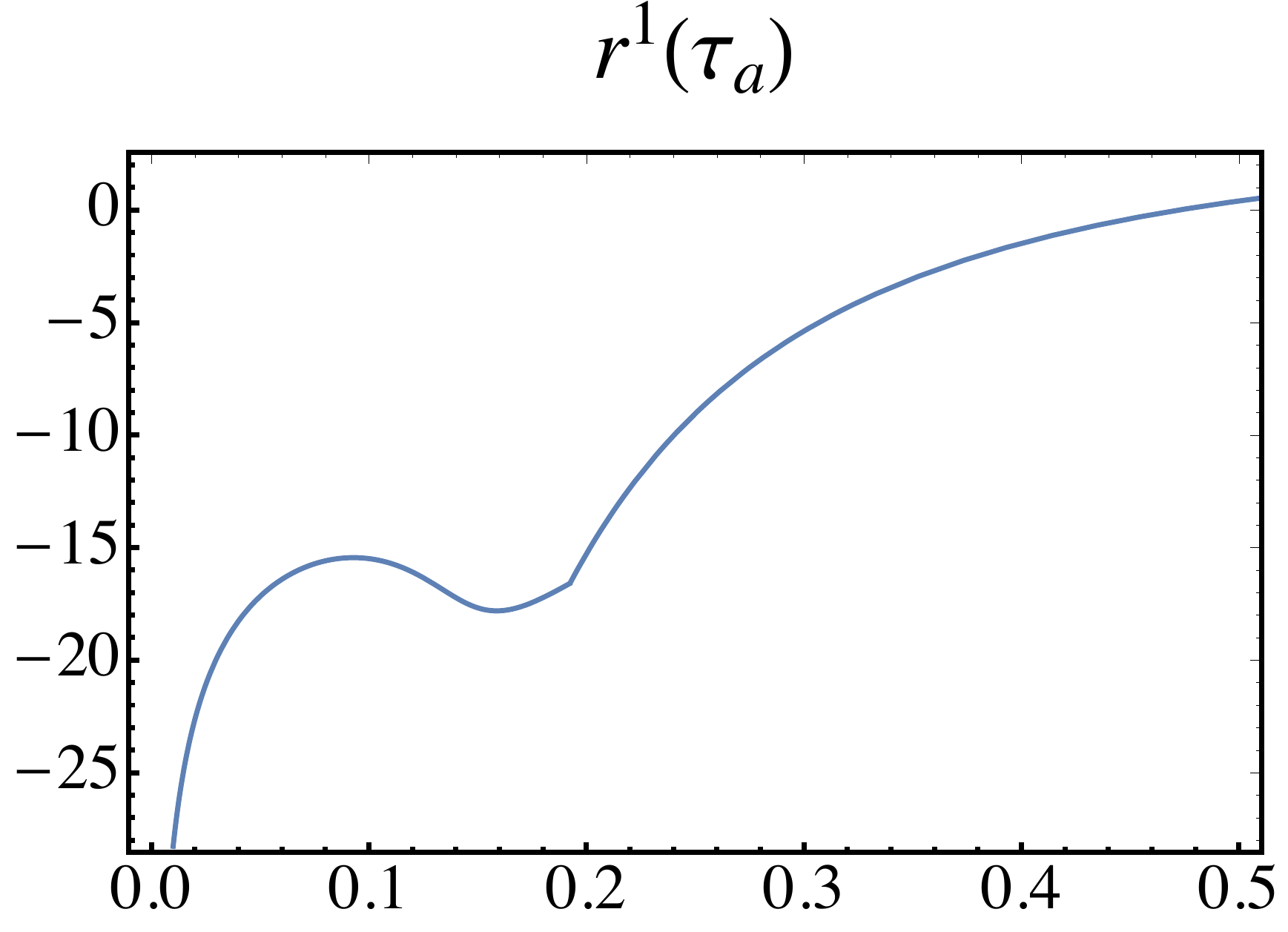}&
\includegraphics[width=.3\columnwidth]{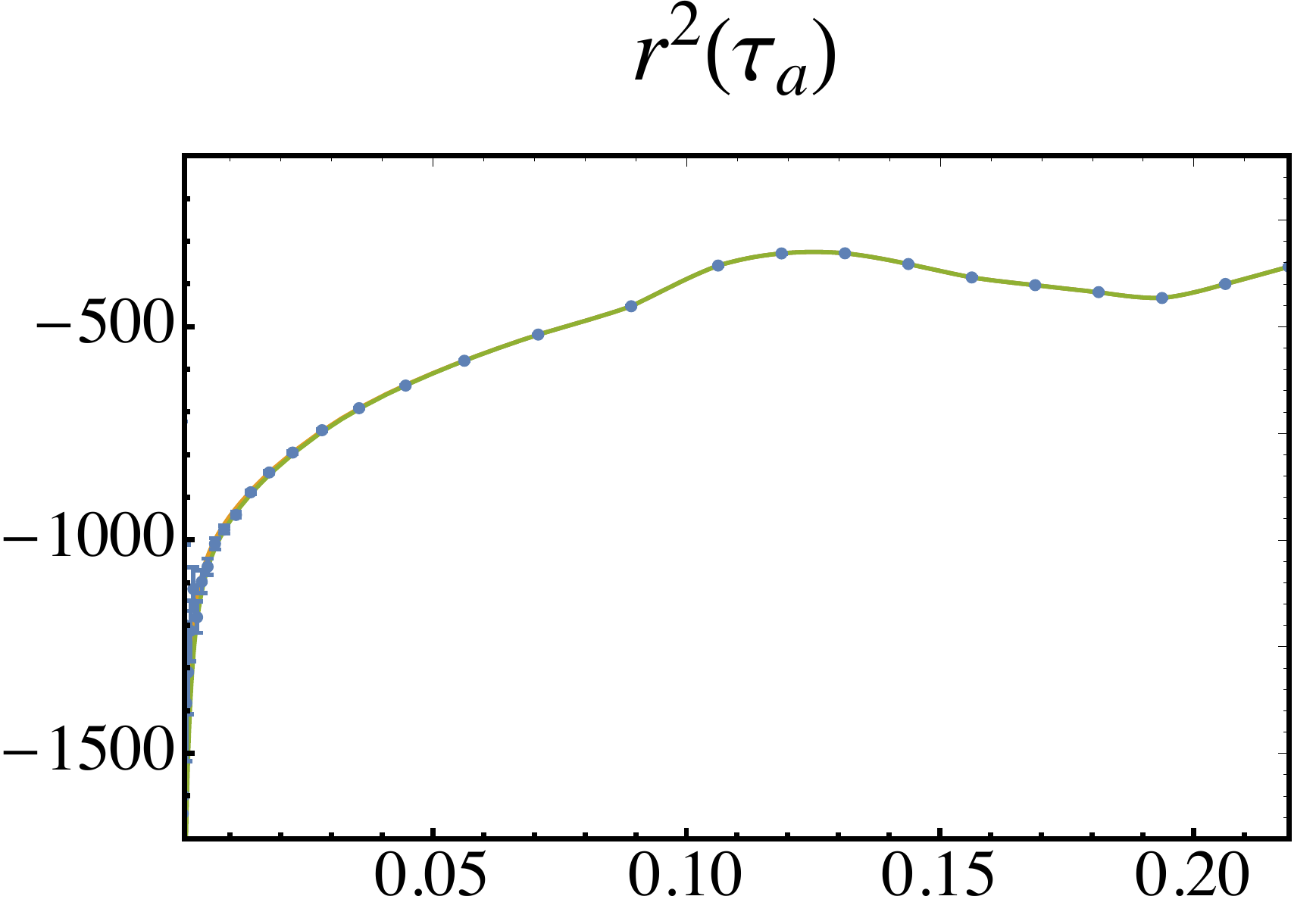}&
\includegraphics[width=.31\columnwidth]{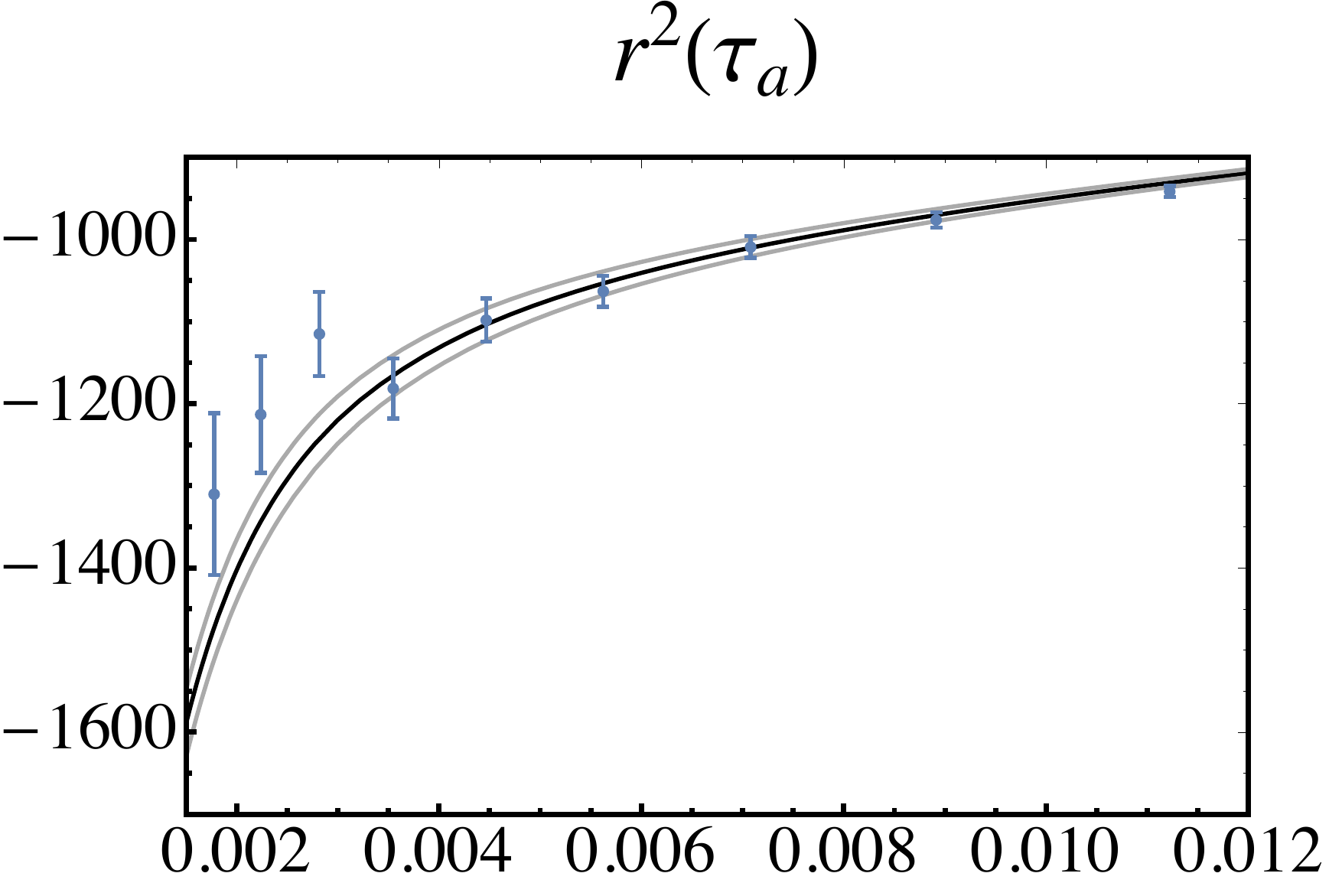} \\
\hspace{-1em}\rotatebox{90}{\quad$a=-0.75$}  &
\includegraphics[width=.285\columnwidth]{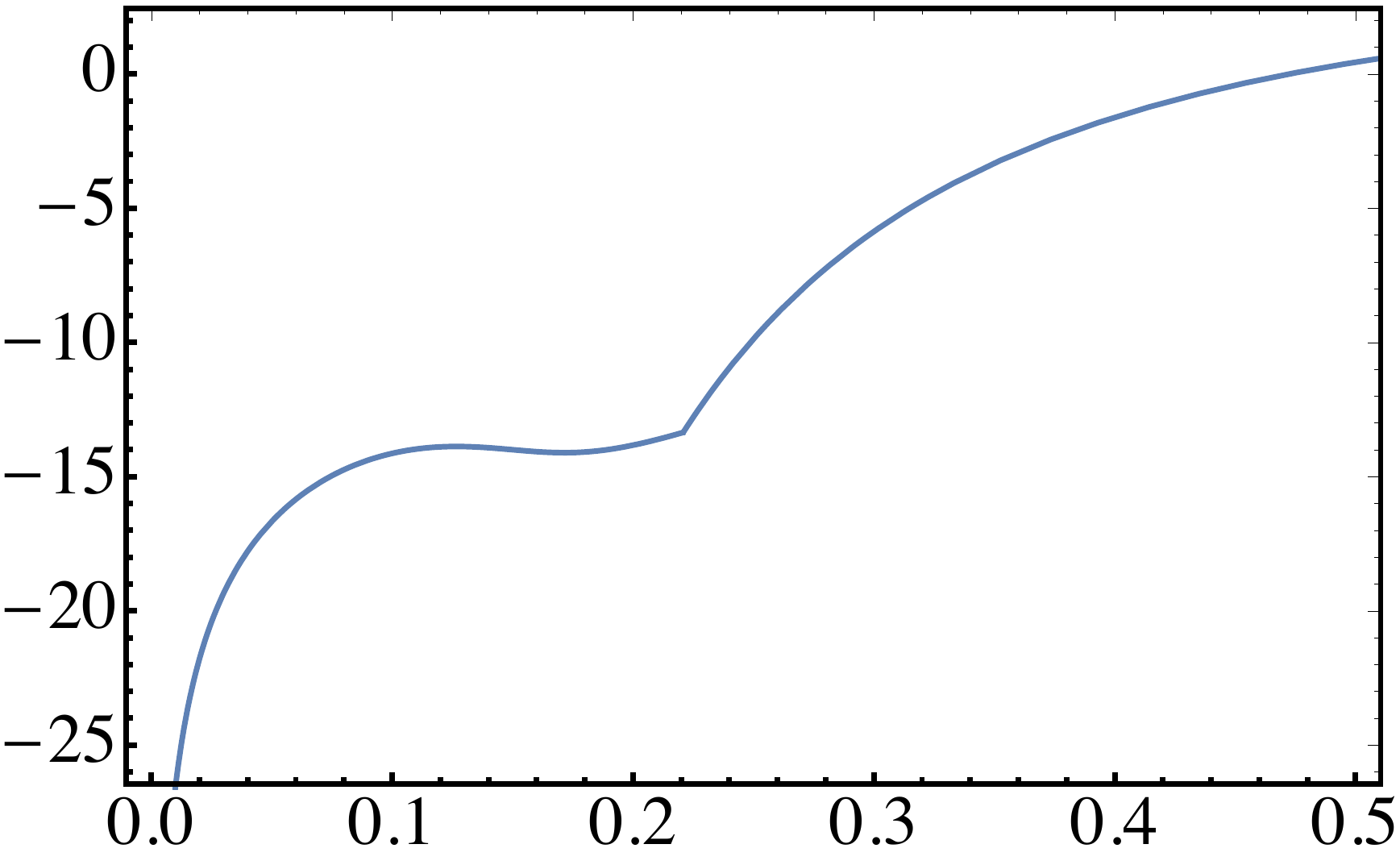}&
\includegraphics[width=.3\columnwidth]{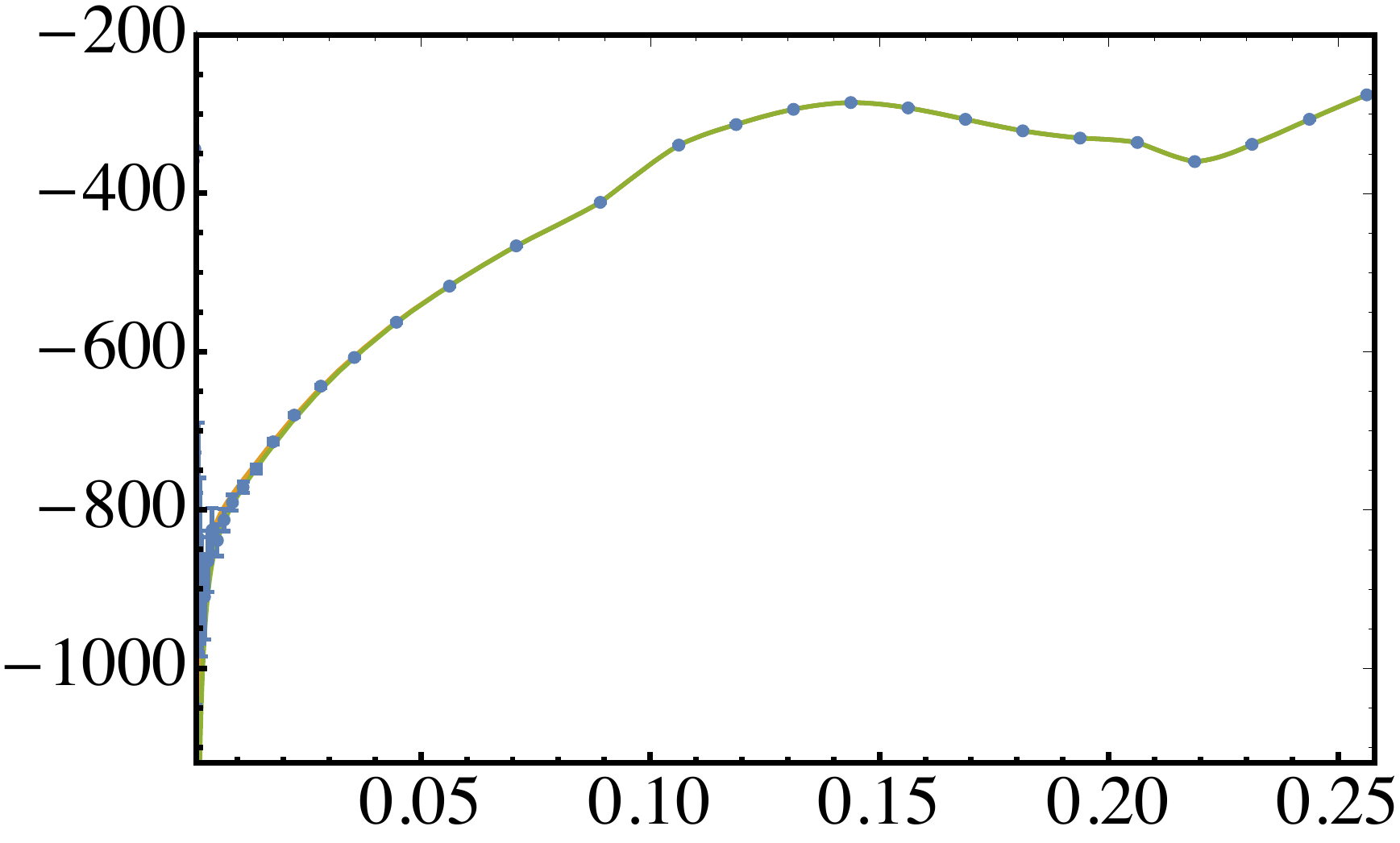}&
\includegraphics[width=.295\columnwidth]{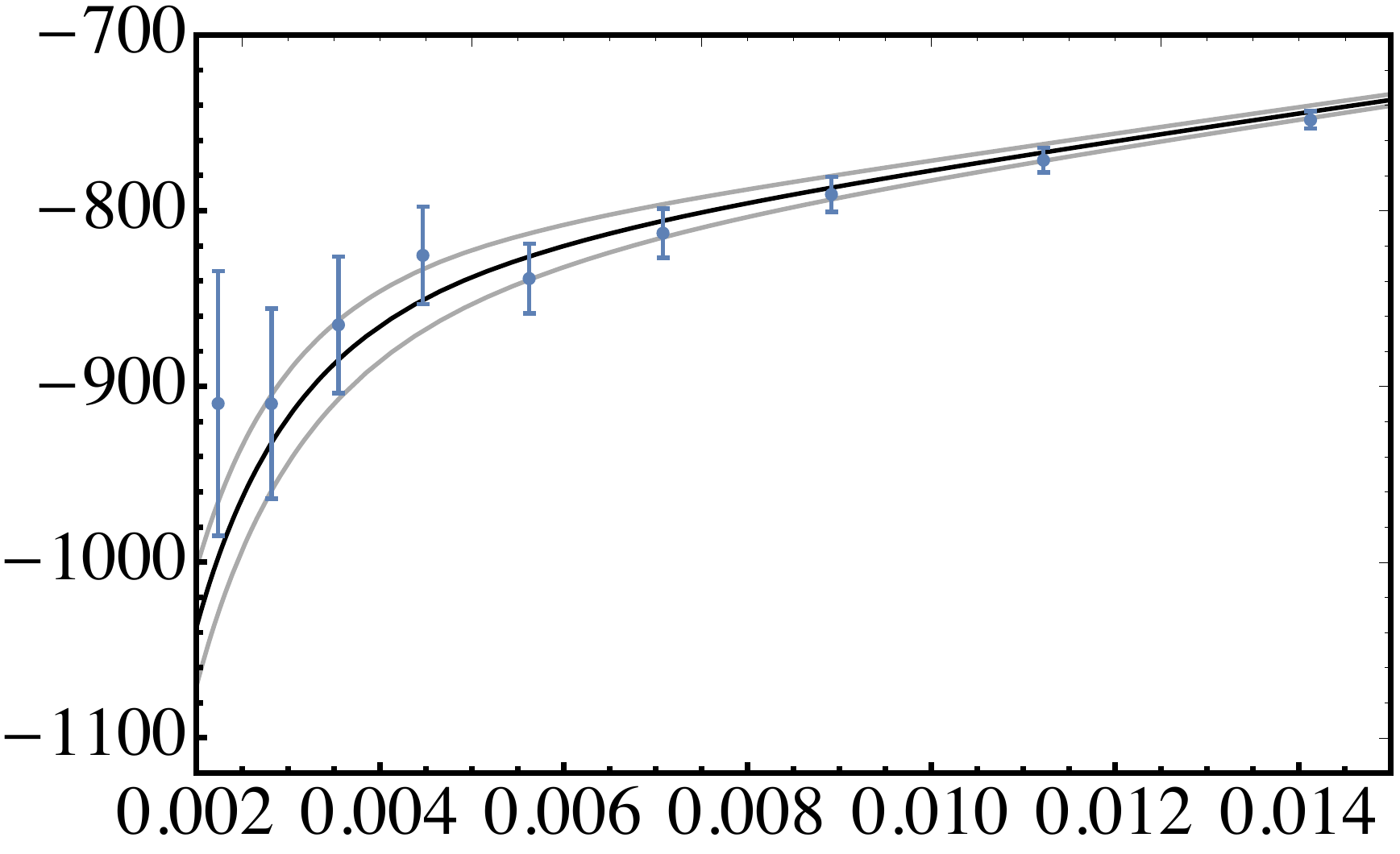}  \\
\hspace{-1em}\rotatebox{90}{\qquad$a=-0.5$}  \hspace{-1em} &
\includegraphics[width=.285\columnwidth]{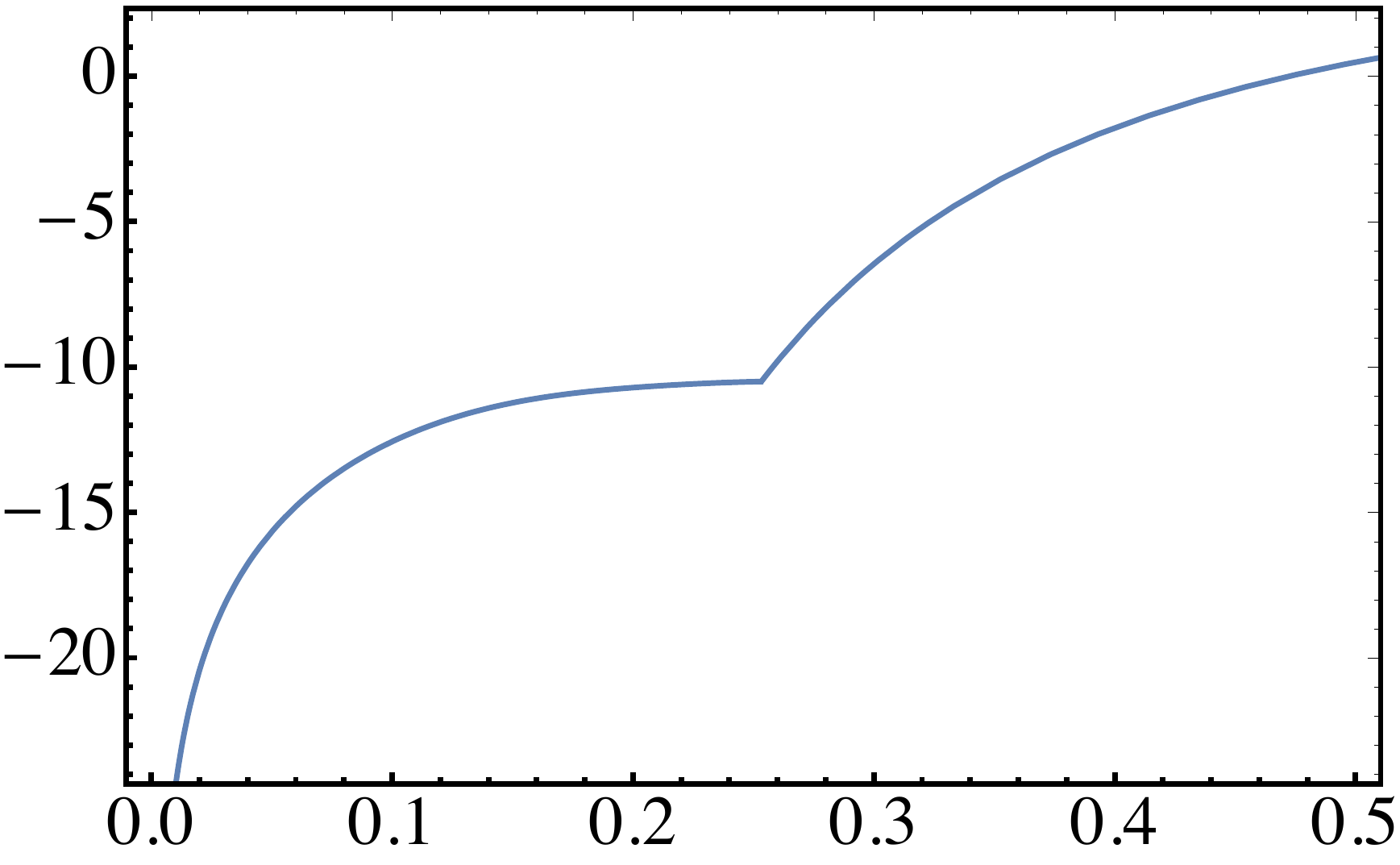}&
\includegraphics[width=.3\columnwidth]{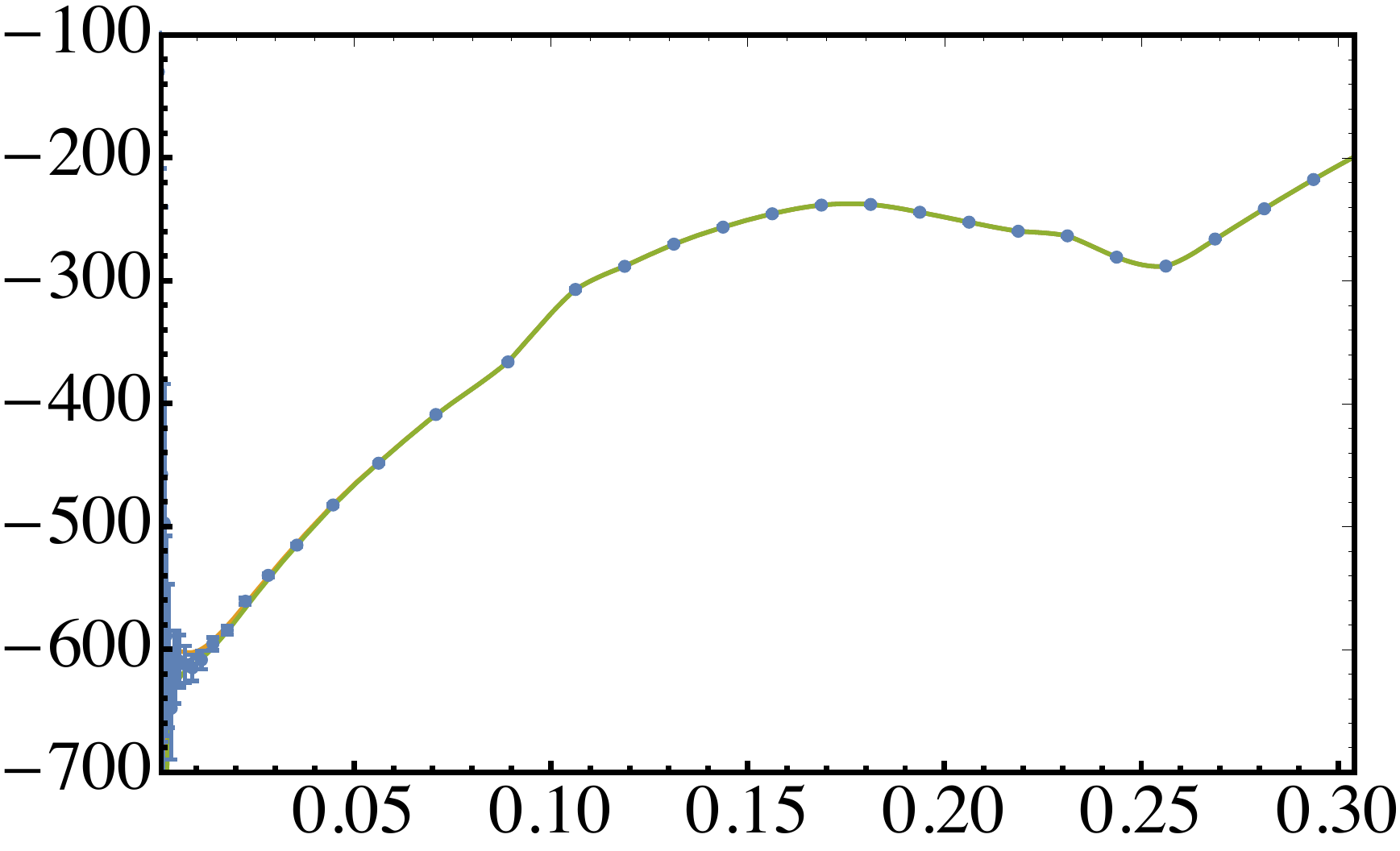}&
\includegraphics[width=.305\columnwidth]{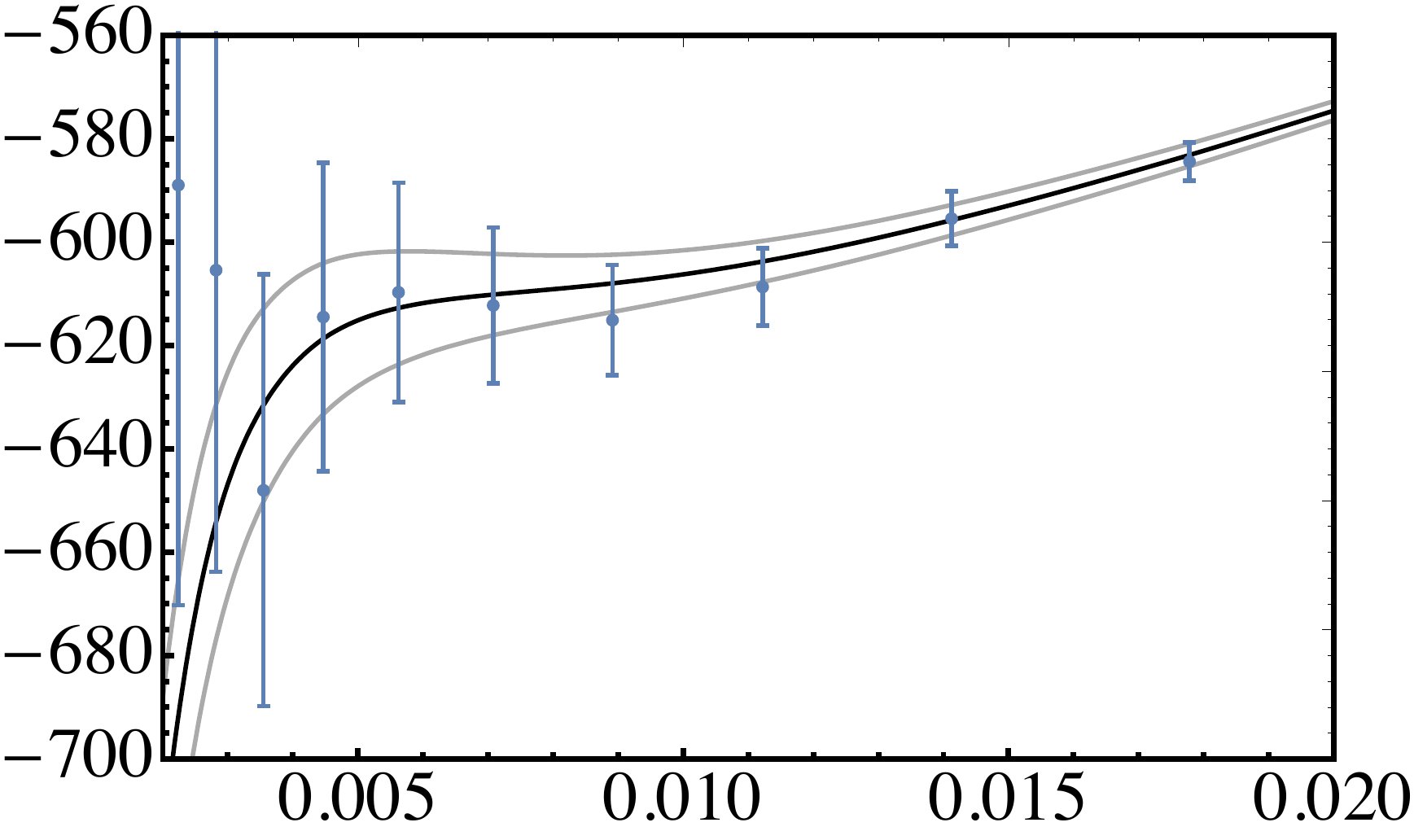} \\
\hspace{-1em}\rotatebox{90}{\quad$a=-0.25$}  \hspace{-1em} &
\includegraphics[width=.285\columnwidth]{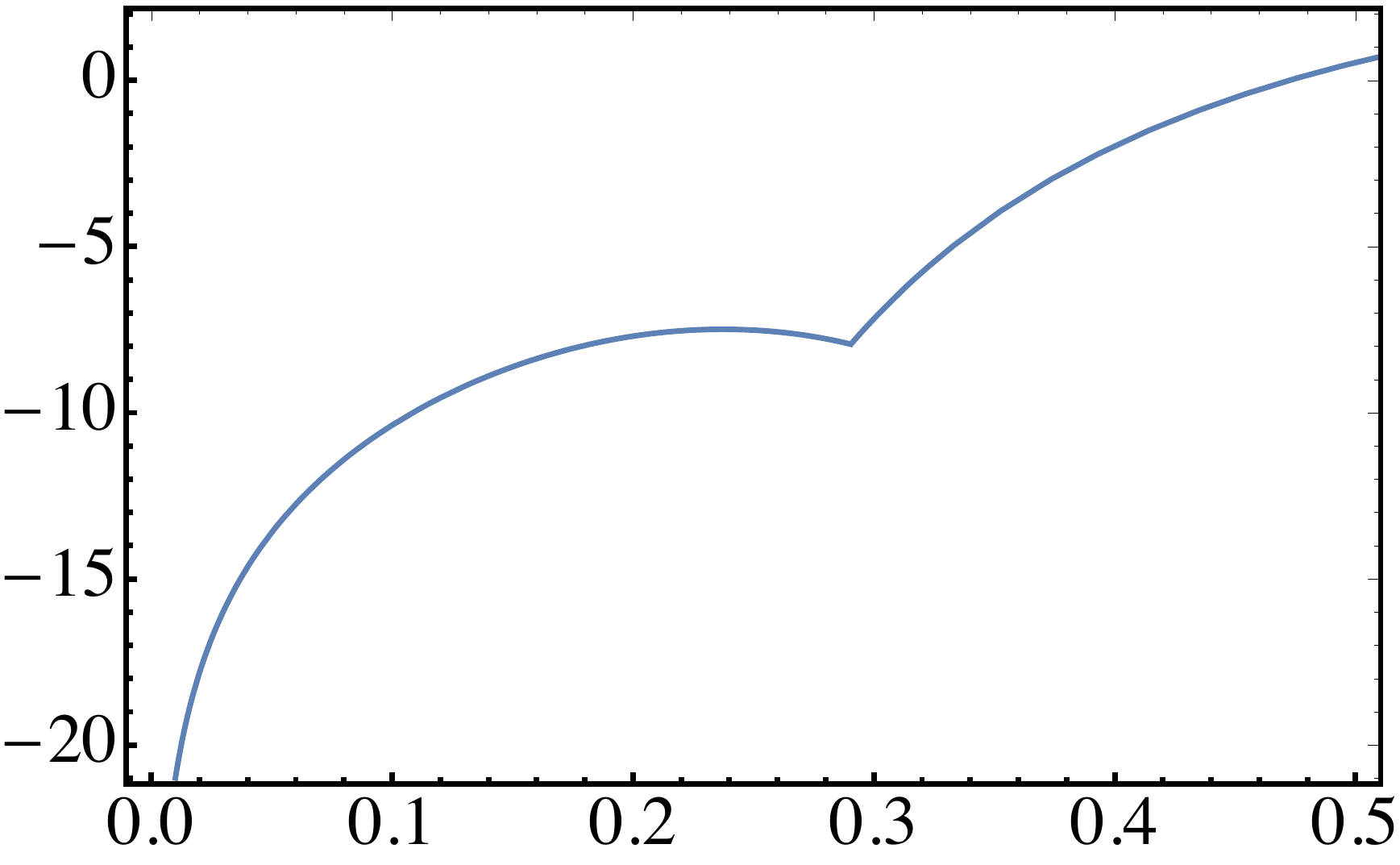}&
\includegraphics[width=.295\columnwidth]{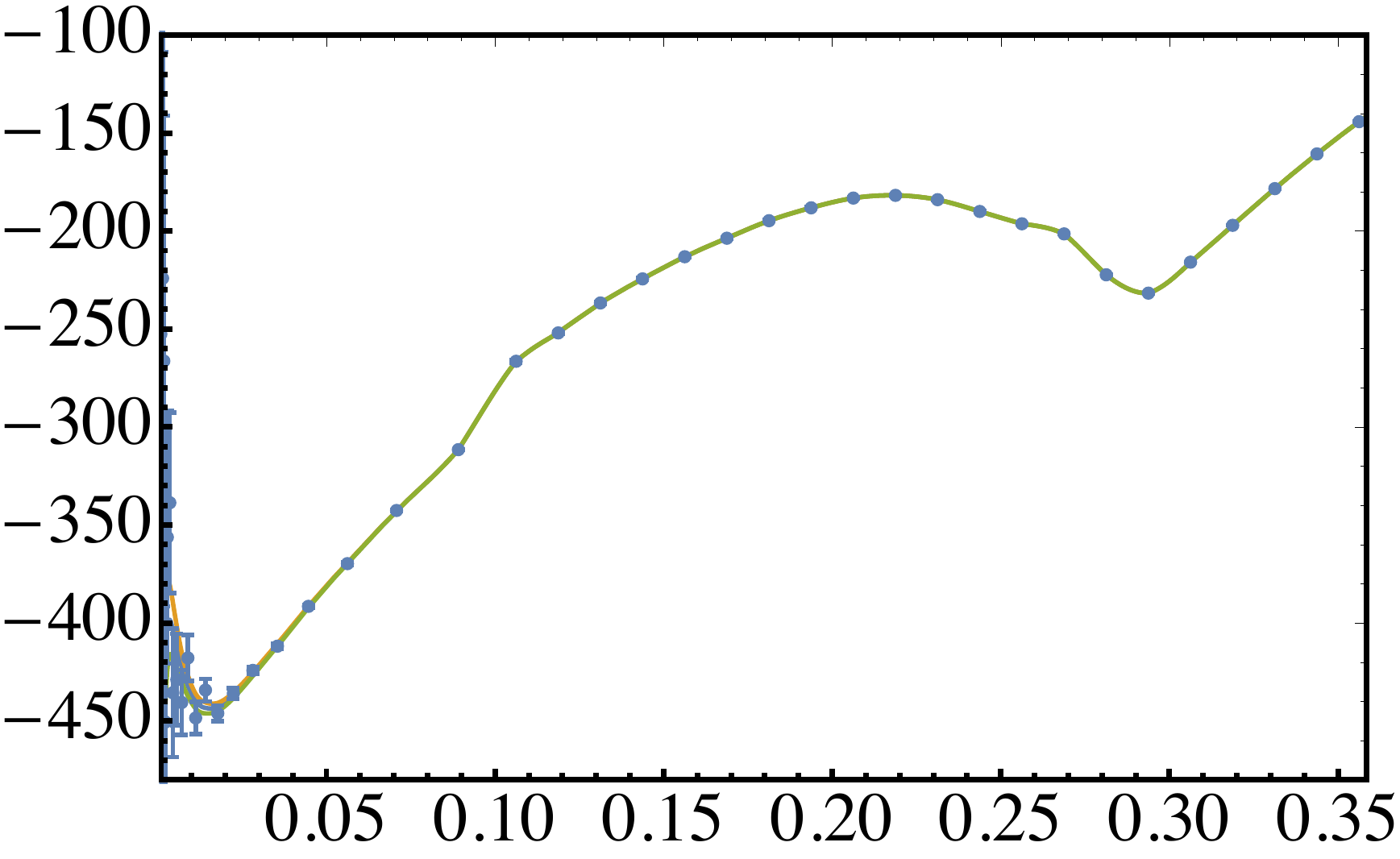}&
\includegraphics[width=.305\columnwidth]{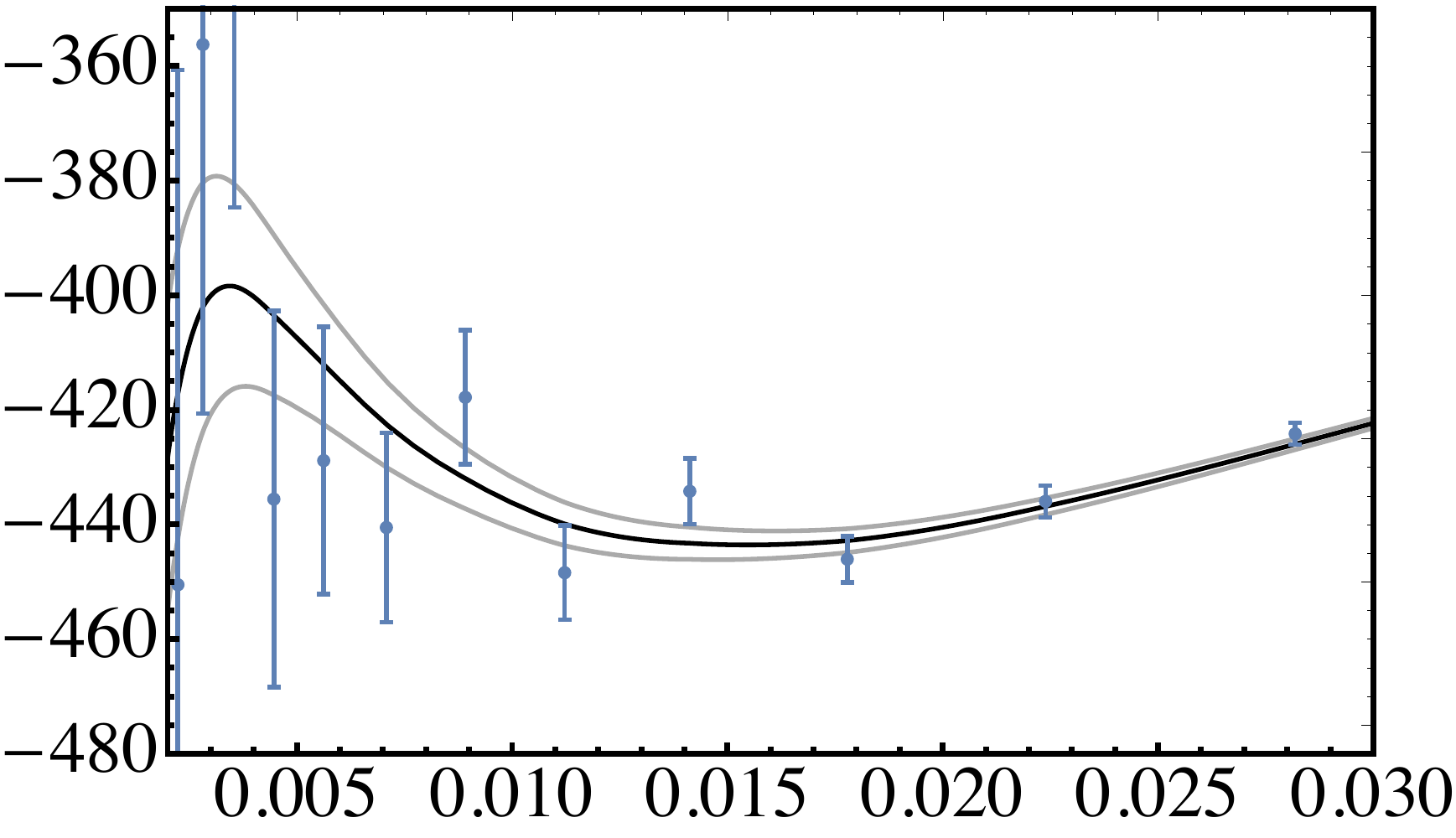} \\
\hspace{-1em}\rotatebox{90}{\qquad\quad$a=0$}  \hspace{-1em} &
\includegraphics[width=.285\columnwidth]{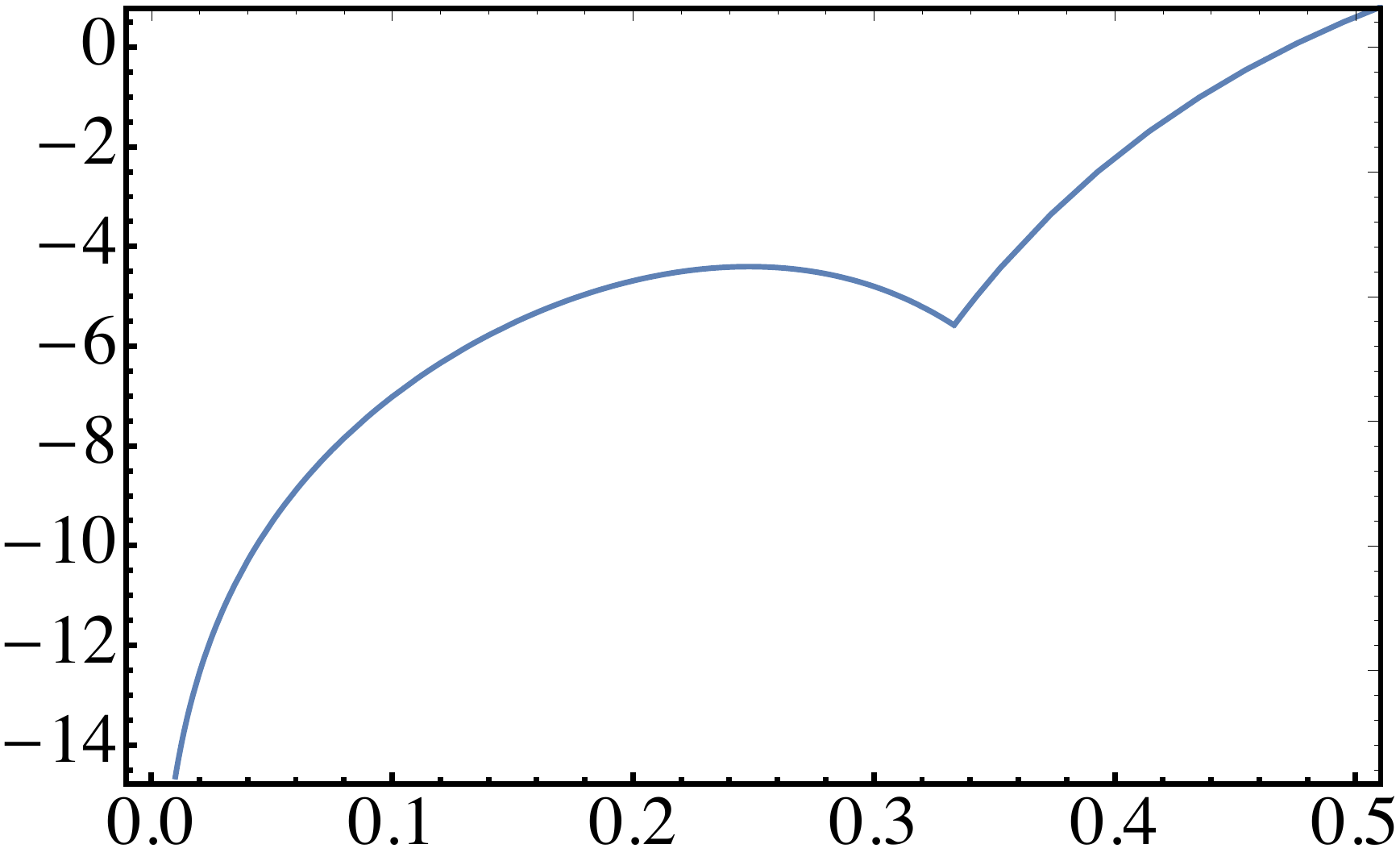}&
\includegraphics[width=.295\columnwidth]{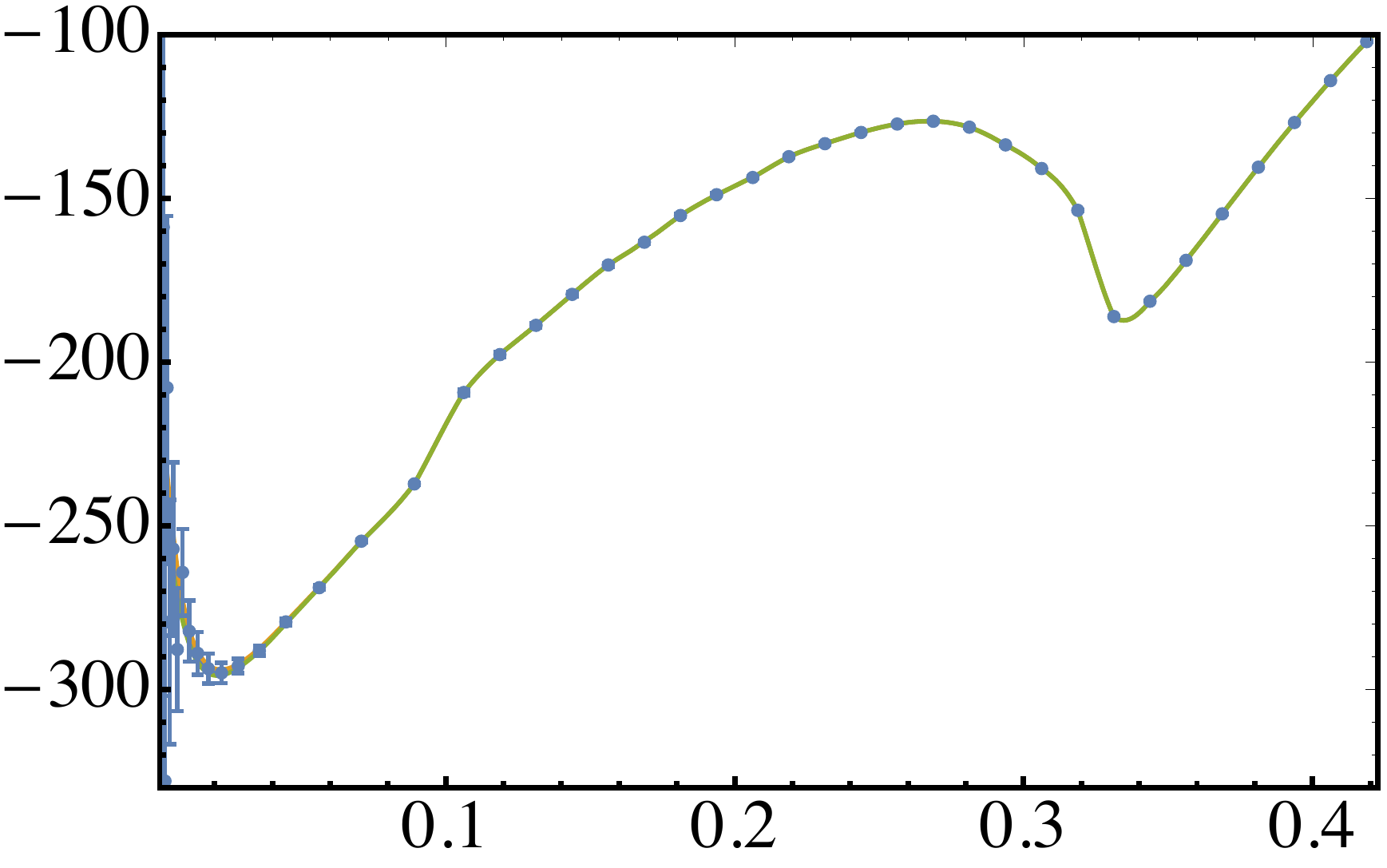}&
\includegraphics[width=.305\columnwidth]{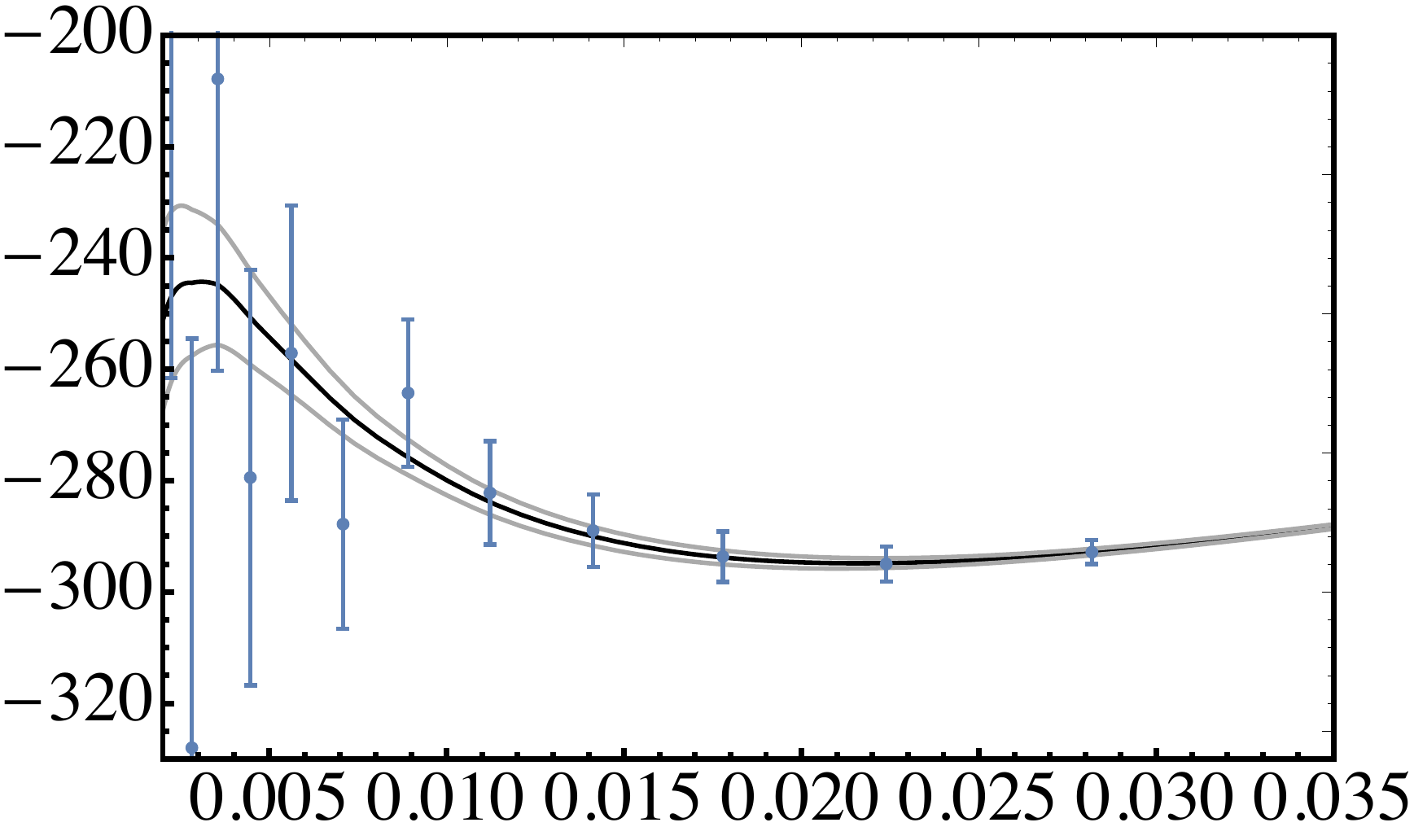} \\
\hspace{-1em}\rotatebox{90}{\qquad$a=0.25$}  \hspace{-1em} &
\includegraphics[width=.285\columnwidth]{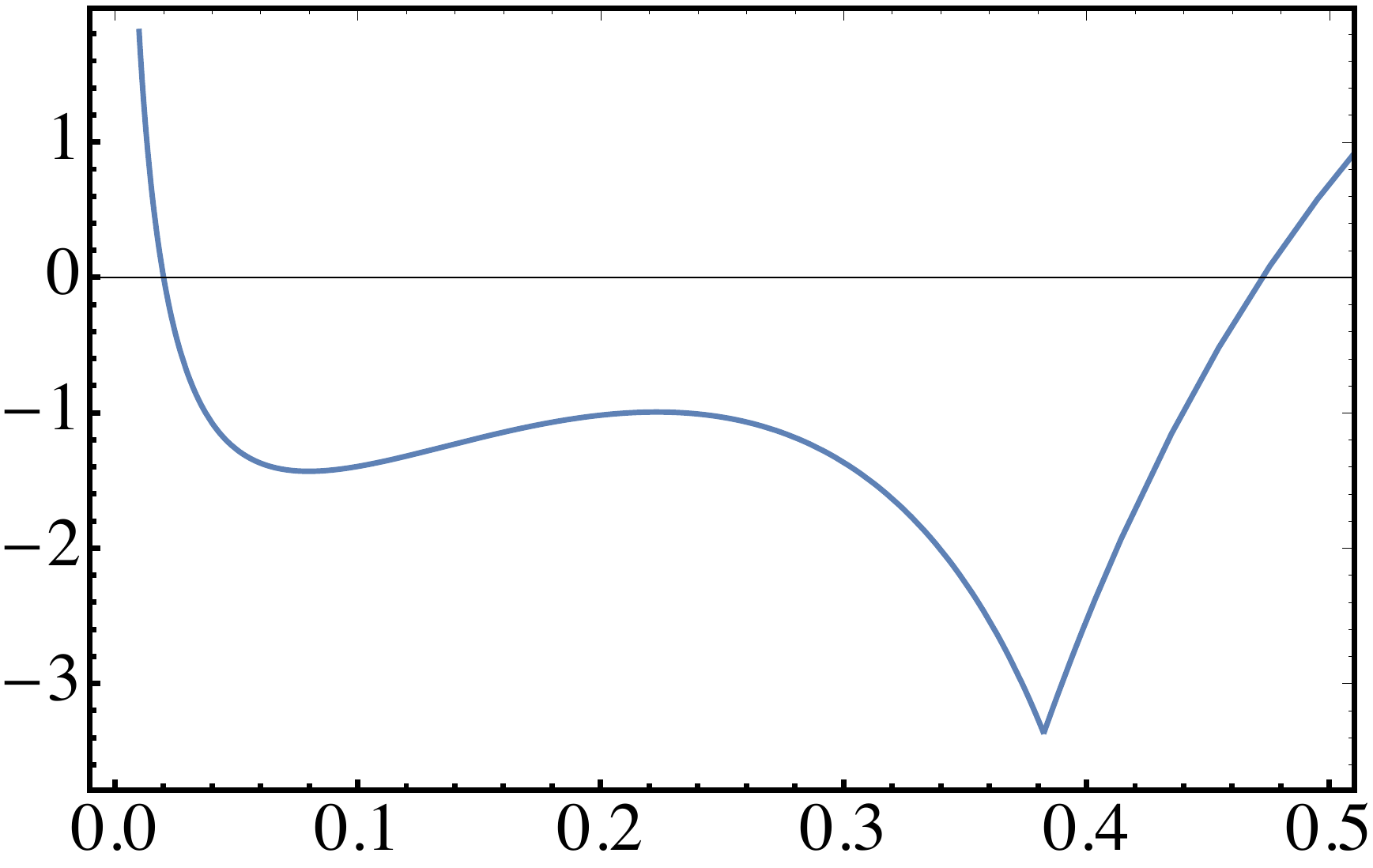}&  \ 
\includegraphics[width=.295\columnwidth]{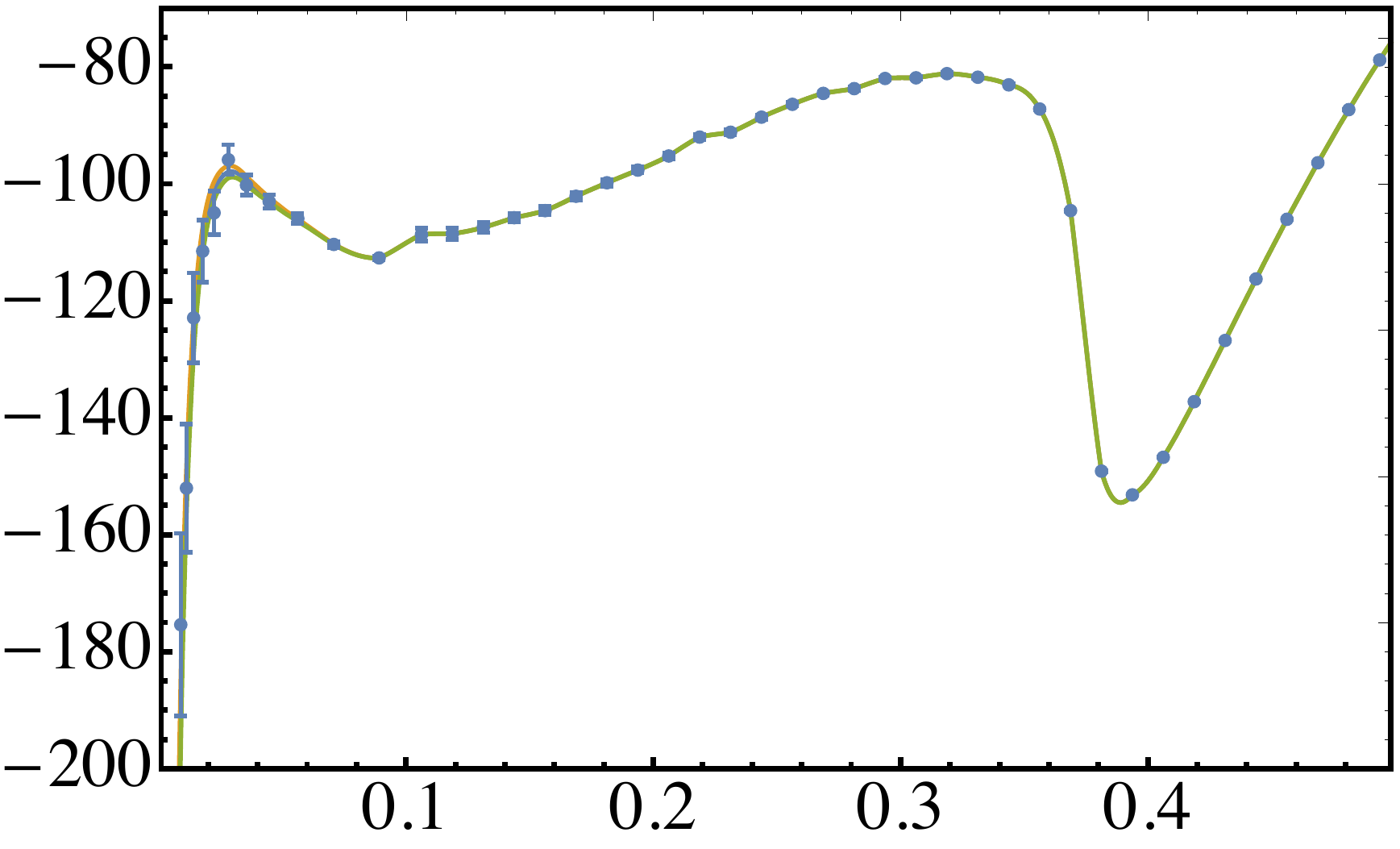}& \
\includegraphics[width=.29\columnwidth]{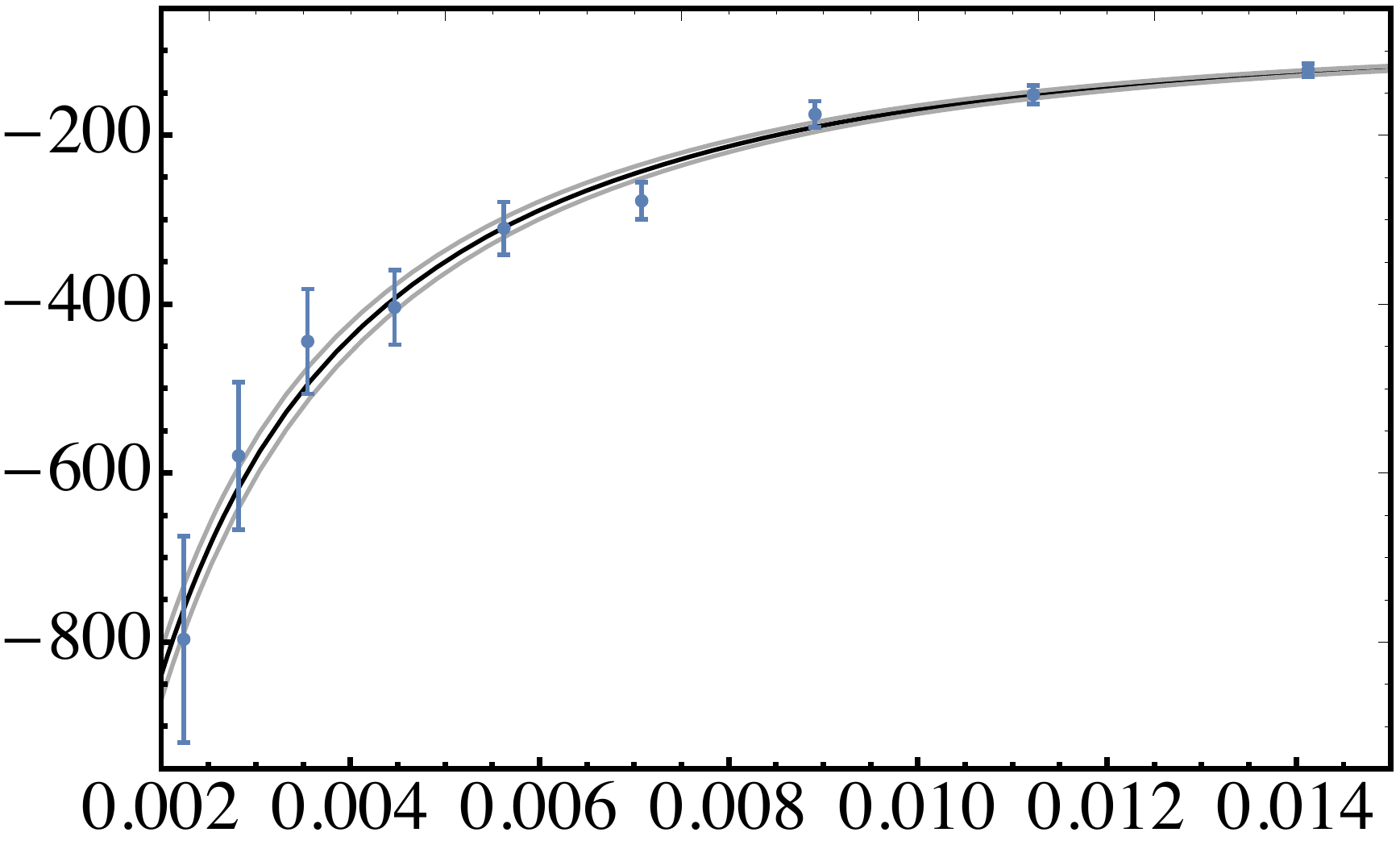} \\
\hspace{-1em}\rotatebox{90}{\qquad\quad$a=0.5$}  \hspace{-1em} &
\includegraphics[width=.285\columnwidth]{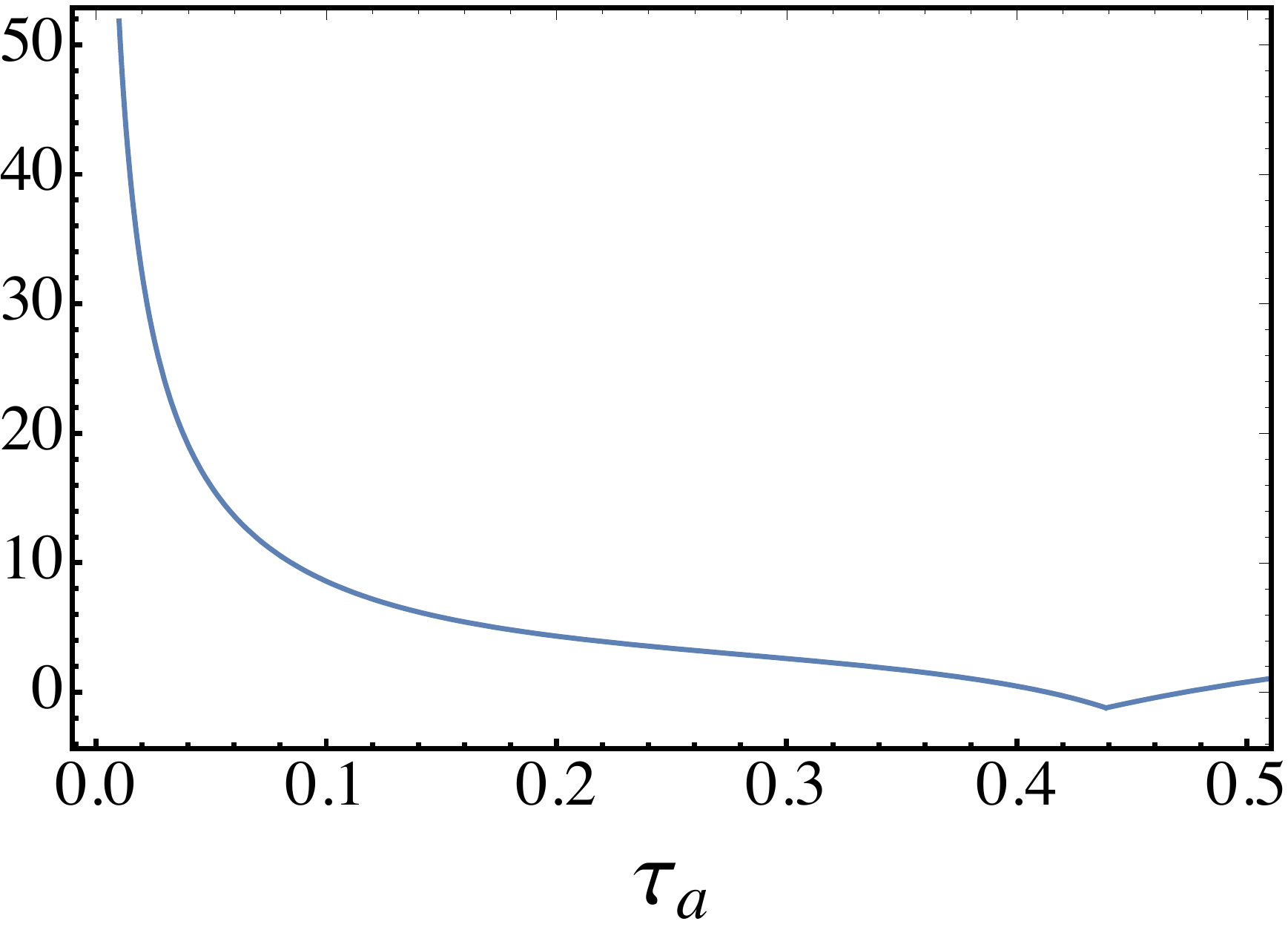}&
\includegraphics[width=.295\columnwidth]{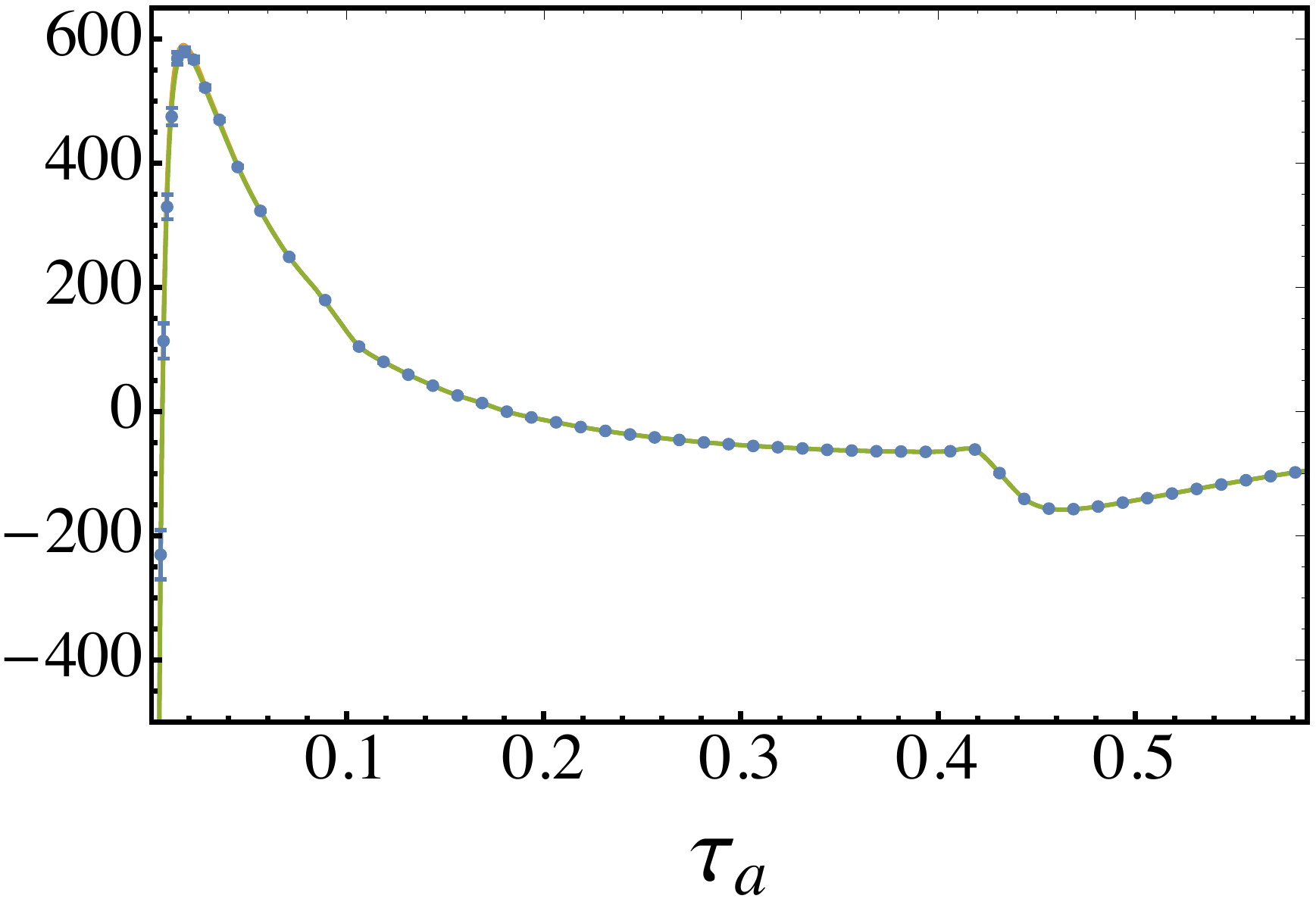}&
\includegraphics[width=.31\columnwidth]{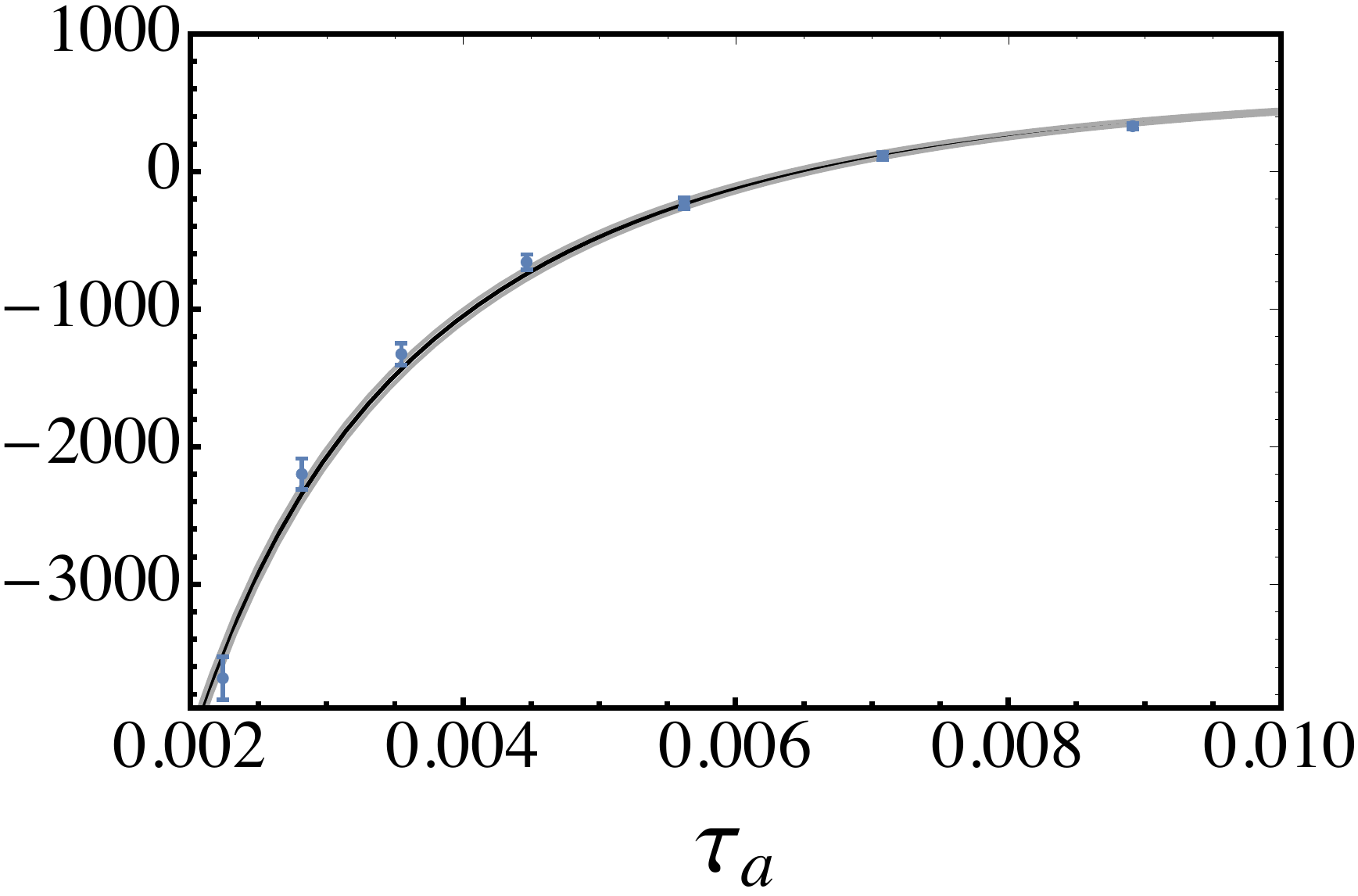}
\end{tabular}
\vspace{-1em}
\caption{
\emph{Left}: $\cO(\as)$ remainder functions from \cite{Hornig:2009vb}. 
\emph{Middle}: $\cO(\as^2)$ remainder functions across the whole physical spectrum.
\emph{Right}: Magnified plots of the latter in the low $\tau_a$ region.
Note that the vertical scales differ significantly across the plots, affecting the visual 
appearance of the uncertainty bands' actual widths.
}
\vspace{- 6em}
\label{fig:r2all}
\end{figure}

We next turn to the functional form of the remainder functions $r^{1}(\tau_a)$ and $r^{2}(\tau_a)$, 
which are displayed in \fig{r2} for $a = 0.25$. For these results we collect \event\ data 
from all events at cutoffs of $10^{-12}$ and $10^{-15}$ as mentioned above. In the left panel the predicted 
curve from \cite{Hornig:2009vb} for the $\cO(\as)$ remainder function is shown, whereas our \event\ results at $\cO(\as^2)$ are given by the data points in the right panel. In this panel we also illustrate how, following \cite{Abbate:2010xh}, 
we divide the $\tau_a$ domain into two regions $\tau_a<0.1$ and $\tau_a>0.1$, where we bin linearly in 
$\log_{10}\tau_a$ or in $\tau_a$ itself, respectively, to accurately capture the shape of the remainder 
function across the whole domain. In the linearly binned region, the \event\ uncertainties are negligible 
and we take a direct interpolation of the data points to be our working remainder function. In the low $\tau_a$ 
region, on the other hand, we perform a fit to a predetermined basis of functions and use the result as our functional 
prediction. We then join these predictions smoothly to determine the final remainder function entering our 
cross section predictions:
\begin{align}
\label{eq:r2joined}
r^2(\tau_a) &=
[1 - f(\tau_a,0.05,0.01)]\, r^2_{\log}(\tau_a) +  f(\tau_a,0.05,0.01)\,r^2_\text{lin}(\tau_a) \,,
\end{align}
with $f(z,z_0,\epsilon) = 1/(1+e^{-(z-z_0)/\epsilon})$. We actually transition to the direct 
interpolation of the \event\ data $r^2_\text{lin}$ at $\tau_a =0.05$, which is a bit below the 
boundary $\tau_a=0.1$ between the logarithmically and linearly binned regions, since the \event\ data 
is so precise well below this point anyway.

The basis of functions we use to fit the data in the low $\tau_a$ region is given by
\be
\label{eq:logfit}
r^2_{\log}(\tau_a) = a_0 + a_1\ln\tau_a + a_2 \ln^2\tau_a + a_3\ln^3\tau_a + b_2\, \tau_a \ln^2\tau_a + b_3\, \tau_a\ln\tau_a\,,
\ee
which is meant to be a representative but not necessarily complete (or even fully accurate) set of functions that can appear in the fixed-order $\cO(\as^2)$ distribution. This basis is simply chosen since it gives a good fit to the growth of $r^2(\tau_a)$ at small values of $\tau_a$ to the accuracy of the \event\ data that we have.
The fit is performed subject to the constraint that the total area $r_c^2(1)$ under the remainder function $r^2(\tau_a)$ is consistent with the values extracted in \tab{d2}. We just use the central values from that  table, as varying it within the ranges shown gives a negligible shift in the extracted fit parameters in \eq{logfit}. The errors on our remainder function $r^2(\tau_a)$ actually come from the 1-$\sigma$ error bands determined by the \texttt{NonlinearModelFit} routine in Mathematica, with the weight of each data point inversely proportional to the square of its \event\ error bar, and with $a_0$ determined by the area condition, leaving five free parameters in the fit. We give the results of these fits in 
\tab{fit}. 

Our final results for the $\cO(\as)$ and $\cO(\as^2)$ remainder functions for all seven values
of the angularity $a$ we consider in this work are shown in \fig{r2all}. The uncertainties of the
aforementioned fit procedure are hardly visible in the plots in the middle panel of \fig{r2all},
but they become noticeable in the low $\tau_a$ region, which is magnified in the right panel.
Although the statistical errors from \event\ at low $\tau_a$ grow substantially, the fits are constrained 
by the highly precise larger $\tau_a$ data and the constraint on the total area under the remainder function. 
More precise results for $r^2(\tau_a)$ at low $\tau_a$, and a more rigorous derivation of the fit functions 
that should appear in \eq{logfit}, are certainly desirable for precision phenomenology in this region, 
but such exercises lie outside the scope of this paper. Here we seek only to obtain 
sufficiently smooth and stable visual results for very low $\tau_a$ values, and focus 
on precision comparisons to data for intermediate-to-large $\tau_a$.

The remainder functions $r^{1}(\tau_a)$ and $r^{2}(\tau_a)$ 
have been extracted at a scale $\mu=Q$, cf. \eq{QCDexpansion}. The singular parts predicted 
by the factorization theorem \eq{differential} and these non-singular parts must separately add up to be
scale independent, if summed to all orders in $\as(\mu)$. Thus we can probe the perturbative uncertainty 
of the non-singular parts by varying them as a function of a separate scale $\mu_{\text{ns}}$:
\begin{align}
\label{eq:rmuns}
r(\tau_a,\mu_{\text{ns}}) &= r^{(1)}(\tau_a,\mu_\text{ns}) + r^{(2)}(\tau_a,\mu_\text{ns})\,, \\
\text{where } r^{(1)}(\tau_a,\mu_\text{ns}) &\equiv \frac{\as(\mu_\text{ns})}{2\pi} \,r^1(\tau_a)\,, \nn \\
r^{(2)}(\tau_a,\mu_\text{ns}) &\equiv \Bigl(\frac{\as(\mu_\text{ns})}{2\pi}\Bigr)^2 \Bigl[r^2(\tau_a) + 
\beta_0\, r^1(\tau_a)\ln\frac{\mu_\text{ns}}{Q}\Bigr]\,, \nn
\end{align}
and similarly for the integrated version $r_c$. These functions 
$r(\tau_a,\mu_\text{ns}),r_c(\tau_a,\mu_\text{ns})$ 
enter our final NLL$'+\cO(\as)$, NNLL$+\cO(\as)$, and NNLL$'+\cO(\as^2)$ resummed and matched predictions 
for the cross sections we will present in \sec{ANGRESULTS}. We have now assembled all the perturbative 
ingredients that we need to make these predictions. But before we do so, we turn our attention 
to the implementation of non-perturbative effects, whose proper treatment can even improve the 
behaviour of the perturbative convergence.
\section{Non-perturbative corrections}
\label{sec:NONPERT}

Having resummed the angularity distributions in \sec{ANGCALC} and matched them to the fixed-order 
predictions in \sec{CONSTANTS}, we will now include a treatment of non-perturbative corrections that will 
influence the overall shape and position of the distributions. 

%%%%%%%%%%%%%%%%%%%%%%%%%%%%%%%%%%%%%%%%%%%%%%%%%%%%%%%
\subsection{Non-perturbative shape function}
\label{sec:NONPERTa}

As with any hadronic observable, event shapes are sensitive to low energetic QCD radiation and effects of confinement. The 
importance of these non-perturbative effects depends on the domain of $\tau_{a}$ considered. For 
angularities with $a <1$, power corrections from the collinear sector are suppressed with respect to 
those from the soft sector \cite{Lee:2006nr,Becher:2013iya}.  
The non-perturbative effects can then be parameterized into a soft \textit{shape function} 
$f_{\text{mod}}(k)$ that is 
convolved with the perturbative distribution~\cite{Korchemsky:1998ev,Korchemsky:1999kt,Hoang:2007vb}:
\be
\label{eq:gap}
S(k,\mu) = 
\int dk' \;S_{\text{PT}}(k-k',\mu) \,f_\text{mod}(k'-2\overline\Delta_a)\,,
\ee
which ultimately leads to \eq{ultimate} for the cross section. Here $S_\text{PT}$ is the soft function 
computed in perturbation theory, and $\overline\Delta_a$ is a \emph{gap parameter}, which we will address in the 
next subsection.
The shape function $f_{\text{mod}}(k)$ is positive definite and normalized.
We follow previous approaches and 
expand the shape function in a complete set of orthonormal basis functions \cite{Ligeti:2008ac}:
\begin{equation}
\label{eq:Smod}
f_{\text{mod}}(k) = 
\frac{1}{\lambda} \left[ \sum_{n=0}^{\infty}\,b_{n} \,f_{n} \left(\frac{k}{\lambda} \right) \right]^{2} \,,
\end{equation}
where 
\begin{align}
\label{eq:fmod}
f_{n}(x) &= 8 \sqrt{\frac{2 x^{3} \left(2n + 1\right)}{3}}\, e^{-2x}\, P_{n}\big(g(x)\big)\,, \\
g(x)&= \frac{2}{3} \left( 3 - e^{-4x} \left(3+12x+24x^{2} + 32 x^{3} \right) \right) -1\,, \nn
\end{align}
and $P_{n}$ are Legendre polynomials. 
The normalization of the shape function implies that the coefficients $b_n$ satisfy
$\sum_{n=0}^{\infty} b_{n}^{2} = 1$. In practice, we only keep one term in the sum \eqref{eq:Smod}, 
setting $b_{n} = 0$ for $n>0$ (cf. \cite{Abbate:2010xh,Hoang:2014wka,Kang:2013nha}). The parameter 
$\lambda$ is then constrained by the first moment of the shape function 
as explained in the next subsection. More terms can in principle be included in \eq{Smod} 
if one wishes to study higher non-perturbative moments beyond the first one. 

This function, when convolved with the perturbative distribution from the previous sections, reproduces the known shift in the tail region~\cite{Korchemsky:1994is,Dokshitzer:1995zt,Dokshitzer:1995qm}, which can be shown rigorously via an operator product expansion (OPE) \cite{Lee:2006nr} to be the dominant non-perturbative effect,\footnote{
In the peak region, the OPE does not apply and the full shape function $f_{\text{mod}}(k)$ is required to capture
the non-perturbative effects. Furthermore, the result in \eqref{eq:NPdistshift} is not only 
leading order in the OPE, it is also subject to other corrections like finite hadron 
masses and perturbative renormalization effects on the quantity $\overline{\Omega}_{1}$, as described 
in~\cite{Mateu:2012nk}.}
\begin{equation}
\label{eq:NPdistshift}
\frac{d\sigma}{d\tau_{a}} (\tau_{a}) \underset{\text{NP}}{\longrightarrow} 
\frac{d\sigma}{d\tau_{a}} \Big(\tau_{a}-c_{\tau_{a}} \frac{\overline\Omega_{1}}{Q}\Big)\,.
\end{equation}
Here $\overline\Omega_{1}$ is a \textit{universal} non-perturbative parameter that is defined as a vacuum matrix element of soft 
Wilson lines and a transverse energy-flow operator (for details, see~\cite{Lee:2006nr}).
On the other hand, $c_{\tau_{a}}$ 
is an exactly calculable \textit{observable-dependent} coefficient which, for the 
angularities, is given by
$c_{\tau_{a}} 
= 2/(1-a)$ \cite{Berger:2003pk,Berger:2004xf,Lee:2006nr}.\footnote{The expression for $c_{\tau_{a}}$ diverges
in the broadening limit $a\to1$, where the SCET$_{\text{I}}$ factorization theorem we use breaks down. A careful analysis revealed that the non-perturbative effects to the broadening
distributions are enhanced by a rapidity logarithm, $c_{B_{T}}=\ln Q/B_T$~\cite{Becher:2013iya}.}

%%%%%%%%%%%%%%%%%%%%%%%%%%%%
\subsection{Gap parameter and renormalon subtraction}

As advocated in \cite{Hoang:2007vb}, we use a soft function with a non-perturbative gap parameter $\overline\Delta_a$, 
as already displayed in \eq{gap}. The gap parameter accounts for the minimum value of $\tau_a$ for a hadronic 
spectrum (the distribution can go down to zero for massless partons, but not for massive hadrons). 
In the tail region of the distributions, \eq{gap} then leads to a shift of the perturbative cross section, 
\eq{NPdistshift}, with
\be
\label{eq:Omegagap}
\frac{2\overline\Omega_1}{1-a} = 
2\overline\Delta_a + \int dk\,k \,f_\text{mod}(k)\,.
\ee
Since the first moment is shifted linearly by $\overline\Delta_a$, this parameter was rescaled 
in \cite{Hornig:2009vb} from its default definition in \cite{Hoang:2007vb} via
\be
\label{eq:Deltaa}
\overline\Delta_a = \frac{\overline\Delta}{1-a}\,,
\ee
where $\overline\Delta\sim \Lqcd$ is an $a$-independent parameter. 
Note that this determines that, with \eq{Smod} truncated at $n=0$, the model function parameter $\lambda = 2(\overline\Omega_1 - \overline\Delta)/(1-a)$.
Up to this point, the barred quantities $\overline\Omega_1,\overline\Delta_{(a)}$ are taken to be defined in a perturbative scheme like $\MSbar$ in which $S_\text{PT}$ has been calculated. In \cite{Hoang:2007vb} it was  pointed out that 
such a definition of the gap parameter $\overline\Delta_a$ has a renormalon ambiguity, shared by the perturbative 
soft function $S_\text{PT}$ in \eq{gap}. This is similar but not identical to the renormalon in the pole 
mass for heavy quarks (see, e.g.~\cite{Beneke:1994sw}). To obtain stable predictions, it is necessary to cancel 
the ambiguity from both $S_\text{PT}$ and $\overline\Delta_a$ in \eq{gap}. This can be done by redefining the gap 
parameter as
\be
\label{eq:Deltabar}
\overline\Delta_a = \Delta_a(\mu) + \delta_a(\mu)\,,
\ee
where $\delta_a$ has a perturbative expansion with the same renormalon ambiguity as $S_\text{PT}$ 
(but opposite sign). The remainder $\Delta_a$ is then renormalon free, but its definition 
depends on the scheme and the scale of the subtraction term $\delta_a$. We adopt here the prescription 
chosen in \cite{Hoang:2008fs} and also later implemented by \cite{Hornig:2009vb,Abbate:2010xh,Hoang:2014wka}, 
which is based on the position-space subtraction for the heavy-quark pole-mass renormalon introduced 
in \cite{Jain:2008gb,Hoang:2008yj}. We will translate this notation to the Laplace-space soft function we have been 
using in this paper---see \eq{Ffixedorder}. The two formulations are completely equivalent with 
$\nu\leftrightarrow ix$.

The subtraction and its evolution equations are easier to formulate in Laplace space than 
in momentum space. For the Laplace-space soft function\footnote{We prefer to use a 
dimensionful Laplace variable in this section, which is related to the one introduced
in \sec{hardjetsoft} by $\nu=\nu_a/Q$.} 
\be
\label{eq:LaplaceS}
\wt S(\nu,\mu) = \int_0^\infty dk\, e^{-\nu k} S(k,\mu)\,,
\ee
the convolution in \eq{gap} becomes
\be
\label{eq:Sproduct}
\wt S(\nu,\mu) = \Bigl[ e^{-2\nu\Delta_a(\mu)}\wt f_\text{mod}(\nu)\Bigr]
\Bigl[ e^{-2\nu\delta_a(\mu)}\wt S_\text{PT}(\nu,\mu)\Bigr]\,,
\ee
where we have grouped the renormalon-free gap parameter $\Delta_a$ with the shape function 
$\wt f_\text{mod}$, and the perturbative subtraction term $\delta_a$ with 
$\wt S_\text{PT}$, rendering each group of terms in brackets renormalon free. There 
are a number of schemes one can choose to define $\delta_a$ that achieve cancellation of the leading 
renormalon. A particularly convenient one found in \cite{Jain:2008gb,Hoang:2008yj,Hoang:2008fs} is 
a condition that fixes the derivative of the soft function at some value of $\nu$ to have an 
unambiguous value:
\be
\label{eq:renormaloncondition}
Re^{\gamma_E} \frac{d}{d\ln\nu} \Bigl[ \ln \widehat S_\text{PT}(\nu,\mu)\Bigr]_{\nu 
= 1/(Re^{\gamma_E})} = 0\,, 
\ee
where $\widehat S_\text{PT}(\nu,\mu) = e^{-2\nu\delta_a(\mu)}\wt S_\text{PT}(\nu,\mu)$, 
which is sufficient to render $\widehat S_\text{PT}$, and thus $\Delta_a$, to be renormalon-free.
This is known as the ``Rgap'' scheme, and its condition determines $\delta_a$ as a function of a new, arbitrary subtraction scale $R$, which should 
be taken to be perturbative, but small enough to describe the characteristic fluctuations in the 
soft function. Explicitly, \eq{renormaloncondition} defines the subtraction term as
\be
\label{eq:deltaa}
\delta_a(\mu,R) = \frac{1}{2}Re^{\gamma_E} \frac{d}{d\ln\nu} 
\Bigl[ \ln \wt S_\text{PT}(\nu,\mu)\Bigr]_{\nu = 1/(Re^{\gamma_E})}\,,
\ee
and we see that $\delta_a$ (and thus $\Delta_a$) depends on two perturbative scales, $\mu$ and $R$. 
Expanding the subtraction terms as
\be
\label{eq:deltaaexp}
\delta_a(\mu,R) = Re^{\gamma_E}\biggl[\frac{\as(\mu)}{4\pi}\delta_a^1(\mu,R) + \Bigl(\frac{\as(\mu)}{4\pi}\Bigr)^2\delta_a^2(\mu,R) + \cdots\biggr]\,,
\ee
we obtain, for the $\MSbar$ Laplace-space soft function $\wt S_\text{PT}$, using its expansion
in \eq{Ffixedorder},
\begin{subequations}
\label{eq:deltaan}
\begin{align}
\delta_a^1(\mu,R) &= \Gamma_S^0 \ln\frac{\mu}{R}\,, \\
\delta_a^2(\mu,R) &= \Gamma_S^0\beta_0 \ln^2\frac{\mu}{R} + \Gamma_S^1\ln\frac{\mu}{R} + \frac{\gamma_S^1(a)}{2} + c_{\tilde S}^1(a)\beta_0\,,
\end{align}
\end{subequations}
where from \eq{cuspproportion}, $\Gamma_S^n = -2 \Gamma_n/(1-a)$, and $\gamma_S^1$ and 
$c_{\tilde S}^1$ are given by \eq{gammaS1a} and \eq{cS1}.

\eq{deltaan} exhibits logarithms of $\mu/R$ that appear in the subtraction term $\delta_a$ and thus the 
renormalon-free gap parameter $\Delta_a$. Since $\mu=\mu_S$ will be chosen in the next section 
to be a function of $\tau_a$, which varies over a large range between $\mu_0\sim 1\text{ GeV}$ and $Q$, 
a fixed value of $R$ can only minimize these logarithms in one region of $\tau_a$, 
but not everywhere. We therefore need to allow $R$ to vary as well to track $\mu_S$, and so we need to know the 
evolution of $\Delta_a$ in both $\mu$ and $R$. The $\mu$-RGE is simple to derive. Since 
$\overline \Delta_a$ in \eq{Deltabar} is $\mu$-independent, we obtain
\begin{align}
\label{eq:DeltaRGE}
\mu\frac{d}{d\mu}\Delta_a(\mu,R) = -\mu\frac{d}{d\mu}\delta_a(\mu,R)  \equiv \gamma_\Delta^\mu[\as(\mu)]\,, 
\end{align}
where from the perturbative expansion of $\delta_a$ in \eq{deltaan}, we can determine 
\begin{align}
\label{eq:gammaDeltamu}
\gamma_\Delta^\mu[\as(\mu)] =  - Re^{\gamma_E} \Gamma_S[\as(\mu)]\,,
\end{align}
explicitly to $\cO(\as^2)$, and which can be shown to hold to all orders \cite{Hoang:2008fs}. The solution 
of this RGE is given by
\be
\label{eq:DeltaRGsol}
\Delta_a(\mu,R) = \Delta_a(\mu_{\Delta},R) + Re^{\gamma_E}\frac{\kappa_S}{2}
\,\eta_\Gamma(\mu,\mu_{\Delta})\,,
\ee
with an initial condition at some scale $\mu_\Delta$, and where $\kappa_S = 4/(1-a)$ 
was given in \eq{jFkF} and the 
kernel $\eta_\Gamma$ was defined in \eq{Keta}.

The evolution of the gap parameter $\Delta_a(\mu,R)$ in $R$ is a bit more involved, and was 
solved in \cite{Hoang:2008yj} for quark masses and applied to the soft gap parameter in \cite{Hoang:2008fs}. 
We follow this derivation here (in our own notation). Since from \eq{DeltaRGsol} we know how to evolve 
$\Delta_a(\mu,R)$ in $\mu$, we just need to derive the evolution of $\Delta_a(R,R)$ in 
$R$. Since $\overline\Delta_a$ in \eq{Deltabar} is also $R$-independent, we can derive from the perturbative 
expansion of $\delta_a$ in \eq{deltaan} 
the ``$R$-evolution'' equation:
\be
\label{eq:RRGE}
\frac{d}{dR}\,\Delta_a(R,R) = -\frac{d}{dR}\,\delta_a(R,R) \equiv - \gamma_R[\as(R)]\,,
\ee
where $\gamma_R$ has a perturbative expansion,
\be
\label{eq:gammaR}
\gamma_R[\as(R)] = \sum_{n=0}^\infty \Bigl(\frac{\as(R)}{4\pi}\Bigr)^{n+1} \gamma_R^n\,,
\ee
whose first two coefficients we read off from \eqs{deltaaexp}{deltaan},
\be
\label{eq:gammaRn}
\gamma_R^0 = 0\,,\quad \gamma_R^1 = \frac{e^{\gamma_E}}{2}\bigl [ \gamma_S^1(a) + 2c_{\wt S}^1 \beta_0\bigr]\,.
\ee
Even though $\gamma_R^0 = 0$ for the soft gap parameter (since $\gamma_S^0(a)=0$), we will keep it 
symbolically in the solution below for generality (and for direct comparison with the quark mass case 
in \cite{Hoang:2008yj}).

To solve \eq{RRGE}, we integrate:
\be
\label{eq:RRGint}
\Delta_a(R_1,R_1) - \Delta_a(R_{\Delta},R_{\Delta}) = -\int_{R_{\Delta}}^{R_1} \frac{dR}{R}\,R \,\gamma_R[\as(R)]\,,
\ee
multiplying and dividing by $R$ in the integrand, anticipating using \eq{dmu} to change integration variables to $\as$. But first we need to invert $\as(R)$ to express $R$. 
To this end, we write \eq{dmu} in the form
\be
\label{eq:lnR}
\ln\frac{R}{R_{\Delta}} = \int_{\as(R_{\Delta})}^{\as(R)}\frac{d\alpha}{\beta[\alpha]} = G[\as(R)] - G[\as(R_{\Delta})]\,,
\ee
where $G[\alpha]$ is the antiderivative of $1/\beta[\alpha]$,
\be
\label{eq:Gprime}
G'[\alpha] = \frac{1}{\beta[\alpha]} = -\frac{2\pi}{\beta_0}\frac{1}{\alpha^2} \frac{1}{1 + \frac{\alpha}{4\pi}\frac{\beta_1}{\beta_0} + \bigl(\frac{\alpha}{4\pi}\bigr)^2\frac{\beta_2}{\beta_0} + \cdots}\,.
\ee
This determines $G$ up to a constant of integration (we  effectively choose it such that $G[\alpha]\to 0$ as 
$\alpha\to\infty$). If $R,R_{\Delta}$ are scales for which $\as$ is perturbative, we can determine $G$ 
explicitly order by order, 
\be
\label{eq:G}
G[\alpha] = \frac{2\pi}{\beta_0}\biggl[\frac{1}{\alpha} + \frac{\beta_1}{4\pi\beta_0} \ln\alpha - \frac{B_2}{(4\pi)^2} \,\alpha + \cdots\biggr]\,,
\ee 
with $B_2$ from \eq{rB}. Then \eq{lnR} gives $R$ in terms of any other perturbative scale $R_{\Delta}$ as
\be
R = e^{G[\as(R)]} \Bigl[ R_{\Delta} e^{-G[\as(R_{\Delta})]}\Bigr]\,,
\ee
and we can use this relation in \eq{RRGint} to achieve the change of variables from $R$ to $\as$,
\be
\Delta_a(R_1,R_1) - \Delta_a(R_{\Delta},R_{\Delta}) = -R_{\Delta} e^{-G[\as(R_{\Delta})]} \int_{\as(R_{\Delta})}^{\as(R_1)} \frac{d\alpha}{\beta[\alpha]} \,e^{G[\alpha]} \,\gamma_R[\alpha]\,.
\ee
Keeping the explicit expression for $G$ in \eq{G} up to the $B_2$ term, and expanding out the integrand in 
powers of $\as$ where we can, we obtain
\begin{align}
&\Delta_a(R_1,R_1) - \Delta_a(R_{\Delta},R_{\Delta}) 
\\
&\quad = \frac{R_{\Delta}}{2\beta_0} \,e^{-G[\as(R_{\Delta})]} \int_{\as(R_{\Delta})}^{\as(R_1)}\frac{d\alpha}{\alpha} \,e^{\frac{2\pi}{\beta_0\alpha}}\;\alpha^{\frac{\beta_1}{2\beta_0^2}} \,
\biggl\{ \gamma_R^0 + \frac{\alpha}{4\pi}\Bigl[ \gamma_R^1 - \frac{\gamma_R^0}{\beta_0} \Bigl(\beta_1 + \frac{B_2}{2}\Bigr)\Bigr] + \cdots \biggr\} \,.\nn
\end{align}
This series of integrals is conveniently evaluated by changing the integration variable to
$t = -2\pi/(\beta_0\alpha)$,
upon which
\begin{align}
&\Delta_a(R_1,R_1) - \Delta_a(R_{\Delta},R_{\Delta}) \\
&\quad = \frac{R_{\Delta}}{2\beta_0}\, e^{-G[\as(R_{\Delta})]} \, 
\Bigl(\frac{2\pi}{\beta_0}e^{i\pi}\Bigr)^{\frac{\beta_1}{2\beta_0^2}}  \int_{t_1}^{t_0}  \frac{dt}{t}\,e^{-t} \, t^{-\frac{\beta_1}{2\beta_0^2}}  
\biggl\{ \gamma_R^0 -  \frac{1}{2\beta_0 t}\Big[ \gamma_R^1 - \frac{\gamma_R^0}{\beta_0}(\beta_1 + \frac{B_2}{2}\Bigr)\Bigr]+ \cdots\biggr\} \,.\nn
\end{align}
These integrals can then be expressed in terms of the incomplete gamma function
$\Gamma(c,x) = \int_x^\infty dt\, t^{-1+c} \,e^{-t}$, such that
\begin{align}
\label{eq:DeltaRRGsol}
&\Delta_a(R_1,R_1) - \Delta_a(R_{\Delta},R_{\Delta})\\
&\quad =  
\frac{R_{\Delta}}{2\beta_0}\,e^{-G[\as(R_{\Delta})]}\,
\Bigl(\frac{2\pi}{\beta_0}e^{i\pi}\Bigr)^{\frac{\beta_1}{2\beta_0^2}} \,
\biggl\{ \gamma_R^0 \,\Bigl[ \Gamma\Bigl(-\tfrac{\beta_1}{2\beta_0^2},-\tfrac{2\pi}{\beta_0\as(R_1)}\Bigr) - \Gamma\Bigl(-\tfrac{\beta_1}{2\beta_0^2},-\tfrac{2\pi}{\beta_0\as(R_{\Delta})}\Bigr)\Bigr] \nn \\
&\qquad\! - \frac{1}{2\beta_0}\Bigl[ \gamma_R^1 - \frac{\gamma_R^0}{\beta_0} \Bigl(\beta_1 \plus \frac{B_2}{2}\Bigr)\Bigr]\Bigl[ \Gamma\Bigl(-\tfrac{\beta_1}{2\beta_0^2}\minus 1,-\tfrac{2\pi}{\beta_0\as(R_1)}\Bigr) - \Gamma\Bigl(-\tfrac{\beta_1}{2\beta_0^2} \minus 1,-\tfrac{2\pi}{\beta_0\as(R_{\Delta})}\Bigr)\Bigr]  \!+ \cdots\! \biggr\}\,,\nn
\end{align}
which is consistent with the solution of the $R$-evolution equation given in \cite{Hoang:2008yj,Hoang:2008fs}. 
The solution of both $\mu$- and $R$-evolution equations in \eqs{DeltaRGsol}{DeltaRRGsol} finally determines the renormalon-subtracted gap 
parameter $\Delta_a(R,\mu)$ at any perturbative scales $R,\mu$ in terms of an input value 
$\Delta_a(R_{\Delta},R_{\Delta})$ at a reference scale $R_{\Delta}$:
\begin{align}
\label{eq:DeltaFullSolution}
&\Delta_a(\mu,R) = \Delta_a(R_{\Delta},R_{\Delta}) + Re^{\gamma_E}\frac{\kappa_S}{2}
\,\eta_\Gamma(\mu,R)  \\
&\quad + \frac{R_{\Delta}}{2\beta_0}\, e^{-G[\as(R_{\Delta})]}\Bigl(\frac{2\pi}{\beta_0}e^{i\pi}\Bigr)^{\frac{\beta_1}{2\beta_0^2}}\, \biggl\{ \gamma_R^0 \Bigl[ \Gamma\Bigl(-\tfrac{\beta_1}{2\beta_0^2},-\tfrac{2\pi}{\beta_0\as(R)}\Bigr) - \Gamma\Bigl(-\tfrac{\beta_1}{2\beta_0^2},-\tfrac{2\pi}{\beta_0\as(R_{\Delta})}\Bigr)\Bigr] \nn \\
&\qquad  - \frac{1}{2\beta_0}\Bigl[ \gamma_R^1 - \frac{\gamma_R^0}{\beta_0} \Bigl(\beta_1 \plus \frac{B_2}{2}\Bigr)\Bigr]\Bigl[ \Gamma\Bigl(-\tfrac{\beta_1}{2\beta_0^2}\minus 1,-\tfrac{2\pi}{\beta_0\as(R)}\Bigr) - \Gamma\Bigl(-\tfrac{\beta_1}{2\beta_0^2}  \minus1,-\tfrac{2\pi}{\beta_0\as(R_{\Delta})}\Bigr)\Bigr] \!+ \cdots \!\biggr\}\,.\nn
\end{align}
This expression is real, and we recall that  $\gamma_R^0 = 0$ in our case.

In the solution of the $\mu$- and $R$-evolution of $\Delta_a$ in \eq{DeltaFullSolution}, we truncate at each order of logarithmic accuracy as follows: at NLL($'$) we keep 
$\gamma_\Delta^\mu,\gamma_R$ to $\cO(\as)$ (i.e. up to the second line of \eq{DeltaFullSolution}), and at NNLL($'$) we keep $\gamma_\Delta^\mu,\gamma_R$ to $\cO(\as^2)$ (i.e up to the last line). These rules are summarized with all other truncation rules in \tab{LogAcc} below. Note that this means that $\eta_\Gamma$ is actually kept to one order of accuracy lower than indicated by \eq{etaclosedform}. This is because the $\mu$-evolution of $\Delta_a$ in \eq{DeltaRGsol} is not multiplied by an extra logarithm as for the hard, jet, and soft functions in the full factorized cross section.\footnote{In comparing to the $R$-evolution for quark masses in \cite{Hoang:2008yj}, it may also appear that we keep one fewer order at N$^k$LL accuracy than in that paper. But the counting for the logarithms is different in the two cases, since the logarithms appear as single logarithms for quark masses, but as double logarithms for
event shapes;
so the terms we call N$^k$LL correspond to terms that are called N$^{k-1}$LL in \cite{Hoang:2008yj}. Also, our truncation scheme seems to differ 
from the one applied for the gap parameter in \cite{Abbate:2010xh,Hoang:2014wka},
as described in Eq.~(A31) of \cite{Abbate:2010xh} or Eq.~(56) of \cite{Hoang:2014wka}. However, it is consistent with the corresponding tables in these papers and with the actual numerical implementations used by these authors in their results \cite{VM}.
}

Transforming the renormalon-free soft function in \eq{Sproduct} back to momentum space, we obtain the shifted version of \eq{gap},
\be
\label{eq:Sfree}
S(k,\mu) = 
\int dk' \;S_{\text{PT}}\bigl(k-k' -2\delta_a(\mu,R),\mu\bigr) \,
f_\text{mod}\bigl(k' - 2\Delta_a(\mu,R)\bigr)\,.
\ee
Then the parameter $\overline\Omega_1$ in \eq{Omegagap}, describing the non-perturbative shift of the perturbative cross section induced by the shape function, turns into a renormalon-free shift:
\be
\label{eq:Omega1sub}
\frac{2\Omega_1(\mu,R)}{1-a} = 2\Delta_a(\mu,R) + \int dk\,k\,f_\text{mod}(k)\,,
\ee
We will take the input gap parameter at a reference scale $R_{\Delta}= 1.5 \text{ GeV}$ to be 
$\Delta(R_{\Delta},R_{\Delta}) = 0.1$~GeV in our phenomenological analysis below.  The exact value of this 
parameter is not particularly relevant to the tail region in which we focus our comparisons to 
data \cite{Abbate:2010xh}.

The shift in the perturbative part of \eq{Sfree} can also be expressed in terms of a differential translation operator that acts on the perturbative soft function $S_\text{PT}$:
\be
S(k,\mu) = 
\int dk' \,\Bigl[e^{-2\delta_a(\mu,R)
\frac{d}{dk}} \,
S_{\text{PT}}\bigl(k-k',\mu\bigr)\Bigr] \,
f_\text{mod}\bigl(k' - 2\Delta_a(\mu,R)\bigr)\,,
\ee
which, after integrating by parts, gives
\be
\label{eq:Srenormalonfree}
S(k,\mu) = 
\int dk' \,
S_{\text{PT}}\bigl(k-k',\mu\bigr)\,\Bigl[e^{-2\delta_a(\mu,R)\frac{d}{d k'}} \,
f_\text{mod}\bigl(k' - 2\Delta_a(\mu,R)\bigr) \Bigr]  \,.
\ee
In the final cross section, the renormalon-subtracted shape function then enters as a convolution against the perturbative distribution,
\be
\label{eq:sigmaconvolved}
\frac{1}{\sigma_0}\,\sigma(\tau_a) = 
\int dk\, \sigma_\text{PT}\Bigl( \tau_a - \frac{k}{Q}\Bigr) \Bigl[e^{-2\delta_a(\mu_S,R)\frac{d}{d k}} f_\text{mod}\bigl(k - 2\Delta_a(\mu_S,R)\bigr) \Bigr]\,,
\ee
which is the expression we anticipated in \eq{ultimate}. This implies we convolve both the singular and nonsingular parts of the cross section \eq{singandns} with the same, renormalon-subtracted shape function. In doing this we follow \cite{Abbate:2010xh} and ensure a smooth transition from the resummation to fixed-order regime even after non-perturbative effects are included.

In practice we expand out the shape function terms to the order we work in $\as$,
\begin{align}
\label{eq:modelexpansion}
e^{-2\delta_a(\mu_S,R)\frac{d}{d k}} f_\text{mod}(k-2 \Delta_a(\mu_S,R))  &= f_\text{mod}^{(0)}(k-2\Delta_a(\mu_S,R)) + f_\text{mod}^{(1)}(k-2\Delta_a(\mu_S,R)) \nn \\
&\quad + f_\text{mod}^{(2)}(k-2\Delta_a(\mu_S,R)) \,, 
\end{align}
where
\begin{subequations}
\label{eq:fmodn}
\begin{align}
f_\text{mod}^{(0)}(k-2\Delta_a(\mu_S,R)) &= f_\text{mod}(k-2\Delta_a(\mu_S,R))\,,\\
f_\text{mod}^{(1)}(k-2\Delta_a(\mu_S,R)) &= - \frac{\as(\mu_S)}{4\pi}\,
2\delta_a^1(\mu_S,R) R e^{\gamma_E} f'_\text{mod}(k-2\Delta_a(\mu_S,R)) \,,\\
f_\text{mod}^{(2)}(k-2\Delta_a(\mu_S,R)) &=  \Bigl(\frac{\as(\mu_S)}{4\pi}\Bigr)^2 \Bigl[ - 2  \delta_a^2(\mu_S,R) Re^{\gamma_E} f'_\text{mod}(k-2\Delta_a(\mu_S,R)) \\
&\qquad \qquad + 2(\delta_a^1(\mu_S,R)Re^{\gamma_E})^2f''_\text{mod}(k-2\Delta_a(\mu_S,R))\Bigr]\,, \nn
\end{align}
\end{subequations}
with $\delta_a^{1,2}(\mu_S,R)$ from \eq{deltaan}. The order to which these terms are kept at each accuracy are included in \tab{LogAcc}.

%------------------------------------------------------------------------------------------------------------------------------------------------
\begin{table}[t]
\centering
\scalebox{.85}{\begin{tabular}{|c|c|c|c|c|}
\hline \hline
\textbf{Accuracy} & \boldmath $\Gamma_{\text{cusp}}$ & \boldmath $\gamma_{F},\gamma_\Delta^\mu,\gamma_R$ & \boldmath $\beta$ & $H,\tilde{J}, \tilde{S},\delta_a$   \\[8pt]
\hline \hline
LL & $\alpha_{s}$ & 1 & $\alpha_{s}$ &  1 \\[8pt]
\hline
NLL &  $\alpha^{2}_{s}$ & $\alpha_{s}$ & $\alpha^{2}_{s}$& 1  \\[8pt]
\hline
NNLL & $\alpha^{3}_{s}$ & $\alpha^{2}_{s}$& $\alpha^{3}_{s}$  & $\alpha_{s}$ \\[8pt]
\hline
N$^{3}$LL & $\alpha^{4}_{s}$ & $\alpha^{3}_{s}$ & $\alpha^{4}_{s}$  &$\alpha^{2}_{s}$  \\[8pt]
\hline \hline
\end{tabular}}
\;
\scalebox{.85}{\begin{tabular}{|c|c| }
\hline \hline
\textbf{Accuracy} & $H, \tilde{J}, \tilde{S},\delta_a$ \\[8pt]
\hline \hline
&  \\[8pt]
\hline
NLL$^{\prime}$ &  $\alpha_{s}$\\[8pt]
\hline
NNLL$^{\prime}$ & $\alpha^{2}_{s}$\\[8pt]
\hline
N$^{3}$LL$^{\prime}$ & $\alpha^{3}_{s}$\\[8pt]
\hline \hline
\end{tabular}}
\;
\scalebox{.85}{\begin{tabular}{|c|c| }
\hline \hline
\textbf{Matching} & $r^n(\tau_a)$ \\[8pt]
\hline \hline
&  \\[8pt]
\hline
$+\cO(\as)$ &  $\alpha_{s}$\\[8pt]
\hline
$+\cO(\as^2)$ & $\alpha^{2}_{s}$\\[8pt]
\hline
$+\cO(\as^3)$ & $\alpha^{3}_{s}$\\[8pt]
\hline \hline
\end{tabular}}
\caption{Ingredients we include at various orders of unprimed N$^k$LL (\emph{Left}), primed N$^k$LL$'$ (\emph{Middle}), and matched (\emph{Right}) accuracies, up to a given fixed order $\cO(\as^n)$. The tables apply to the integrated distribution in \eq{sigmaPTexpansion} and the Laplace-transformed distribution, but not, for unprimed accuracies, directly to the differential form in \eq{differential}---see \cite{Almeida:2014uva} for details. We have included a counting for the renormalon subtractions terms $\delta_a$ in \eq{deltaaexp} and 
the $\mu$- and $R$-evolution anomalous dimensions $\gamma_\Delta^\mu,\gamma_R$ in \eqs{gammaDeltamu}{gammaR} as described in the text.}
\label{tab:LogAcc}
\end{table}
%------------------------------------------------------------------------------------------------------------------------------------------------

\subsection{Final resummed, matched, and renormalon-subtracted cross section}
\label{ssec:finalcs}

We now collect all pieces described above, giving our final expressions for the resummed cross section, matched to fixed-order and convolved with a renormalon-free shape function.

In evaluating the convolution in \eq{sigmaconvolved}, we must  truncate the product of the fixed-order perturbative pieces contained in \eqs{cumulant2}{remainder} along with the non-perturbative pieces in \eq{modelexpansion} to the appropriate order in $\as$ for N$^k$LL$^{(')}$ accuracy. Namely, starting with \eq{cumulant2} for the integrated distribution, we expand the fixed-order coefficients in powers of $\as$:
\be
\frac{\sigma_{c,\text{PT}}(\tau_a)}{\sigma_0} = \sigma_c^{\text{LO}}(\tau_a) + \sigma_c^{\text{NLO}}(\tau_a)  + \sigma_c^{\text{NNLO}}(\tau_a) \,,
\ee
where 
\begin{align}
\label{eq:sigmaPTexpansion}
 \sigma_c^{\text{N$^k$LO}}(\tau_a;\mu_H,\mu_J,\mu_S,\mu_{\text{ns}}) &= e^{\wt K (\mu_H,\mu_J,\mu_S;Q) + K_\gamma(\mu_H,\mu_J,\mu_S)}\Bigl(\frac{1}{\tau_a}\Bigr)^{\Omega(\mu_J,\mu_S)} \\
 &\quad \times  \wt F_k(\tau_a,\partial_\Omega,\mu_H,\mu_J,\mu_S)\;\frac{e^{\gamma_E\Omega}}{\Gamma(1-\Omega)} 
\ +  \ r_c^{(k)}(\tau_a,\mu_\text{ns}) 
\,,  \nn
\end{align}
with the fixed-order operators $\wt F_k$ at each order given by
\begin{subequations}
\label{eq:fixedorderoperators}
\begin{align}
\wt F_0(\tau_a,\partial_\Omega,\mu_H,\mu_J,\mu_S) &= 1\,, \\
\wt F_1(\tau_a,\partial_\Omega,\mu_H,\mu_J,\mu_S) &= \frac{\alpha_H}{4\pi}H_1(Q^2,\mu_H) + 2\frac{\alpha_J}{4\pi}\wt J_1\Bigl(\partial_\Omega + \ln\frac{\mu_J^{2-a}}{Q^{2-a}\tau_a},\mu_J\Bigr) \\
&\quad + \frac{\alpha_S}{4\pi}\wt S_1\Bigl(\partial_\Omega + \ln\frac{\mu_S}{Q\tau_a},\mu_S\Bigr) \,, \nn \\
\wt F_2(\tau_a,\partial_\Omega,\mu_H,\mu_J,\mu_S) &= \Bigl(\frac{\alpha_H}{4\pi}\Bigr)^2 H_2(Q^2,\mu_H) + \Bigl(\frac{\alpha_J}{4\pi}\Bigr)^2 (2\wt J_2 + \wt J_1^2)\Bigl(\partial_\Omega + \ln\frac{\mu_J^{2-a}}{Q^{2-a}\tau_a},\mu_J\Bigr) \nn \\
&\quad + \Bigl(\frac{\alpha_S}{4\pi}\Bigr)^2 \wt S_2\Bigl(\partial_\Omega+\ln\frac{\mu_S}{Q\tau_a},\mu_S\Bigr) + \frac{\alpha_H}{4\pi}H_1(Q^2,\mu_H)\\
&\qquad \times \biggl[2 \frac{\alpha_J}{4\pi}\wt J_1\Bigl(\partial_\Omega \plus \ln\frac{\mu_J^{2-a}}{Q^{2-a}\tau_a},\mu_J\Bigr) + \frac{\alpha_S}{4\pi}\wt S_1\Bigl(\partial_\Omega \plus \ln\frac{\mu_S}{Q\tau_a},\mu_S\Bigr) \biggr] \nn \\
&\quad + 2 \frac{\alpha_J}{4\pi} \frac{\alpha_S}{4\pi}\wt J_1\Bigl(\partial_\Omega \plus \ln\frac{\mu_J^{2-a}}{Q^{2-a}\tau_a},\mu_J\Bigr)\wt S_1\Bigl(\partial_\Omega \plus \ln\frac{\mu_S}{Q\tau_a},\mu_S\Bigr) \,, \nn
\end{align}
\end{subequations}
where $\alpha_F \equiv \as(\mu_F)$, $H_n,\wt J_n,\wt S_n$ are given by \eq{Ffixedorder}, and $r_c^{(k)}$ are the integrated versions of the remainder functions in \eq{rmuns}
with $r_c^{(0)}(\tau_a,\mu_\text{ns})=0$.

In the convolved cross section \eq{sigmaconvolved}, the proper fixed-order expansion up to NNLL($'$) accuracy is then given by:
\be
\label{eq:convolvedexpanded}
\frac{\sigma_{c}(\tau_a)}{\sigma_0} = \sigma_c^{[0]}(\tau_a) 
+ \sigma_c^{[1]}(\tau_a)  + \sigma_c^{[2]}(\tau_a) \,,
\ee
where 
\begin{subequations}
\label{eq:convolvedterms}
\begin{align}
\sigma_c^{[0]}(\tau_a) &= 
\int dk\; \sigma_c^{\text{LO}}\Bigl(\tau_a - \frac{k}{Q};\mu_H,\mu_J,\mu_S\Bigr) 
\,f_\text{mod}^{(0)}(k-2\Delta_a(\mu_S,R))\,, \\
\sigma_c^{[1]}(\tau_a) &= \int dk\; \Bigl\{ \sigma_c^{\text{NLO}}
\Bigl(\tau_a - \frac{k}{Q};\mu_H,\mu_J,\mu_S,\mu_\text{ns}\Bigr) \,
f_\text{mod}^{(0)}(k-2\Delta_a(\mu_S,R)) \\
&\qquad\quad + \sigma_c^{\text{LO}}\Bigl(\tau_a - \frac{k}{Q};\mu_H,\mu_J,\mu_S\Bigr) \,
f_\text{mod}^{(1)}(k-2\Delta_a(\mu_S,R))\Bigr\}\,, \nn \\
\sigma_c^{[2]}(\tau_a) &= \int dk\; \Bigl\{\sigma_c^{\text{NNLO}}
\Bigl(\tau_a - \frac{k}{Q};\mu_H,\mu_J,\mu_S,\mu_\text{ns}\Bigr) \,
f_\text{mod}^{(0)}(k\minus 2\Delta_a(\mu_S,R)) \\
&\qquad\quad + \sigma_c^{\text{NLO}}\Bigl(\tau_a - \frac{k}{Q};\mu_H,\mu_J,\mu_S,\mu_\text{ns}\Bigr) \,
f_\text{mod}^{(1)}(k\minus 2\Delta_a(\mu_S,R)) \nn \\
&\qquad\quad + \sigma_c^{\text{LO}}\Bigl(\tau_a - \frac{k}{Q};\mu_H,\mu_J,\mu_S\Bigr) \,
f_\text{mod}^{(2)}(k-2\Delta_a(\mu_S,R))\Bigr\} \,. \nn
\end{align}
\end{subequations}
\eq{convolvedexpanded} is to be truncated at the fixed order demanded by \tab{LogAcc} (first term up to NLL, second term for NLL$'$ and NNLL, and up to the last term for NNLL$'$), while the evolution kernels contained in each expression are to be evaluated to the appropriate resummed logarithmic accuracy as described in \sec{ANGCALC}.

%----------------------------------------------------------------------------
\begin{figure}[t]
\centering
\includegraphics[scale=0.28]{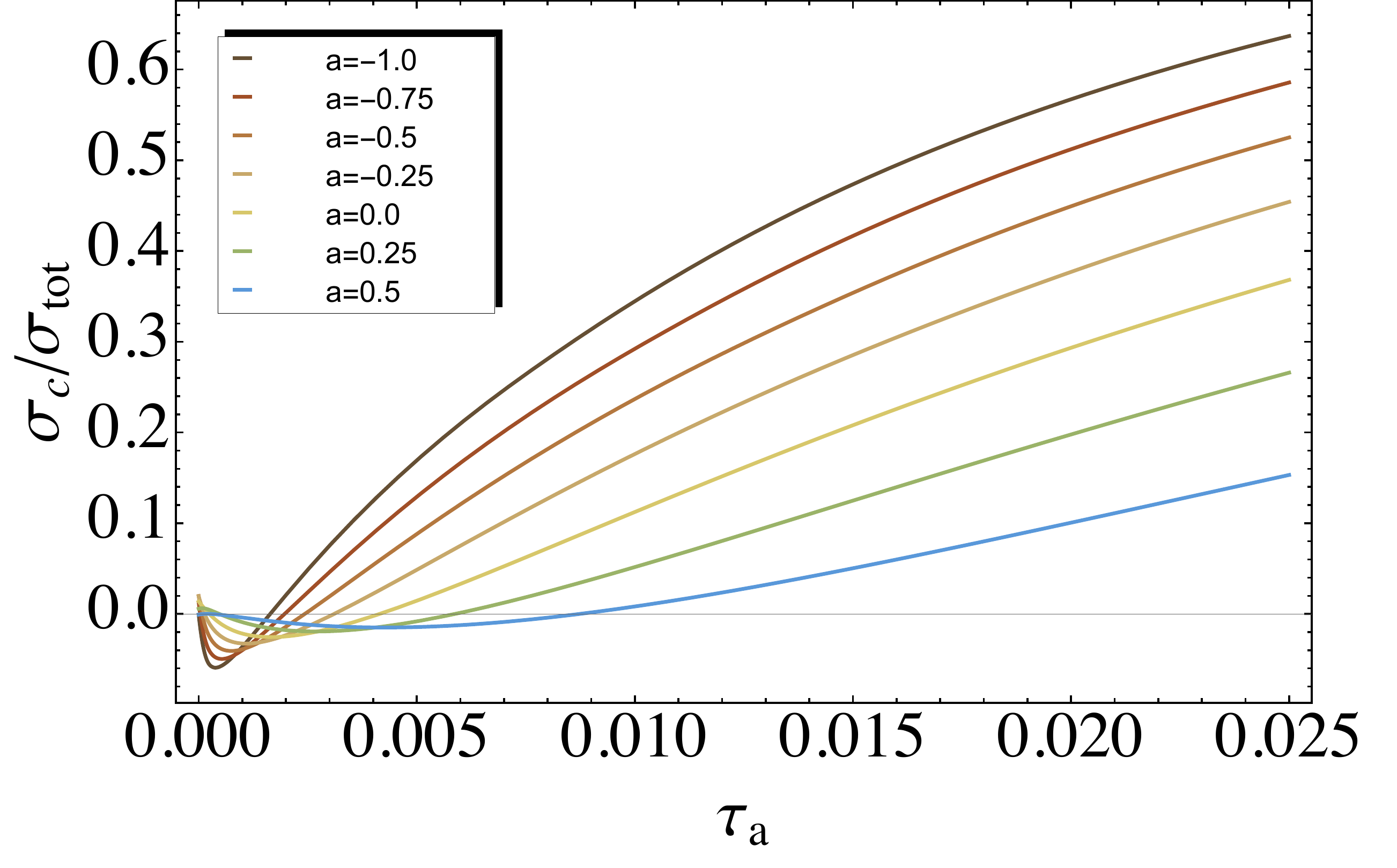}
\includegraphics[scale=0.28]{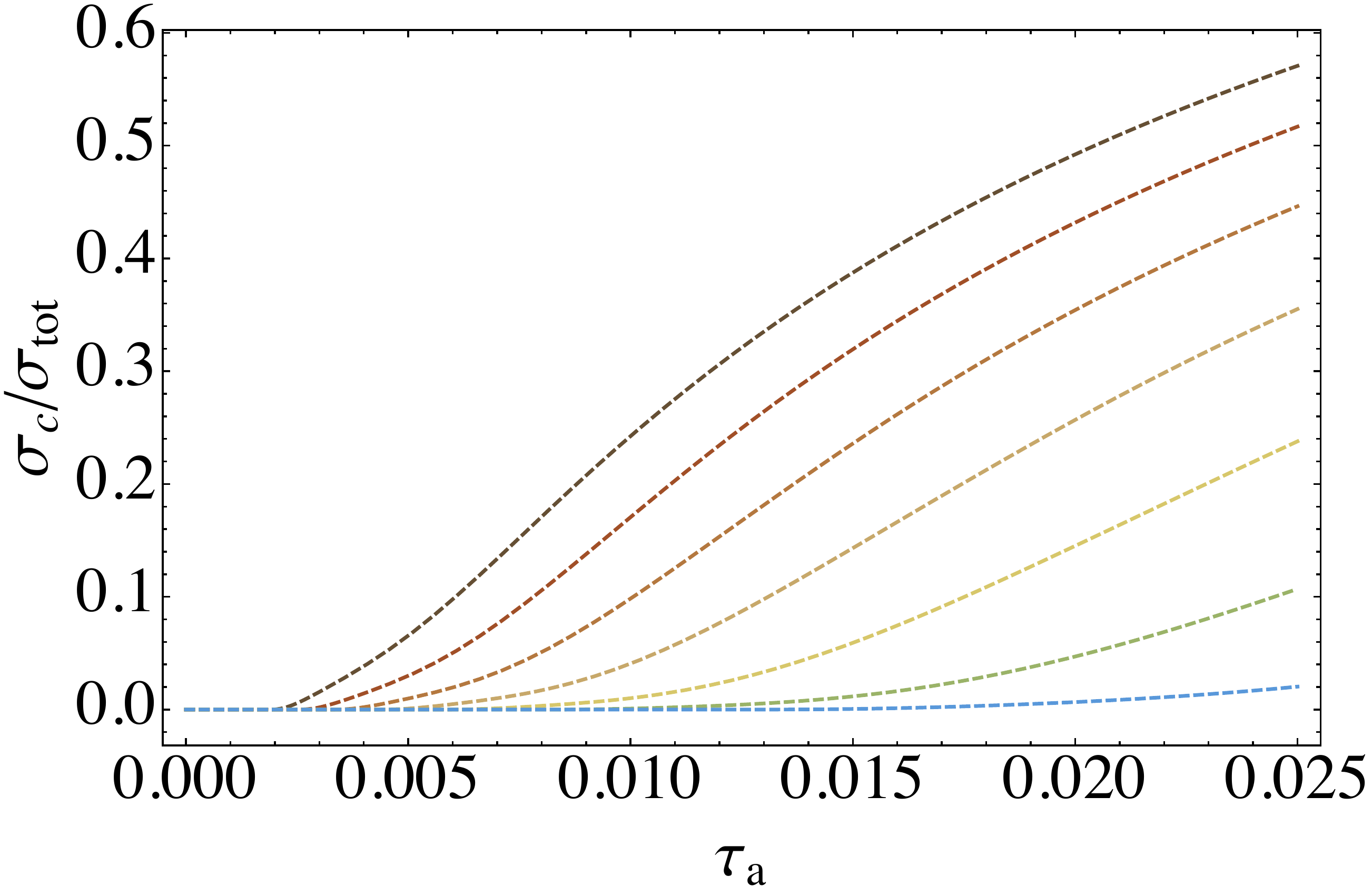}
\vspace{-1em}
\caption{Predictions for the central values of the integrated cross sections at 
NNLL$^{\prime}$ accuracy in the low $\tau_{a}$ domain, shown here for seven values of the angularity parameter $a$.  \emph{Left}: Predictions from the purely perturbative cross section.
\emph{Right}: Predictions after renormalon subtraction as implemented in \eq{convolvedterms}.
}
\label{fig:renormalon}
\end{figure}
%-----------------------------------------------------------------------------
\eqs{convolvedexpanded}{convolvedterms} represent our final expression for the renormalon-free resummed and matched cross section that is convolved with a non-perturbative shape function.
We should point out that we will perform the convolution in $k$ prior to choosing particular values for the scales $\mu_{H,J,S,\text{ns}}$ and $R$. We have clearly exhibited the dependence on these scales versus the explicit dependence on $\tau_a$ appearing in \eqs{sigmaPTexpansion}{fixedorderoperators}. In the next section we will describe how we choose these scales, but for now it suffices to say that they are functions of the measured $\tau_a$ in the cross section, and not functions of the convolution variable $\tau_a - k/Q$. 
Thus the $\tau_a$ dependence inside the scales $\mu_{H,J,S,\text{ns}}$ and $R$ are not convolved over in \eq{convolvedterms}. In our numerical code, the convolution between the explicit $\tau_a-k/Q$ dependence from \eqs{sigmaPTexpansion}{fixedorderoperators} and the $k$ dependence in $f_\text{mod}$ is then computed for a given set of scales 
$\mu_i(\tau_a),R(\tau_a)$. 

\fig{renormalon} illustrates the practical effect of implementing the renormalon subtraction and the convolution with the non-perturbative shape function as described above.  We present the central theory curves at NNLL$^{\prime}$ accuracy for the cumulant cross sections at seven values of the parameter $a$, all in the low-$\tau_{a}$ domain. The curves in the left panel reflect the purely perturbative calculation, which exhibit unphysical (negative) values for the cross section. The curves in the right panel, on the other hand, represent the calculation performed with \eq{convolvedterms}.  It is clear that the renormalon cancellation is successful and no unphysical behavior is observed.

\section{Scale choices}
\label{sec:SCALES}

\subsection{Profile functions}
\label{ssec:profile}

From the arguments of the logarithms in the fixed-order hard, jet, and soft functions appearing in 
\eq{cumulant2}, one can identify the natural scales at which these logarithms are minimized, 
finding
\begin{equation}
\label{eq:angscales}
\mu_{H}^{\text{nat}} = Q, \,\,\,\,\,\,\, \mu_{J}^{\text{nat}} = Q \tau_{a}^{1/(2-a)}, \,\,\,\,\,\,\, \mu_{S}^{\text{nat}} = Q \tau_{a} \,.
\end{equation}
In the tail region, where the resummation is critical, we want to evaluate the distributions near 
these scales. However, we ultimately predict the distributions over a domain of $\tau_{a}$ that can 
be roughly broken into three regions where the comparative scalings differ (see, e.g.~\cite{Abbate:2010xh}):
\begin{itemize}
\item \emph{Peak Region}: \hspace{3.0mm}
$\mu_{H} \gg \mu_{J} \gg \mu_{S} \sim \Lambda_{QCD}$\,,
\item \emph{Tail Region}: \hspace{4.2mm}
$\mu_{H} \gg \mu_{J} \gg \mu_{S} \gg \Lambda_{QCD}$\,,
\item \emph{Far-tail Region}:   
$\mu_{H} = \mu_{J} = \mu_{S} \gg \Lambda_{QCD}$\,.
\end{itemize} 
In the peak region the soft scale is non-perturbative, and it is here that the full model shape function 
described in \sec{NONPERT} becomes necessary for making reliable predictions.  In this region 
we will adjust the scales to plateau at a constant value just above $\Lqcd$.  On the other hand, the scales 
are well separated in the tail region where the resummation is most important.  We want to minimize the 
logarithms in the resummed distributions, and hence the scalings are close to the natural values in 
\eq{angscales}.  Finally, our predictions should match onto fixed-order perturbation theory in the 
far-tail region.  The resummations should therefore be switched off, and the scales should merge at 
$\mu_{H,J,S} \sim Q$.  

Getting the scales to merge near $\mu_{H,J,S}\sim Q$ in the far-tail region will require $\mu_{J,S}$ to rise 
faster with $\tau_a$ than the natural scales in \eq{angscales}, since the physical maximum value of $\tau_a$ 
is less than 1. We will achieve this below by defining a smooth function to transition between the resummation 
and fixed-order regions.  But the transition can be made less sudden by increasing the rate of change of 
$\mu_{J,S}$ even in the resummation region.  Such an increased slope was used for the $C$-parameter and 
thrust distributions in \cite{Hoang:2014wka}.  For thrust, i.e. $a=0$, the authors used the central values
\be
\label{eq:rs}
\mu_S = r_s \mu_H \tau_0\,,\qquad \mu_J = (\mu_H \mu_S)^{1/2}\,,
\ee
with $r_s = 2$ in the resummation region. We will follow this strategy here and give a physical interpretation 
to the slope parameter $r_s$. The maximal value for thrust is $\tau_0 = 1/2$, which is achieved for a 
perfectly spherically symmetric distribution of particles in the final state. The slope $r_s = 2$ thus ensures 
that $\mu_{J,S}$ merge with $\mu_H$ at this maximum value $\tau_0^\text{sph}$, 
instead of at $\tau_a=1$ as the natural scales \eq{angscales} do. For arbitrary $a$, the 
angularity of the spherically symmetric configuration is
\begin{align}
\label{eq:tausph}
\tau_a^\text{sph} &= \frac{1}{4\pi}\int_{0}^{2\pi}d\phi \int_{-1}^1 d\cos\theta\, \sin^a\theta (1-\abs{\cos\theta})^{1-a} 
= \frac{1}{2-\frac{a}{2}}\;\HypF\Bigl(1,-\frac{a}{2};3-\frac{a}{2};-1\Bigr) \,, 
\end{align}
which ranges from $\tau_{-1}^\text{sph} \approx 0.356$ to $\tau_{1/2}^\text{sph}\approx 0.616$ for the values 
of $a$ we consider in this work. 
These may be compared to the maximum values of a three- and four-particle configuration in \fig{taumax} 
in \appx{tetra}. We will then choose a default slope $r_s = 1/\tau_a^\text{sph}$ in \eq{rs} that ensures 
that the scales $\mu_{H,J,S}$ meet at $\tau_a=\tau_a^\text{sph}$.
We actually want the scales to merge a bit before $\tau_a^\text{sph}$, so that there is a non-vanishing 
region where the predicted distributions are purely fixed order. 

We have designed a set of \emph{profile functions} (see, 
e.g.~\cite{Ligeti:2008ac, Abbate:2010xh,Kang:2014qba,Hornig:2016ahz}) that fulfill all of the criteria discussed 
above while smoothly interpolating between the various regions.  
The precise form of our profiles depends on a running scale defined by
\begin{equation}
\label{eq:murun}
\mu_{\text{run}}(\tau_a) = 
\begin{cases}
\mu_{0} & \tau_a \le t_{0} \\
\zeta\Big(\tau_a; \lbrace t_{0}, \mu_{0}, 0 \rbrace, \lbrace t_{1}, 0, \frac{r}{\tau_{a}^\text{sph}}\mu_H  \rbrace \Big) & t_{0} \le \tau_a \le t_{1}\\
\frac{r}{\tau_{a}^\text{sph}}\,\mu_H\tau_a & t_{1} \le \tau_a \le t_{2} \\
\zeta\Big(\tau_a; \lbrace t_{2},  0, \frac{r}{\tau_{a}^\text{sph}}\mu_H \rbrace, \lbrace t_{3}, \mu_H, 0 \rbrace \Big) & t_{2} \le \tau_a \le t_{3} \\
\mu_H & \tau_a \ge t_{3}
\end{cases}\,.
\end{equation}
The function $\zeta$ ensures that $\mu_\text{run}$ and its first derivative are smooth. 
Specifically, we adopt the functional form from \cite{Hoang:2014wka},
which connects a straight line in the region $\tau_a<t_0$ with slope $r_0$ and intercept $y_0$ with another 
straight line in the region $\tau_a>t_1$ with slope $r_1$ and intercept $y_1$ via
\begin{equation}
\label{eq:zeta}
\zeta\left(\tau_a; \lbrace t_{0}, y_{0}, r_{0} \rbrace, \lbrace t_{1}, y_{1}, r_{1} \rbrace \right) =
\begin{cases} 
a + r_0(\tau_a-t_0) + c(\tau_a-t_0)^2 & \tau_a \le \frac{t_{0} + t_{1}}{2} \\
A + r_1(\tau_a-t_1) + C(\tau_a-t_1)^2 & \tau_a \ge \frac{t_{0} + t_{1}}{2} 
\end{cases}\,,
\end{equation}
where the coefficients of the polynomials are determined by 
continuity of the function and its first derivative:
\begin{align}
\label{eq:CPAR}
a &= y_0 + r_0 t_0\,, &  A &= y_1 + r_1 t_1\,, \\
c&= 2\,\frac{A-a}{(t_0-t_1)^2} + \frac{3r_0+r_1}{2(t_0-t_1)}\,, &  
C &= 2\,\frac{a-A}{(t_0-t_1)^2} + \frac{3r_1 + r_0}{2(t_1-t_0)} \,. \nn
\end{align}
The parameters $t_{i}$ control the transitions between the non-perturbative, resummation, and fixed-order regions of the distributions,
and can be varied as well as part of the estimation of the theoretical uncertainties.
We will set these parameters to
\begin{align}
\label{eq:tparameters}
t_{0} &=  \frac{n_{0}}{Q} \,3^a,\,\,\, &&t_{2} = n_2\times0.295^{1-0.637\,a}\,,\\
t_{1} &= \frac{n_{1}  }{Q} \, 3^a,\,\,\,&&t_{3} = n_3\, \tau_{a}^\text{sph} \,, \nn
\end{align}
with coefficients $n_i$ that we can adjust.
The design of profile functions is somewhat of an art. The chosen $a$-dependence of $t_{0,1,2}$ in 
\eq{tparameters} is based on some empirical observations about the theory distributions ultimately predicted.
The first two, $t_{0,1}$, control the transition between the non-perturbative and resummation regions, 
and we have chosen them to track the location of the peak of the differential $\tau_a$ distributions. Very 
roughly, this location scales like $3^a$. The parameter $t_{2}$ was determined  as a numerical approximation 
to the point where singular and nonsingular contributions become equal in magnitude, since the 
resummation should be turned off once the latter become as sizable as the former. 
The formula for $t_2$ in \eq{tparameters} is our rough empirical fit to this crossing point in $\tau_a$.
Finally, the parameter $t_3$ is chosen so that our predictions for the distributions revert to those of 
fixed-order perturbation theory a bit below the maximum value $\tau_a^\text{sph}$, as described above,
which we will achieve by choosing $n_3\lesssim 1$.

The parameters $n_i$ in \eq{tparameters}, and $r$ and $\mu_{0}$ in \eq{murun}, will be treated differently 
in the discussion of \sec{scalevariations}, where we probe the theoretical uncertainty of our 
predictions in two different ways:  the \emph{band} and \emph{scan} methods.  In the former, certain parameters 
including $n_{i}$, $r$ and $\mu_{0}$ will be taken as constants, whereas in the latter these will 
be varied over.  The central values and/or scan ranges for all parameters we consider are given for 
both methods in \tab{scan}, and the details of the two procedures are described in 
\sec{scalevariations}.

%-----------------------------------------------------------------------------------------------------------------------------------------------------
\begin{table}[t]
\small
\begin{center}
\begin{tabular}{|c|c|c|}
\hline
& Band Method & Scan Method \\
\hline
\hline 
$e_H$ & 
$\{0.5,2\}$ & $0.5\leftrightarrow 2$ \\ 
\hline 
$e_J$ & $\{-0.5,0.5\}$ &
$-0.75\leftrightarrow 0.75$ 
\\ 
\hline 
$e_S$ & $\{-0.5,0.5\}$ &
0 \\ 
\hline
$n_0$ & $2~\text{GeV}$ &
$1\leftrightarrow 2$ GeV
\\
 \hline
$n_1$ & $10~\text{GeV}$ &
$8.5\leftrightarrow 11.5$ GeV
\\ 
\hline
$n_2$ & $0.85$ &
$0.9\leftrightarrow 1.1$
\\ 
\hline
$n_3$ & $0.8$ &
$0.8\leftrightarrow 0.9$
\\ 
\hline
\end{tabular}
\qquad
\begin{tabular}{|c|c|c|}
\hline
& Band Method & Scan Method \\
\hline
\hline 
$\mu_0$ & $1~\text{GeV}$ &
$0.8\leftrightarrow 1.2$ GeV
\\ 
\hline
$R_0$ & $0.6~\text{GeV}$ &
$\mu_0 - 0.4$ GeV \\ 
\hline
$r$ & $1$ &
$0.75 \leftrightarrow 1.33$
\\
 \hline
$\delta c_{\tilde J}^2$ &  $\{-1, 1\}$&
$-1\leftrightarrow 1$
\\
 \hline
$\delta r^2$ & $\{-1, 1\}$ &
$-1\leftrightarrow 1$
\\
 \hline
$n_s$ & $\{-1, 1\}$ & $\{-1,0,1\}$ \\ 
\hline
\end{tabular}
\end{center}
\vspace{-1.2em}
\caption{
Central values and/or parameter ranges we use in the band and scan methods described in 
\sec{scalevariations}. For the band method, the lists of parameters are discrete variations whose 
results are added in quadrature. For the scan method, all parameters are chosen randomly within the 
ranges shown. It is important to remember that the range of values chosen for any given parameter 
need not be identical in the two independent approaches we use to estimate the theoretical
uncertainties.
}
\label{tab:scan}
\vspace{-3mm}
\end{table}
%-----------------------------------------------------------------------------------------------------------------------------------------------------

Finally then, we use the following forms for the hard, jet, and soft profiles:
\begin{subequations}
\label{eq:muHJS}
\begin{align}
\label{eq:muH}
\mu_{H} &= e_H Q\,, \\
\label{eq:muS}
\mu_{S}(\tau_a) &= \left[1+ e_{S} \, \theta(t_{3} - \tau_a) 
\left( 1 - \frac{\tau_a}{t_{3}} \right)^{2}\right] \mu_\text{run}(\tau_a)\,,\\
\label{eq:muJ}
\mu_{J}(\tau_a) &= \left[1+ e_{J} \, \theta(t_{3} - \tau_a) 
\left( 1 - \frac{\tau_a}{t_{3}} \right)^{2}\right]\, \mu_H^{\frac{1-a}{2-a}} \, \mu_\text{run}(\tau_a)^{\frac{1}{2-a}}\,,
\end{align}
\end{subequations}
where varying the parameter $e_H$ will adjust the overall magnitude for all the scales together 
(since $\mu_H$ enters $\mu_\text{run}$ in \eq{murun}), and varying $e_{S,J}$ controls the width of the 
respective soft and jet bands, thereby allowing a variation about the default shape and the canonical 
relation $\mu_J^{2-a} = \mu_H^{1-a}\mu_S$.

In addition to \eq{muHJS}, our predictions depend on the scale $R$ associated with the renormalon subtraction 
in the soft function, see \eqs{Srenormalonfree}{sigmaconvolved}.  
For this scale, following \cite{Abbate:2010xh,Hoang:2014wka}, we use a profile that 
mimics the soft scale $\mu_S$ in \eq{muS} but that starts at an initial value of $R_0<\mu_0$ for $\tau_a<t_0$:
\begin{equation}
\label{eq:muR}
R(\tau_a) = \mu_S(\tau_a)\quad\text{ with }\,\mu_0 \to R_0\,,
\end{equation}
where our choice for $R_{0}$ is given in \tab{scan}.  As pointed out in  \cite{Abbate:2010xh}, 
a choice like this ensures that the hierarchy between $R$ and $\mu_{S}$ is never large enough to generate 
large logarithms in the subtraction terms $\delta_a^n$ in \eq{deltaaexp},
while also keeping a nonzero $\cO(\as)$ subtraction $\delta_a^1$ (with the right sign) below $t_1$ 
where the effect is most pronounced, by strictly taking $R<\mu_S$.

Finally, our predictions depend on the scale $\mu_{\text{ns}}$ that enters the nonsingular fixed-order 
contribution in \eq{sigmaPTexpansion}. A variation of this scale probes missing higher-order terms in the 
fixed-order prediction. For small values of $\tau_a$, this includes subleading logarithms of the form
$\tau_a\ln^k\tau_a$ (in the integrated distribution) (see, e.g., \cite{Freedman:2013vya})
which are not resummed by our leading-power factorization theorem in \eq{cumulant2} (see \cite{Moult:2018jjd} 
for a resummation of subleading logarithms of thrust in $H\to gg$). 
To this end, we again adopt the strategy of \cite{Abbate:2010xh,Hoang:2014wka} (cf. 
also \cite{Kang:2013nha}) and consider three choices of $\mu_{\text{ns}}$ as a function of $\tau_a$:
\be
\label{eq:muns}
\mu_{\text{ns}}(\tau_a)= 
\begin{cases}
\frac{1}{2}\big(\mu_H+\mu_J(\tau_a)\big) & n_s = 1 \\
\mu_H & n_s = 0 \\
\frac{1}{2}\big(3\mu_H-\mu_J(\tau_a)\big) & n_s = -1 \\
\end{cases} \,,
\ee
where $n_s$ is a discrete parameter, which we will vary to select between the three
different choices of $\mu_{\text{ns}}$.

In order to obtain a comprehensive theory error estimate, we must consider uncertainties associated with all 
of the scale and profile parameters that contribute to our calculation. Far from adding 
arbitrariness to our predictions, these scales and parameters have meanings that reflect the physics contributing 
to the different regions of the event shape distributions, while the observation of reduced dependence on these parameters from 
one order of accuracy to the next provides a stringent test that we have organized the perturbative expansion 
correctly. We have carried out these variations with two different methods, which we discuss next.

%%%%%%%%%%%%%%%%%%%%%%%%%%%%%%%%%%%%%%%%%%%%%%%%
\subsection{Scale variations}
\label{sec:scalevariations}

There exist numerous approaches to varying the different scale and profile parameters that enter the 
theoretical predictions in order to gauge their respective uncertainties. Two methods were contrasted 
in \cite{Abbate:2010xh}, the so-called \emph{band} and \emph{scan} methods. In the first, one parameter 
is varied at a time and the envelope of the resulting predictions is taken to be the total uncertainty band.  
In the second, many random choices for all the parameters are taken at once, within predefined ranges for each, 
and the resulting envelope of the predictions is again taken to be the total uncertainty band. The analysis in 
\cite{Abbate:2010xh} concluded that the band method implemented this way
tends to underestimate uncertainties, and the authors therefore advocated the use of the scan method. We will 
apply both methods to all seven angularity distributions treated in this study, but we will purposefully make 
the band method more conservative by adding uncertainties from each individual variation in quadrature 
(cf.~also \cite{Kang:2013nha}). We  furthermore adjust the ranges of the parameter variation in each method to 
achieve convergence plots for the final cross section that exhibit both good convergence and display 
sufficiently conservative error estimates. In the end, the two methods can be made to give similar results. 
We undertake this comparison not to advocate one or the other method, but more as a conservative test of 
the reliability of our displayed uncertainties.

\subsubsection*{Band method}

A simple and intuitive procedure for obtaining uncertainty estimates is to execute independent variations 
about the relevant scales $\mu_{i}$ ($i=H,J,S,R$ with $\mu_{R} \equiv R$) and 
$\mu_\text{ns}$, with each variation designed to probe missing orders of perturbative 
accuracy not calculated in our study. As mentioned above, we then add the 
uncertainties from the individual variations in quadrature to obtain our final theory error
estimate. We have performed six independent variations, with the particular scale settings in each one given by:
\begin{itemize}
\item a 
jet-scale variation with $\mu_{H,S,R} = \mu_{H,S,R}^{c}$, $\mu_{J} \in \lbrace \mu_{J}^{J,low},  \mu_{J}^{J,high} \rbrace$ and $\mu_{\text{ns}} = \mu_H^{c}$,
\item a 
soft-scale variation with $\mu_{H,J} = \mu_{H,J}^{c}$, $\mu_{R,S} \in \lbrace \mu_{R,S}^{(R,S),low},  \mu_{R,S}^{(R,S),high} \rbrace$ and $\mu_{\text{ns}} =  \mu_H^{c}$,
\item a 
hard-scale variation with $\mu_{i} \in \lbrace \mu_{i}^{low},  \mu_{i}^{high} \rbrace$ and 
$\mu_{\text{ns}} \in \lbrace \mu_H^{low}, \mu_H^{high} \rbrace$,
\item a nonsingular-scale variation with $\mu_{i} = \mu_{i}^{c}$ and 
$\mu_{\text{ns}} \in \lbrace 
\frac{1}{2}(\mu^{c}_H+\mu^{c}_J), \frac{1}{2}(3\mu^{c}_H-\mu^{c}_J)\rbrace$, 
\item a variation of 
$c_{\tilde{J}}^{2}$ as described below, with $\mu_{i} = \mu_{i}^{c}$ and 
$\mu_{\text{ns}} =  \mu_H^{c}$, 
\item a variation of $r^{2}$ as described below, with $\mu_{i} = \mu_{i}^{c}$ and 
$\mu_{\text{ns}} =  \mu_H^{c}$. 
\end{itemize}
Our notation is such that in the hard-scale variation
$\lbrace \mu_{i}^{low}, \mu_{i}^{high} \rbrace \equiv \lbrace \mu_{i}(\mu_H = Q/2), \mu_{i}(\mu_H = 2Q)  \rbrace$ 
whereas in the jet- and soft-scale variations $\mu_{i}^{i,low}$, $\mu_{i}^{c}$, and 
$\mu_{i}^{i,high}$ ($i = J,S$) correspond to $e_{J,S} = \lbrace -0.5, 0.0, 0.5 \rbrace$, respectively.  
Note that, while we have not included an independent variation of the renormalon scale (due to its negligible 
difference with the soft scale), we have maintained the relative splitting and hierarchy between 
$\mu_{S}$ and $R$ (and therefore the size of the logarithms of their ratios) by defining $R^{high}$ and 
$R^{low}$ as
\begin{equation}
R^{high} = \phi_{RS} \, \mu_{S}^{high}, \qquad 
R^{low} =  \phi_{RS} \, \mu_{S}^{low}, \qquad 
\text{with} \qquad
\phi_{RS} = \frac{R^{c}}{\mu_{S}^{c}}\,,
\end{equation}
and similarly for the functions $R^{R,low}$ and $R^{R,high}$.

%------------------------------------------------
\begin{figure}[h]
\centering
\vspace{-1.2em}
\begin{tabular}{ccc}
 & \hspace{-16.8em} \scriptsize  GeV & \hspace{-16.7em}  \scriptsize  GeV  \\
\rotatebox{90}{\qquad\qquad\quad$a=-1$} &
\includegraphics[width=0.45\columnwidth]{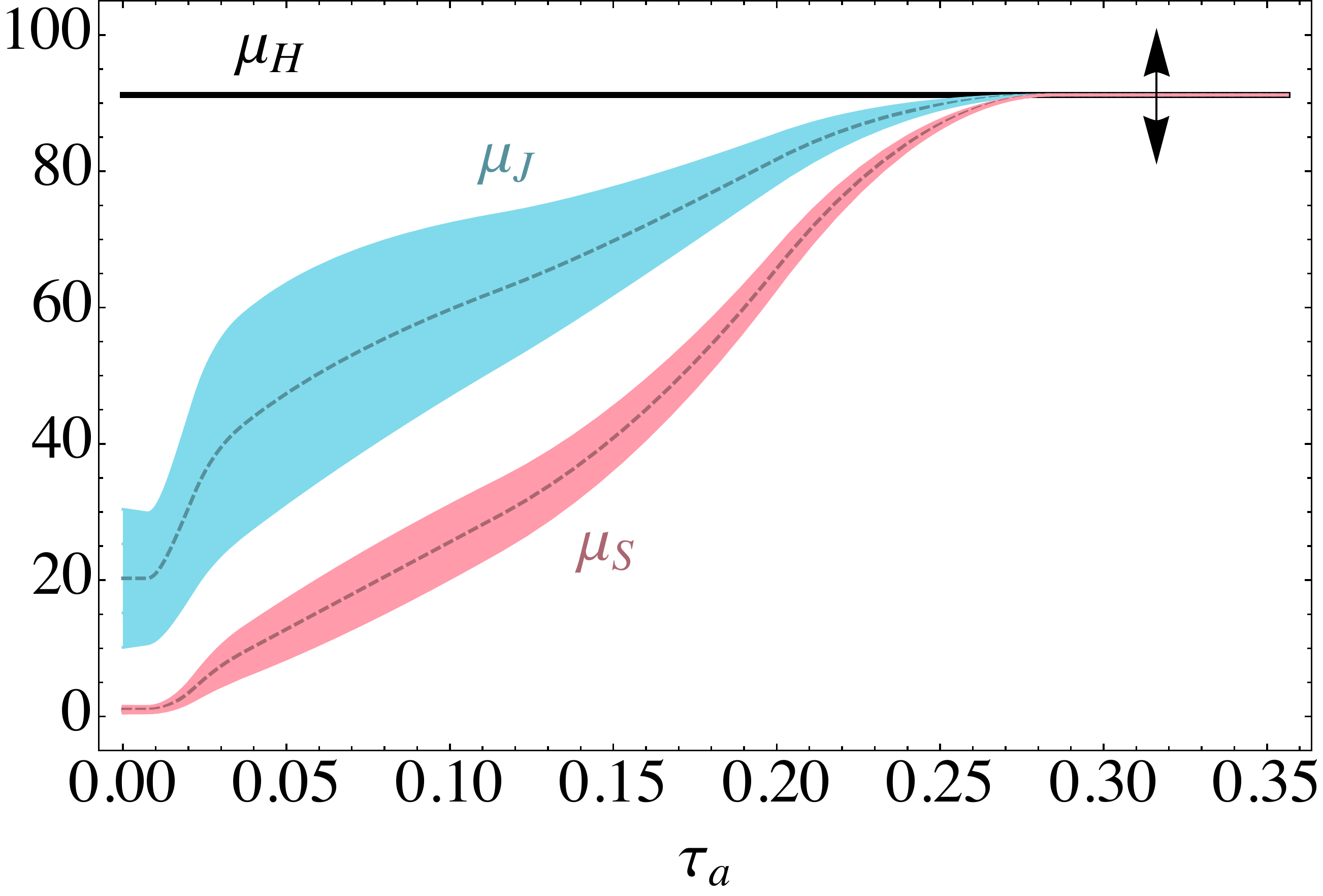} &
\includegraphics[width=0.45\columnwidth]{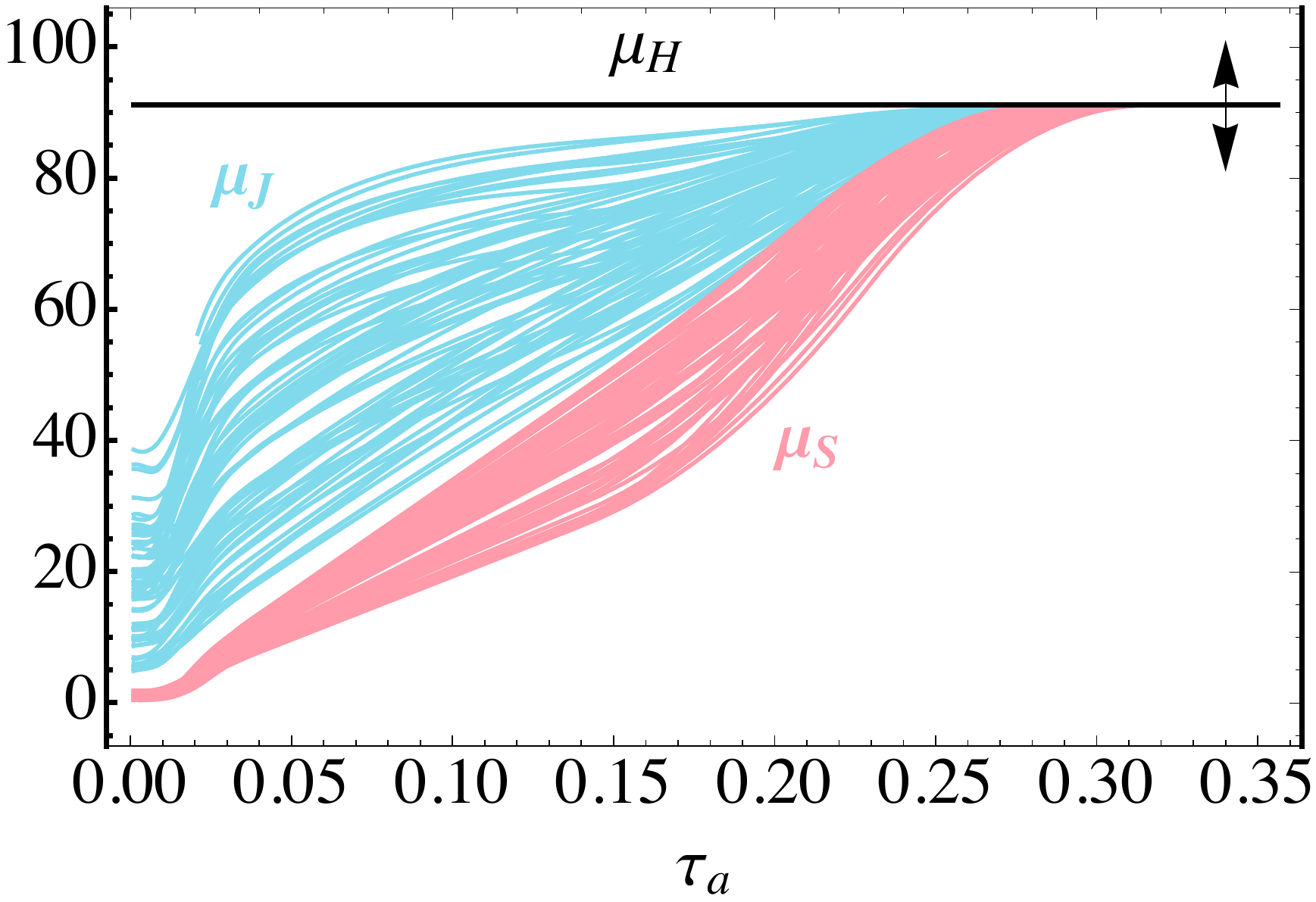} \\
\rotatebox{90}{\qquad\quad\quad$a=-0.5$} &
\includegraphics[width=0.45\columnwidth]{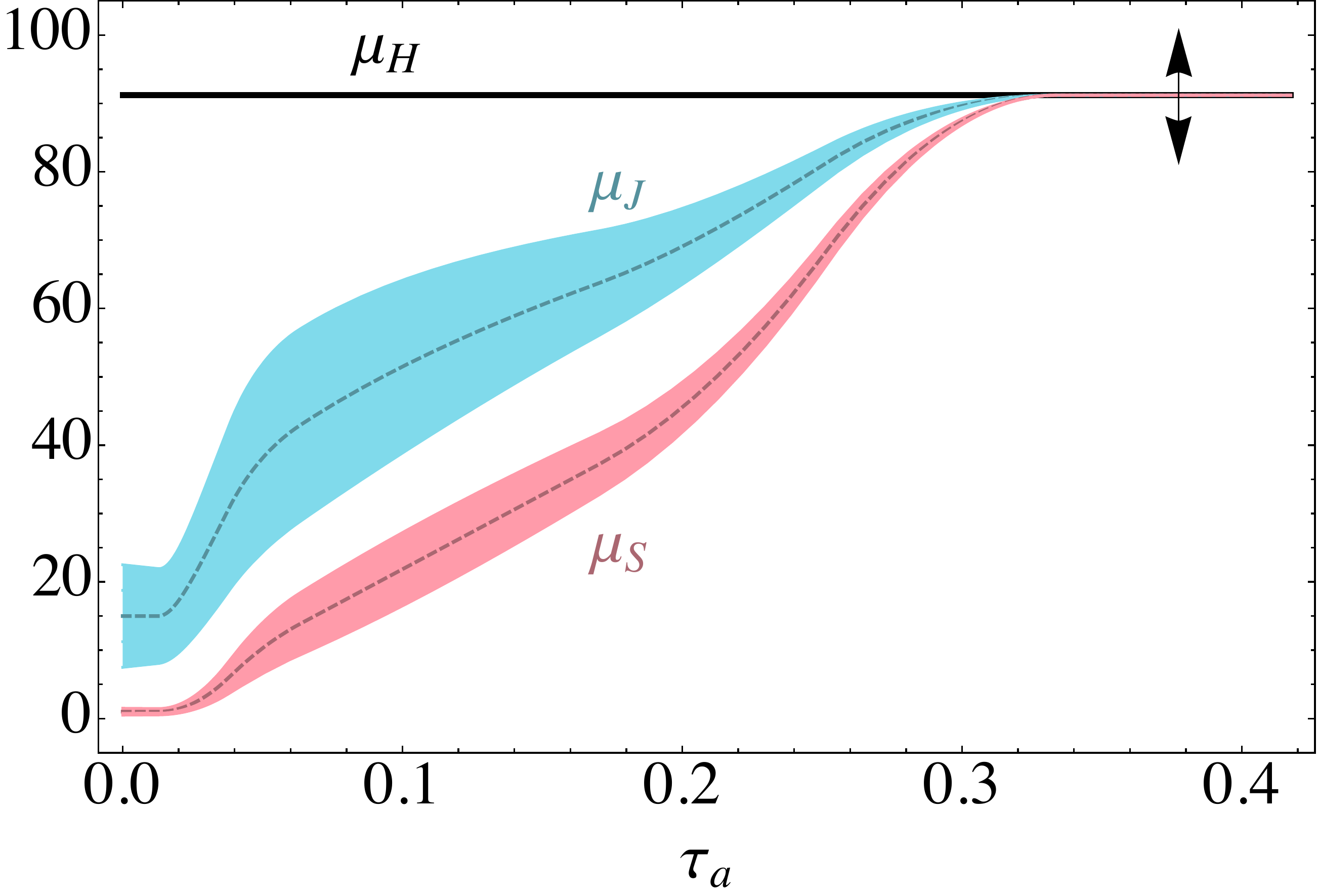} &
\includegraphics[width=0.45\columnwidth]{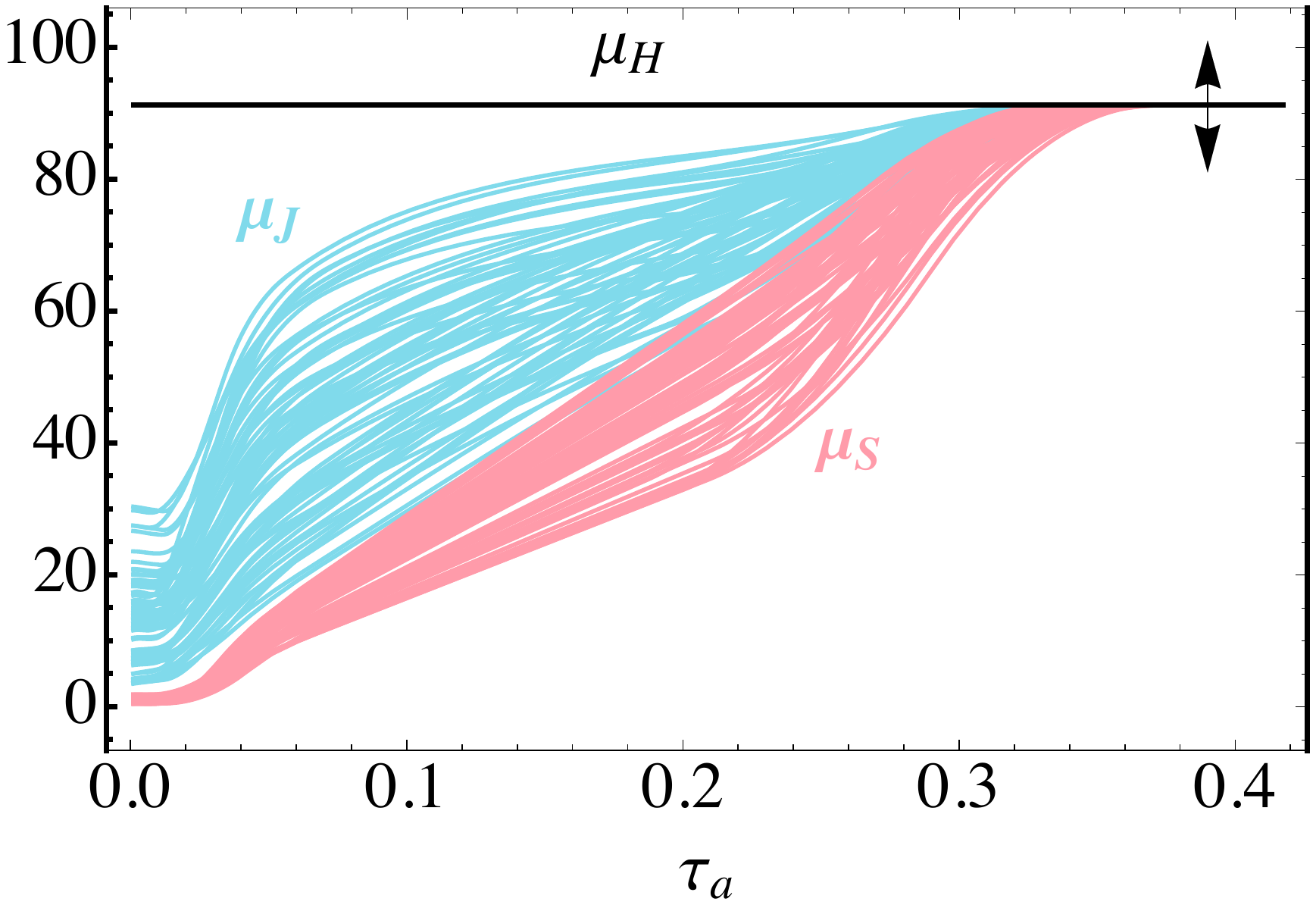} \\
\rotatebox{90}{\qquad\qquad\quad$a=0$} &
\includegraphics[width=0.45\columnwidth]{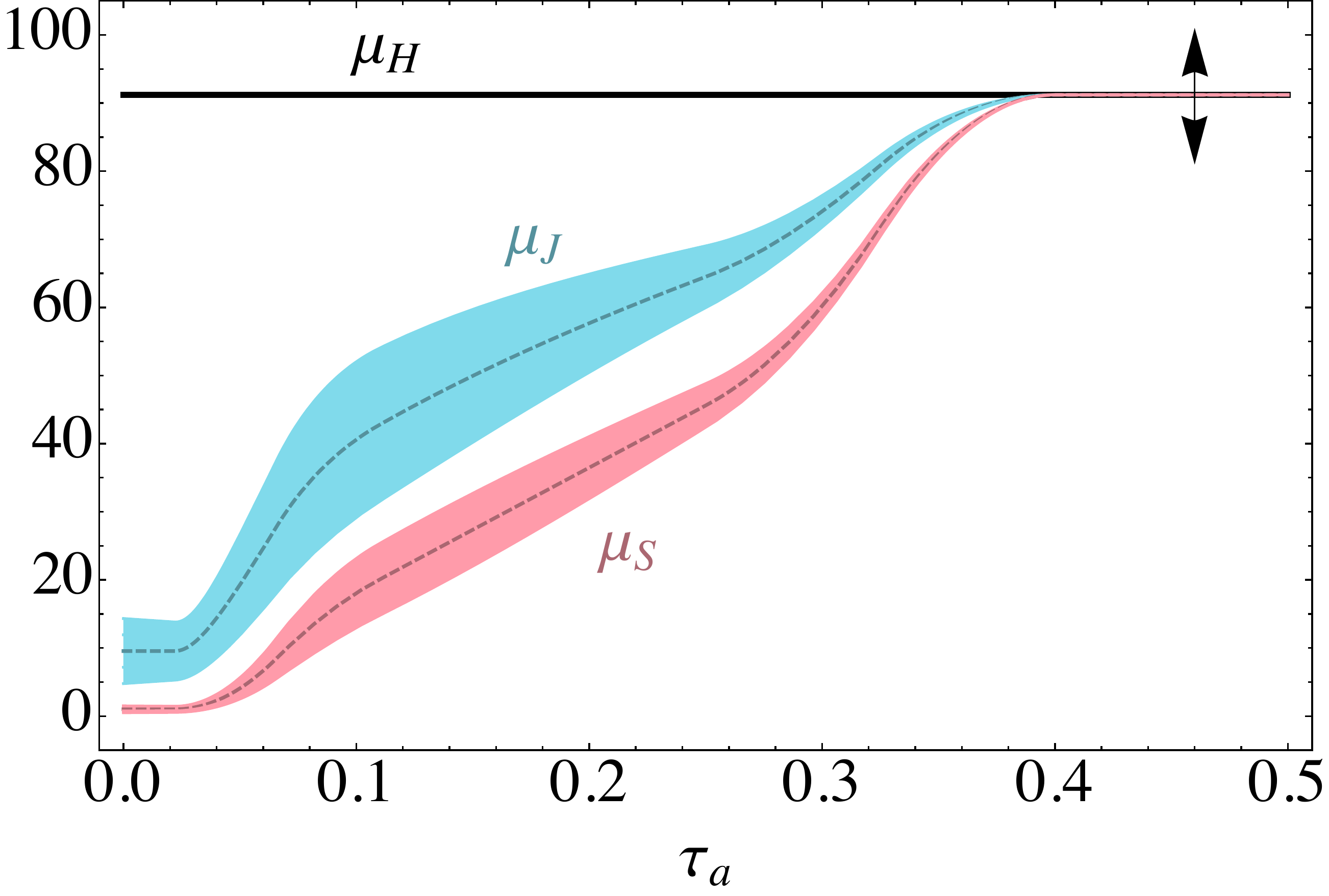} &
\includegraphics[width=0.45\columnwidth]{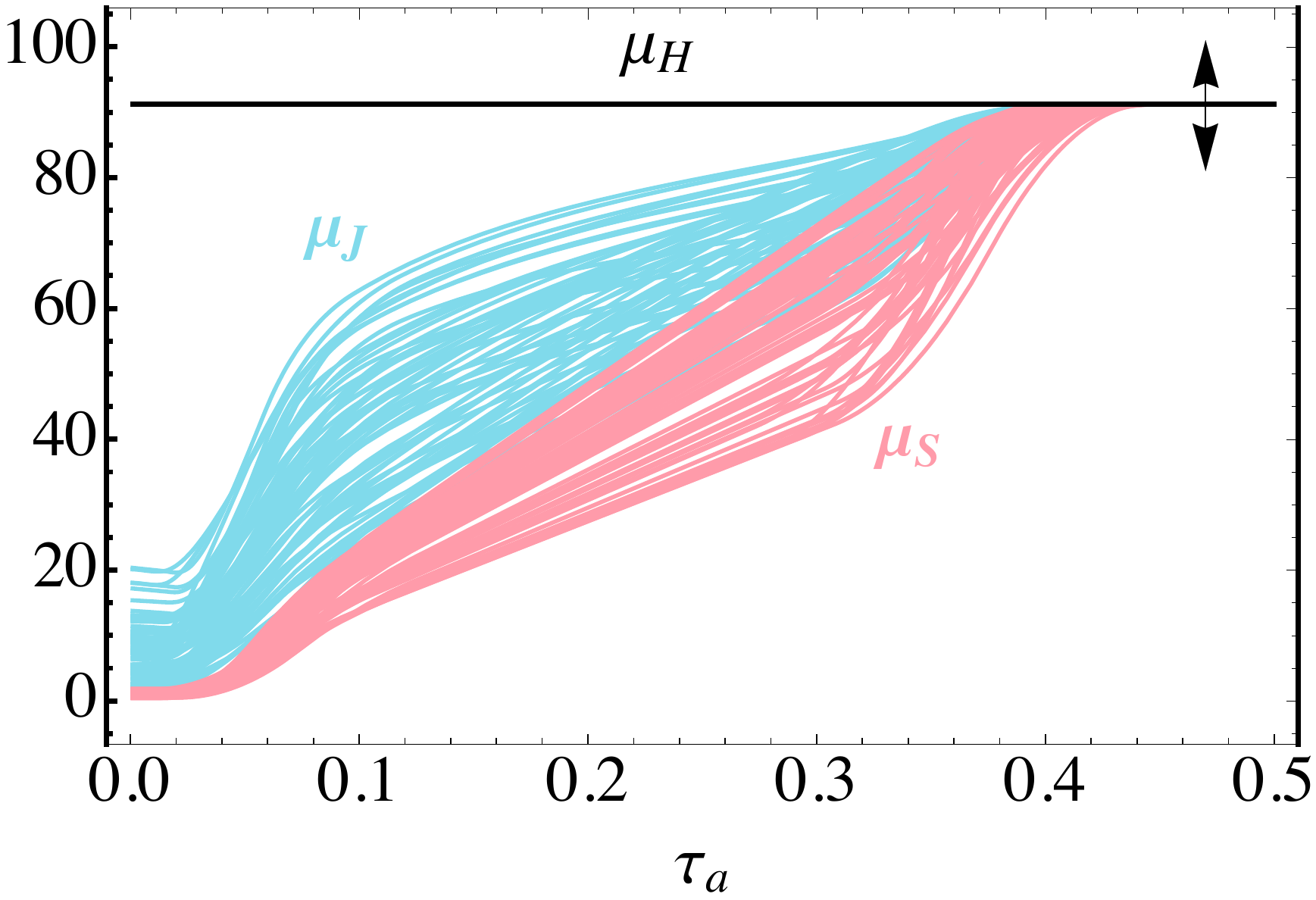} \\
\rotatebox{90}{\qquad\qquad\quad$a=0.5$} &
\includegraphics[width=0.45\columnwidth]{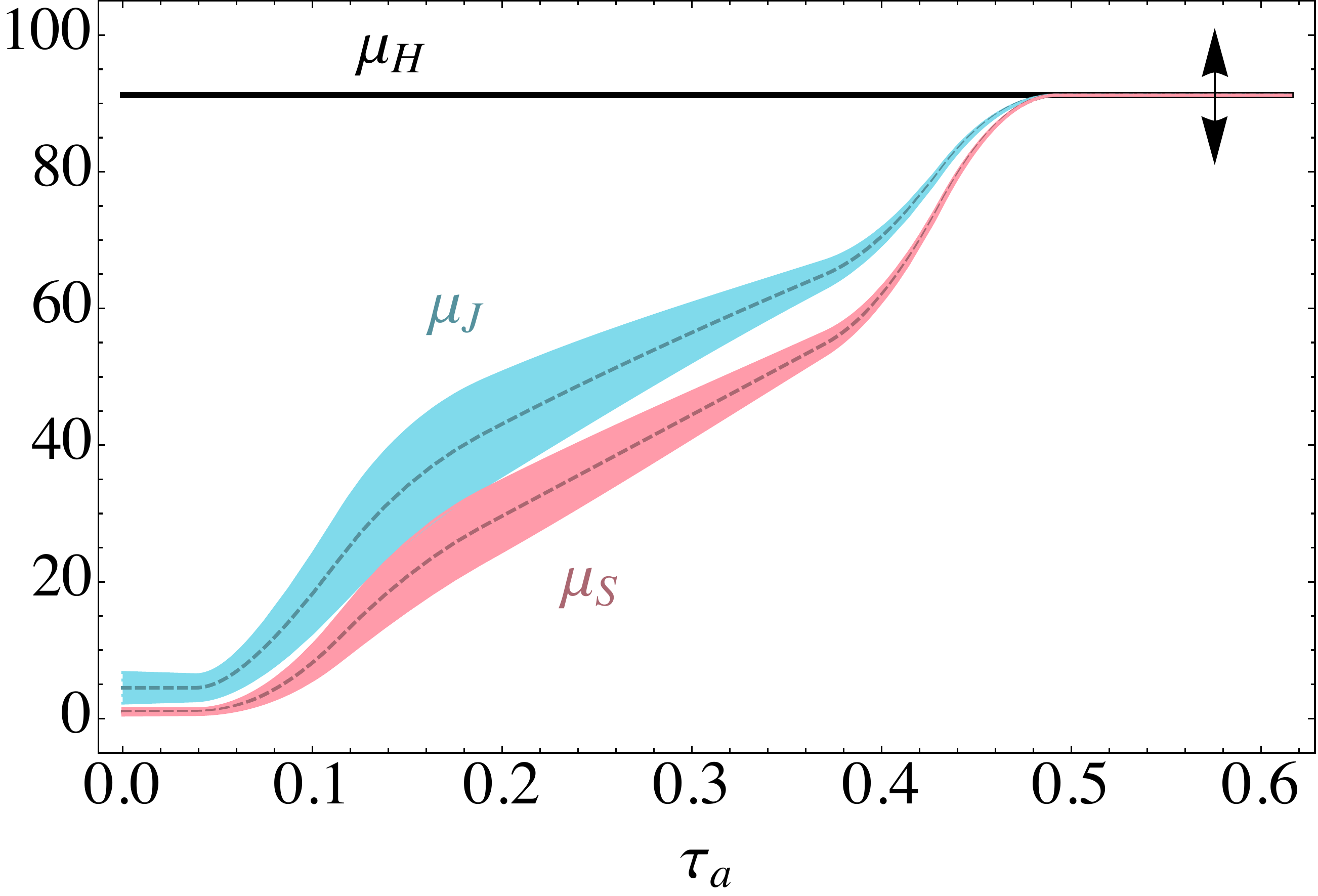} &
\includegraphics[width=0.45\columnwidth]{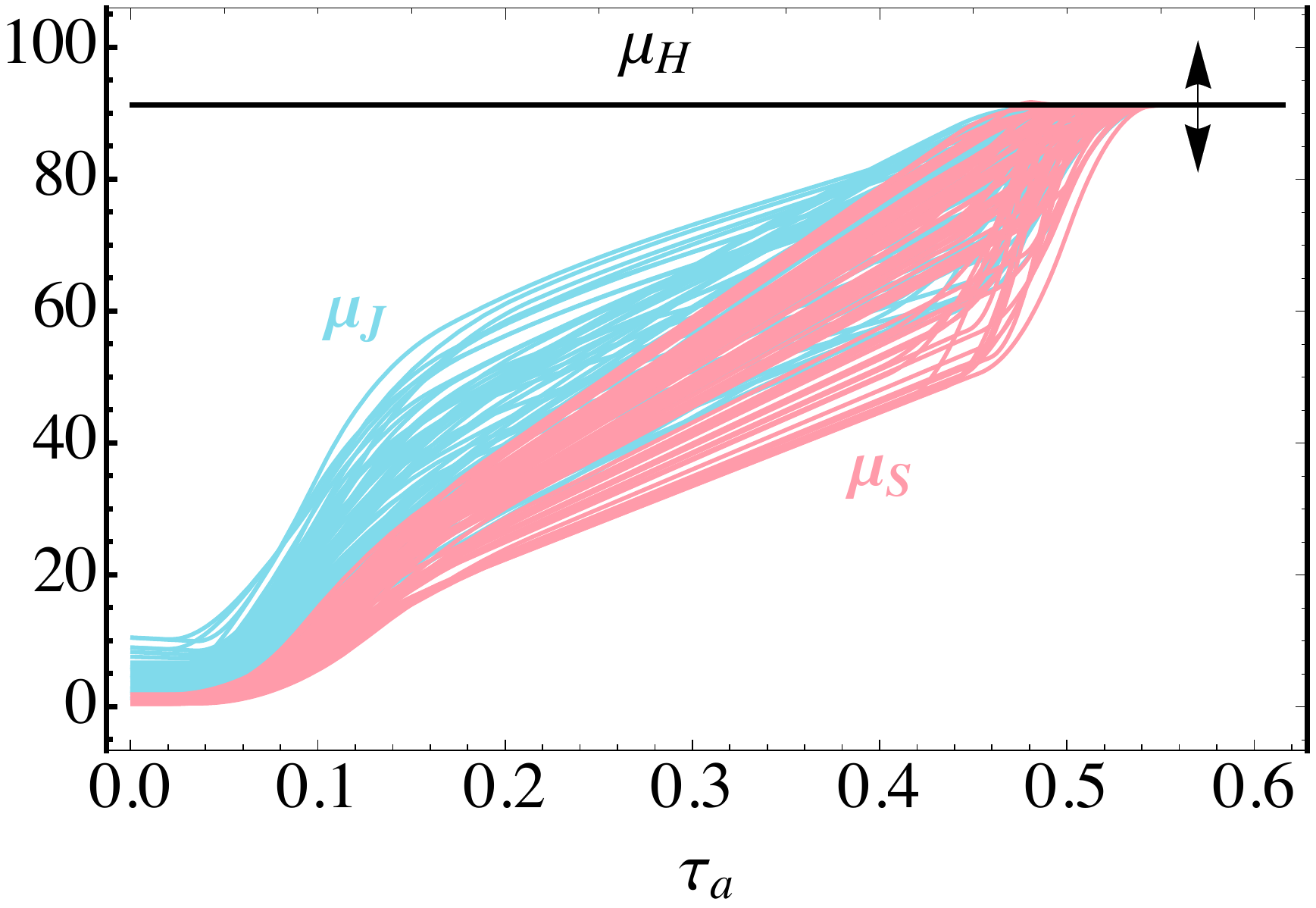}
\end{tabular}
\vspace{-1em}
\caption{Profile functions implemented in our resummation, shown for 
$a = \lbrace -1.0, -0.5, 0.0, 0.5 \rbrace$ and $Q = m_{Z}$.  
\emph{Left}: Band method.
\emph{Right}: Scan method.
Each set of profiles $\mu_{H,J,S}$ actually has a different absolute vertical scale set by $\mu_H = e_H Q$, 
which has been rescaled for this illustration to $e_H=1$. Also, for each individual set of scales,  
it is the case that $\mu_S<\mu_J$, although the overall bands from the scan overlap.}
\vspace{-1em}
\label{fig:ScalePlot}
\end{figure}
%------------------------------------------------

In the left panel of \fig{ScalePlot}, we illustrate the width of the jet- and soft-scale variations 
for $a = \lbrace -1.0,\, -0.5,\, 0.0,\, 0.5 \rbrace$ and $Q=m_{Z}$. 
One clearly observes the leveling off in the low $\tau_{a}$ region, the natural behavior in the mid  
$\tau_{a}$ region, and finally the convergence of all three scales in the far-tail region.  Again, the 
coloured bands represent independent variations of the jet and soft scales through 
$e_{J,S} \in \lbrace -0.5, 0.0, 0.5 \rbrace$.  The third 
hard-scale variation then corresponds to shifting the overall scale of the plots in  
\fig{ScalePlot} up and down, and we do so through $Q/2 \le \mu_H \le 2 Q$.  

The last three variations are not illustrated in \fig{ScalePlot}. The nonsingular scale variation is designed 
to probe missing fixed-order terms not accounted for in our calculation, and the variation of 
$c_{\tilde J}^{2}$ and $r^{2}$ estimate the systematic uncertainty of our {\tt{EVENT2}} extraction of both 
the two-loop jet-function constant and
the $\mathcal{O}(\alpha_{s}^{2})$ remainder function.  For the former, differences from central values are 
obtained by maximizing and minimizing the cross section 
over the listed uncertainties of $c_{\tilde J}^{2}$ in Table \ref{tab:cJ2}. For the latter, differences from 
central values are obtained by varying $r^{2}(\tau_a)$ across the 1-$\sigma$ error bands obtained in the 
{\tt{NonLinearModelFit}} described in Section 
\ref{sec:remainder} (these procedures correspond to setting the parameters $\delta c^{2}_{\tilde{J}}$ and $\delta r^{2}$ to $\pm1$ in \eqs{cj2def}{r2def} below).  The errors induced by these variations are minimal in comparison to the other four 
variations.

%%%%%%%%%%%%%%%%%%%%%%%%%%%%%%%%%%%%%%%%%%%%%%%%%
\subsubsection*{Scan method}

For the scan method we take 64 random selections of profile and scale parameters within the ranges 
shown in \tab{scan}. Note that in this method we have chosen not to vary $e_S$ away from 0, thus allowing 
the variations of the other relevant parameters (e.g.~$n_0,n_1,\mu_0,r$) to probe the relevant range of the 
soft scale $\mu_S$.
Note also that we fix $R_0$ to always be \mbox{($\mu_0-0.4$~GeV)} so that while $\mu_0$ varies, the difference 
between $\mu_0$ and $R_0$ remains fixed, to preserve the scheme choice for the gap parameter 
$\Delta_a$ (cf. \cite{Abbate:2010xh,Hoang:2014wka}). The parameters $\delta c_{\tilde J}^2$ and 
$\delta r^2$ are defined to probe the uncertainties coming from scanning over the 1-$\sigma$ ranges of 
$c_{\tilde J}^2$ and $r^2(\tau_a)$ determined in \sec{CONSTANTS}. That is, we take
\be
\label{eq:cj2def}
c_{\tilde J}^2 = c_{\tilde J}^{2,\text{central}} + 
\begin{cases}
\delta c_{\tilde J}^2 \; \Delta c_{\tilde J}^{2,\text{upper}} & \delta c_{\tilde J}^2 > 0 \\
\delta c_{\tilde J}^2 \;  \Delta c_{\tilde J}^{2,\text{lower}} & \delta c_{\tilde J}^2 <0
\end{cases} \,,
\ee
where ``central, upper, lower'' refer to the central values, upper, and lower uncertainties on 
$c_{\tilde J}^2$ given in \tab{cJ2}.   Similarly,
\be
\label{eq:r2def}
r^2(\tau_a) = r^{2,\text{central}}(\tau_a) + \delta r^2\;\Delta r^2(\tau_a)\,,
\ee
where $r^{2,\text{central}}$ and $\Delta r^2$ are the central values and 1-$\sigma$ uncertainty function 
illustrated in \fig{r2all}.

In the right panel of \fig{ScalePlot}, we show the analogous width of the jet- and soft-scale variations 
probed with the scan method. As is clear, the error bands between the two methods are of a similar form 
and they exhibit all of the qualities we demand of the profile functions across the whole $\tau_{a}$ domain.  
Given the current scan ranges and parameter choices, the errors in the scan method indeed appear to be 
larger than those of the band method, 
but this is somewhat compensated for by taking the simple envelope of resulting variations rather than
adding in quadrature.  In the next section we will apply both methods to 
our final predictions for the resummed and matched cross sections,
and we will compare the quality of their convergence both for integrated and differential distributions.

\section{Results and data comparison}
\label{sec:ANGRESULTS}

%----------------------------------------------------------------------------
\begin{figure}[t]
\centering
{ \hspace{.07\columnwidth} \large\bf Band Method \hspace{.28\columnwidth} Scan Method} \\[.5em]
\begin{tabular}{ccc}
\rotatebox{90}{\qquad\qquad\quad$a=-1$} &
\includegraphics[width=.45\columnwidth]{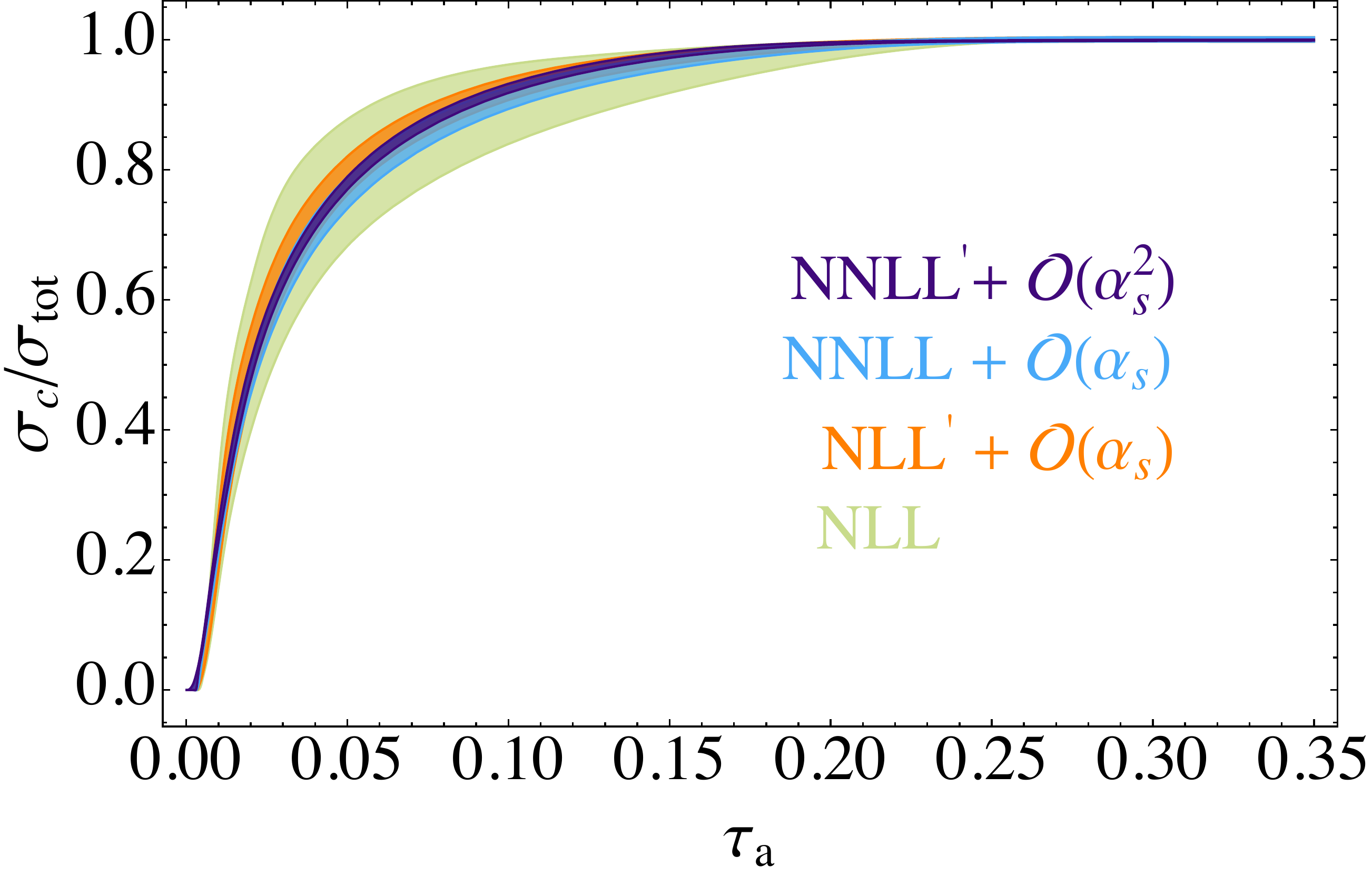} &
\includegraphics[width=.45\columnwidth]{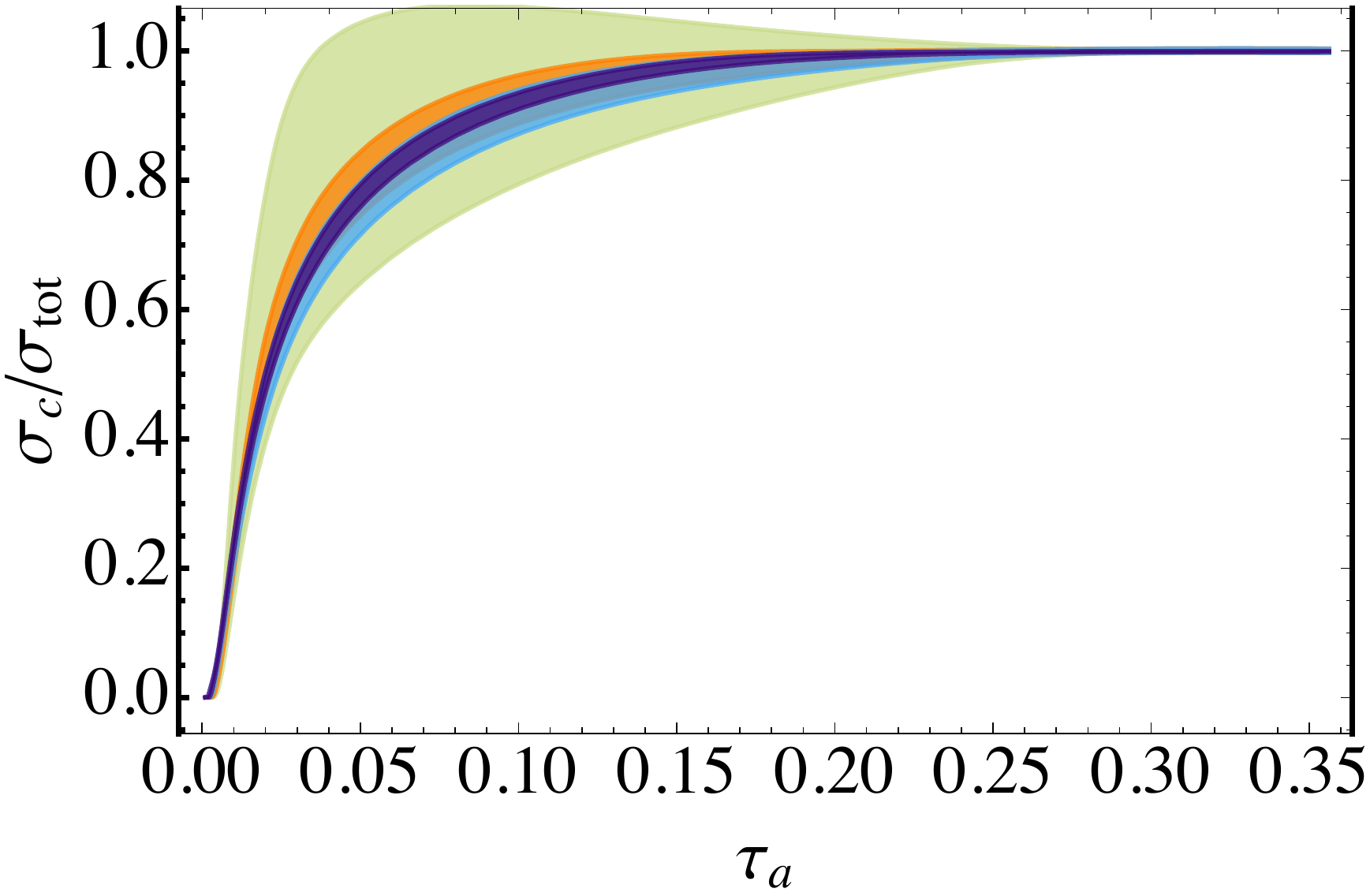} \\
\rotatebox{90}{\qquad\qquad\quad$a=-0.5$} &
\includegraphics[width=.45\columnwidth]{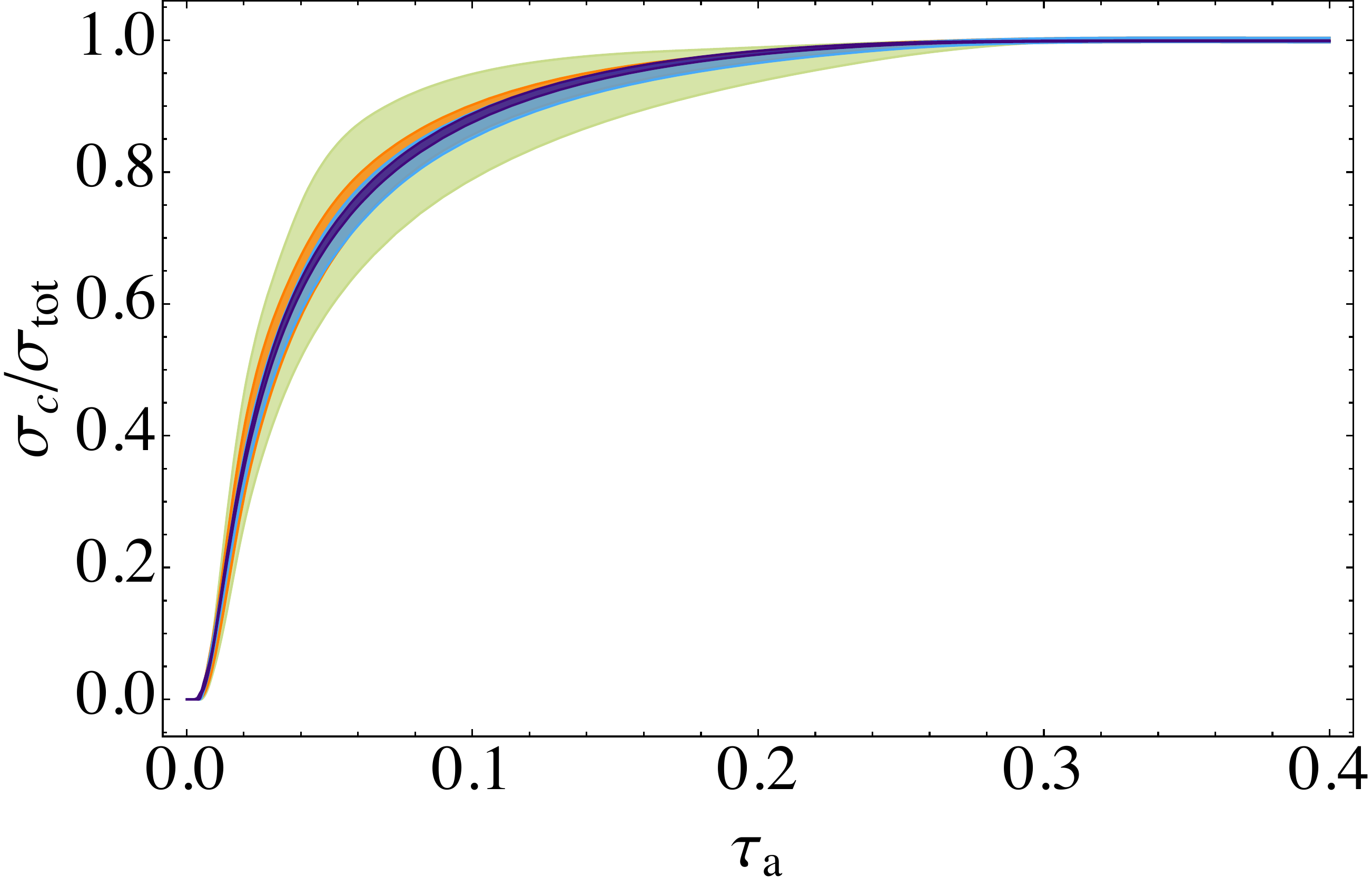} &
\includegraphics[width=.45\columnwidth]{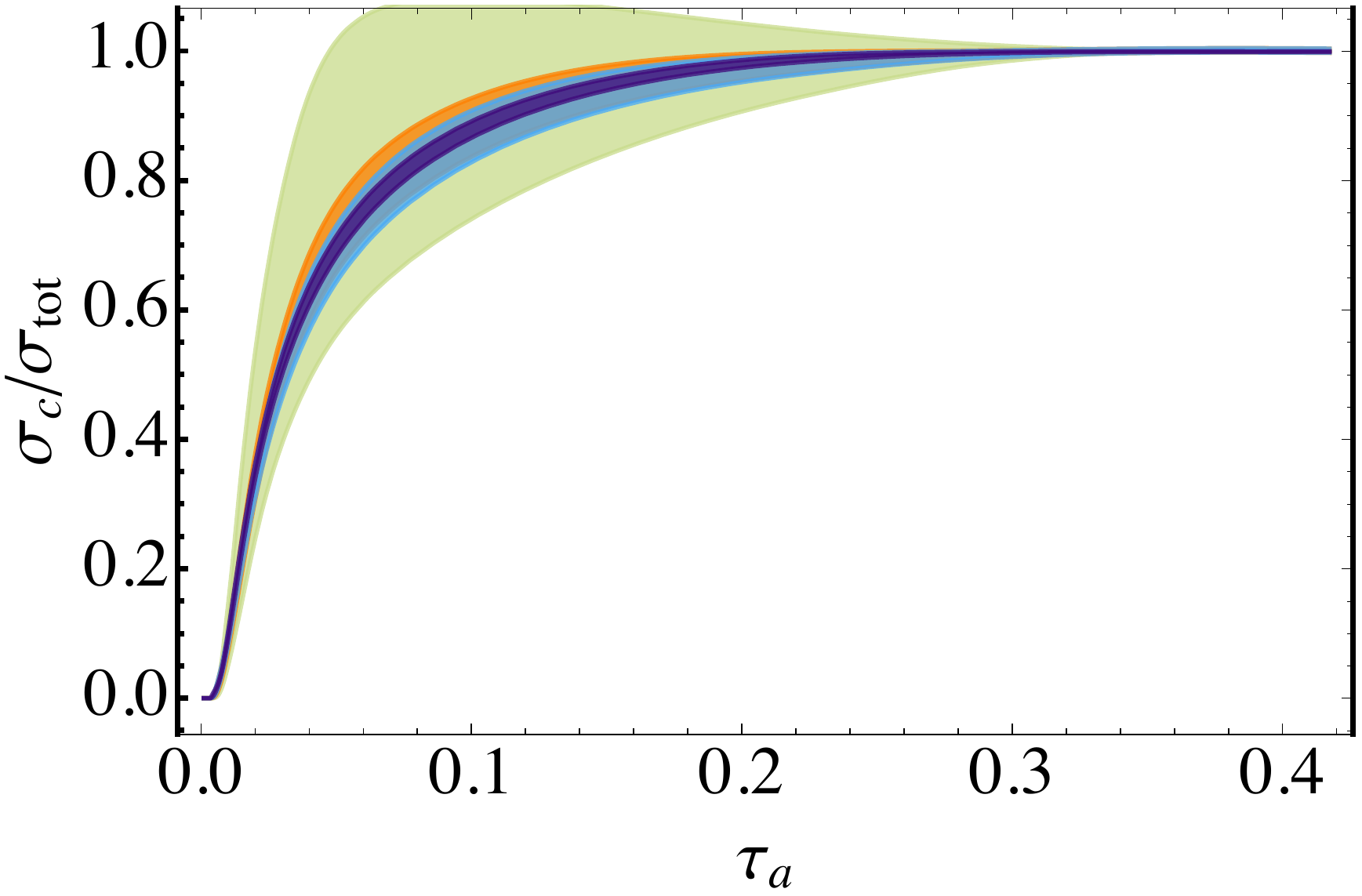} \\
\rotatebox{90}{\qquad\qquad\quad$a=0$} &
\includegraphics[width=.45\columnwidth]{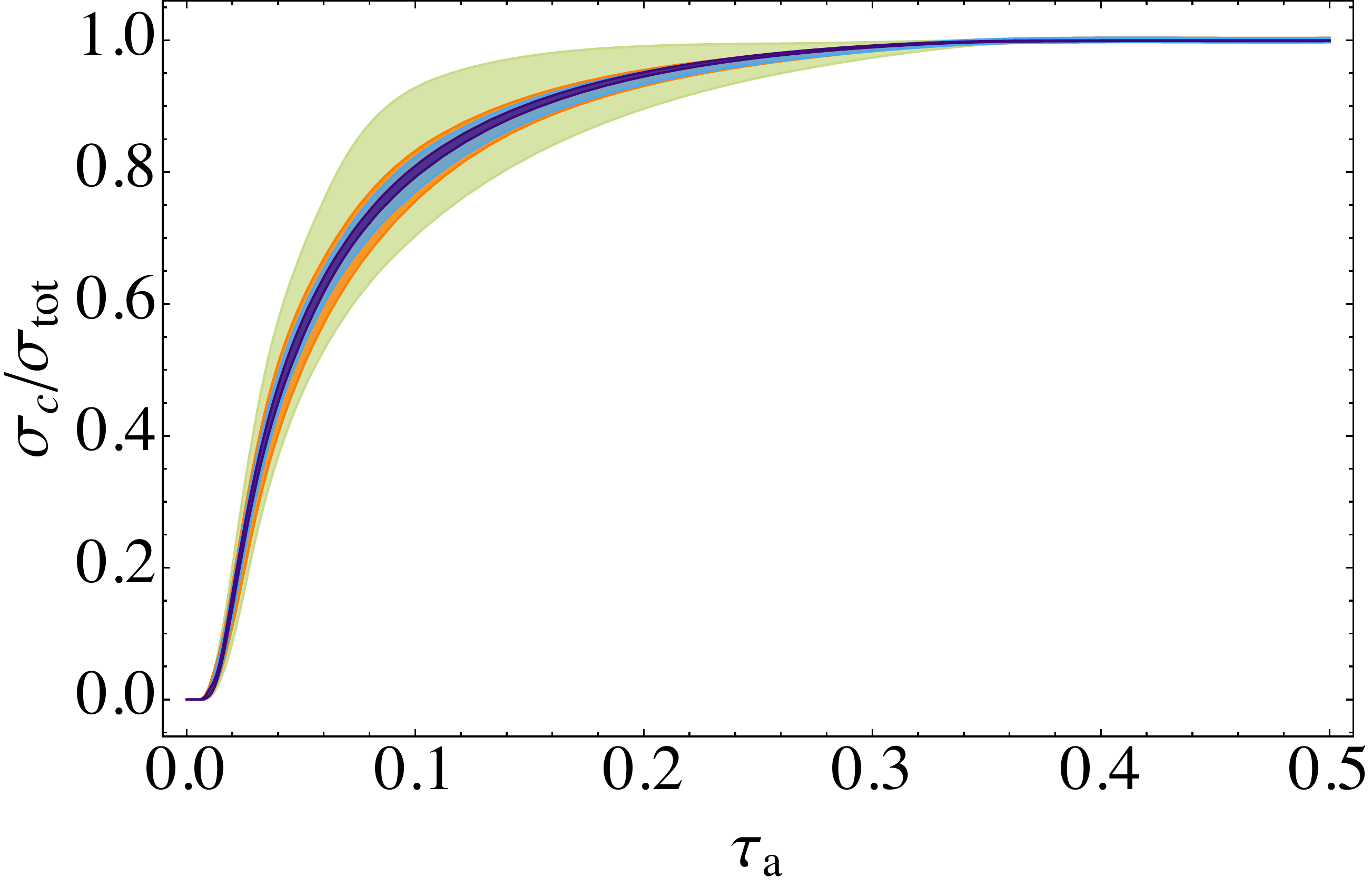} &
\includegraphics[width=.45\columnwidth]{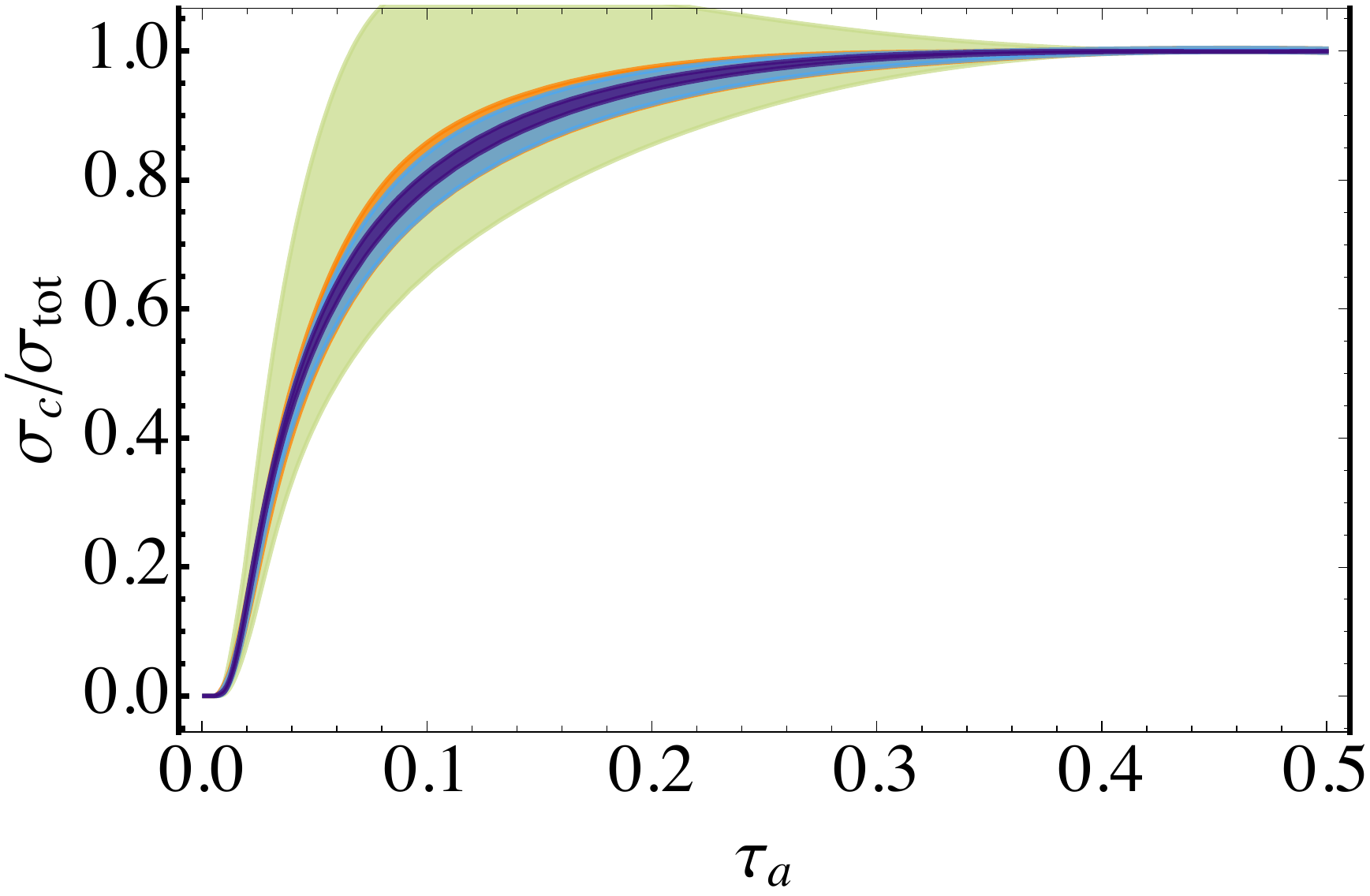} \\
\rotatebox{90}{\qquad\qquad\quad$a=0.5$} &
\includegraphics[width=.45\columnwidth]{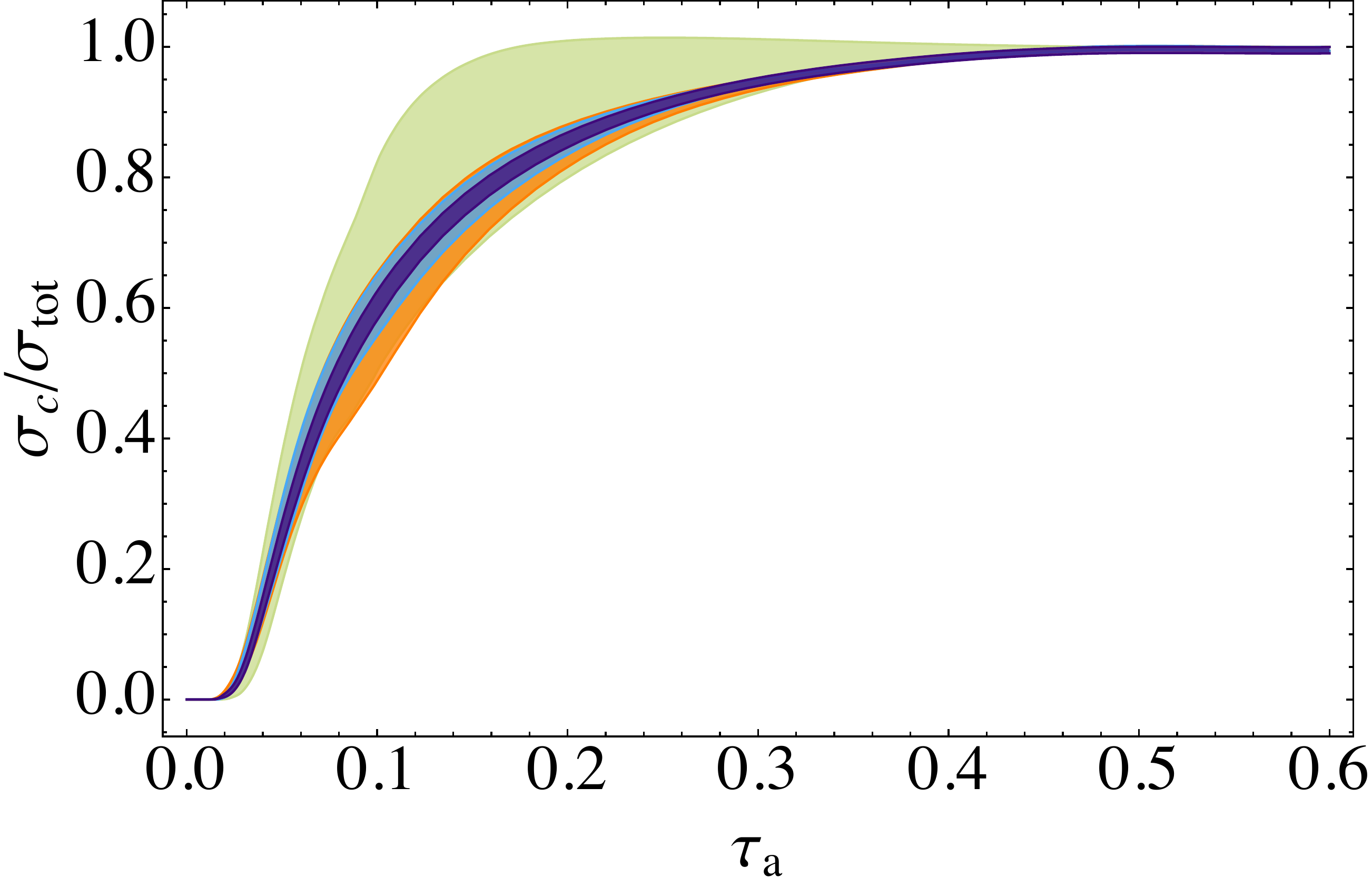} &
\includegraphics[width=.45\columnwidth]{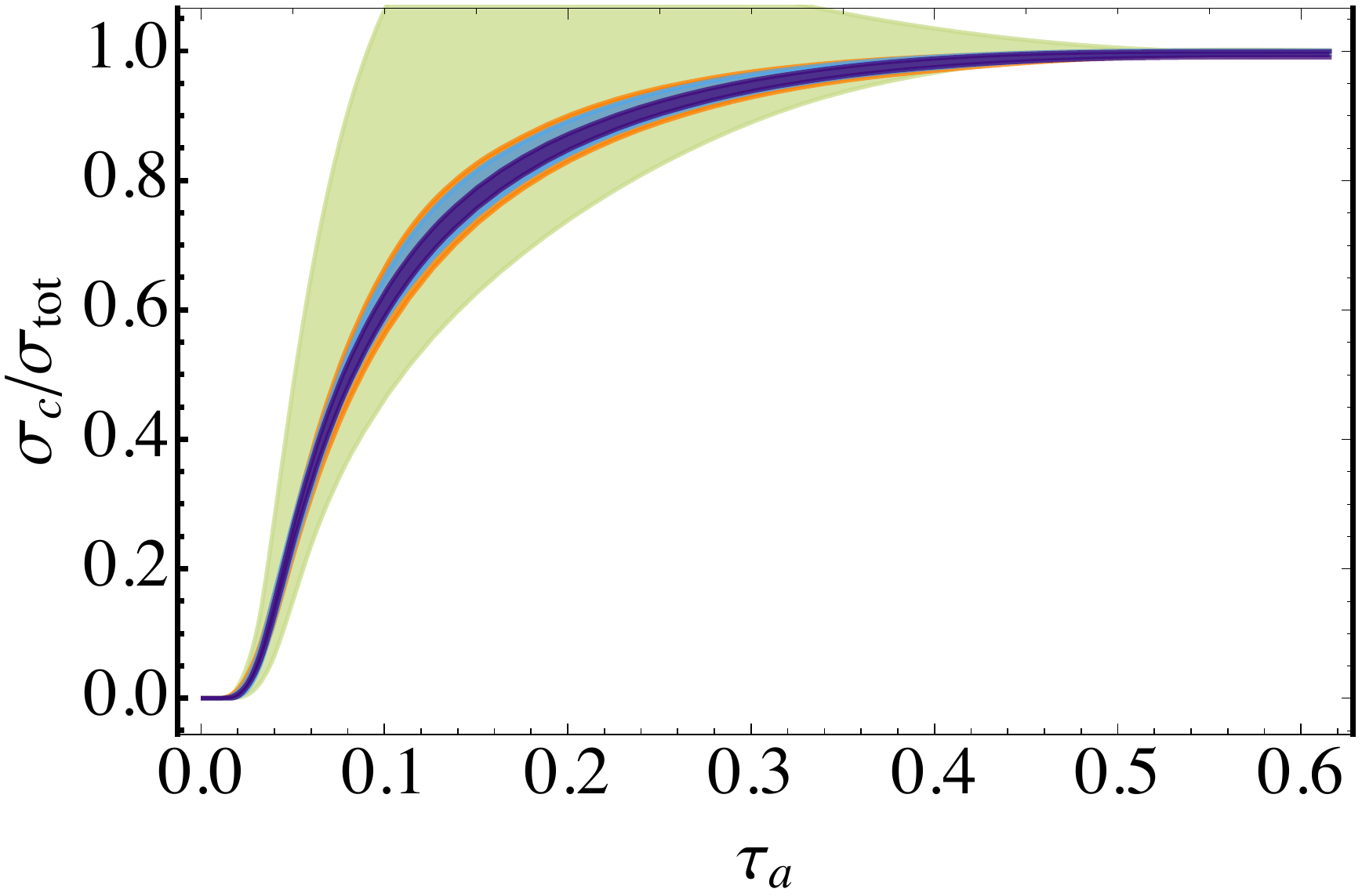}
\end{tabular}
\vspace{-1em}
\caption{
Integrated angularity distributions for four values of 
$a=\{-1.0,-0.5,0.0,0.5\}$ and $Q=m_Z$ at NLL (green), NLL$^{\prime}+\cO(\as)$ (orange), NNLL$+\cO(\as)$ (blue), 
and NNLL$^{\prime}+\cO(\as^2)$ (purple) accuracy, with renormalon subtractions to the corresponding orders.
The theoretical uncertainties have been estimated with the
band method (left) and the scan method (right) as discussed in \sec{scalevariations}.
}
\label{fig:cumulantconvergence}
\end{figure}
%-----------------------------------------------------------------------------

%----------------------------------------------------------------------------
\begin{figure}[t]
\centering
\includegraphics[width=.48\columnwidth]{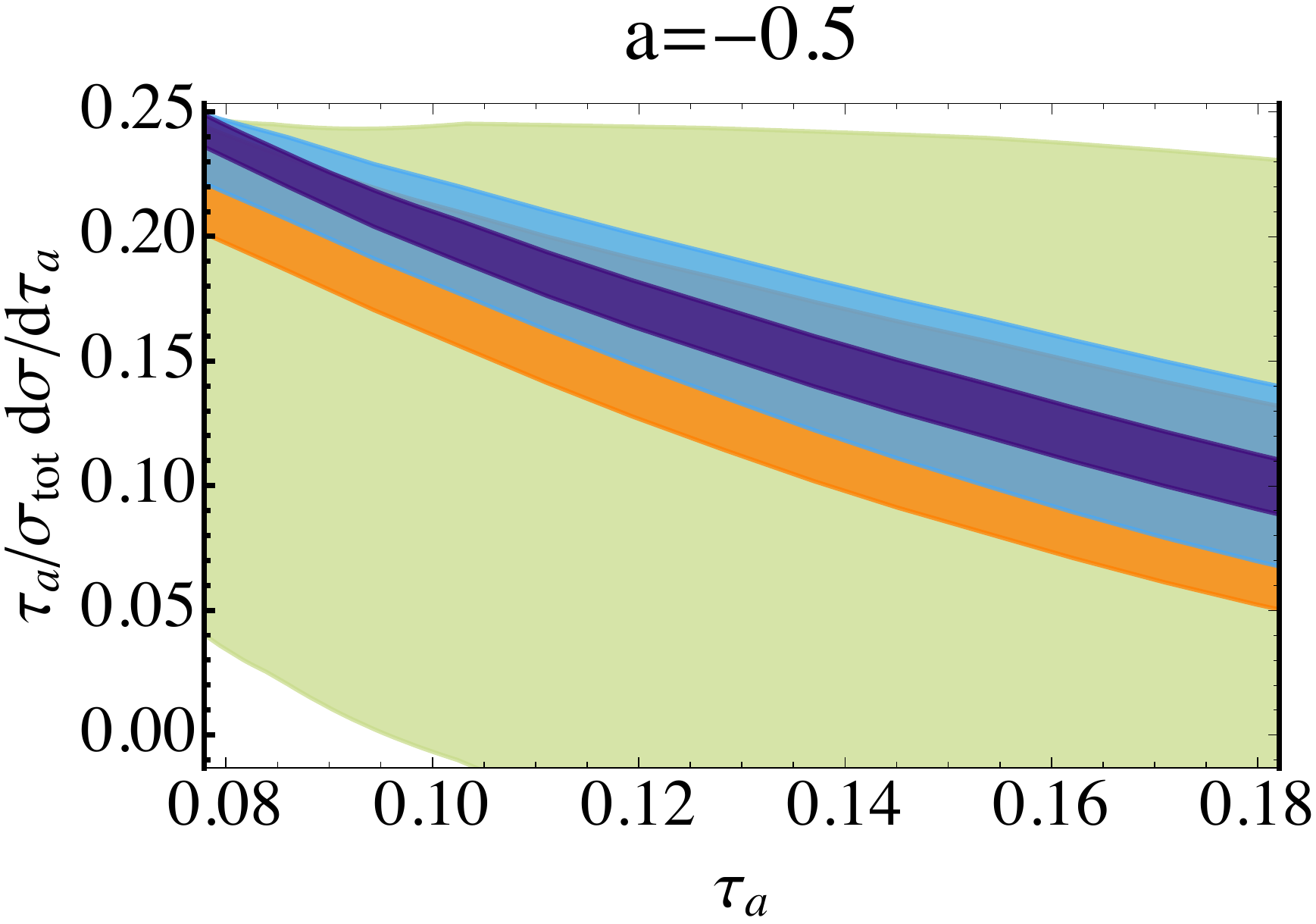} \quad
\includegraphics[width=.48\columnwidth]{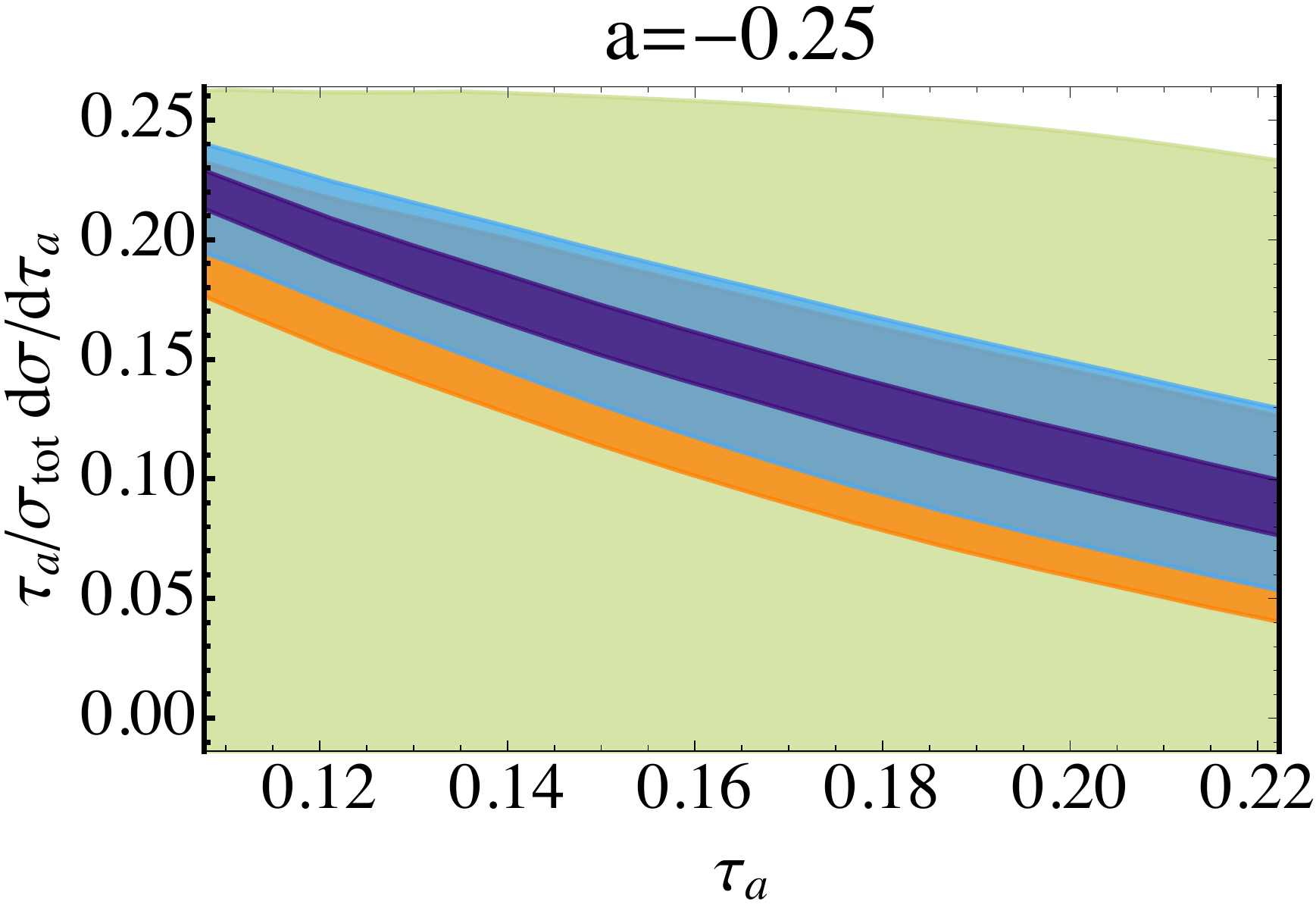} \\[1ex]
\includegraphics[width=.48\columnwidth]{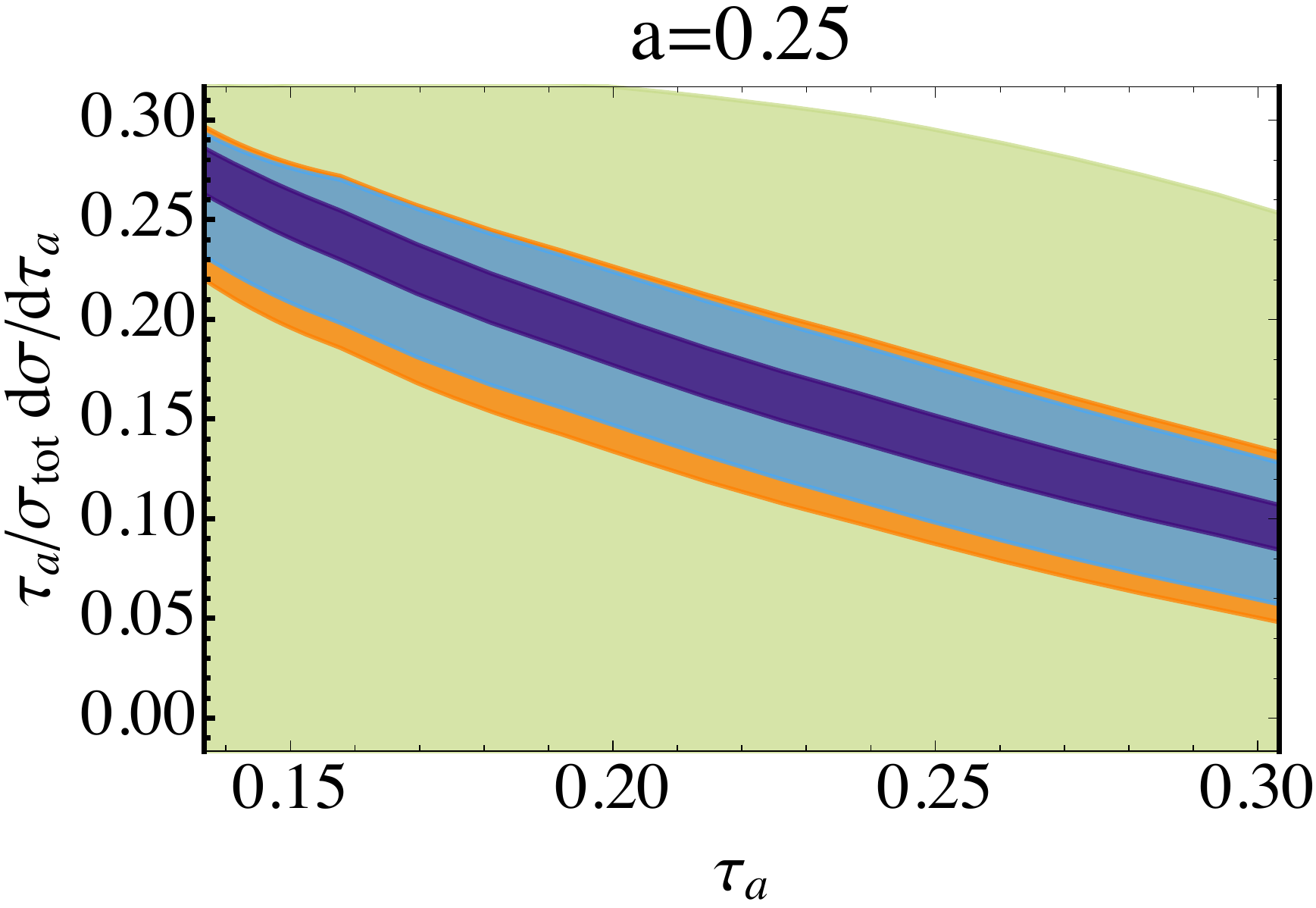} \quad
\includegraphics[width=.48\columnwidth]{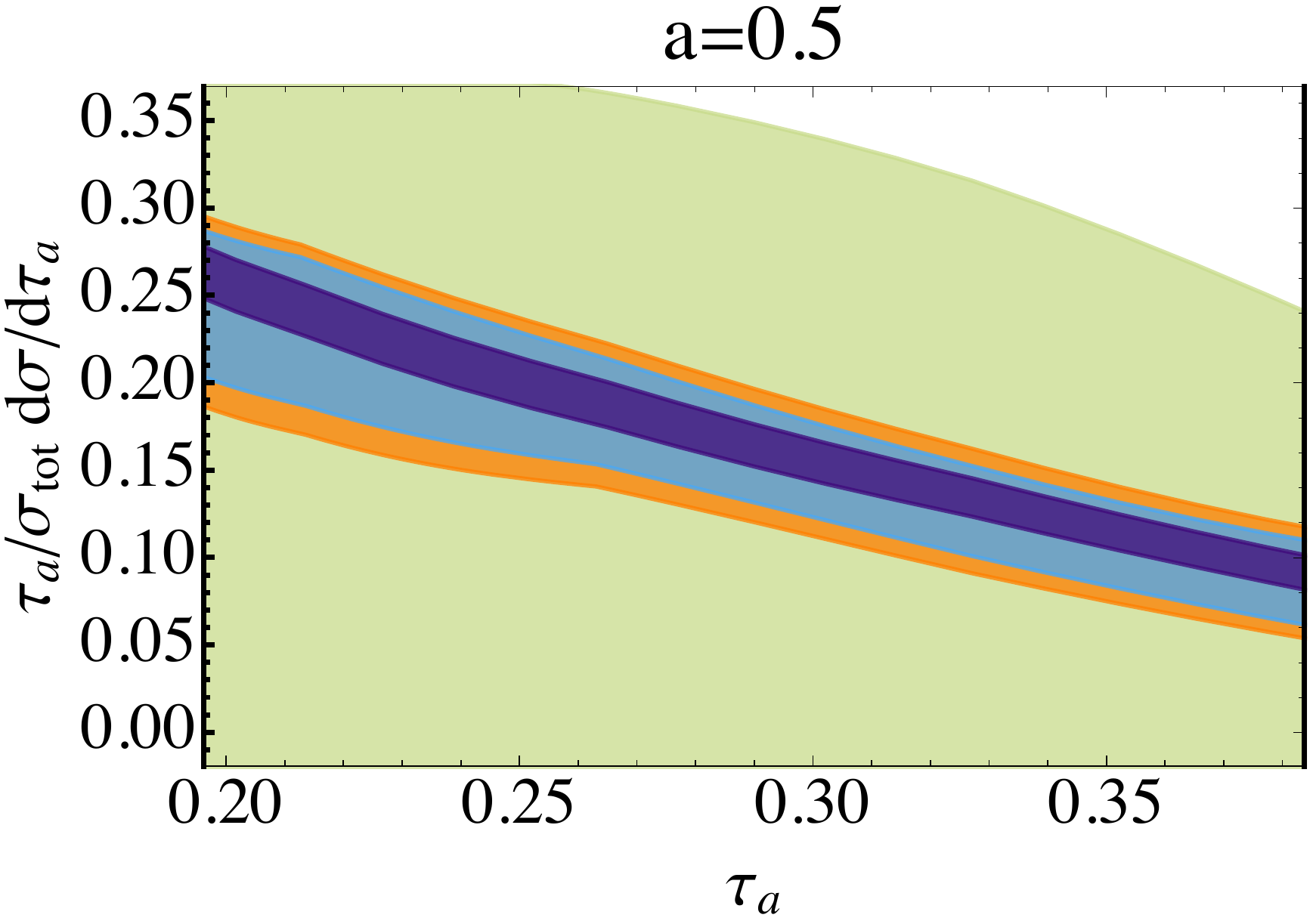}
\vspace{-2em}
\caption{
Differential angularity distributions 
for four values of 
$a=\{-0.5,-0.25,0.25,0.5\}$ and $Q=m_{Z}$ in the central $\tau_a$ region, resummed and matched to NLL (green), NLL$^{\prime}+\cO(\as)$ (orange), NNLL$+\cO(\as)$ (blue), 
and NNLL$^{\prime}+\cO(\as^2)$ (purple) accuracy, with renormalon subtractions to the corresponding orders, and 
uncertainties estimated with the scan method.
}
\label{fig:convergence}
\end{figure}
%-----------------------------------------------------------------------------

We now present our predictions for the angularity distributions for seven values of
the angularity parameter 
$a \in \lbrace-1.0,\,-0.75,\,-0.5,\,-0.25,\,0.0,\,0.25,\,0.5\rbrace$
at $Q = 91.2$ GeV and $Q=197$ GeV, and compare them to the L3 data presented 
in~\cite{Achard:2011zz}.\footnote{We do not include the case $a=0.75$ in our study, since 
$a\to 1$ corrections to the \SCETa factorization theorem \eq{cumulant2}
cannot be neglected for such large values of $a$.}
In our analysis we will use the $\overline{\mathrm{MS}}$ coupling constant
$\alpha_{s}(m_{Z}) = 0.11$ and 
the non-perturbative shift parameter, defined through \eq{Omega1sub},
$\Omega_1(R_\Delta,R_\Delta)= 0.4$~GeV
at $R_\Delta = 1.5$~GeV.
These values are chosen to be consistent with the central fit values from \cite{Hoang:2015hka} for $\as(m_Z)$ 
(to two signficant digits)  and $\Omega_1(R_\Delta,R_\Delta)$ (to one signficant digit) at NNLL$'$ accuracy.
Some discussion on the phenomenological impact of choosing different values for these input parameters will be
given in \sec{CONCLUSIONS}.

We first show  in \fig{cumulantconvergence} our predictions for the integrated distributions $\sigma_c(\tau_a)$ 
given by \eqs{convolvedexpanded}{convolvedterms} for four values of $a$ at $Q=m_Z$ up to NNLL$'+\cO(\as^2)$ 
accuracy, including renormalon subtractions. At the same time we
compare the two methods discussed in \sec{scalevariations} to estimate the overall
uncertainties of our analysis. 
The
band method has been applied in the left panel and the scan method in the right panel.
One clearly observes that moving to primed accuracies not only dramatically reduces the scale uncertainties, 
but that also the variations converge across the entire spectra as we move to higher accuracies. 
One also sees that the two methods that have been applied to estimate the theory uncertainties are consistent 
with one another, given the parameter ranges and variations chosen. However, when computing differential 
distributions by taking the derivative of \eq{convolvedexpanded}, we notice a slight improvement in numerical 
stability when using the scan method.
This is partially due to the envelope of the many (64) variations smoothing out wiggles coming from 
derivatives of profile functions $\mu_i(\tau_a)$, especially in transition regions.
\fig{convergence} illustrates this convergence of the differential distribution (multiplied by $\tau_a$) in the central $\tau_{a}$ domain for four values of the angularity parameter $a$. For this reason, and to allow for a more direct comparison of our results to those of \cite{Abbate:2010xh,Hoang:2014wka,Hoang:2015hka}, we implement theory uncertainties as obtained with the 
scan method in the following. 

%----------------------------------------------------------------------------
\begin{figure}[t]
\centering
\begin{tabular}{ccc}
\rotatebox{90}{\qquad\qquad\quad$a=-0.5$} &
\includegraphics[width=.45\columnwidth]{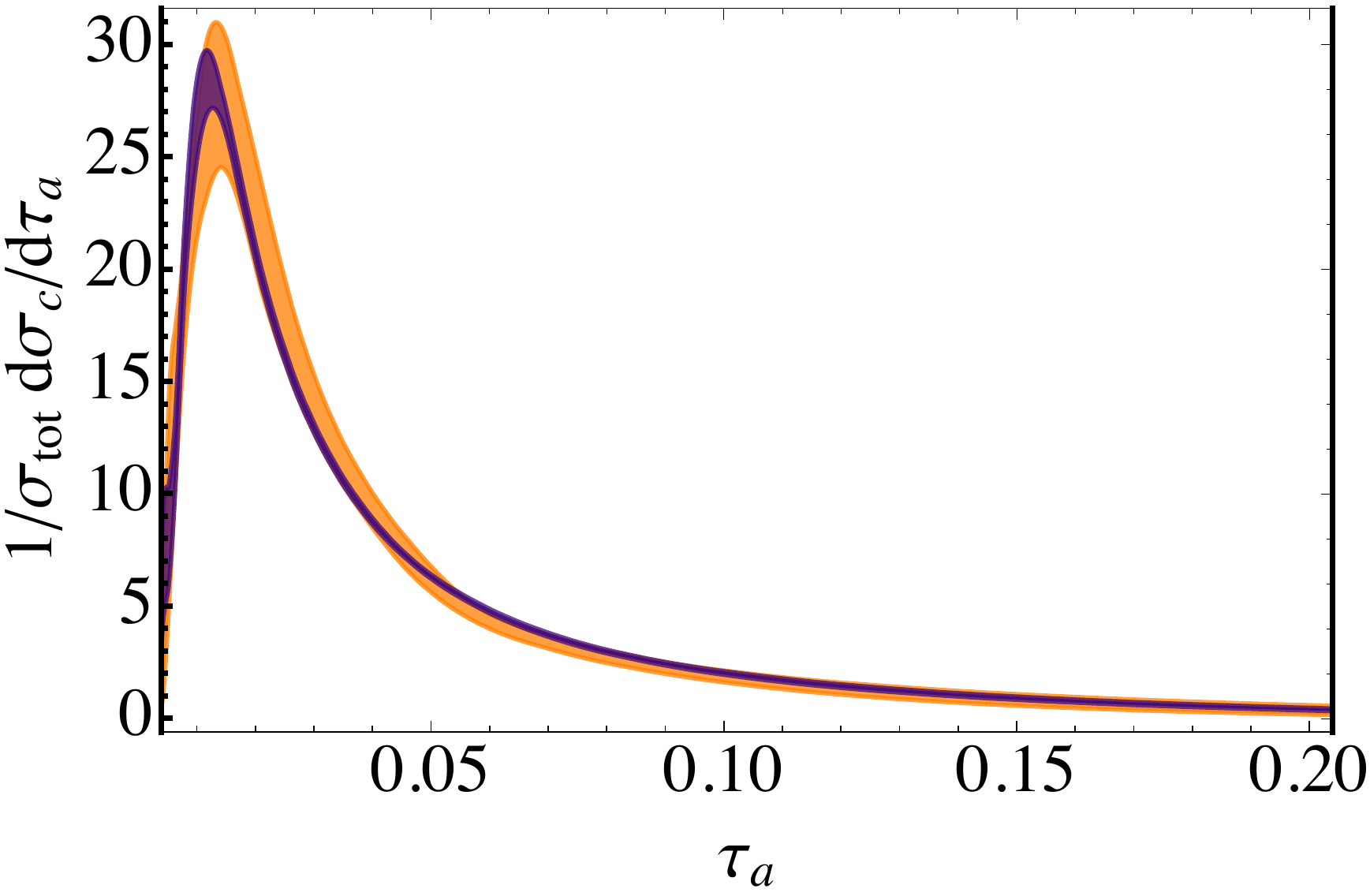} &
\includegraphics[width=.45\columnwidth]{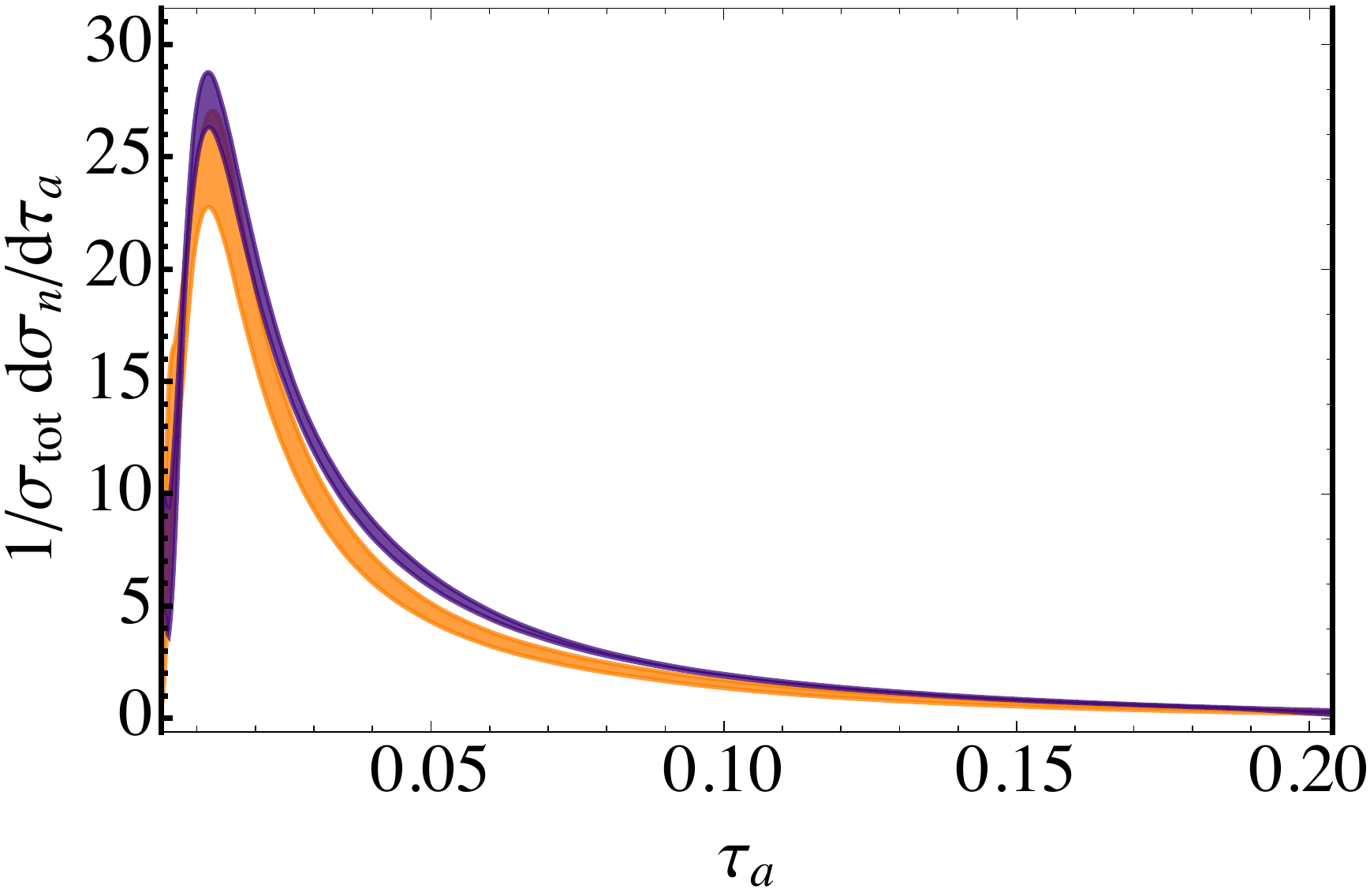} \\
\rotatebox{90}{\qquad\qquad\quad$a=0.25$} &
\includegraphics[width=.45\columnwidth]{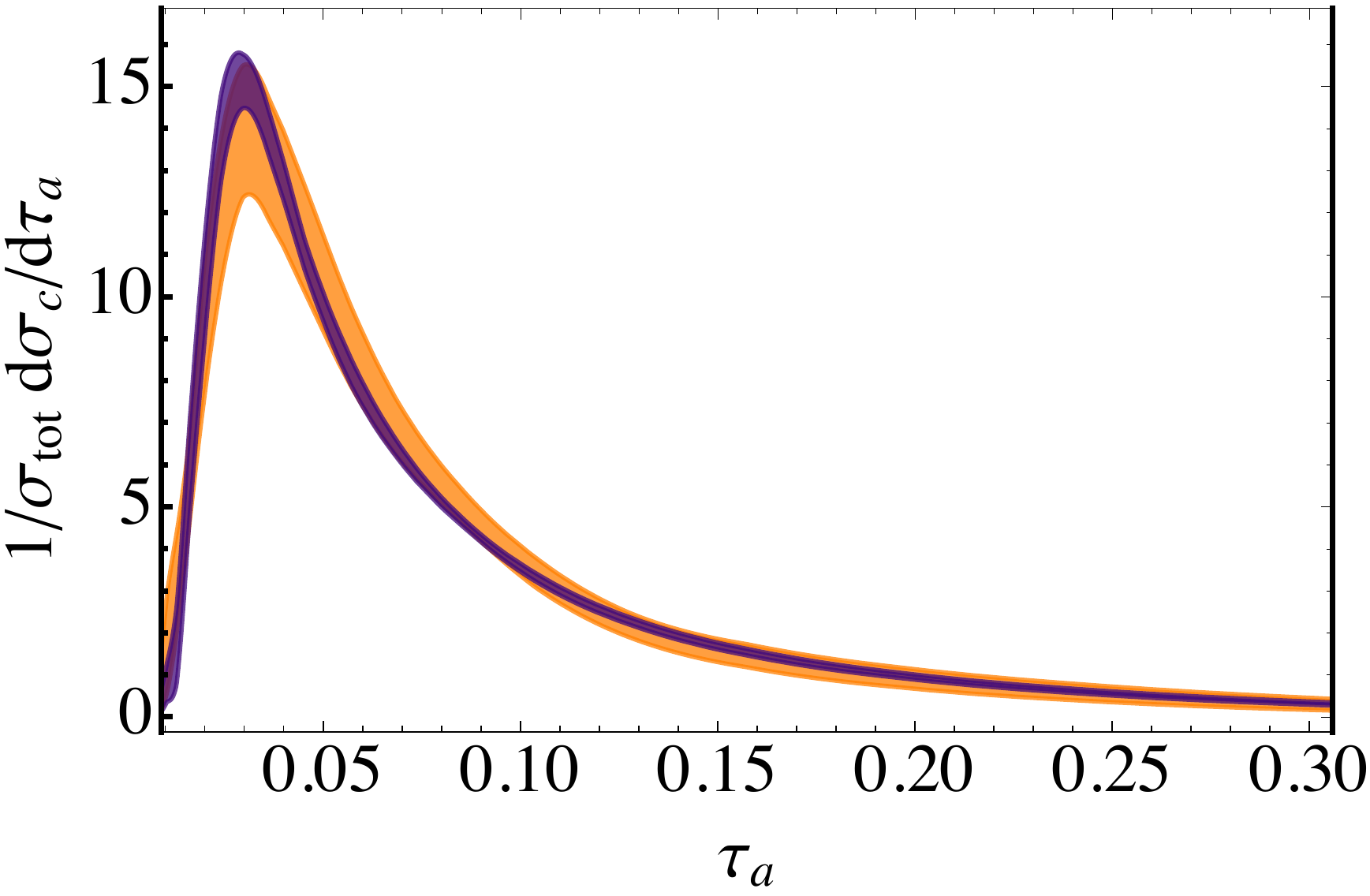} &
\includegraphics[width=.45\columnwidth]{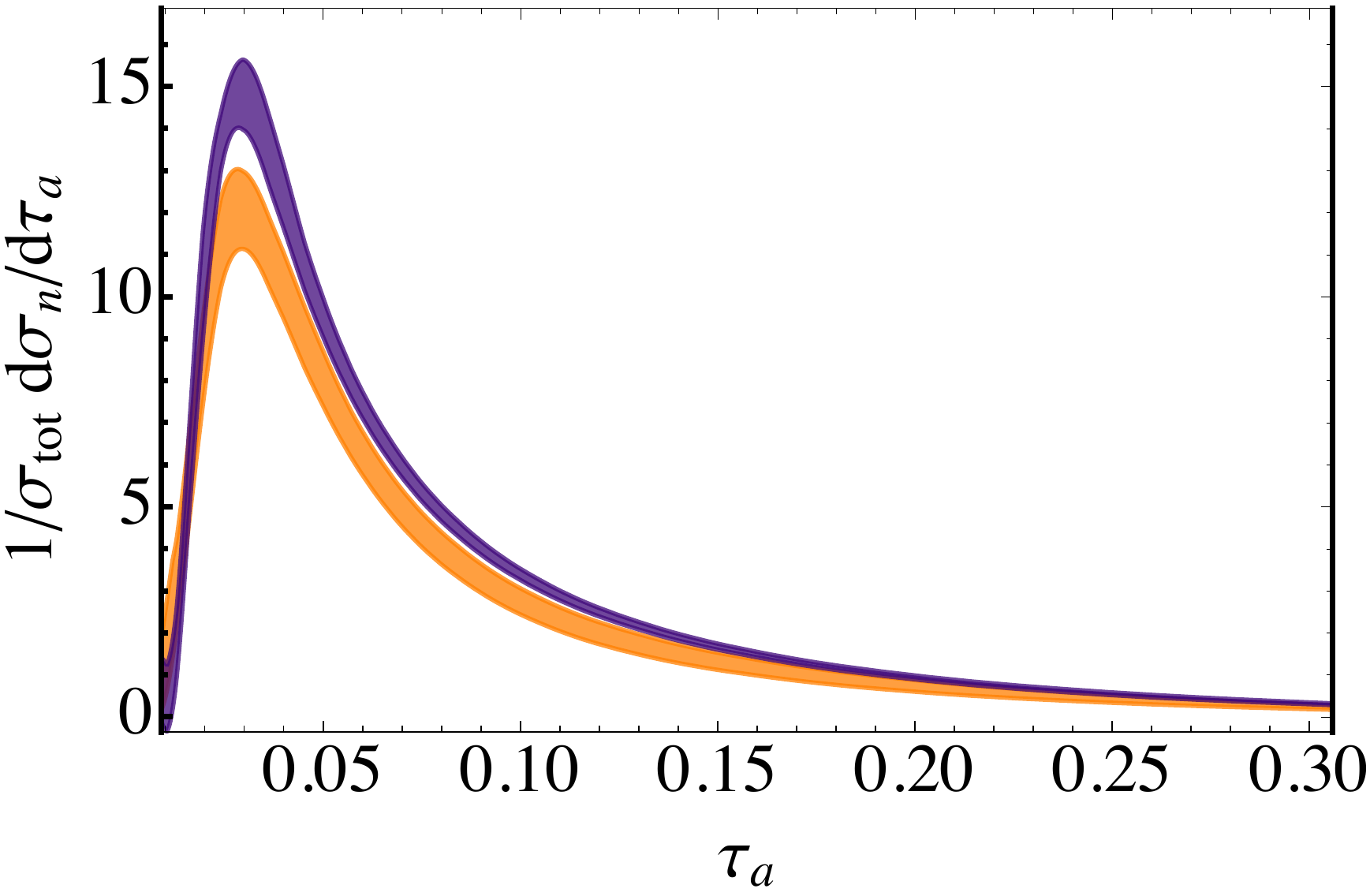}
\end{tabular}
\vspace{-1.5em}
\caption{
Differential angularity distributions for $a=-0.5$ and $a=0.25$ at $Q=m_{Z}$ over the entire 
$\tau_a$ domain. The distributions correspond to NLL$' + \mathcal{O}(\alpha_{s})$ (orange)
and NNLL$' +\mathcal{O}(\alpha^{2}_{s})$ (purple) accuracy with renormalon subtractions,
and they are either obtained
as the derivative of the integrated distributions  (left) or 
directly resummed as differential distributions (right).
As expected from the analysis in \cite{Almeida:2014uva}, the former show a bit better 
convergence than the latter.}
\vspace{-1em}
\label{fig:fullspec}
\end{figure}
%-----------------------------------------------------------------------------

In order to demonstrate the improvement in precision that we achieve for the differential 
distributions in moving from NLL$' + \mathcal{O}(\alpha_{s})$ \cite{Hornig:2009vb} to 
NNLL$' +\mathcal{O}(\alpha^{2}_{s})$ accuracy, we present our (renormalon-subtracted) 
predictions for $a = \lbrace -0.5, 0.25 \rbrace$ and $Q=m_{Z}$ across the entire $\tau_{a}$ domain in 
\fig{fullspec}.  This figure also illustrates the differences between taking the derivative of the integrated 
distributions (which we call $d\sigma_c/d\tau_a$ here) in the left panel, and directly evaluating the 
resummed (``na\"{i}ve'') differential distributions (which we call $d\sigma_n/d\tau_a$ here) in the right panel. 
We see that the former give better convergence and
that they better preserve the total integral under the distributions. 
These issues with the na\"{i}ve distributions were extensively discussed in 
\cite{Almeida:2014uva}. In fact, as shown there, at unprimed orders the na\"{i}ve formulas do not even 
preserve the correct order of accuracy, and even the primed orders suffer from the illustrated mismatch with 
the total integral under the curve. These issues can be remedied by supplementing the na\"{i}ve formula with 
additional terms that both preserve accuracy (at unprimed orders) and maintain agreement with the total integral 
under the curve (at any order). See \cite{Almeida:2014uva,Procura:2018zpn} for such strategies, 
and \cite{Bertolini:2017eui} for a beautiful mathematical solution to this problem. Here, for simplicity, we 
have not implemented such strategies, and we simply stick to the integrated distributions from
\eq{convolvedexpanded} as the basis for all of our predictions. 
 
 %----------------------------------------------------------------------------
\begin{figure}[t]
\centering
\begin{tabular}{ccc}
 & \quad {\large\bf Slope $r_s = 1/\tau_a^\text{sph}$} & \quad {\large\bf Slope $r_s=1$} \\[1ex]
\rotatebox{90}{\qquad\qquad\quad$a=-0.5$} &
\includegraphics[width=.45\columnwidth]{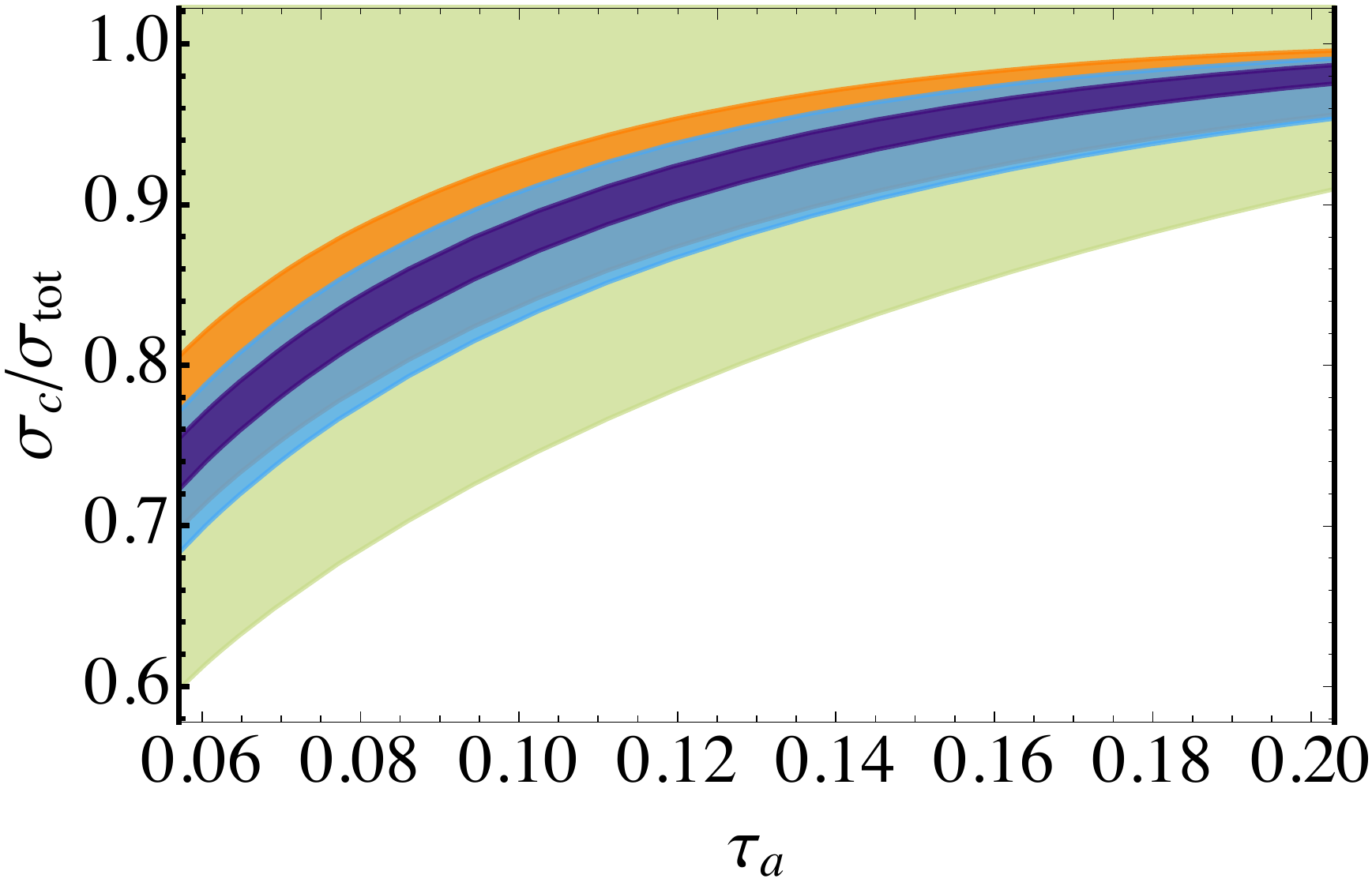} &
\includegraphics[width=.45\columnwidth]{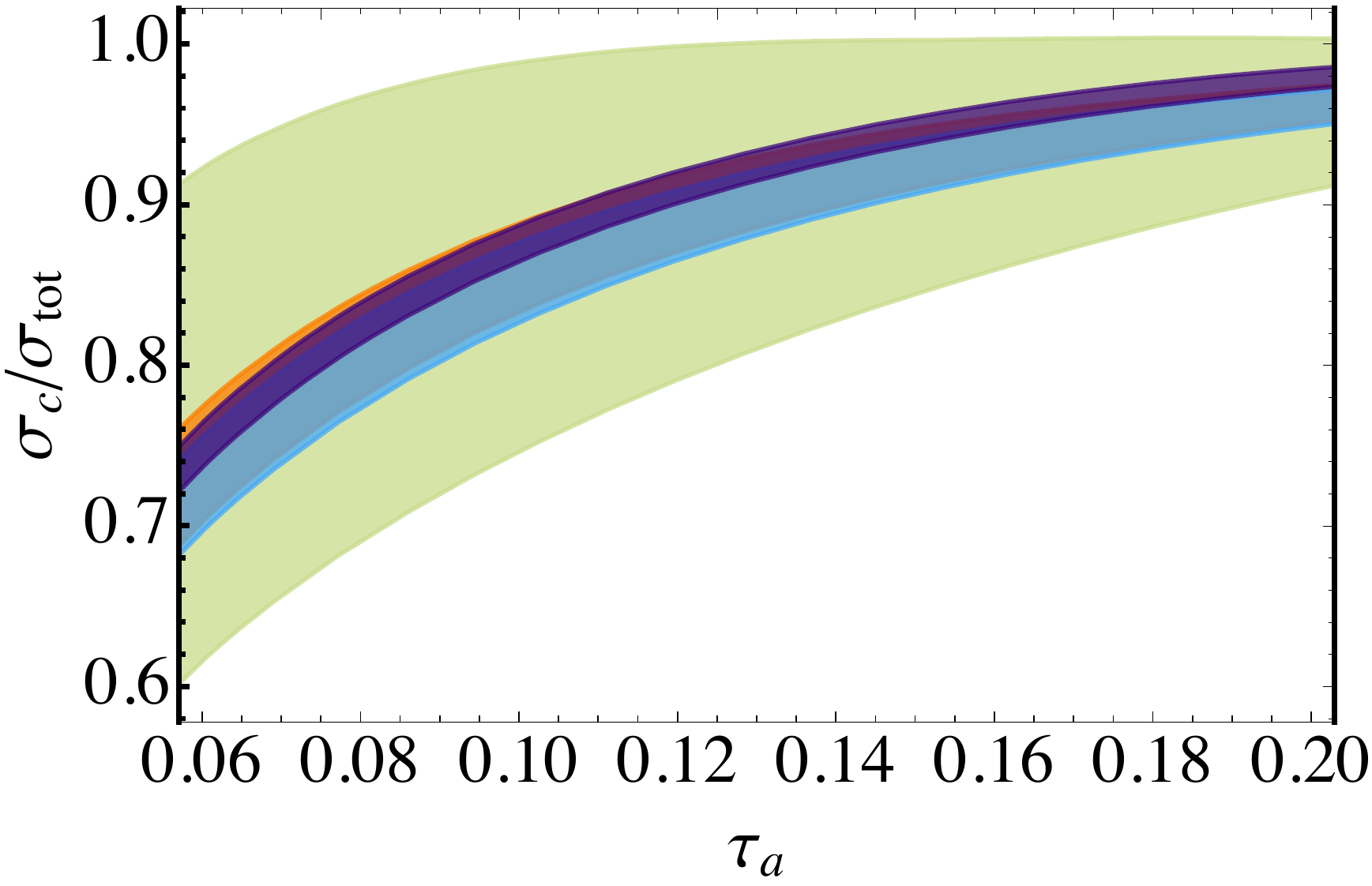} \\
\rotatebox{90}{\qquad\qquad\quad$a=0.25$} &
\includegraphics[width=.45\columnwidth]{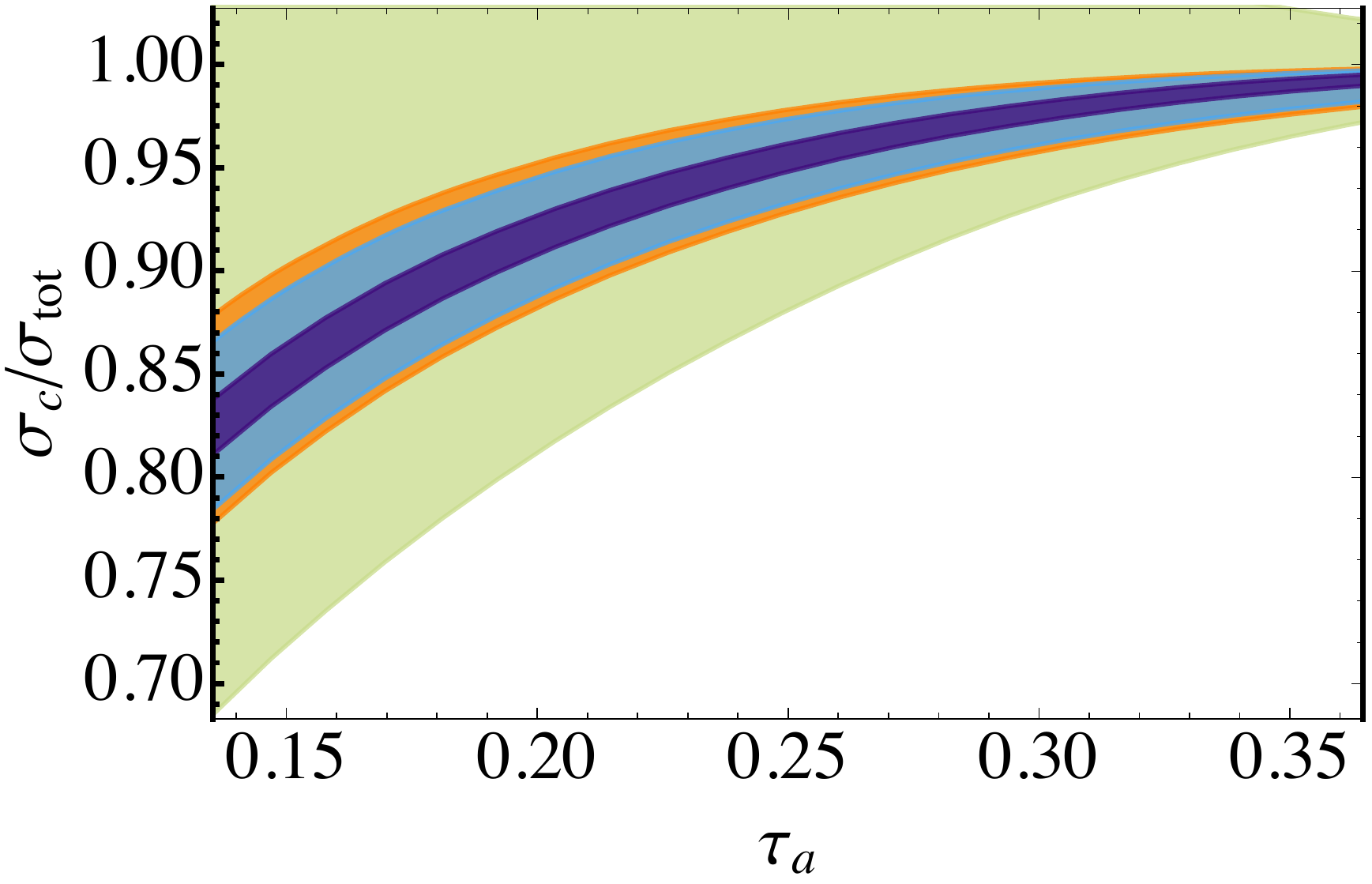} &
\includegraphics[width=.45\columnwidth]{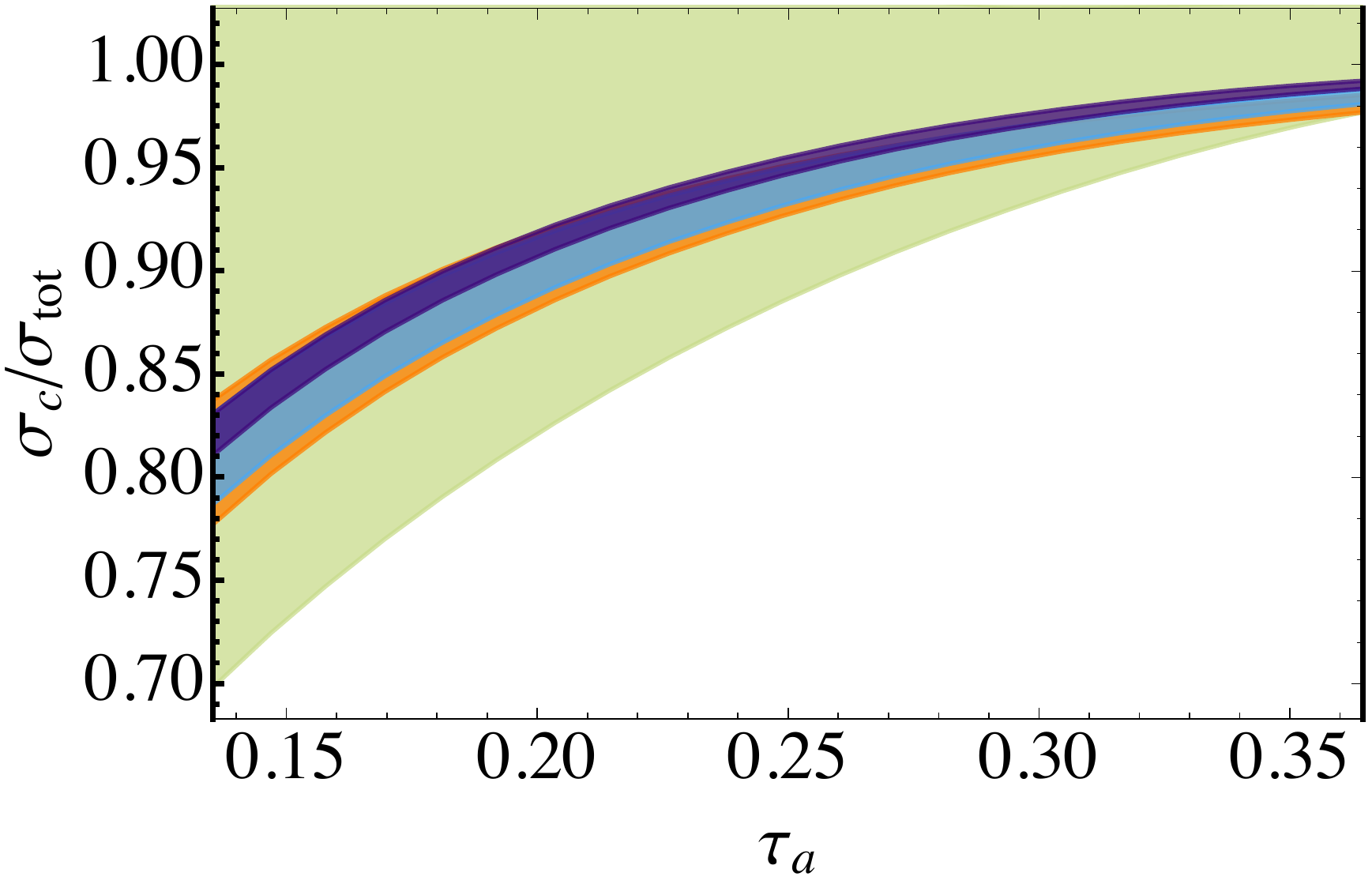}
\vspace{-1em}
\end{tabular}
\caption{
Integrated angularity distributions for $a=-0.5$ and $a=0.25$ at $Q=m_{Z}$ at 
NLL (green), NLL$^{\prime}+\cO(\as)$ (orange), NNLL$+\cO(\as)$ (blue), 
and NNLL$^{\prime}+\cO(\as^2)$ (purple) accuracy, with renormalon subtractions to the corresponding orders.
The distributions have been obtained with the slope 
parameter $r_s = 1/\tau_a^\text{sph}$ (left) and $r_s = 1$ (right).
}
\label{fig:slopecompare}
\end{figure}
%-----------------------------------------------------------------------------

In \fig{slopecompare} we illustrate the effect of choosing a slope parameter $r_s = 1/\tau_a^\text{sph}$
for the running scale in \eq{murun} in the resummation region.
In this figure we compare predictions for integrated distribution for 
$a = \{-0.5,0.25\}$ and $Q=m_Z$ at different logarithmic accuracies.
One observes that the convergence is slightly better for the choice $r_s = 1/\tau_a^\text{sph}$
than for $r_s = 1$, although the differences between both choices become smaller at higher
logarithmic accuracies.
This is consistent with the findings of \cite{Hoang:2014wka}, who made a similar comparison for thrust ($a=0$), and this  also validates our interpretation of the slope parameter $r_s$
as being related to the maximum value of the angularity 
$\tau_a^\text{sph}$.

%----------------------------------------------------------------------------
\begin{figure}[t]
\centering
\includegraphics[scale=0.27]{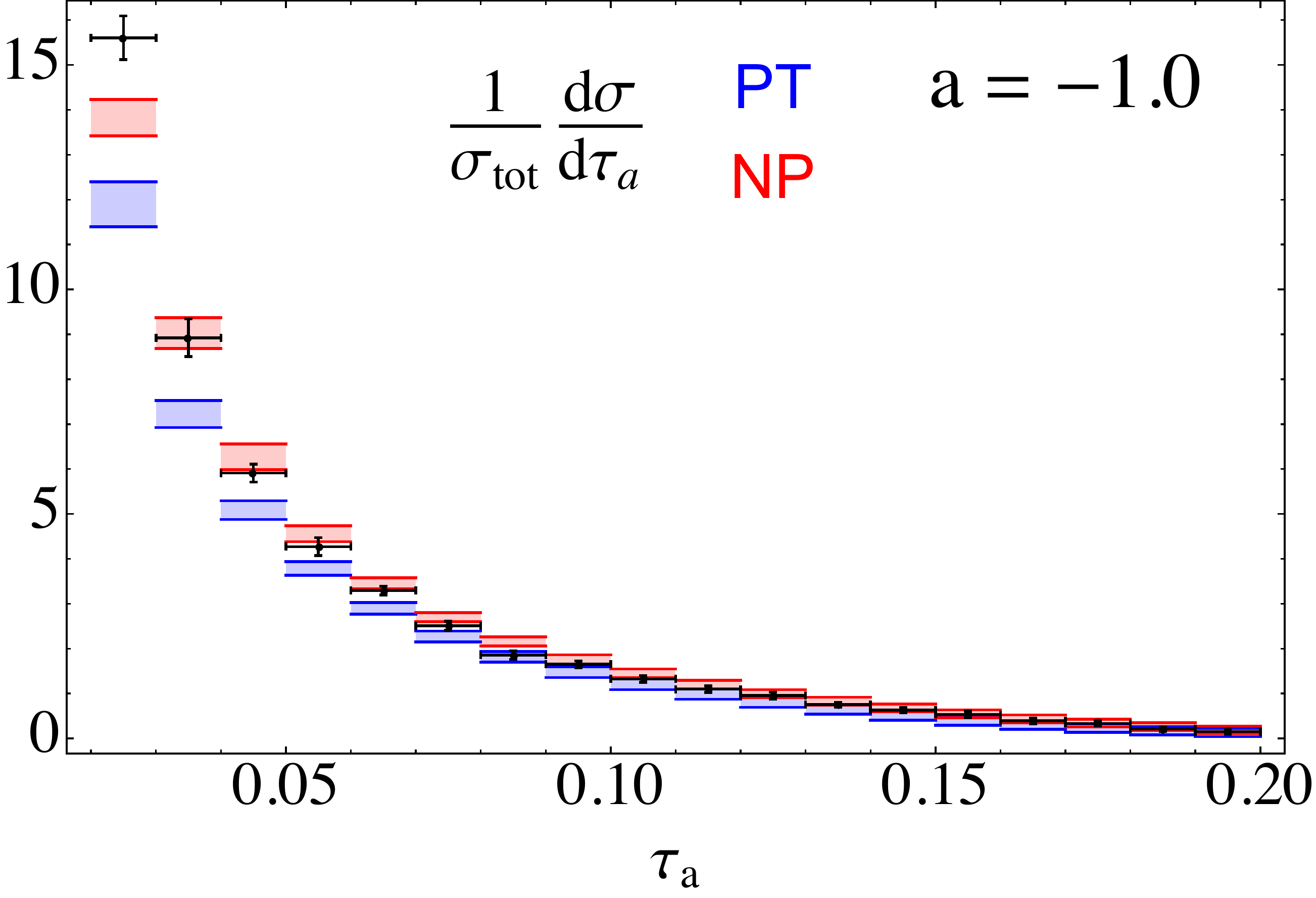}\\[1ex]
\includegraphics[scale=0.27]{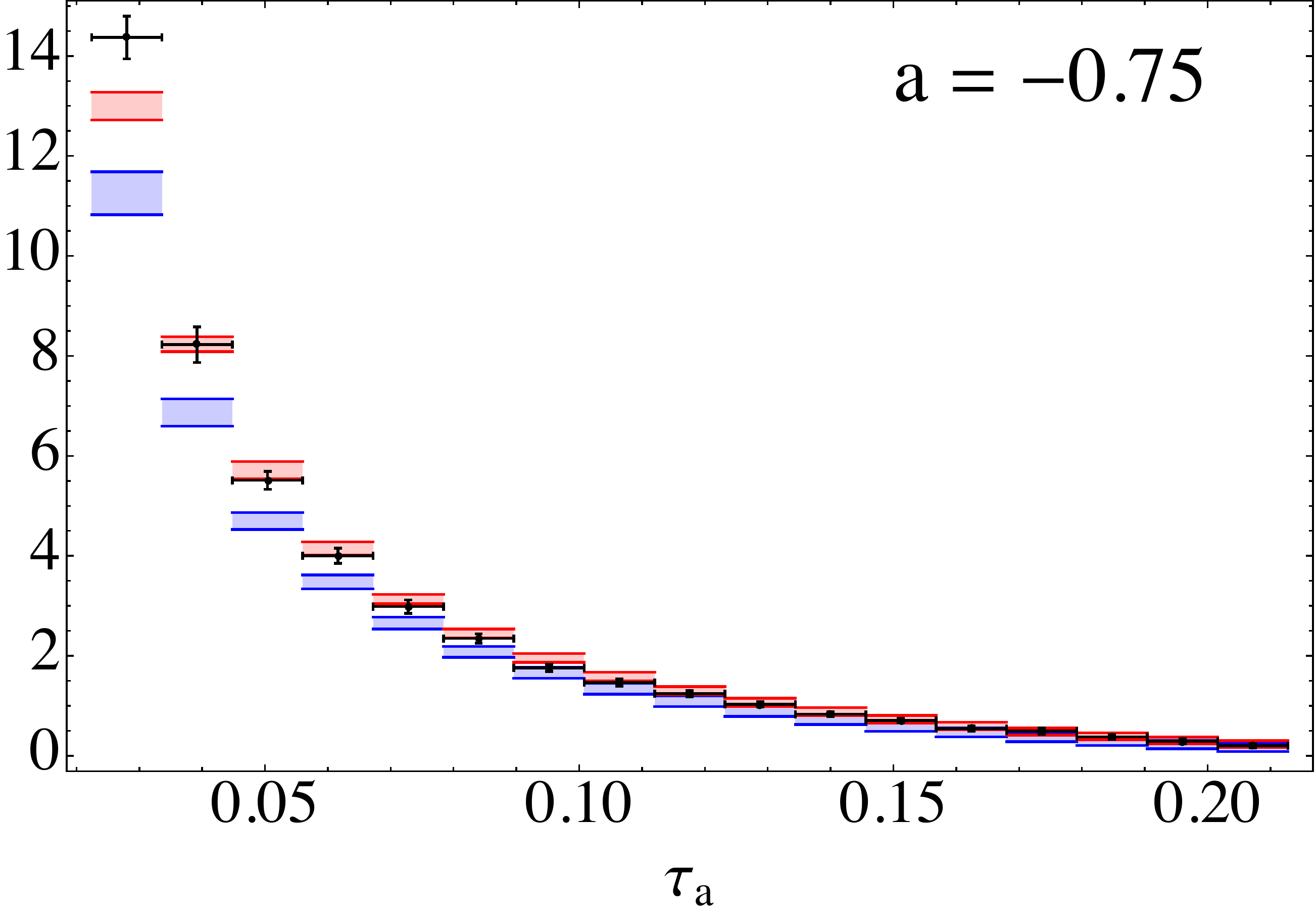}\quad
\includegraphics[scale=0.27]{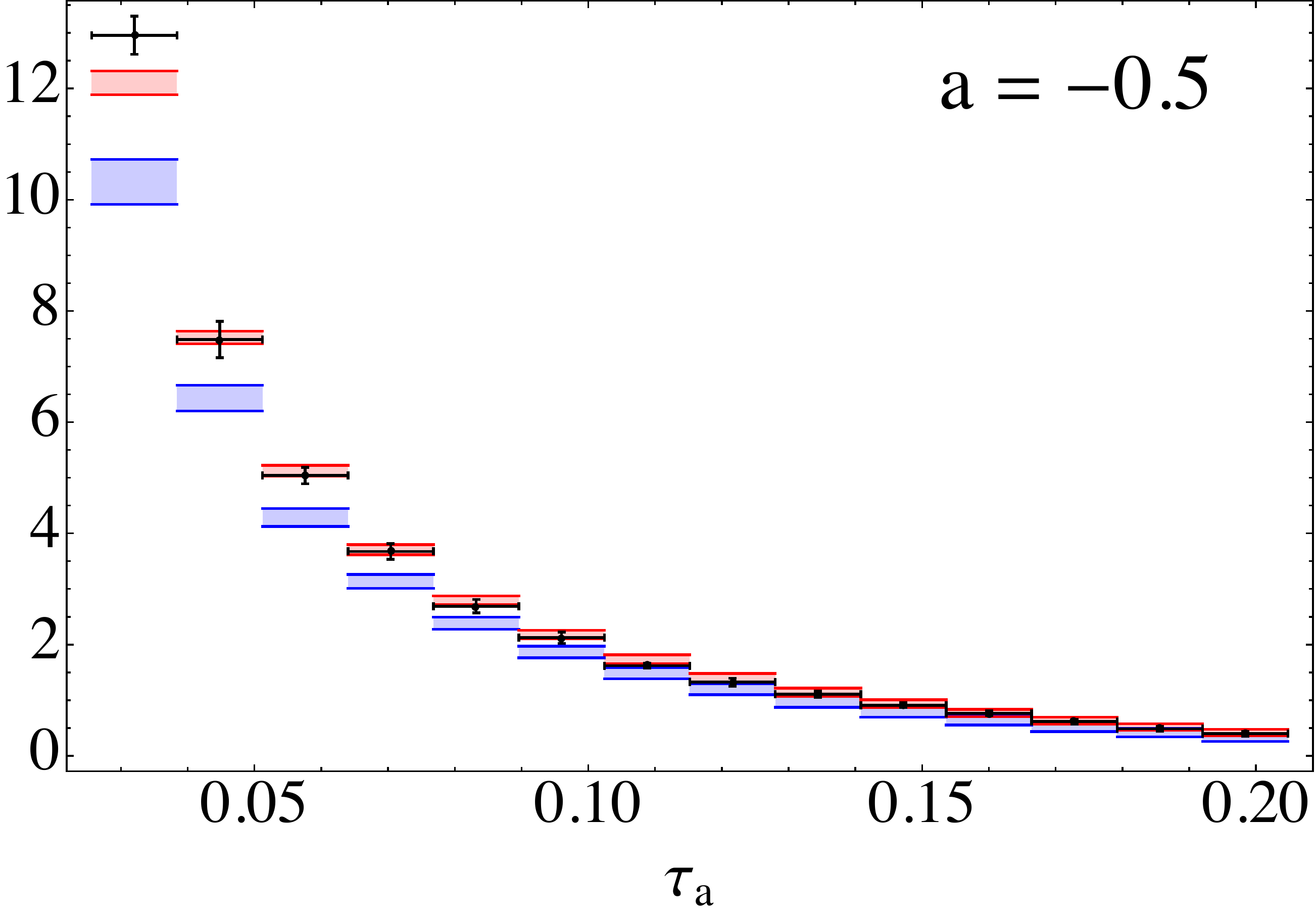}
\includegraphics[scale=0.27]{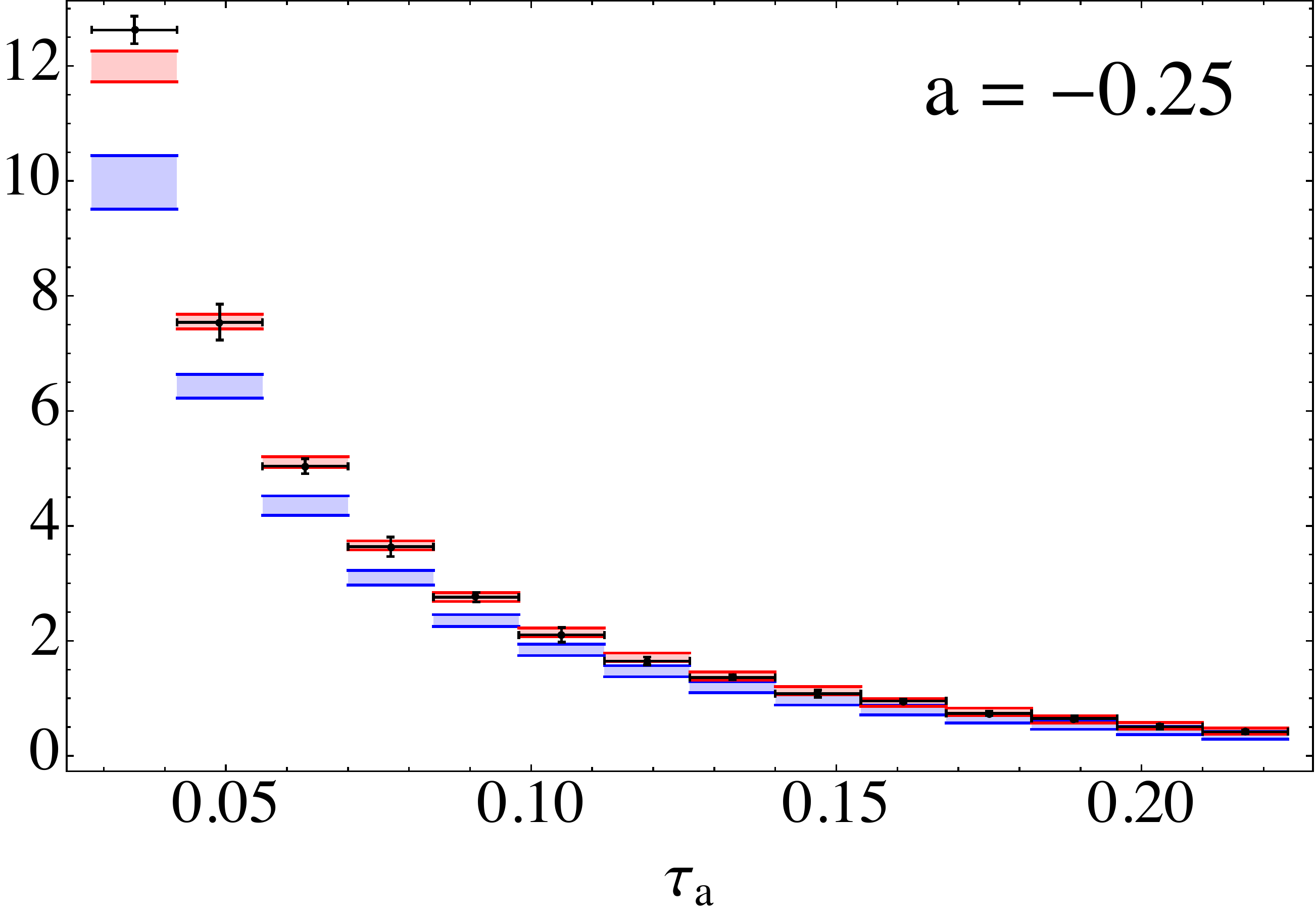}\quad
\includegraphics[scale=0.27]{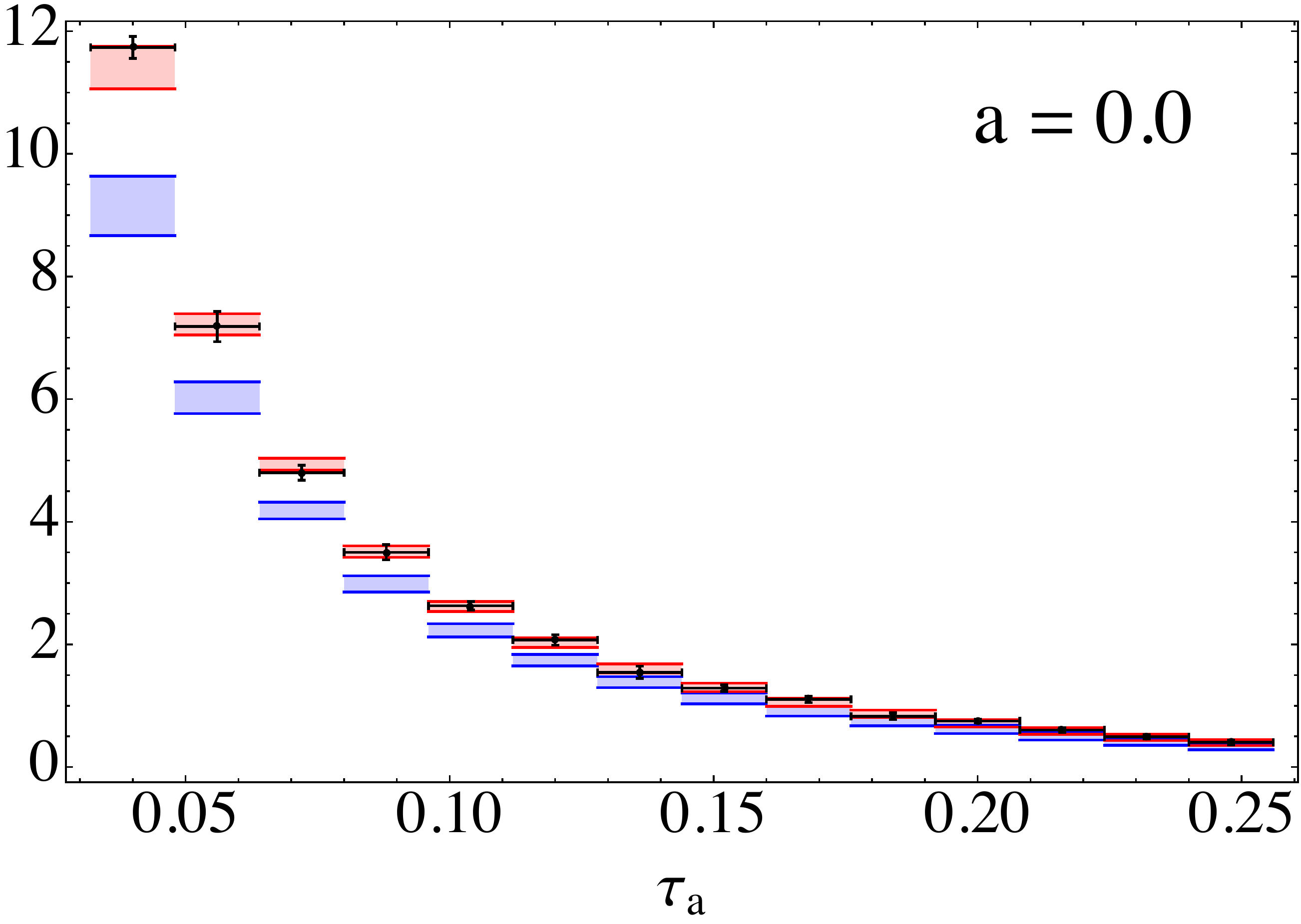}
\includegraphics[scale=0.27]{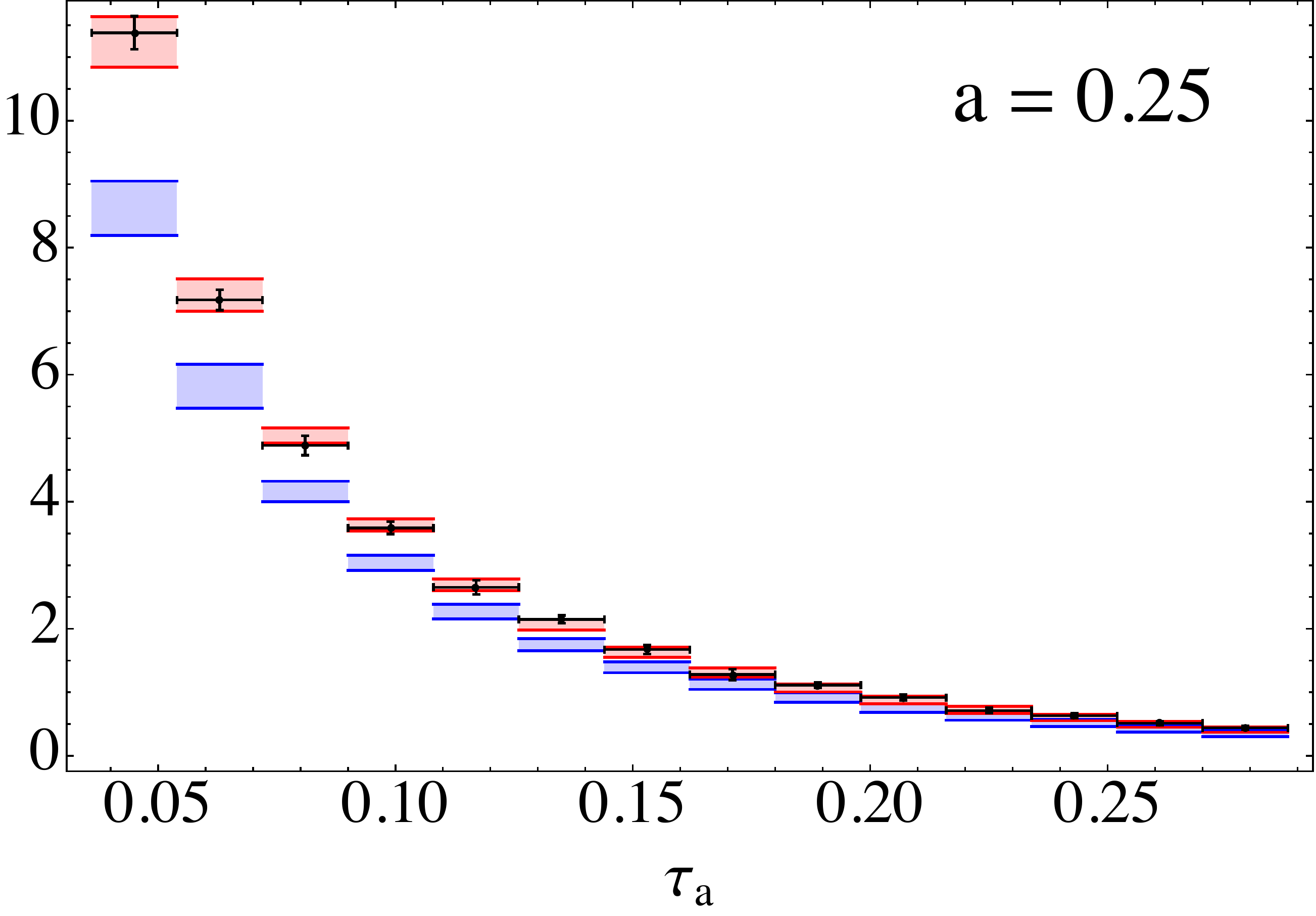}\quad
\includegraphics[scale=0.27]{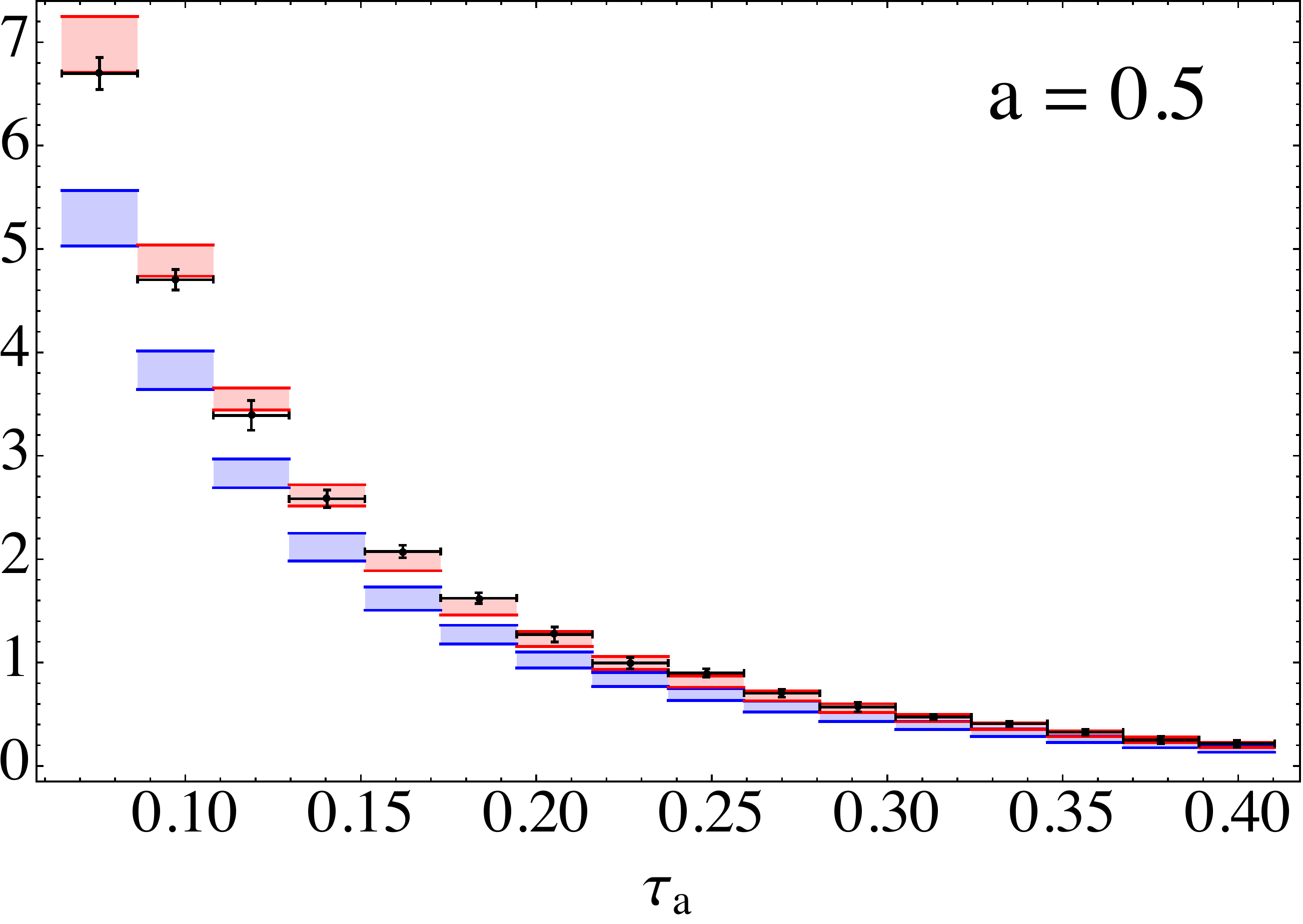}
\caption{
NNLL$^{\prime}$ resummed  and $\mathcal{O}(\alpha_{s}^{2})$ matched angularity distributions for all values of 
$a$ considered in this study, $a \in \lbrace -1.0, -0.75, -0.5, -0.25, 0.0 , 0.25, 0.5 \rbrace$, at 
$Q = m_Z$,
with $\as(m_Z) = 0.11$. The blue bins represent the purely perturbative prediction and the red bins include
a convolution with a gapped and renormalon-subtracted shape function, with a first moment set to 
$\Omega_1(R_\Delta,R_\Delta) = 0.4\text{ GeV}$. Overlaid is
the experimental data from  \cite{Achard:2011zz}.
}
\vspace{-4em}
\label{fig:NPplots1}
\end{figure}
%-----------------------------------------------------------------------------

%----------------------------------------------------------------------------
\begin{figure}[t]
\centering
\includegraphics[scale=0.27]{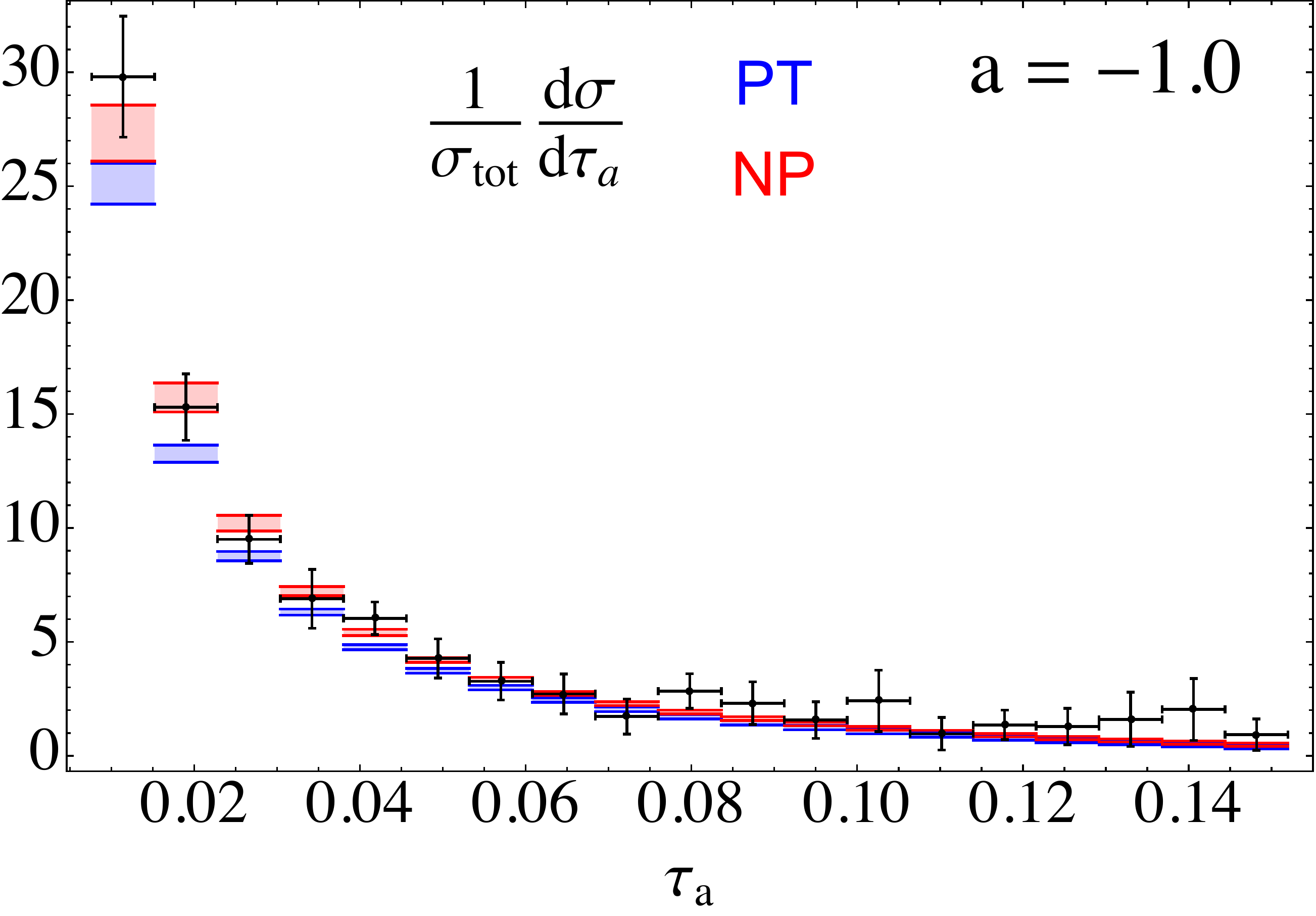}\\[1ex]
\includegraphics[scale=0.27]{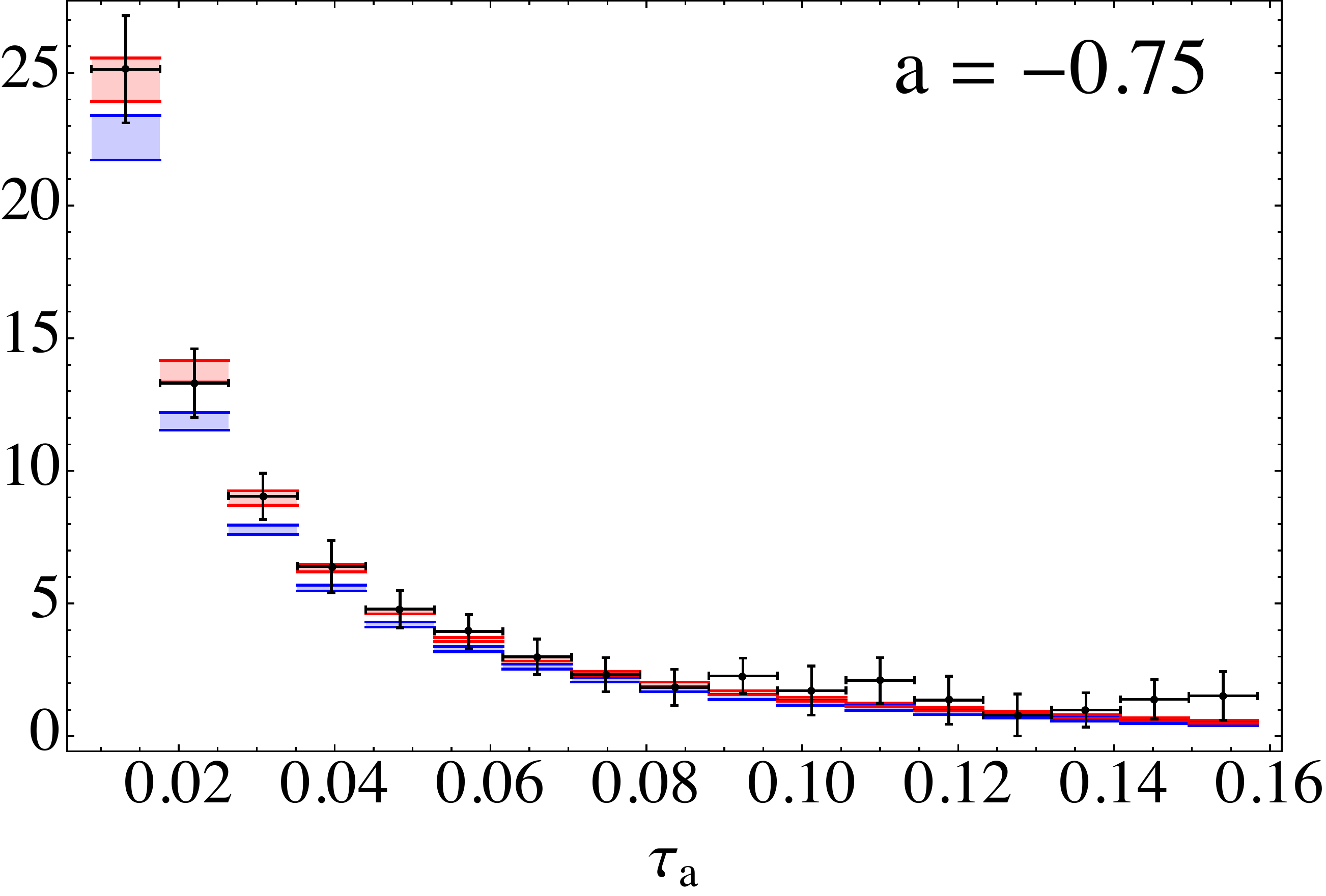}\quad
\includegraphics[scale=0.27]{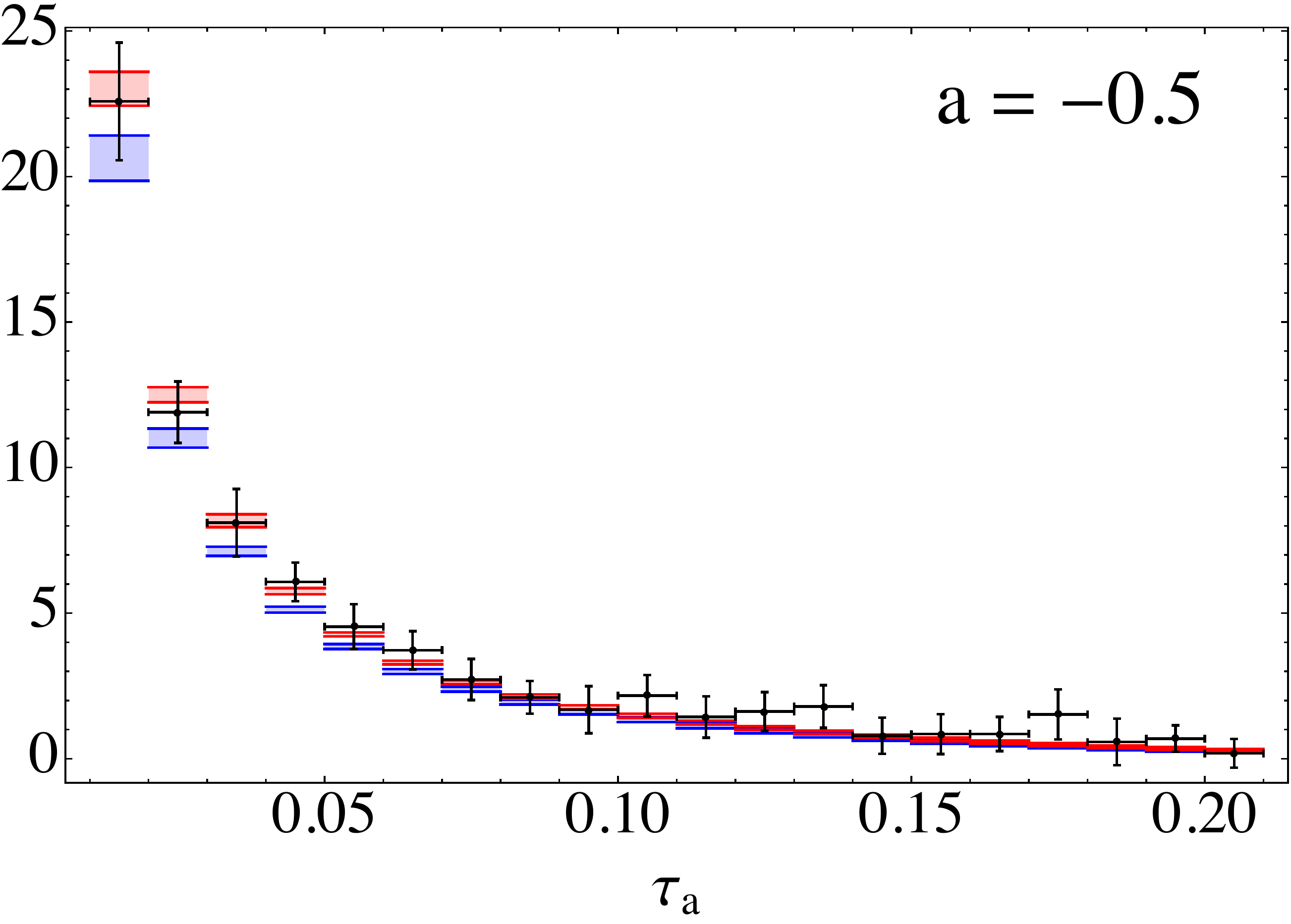}
\includegraphics[scale=0.27]{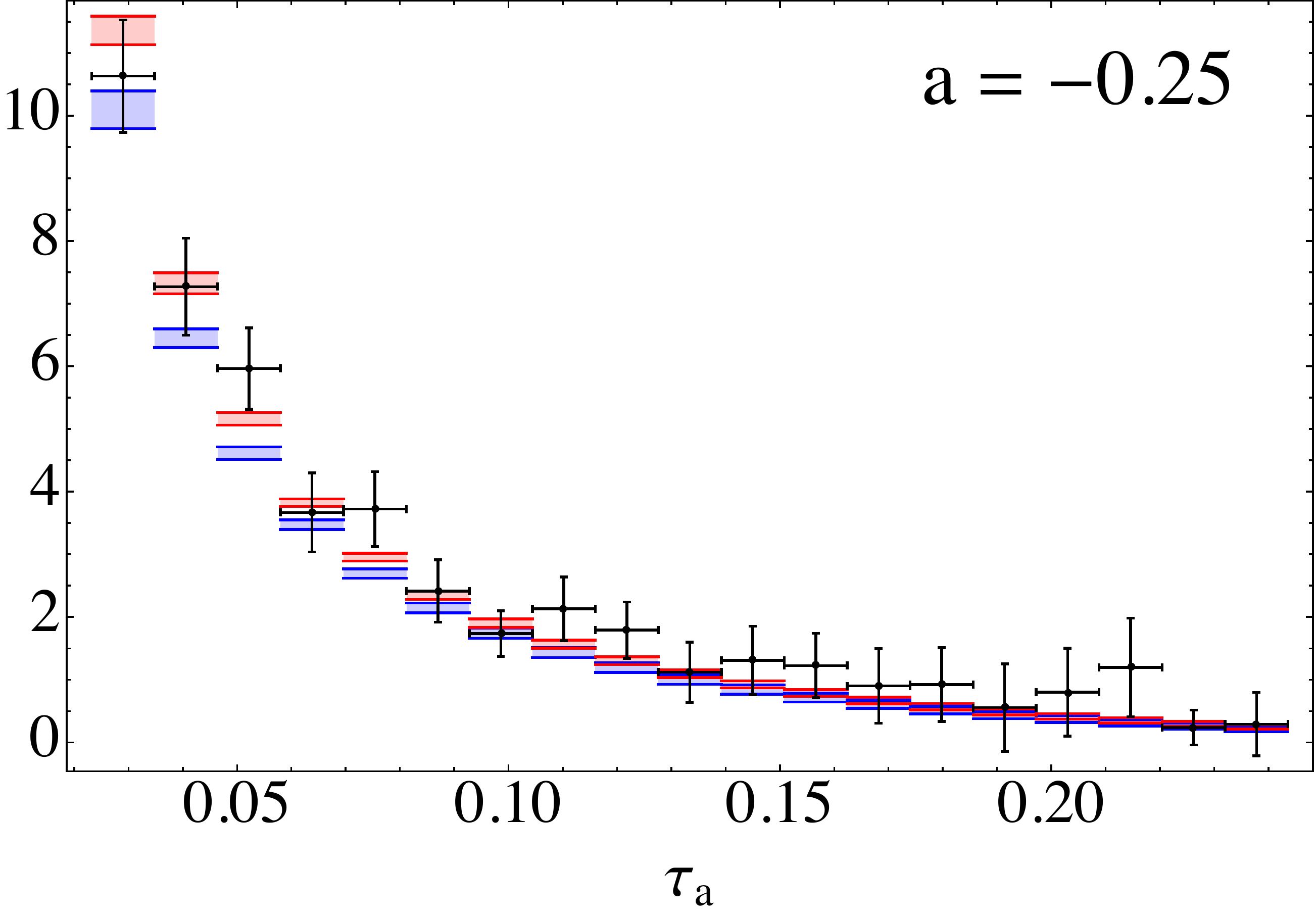}\quad
\includegraphics[scale=0.27]{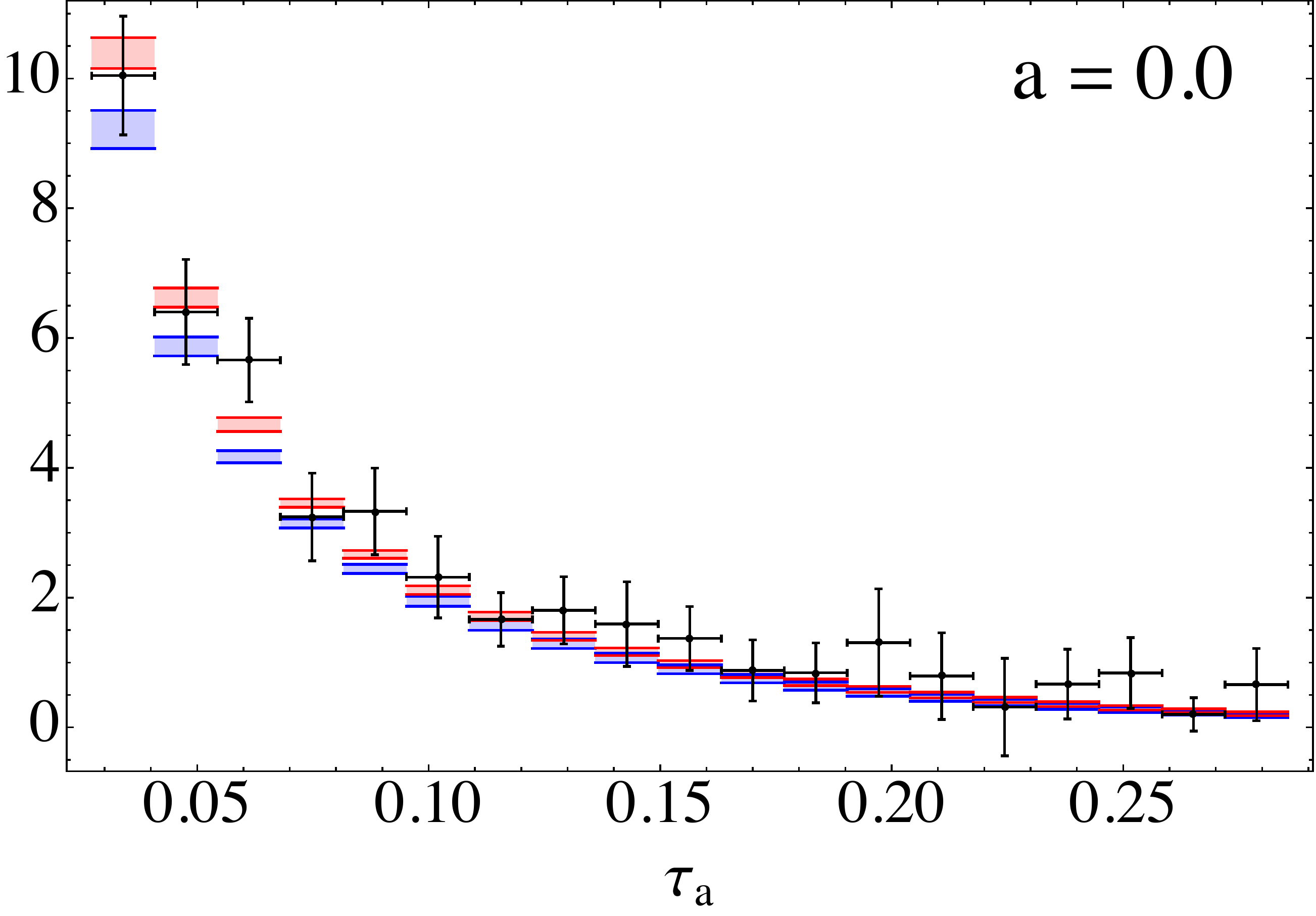}
\includegraphics[scale=0.27]{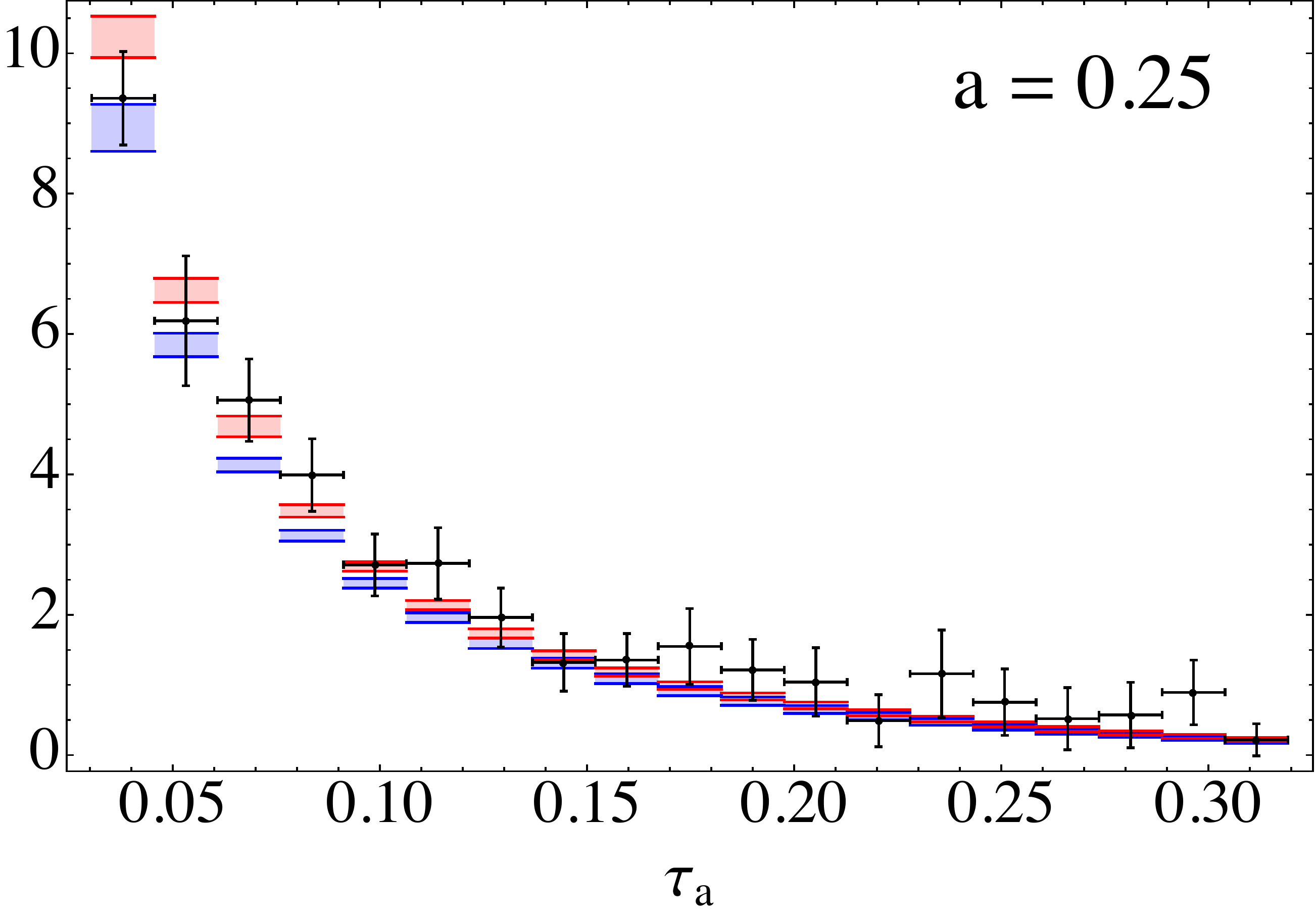}\quad
\includegraphics[scale=0.27]{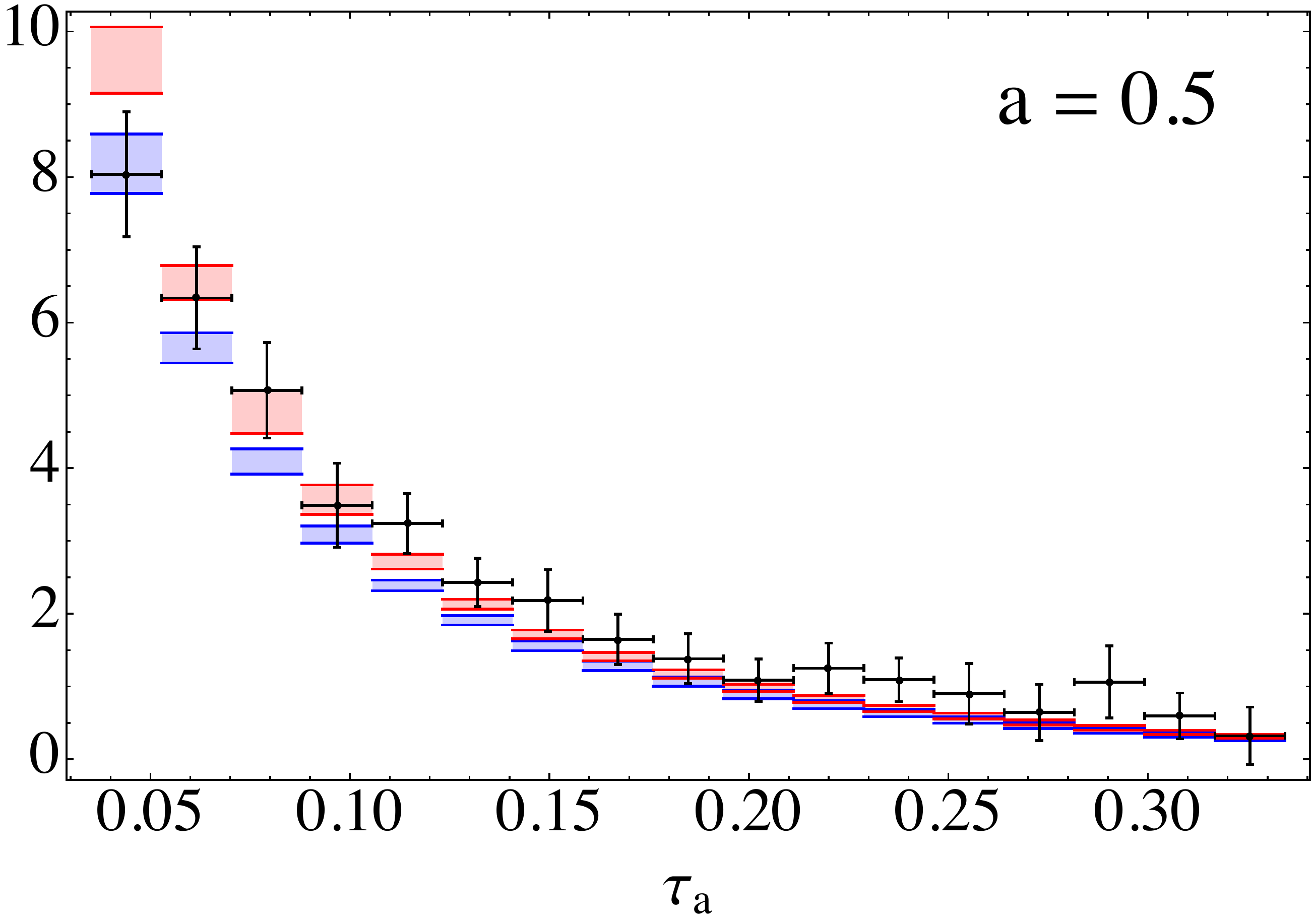}
\caption{The same as in Figure \ref{fig:NPplots1}, but with $Q = 197$ GeV.}
\label{fig:NPplots2}
\end{figure}
%-----------------------------------------------------------------------------

%----------------------------------------------------------------------------
\begin{figure}[t]
\centering
\begin{tabular}{ll}
\includegraphics[width=.48\columnwidth]{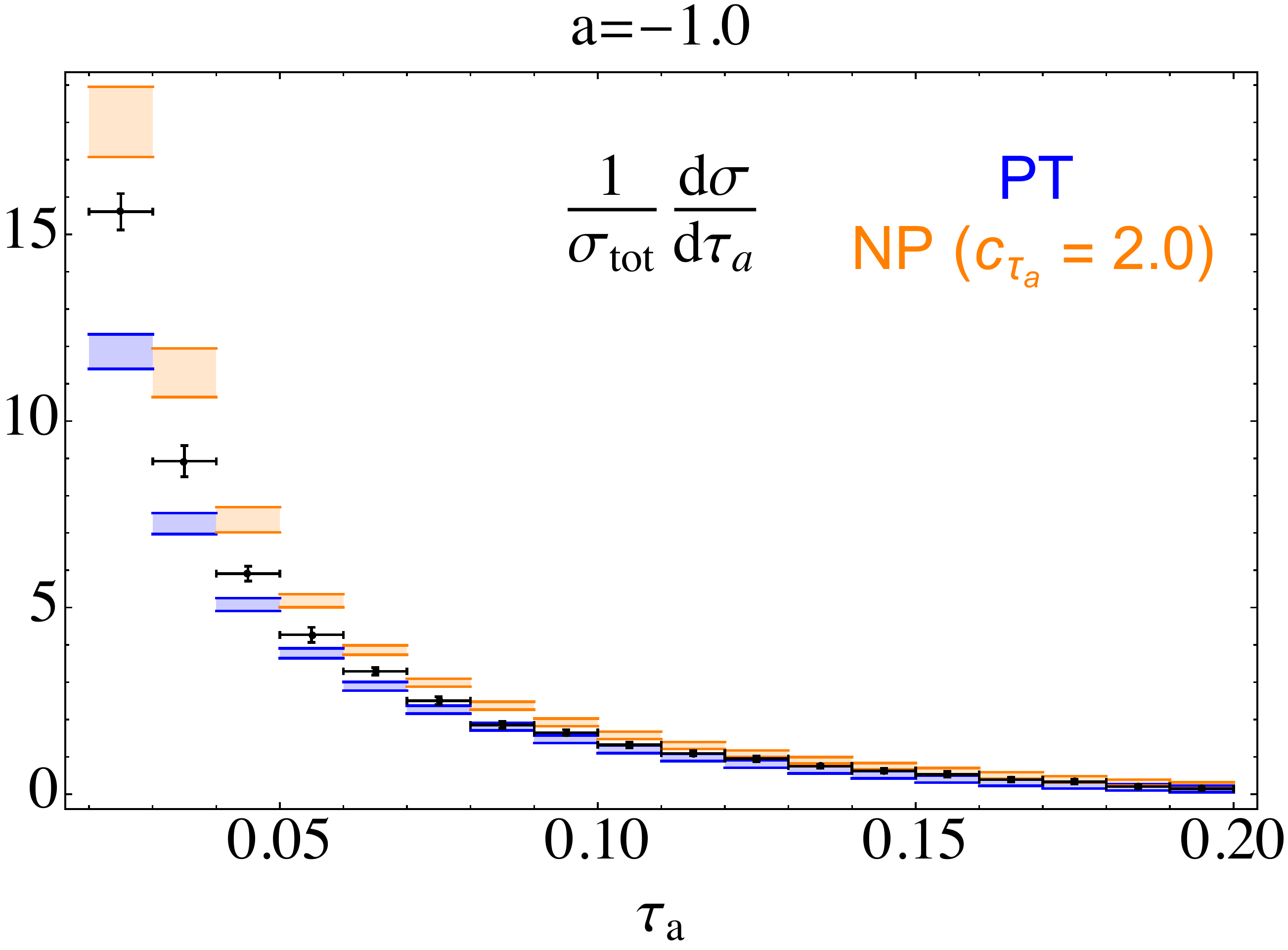} &
\includegraphics[width=.47\columnwidth]{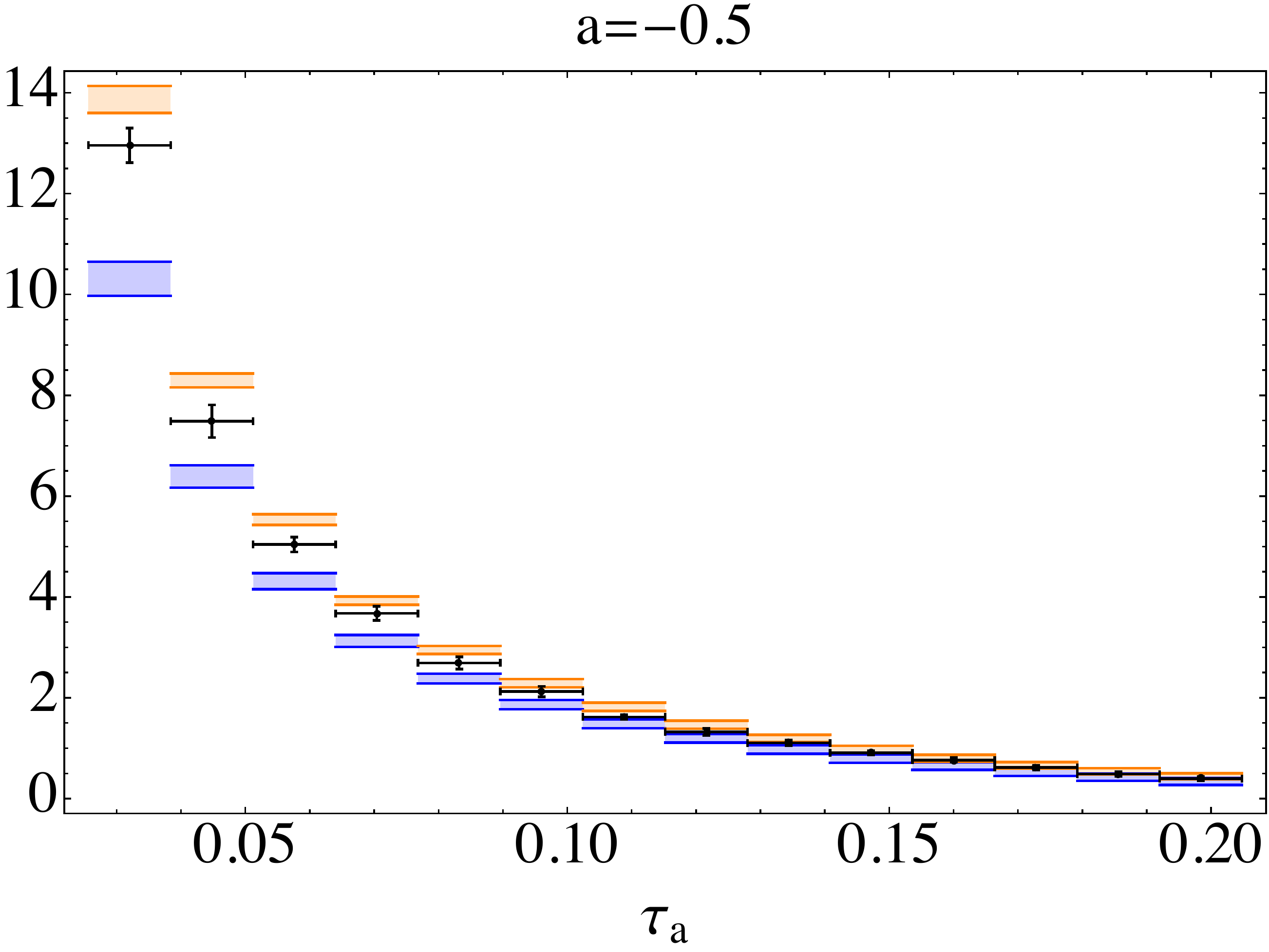} \\
\includegraphics[width=.48\columnwidth]{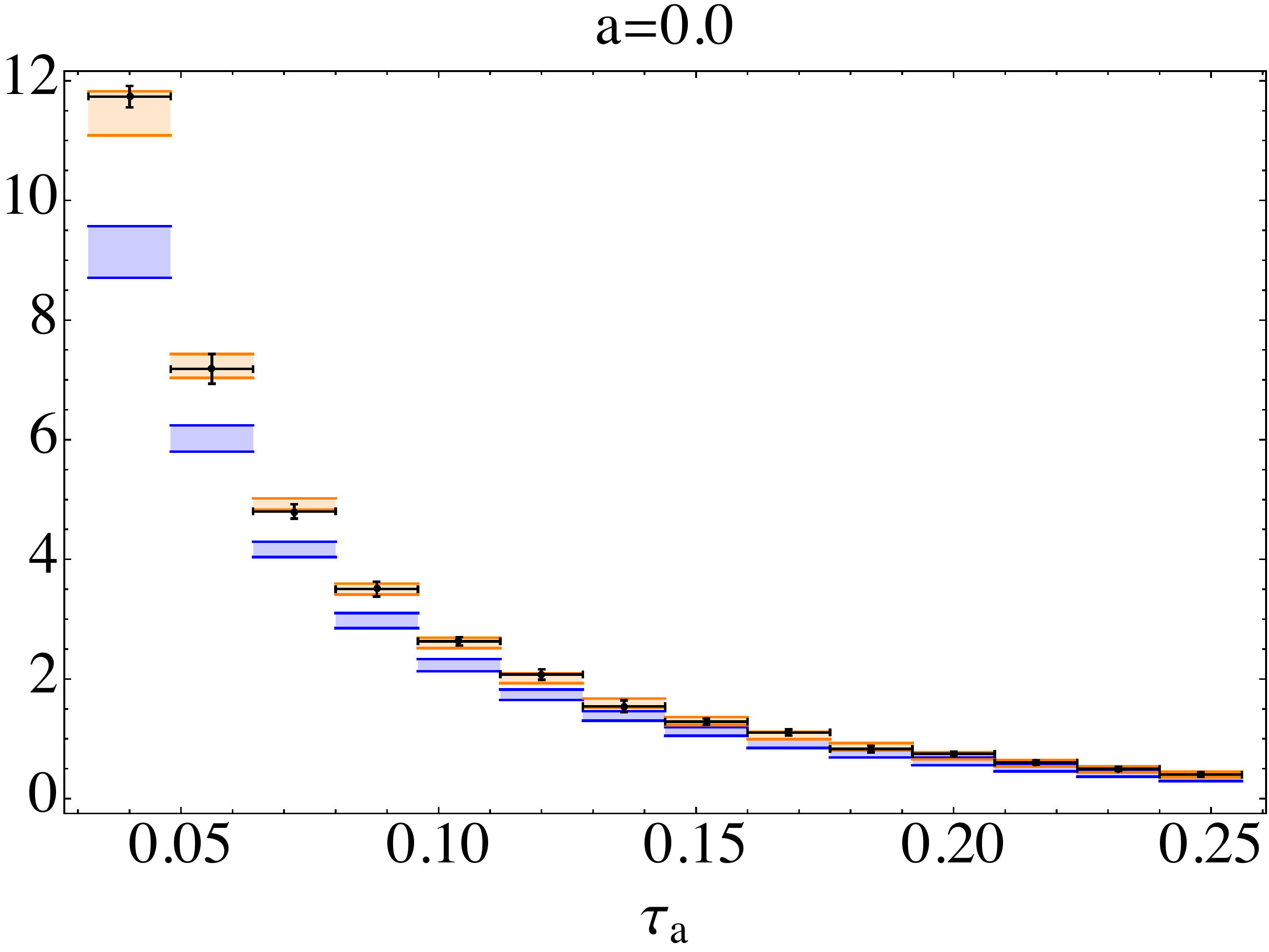} &
\includegraphics[width=.47\columnwidth]{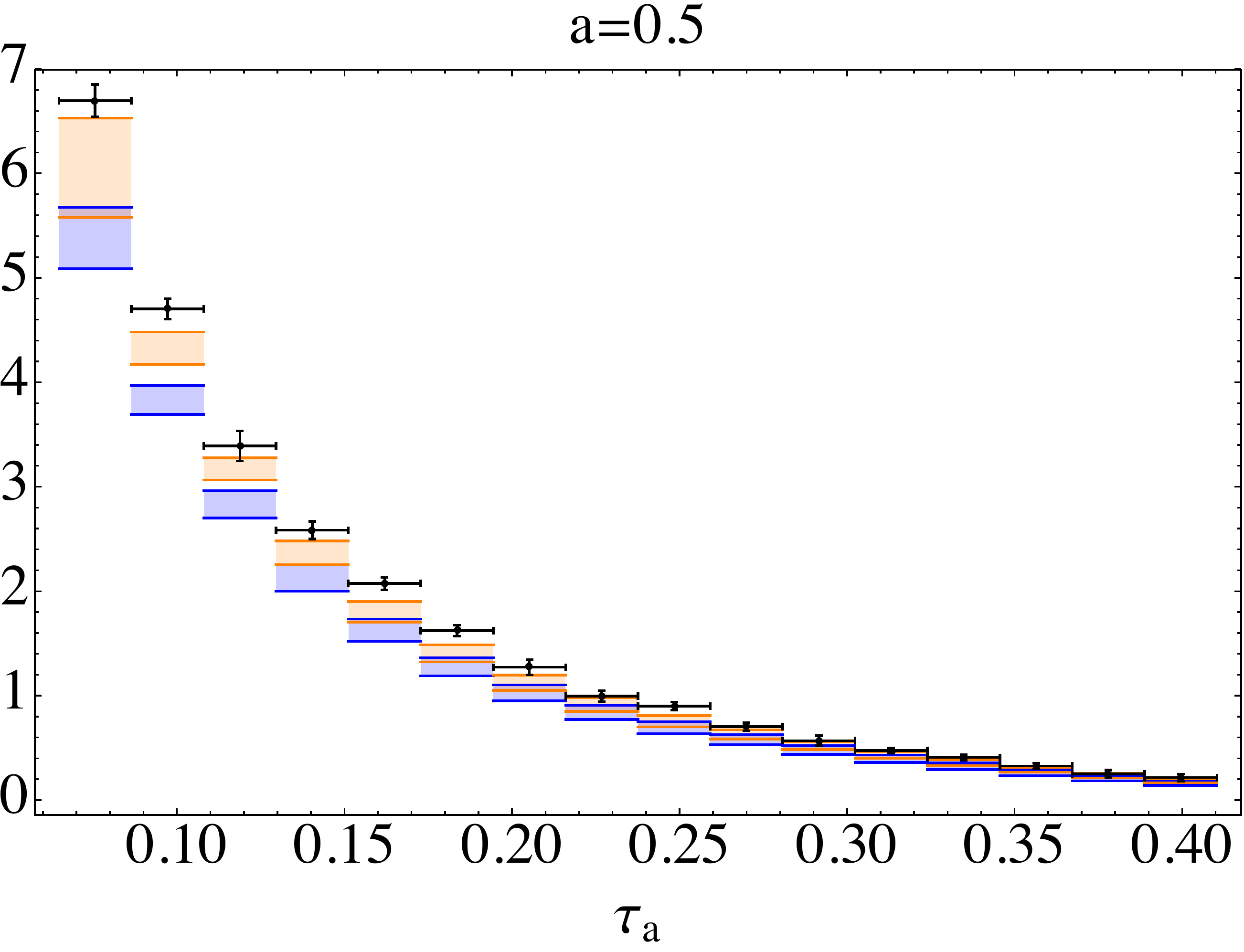}
\end{tabular}
\vspace{-1.5em}
\caption{
The same as in Figure \ref{fig:NPplots1} (for $a=\{-1.0,-0.5,0.0,0.5\}$), but with the non-perturbative scaling 
coefficient purposefully altered to $c_{\tau_{a}} = 2.0$, instead of $c_{\tau_{a}} = 2/(1-a)$.}
\label{fig:NOLS}
\end{figure}
%-----------------------------------------------------------------------------

Finally, we present binned results for NNLL$^{\prime}$ resummed, $\mathcal{O}(\alpha_{s}^{2})$ matched and 
renormalon-subtracted differential distributions for 
$a = \lbrace-1.0, -0.75, -0.5, -0.25, 0.0, 0.25, 0.5 \rbrace$ in \fig{NPplots1} ($Q=m_Z$) and 
\fig{NPplots2} ($Q=197$ GeV). To obtain the binning, we have integrated the upper and lower bounds of the 
differentiated cumulative distribution over each bin, which we then divided by the bin width.  In both cases we 
compare to the data from the L3 collaboration \cite{Achard:2011zz}, and we
explicitly illustrate the effect of including the non-perturbative effects from \eq{convolvedterms} (red bins).  
In the figures we focus on the central 
$\tau_{a}$ domain where the effect of the resummation is most relevant.  It is clear that---given the 
parameters we have applied in our analysis---the data is consistent with the non-perturbative shifted curves 
for the entirety of the angularity spectra.  On the other hand, the bins obtained using only the perturbative 
predictions (blue bins) systematically undershoot the available data throughout the resummation-sensitive 
region.\footnote{Of course, better agreement with the purely perturbative distributions would be obtained at a 
higher value of $\alpha_{s}$, although it 
does not appear that such a fit will be as good as this two-parameter fit that includes a significant 
$\Omega_1$---see Figure \ref{fig:avariation} and the associated discussion in Section \ref{sec:CONCLUSIONS}.} 
These observations are less obvious with the less precise 197 GeV data in \fig{NPplots2}, which
nevertheless provides a decent test of the $Q$ dependence of $\as$ and the $1/Q$ scaling of the leading 
nonperturbative correction.

These results serve as an impressive visual confirmation of the leading non-perturbative correction discussed in 
\sec{NONPERT}, which clearly indicates a scaling with the angularity parameter $a$. To further emphasize that 
this scaling is actually reflected in the data, in \fig{NOLS} we have drawn analogous plots to those in 
\fig{NPplots1}, but we have purposefully altered the non-perturbative scaling coefficient such that $c_{\tau_{a}} 
\rightarrow 2.0$, instead of the predicted $c_{\tau_{a}} = 2/(1-a)$.\footnote{We have also removed the 
$1/(1-a)$ scaling of the gap parameter, cf. \eqref{eq:Deltaa}.}  The associated 
bins for the predictions convolved with a renormalon-free shape function,
now shown in orange, clearly overshoot the data at $a = \lbrace -1.0, -0.5 \rbrace$, 
obviously remain unchanged 
for thrust ($a=0$), and are beginning to slightly undershoot some bins at $a = 0.5$.  This indicates that the scaling 
$c_{\tau_{a}} = 2/(1-a)$ is indeed a good theoretical prediction.

Note that the leading bin depicted in all panels of Figs.~\ref{fig:NPplots1}--\ref{fig:NOLS} is actually the 
first bin after the peak of the distribution, as determined by the experimental data.  It is unsurprising that 
in most cases our theory predictions do not successfully describe this data point, as it is particularly 
sensitive to the form of the non-perturbative shape function. As described in \sec{NONPERTa}, we have only 
included the first term in its expansion in a series of basis functions, and one would need to 
include additional terms in this expansion in order to accurately describe the data in the peak region. 

The above results serve as a powerful visual confirmation of the predictions of factorization and resummation in 
QCD using the SCET framework, the evolution of $\as$, and the scaling and universality of leading 
non-perturbative corrections for $e^+e^-$ event shapes. We have not yet performed a full, robust statistical fit 
to the available L3 data for $\as(m_Z)$ and $\Omega_1(R_\Delta,R_\Delta)$, but visually the fit values from 
\cite{Hoang:2015hka} appear consistent within our calculations when compared to the data in \fig{NPplots1}. 
This motivates carrying out such a fit to the available LEP angularity data 
in the future.

\section{Summary and outlook}
\label{sec:CONCLUSIONS}

We have presented NNLL$^{\prime}$ resummed and $\mathcal{O}(\as^{2})$ matched distributions for the $e^{+}e^{-}$ 
event shape angularities at $Q =  91.2$ GeV and $Q = 197$ GeV. Our results are the most precise achieved 
for this observable to date, and they are made possible by a recent calculation of the two-loop soft anomalous 
dimension $\gamma_{S}^{1}$ \cite{Bell:2015lsf,Bell:2018vaa} and finite constants 
$c_{\tilde S}^{2}$ \cite{Bell:2015lsf,Bell:2018oqa}.  We determined the remaining unknown  
NNLL$^{\prime}$ ingredient $c_{\tilde J}^{2}$ from a fit to the {\tt{EVENT2}} generator, 
from which we also obtained the fixed-order matching to QCD at  $\mathcal{O}(\as^{2})$.  
We further modeled non-perturbative effects with a gapped shape function whose renormalon ambiguity was cancelled using the ``Rgap'' scheme to define the 
appropriate subtractions and obtain the ``$R$-evolution'' of the non-perturbative moment $\Omega_1(\mu,R)$.
Finally, reliable theory uncertainty estimates have been obtained via variations about a suitable profile function 
describing the transition of non-perturbative and perturbative scales throughout the relevant domain.

We have compared the predictions resulting from this analysis to data from the L3 collaboration,
and find excellent agreement for all angularities considered in this study. In particular, we have visually 
confirmed the predicted sensitivity of the distributions to leading non-perturbative scaling effects 
that are simultaneously encoded in the angularity parameter $a$ and the universal shift 
parameter ${\Omega}_{1}$ --- the distributions indeed seem to require shifts proportional to $2/(1-a)$ 
in order to accurately describe the data.  

%----------------------------------------------------------------------------
\begin{figure}[t]
\centering
\begin{tabular}{cc}
\quad {\large\bf $a = -1.0$} & \quad {\large\bf $a = 0.5$} \\ [0.7em]
\hspace{2mm}
\includegraphics[scale=0.24]{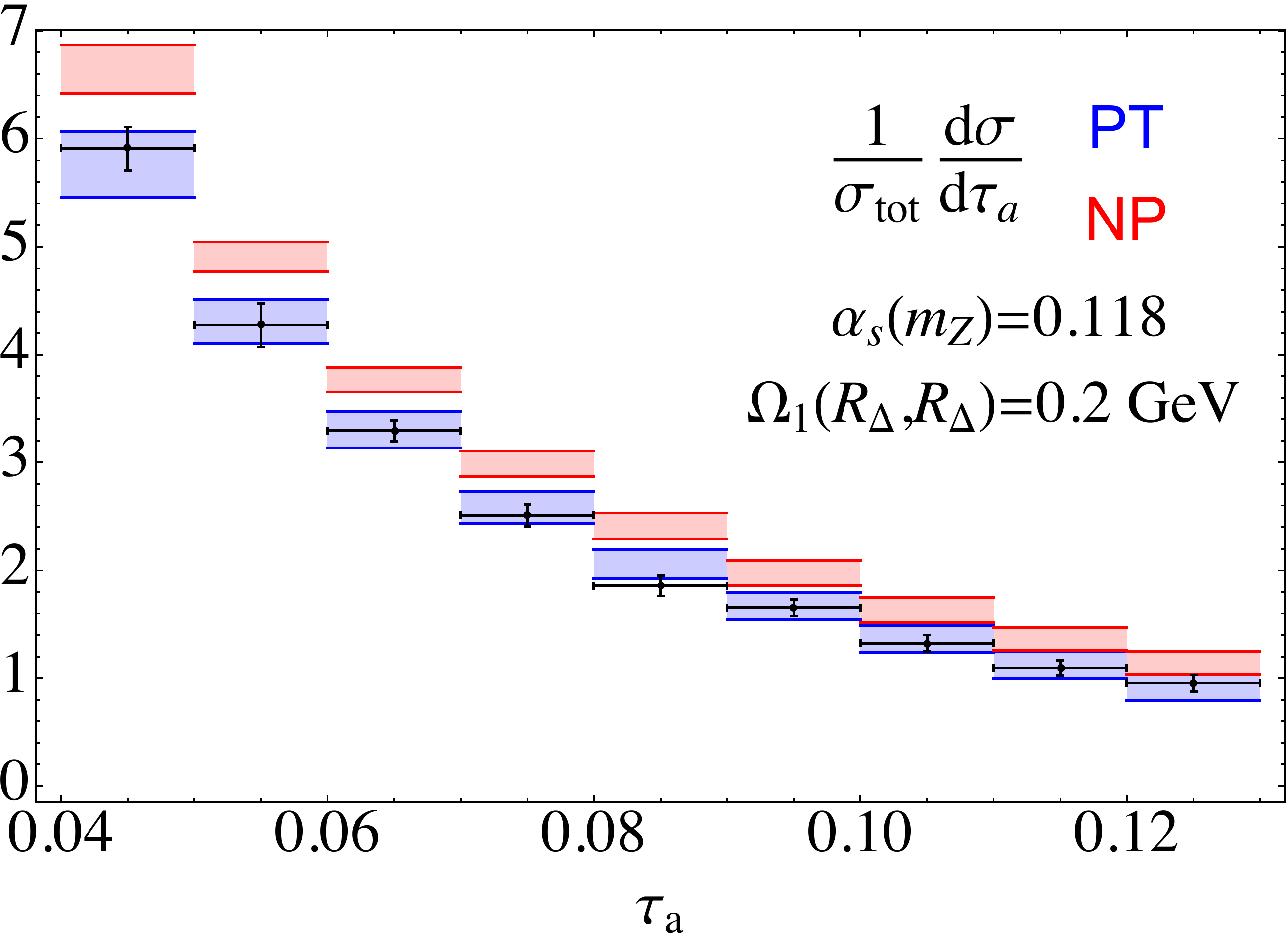}&
\hspace{5mm}
\includegraphics[scale=0.24]{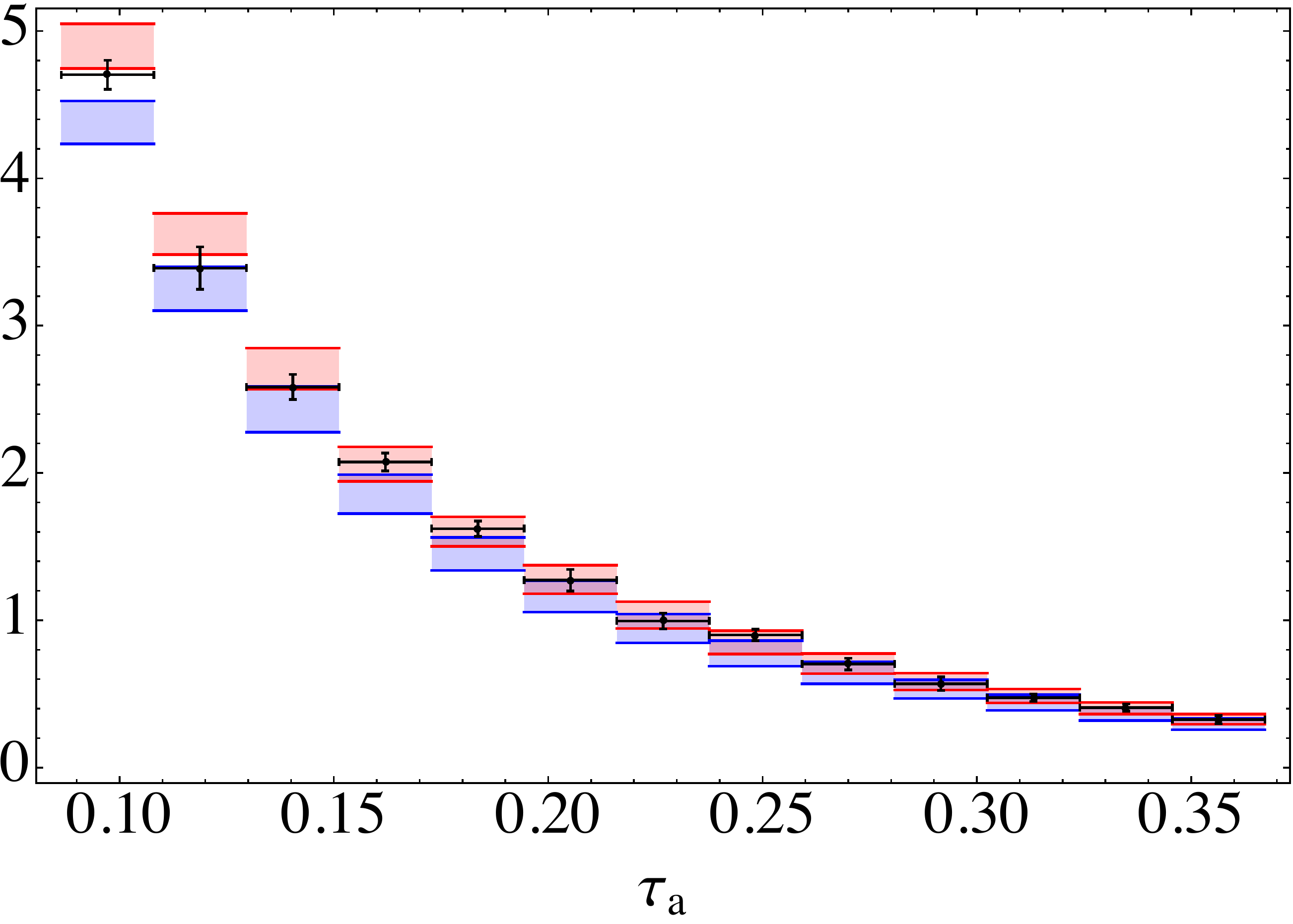}\\
\hspace{-5mm}
\includegraphics[scale=0.28]{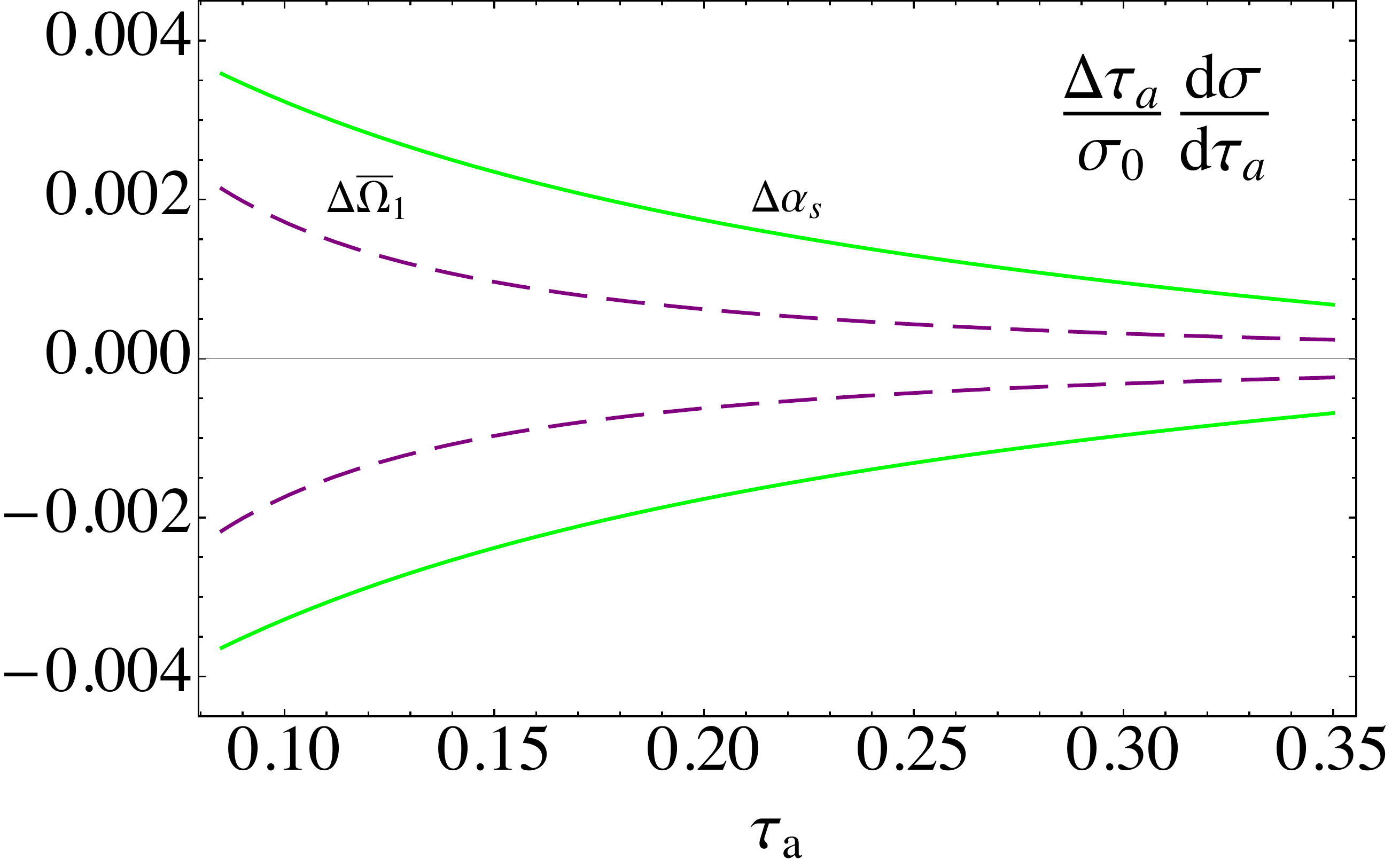} &
\includegraphics[scale=0.28]{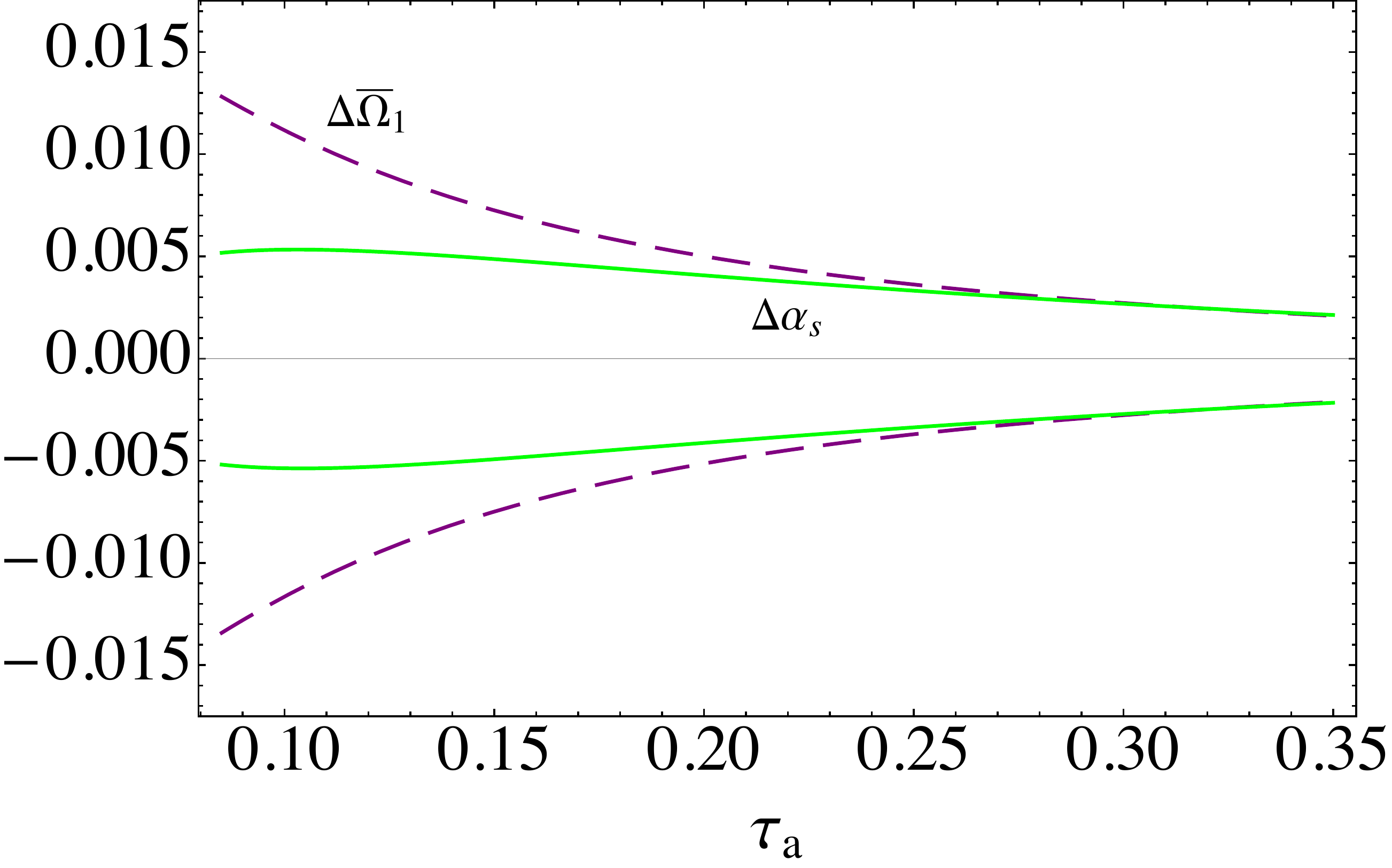}
\end{tabular}
\vspace{-1em}
\caption{\emph{Top}:  A binned data comparison at $a = \lbrace -1.0, \,0.5 \rbrace$ for distributions 
generated with $\alpha_{s}(m_Z) = 0.118$ and $\Omega_1(R_\Delta,R_\Delta) = 0.2$~GeV.  All other settings are the same as in 
Figure \ref{fig:NPplots1}.  As is clear, the final renormalon-corrected (red) curves overshoot the data 
at $a = -1.0$ but, thanks to the small shift parameter, successfully describe the $a = 0.5$ data.  
\emph{Bottom}: Difference between central curves and curves evaluated with single variations of 
either $\overline{\Omega}_1$ (dashed, purple) or $\alpha_{s}(m_{Z})$ (solid, green)
at the same values of $a$ and $Q= m_Z$.  Note that the lower plots are generated with canonical scales, are 
unmatched, and the variation about $\overline\Omega_{1}$ is illustrated using the shift \eqref{eq:shift}, 
rather than the fully shape and renormalon-corrected formulae.
}
\label{fig:avariation}
\end{figure}
%-----------------------------------------------------------------------------

This characteristic non-perturbative scaling strongly motivates an extraction of both $\alpha_{s}$ 
and $\Omega_1$ from a fit to experimental data
on angularities.  As previous extractions of $\as$ from $e^+e^-$ event shapes like thrust and $C$-parameter have shown~\cite{Abbate:2010xh,Abbate:2012jh,Hoang:2015hka}, the disentangling of perturbative and non-perturbative effects is crucial and has a large impact on the determined value of $\as$ itself. Thus angularities, with their tunable relative contributions of these effects, 
can play an important role in confirming (or not) the values coming from these other extractions, which as mentioned before, are considerably lower than the PDG world average~\cite{PhysRevD.98.030001}.
%, as {\color{purple} a conclusive confirmation or resolution} of the aforementioned discrepancies 
%between values of the strong coupling
%extracted from $e^{+}e^{-}$ event-shape data using 
%SCET~\cite{Abbate:2010xh,Abbate:2012jh,Hoang:2015hka} and
%the PDG world average~\cite{PhysRevD.98.030001}
%will require non-perturbative and perturbative effects to be disentangled.  
While we have not performed a 
dedicated extraction in this study, in the top panels of \fig{avariation} we have shown the analogous plots 
to those in \fig{NPplots1} at $a = \lbrace -1.0, 0.5 \rbrace$, but with $\alpha_{s}(m_{Z}) = 0.118$ 
(consistent with the PDG value)
and 
$\Omega_1(R_\Delta,R_\Delta) = 0.2$~GeV, just large enough to consistently implement our gap parameter and provide 
a good fit to the data for the angularity $a=0.5$.  As can be seen, however, at $a = -1.0$ the final theory 
bins are then well above the data in the tail region.  As the practical effect of lowering (raising) 
$\alpha_{s}(m_{Z})$ (${\Omega}_{1}$) away from the best values used in \fig{NPplots1} is to the lower (raise) 
the theory prediction, this na\"ive exercise leads one to provisionally expect that such parameter values will 
be difficult to reconcile with the available data.

In the bottom panels of \fig{avariation} we finally show the difference
$(d\sigma / d \tau_{a})_{\text{central}} - d\sigma / d \tau_{a} $ over the range 
$0.085 \le \tau_{a} \le 0.35$ for $a = \lbrace -1.0, \,0.5 \rbrace$, where 
$(d\sigma / d \tau_{a})_{\text{central}}$ is an (unmatched, non-renormalon corrected) NNLL$^{\prime}$ 
resummed curve. For $(d\sigma / d \tau_{a})$ we have varied 
2$\overline\Omega_1$ by $\pm\, 0.1$ GeV 
and $\alpha_{s}(m_{Z})$ by $\pm\, 0.001$, corresponding to the purple and green curves, respectively.  
Our plots are to be compared to Fig.~10 in \cite{Abbate:2010xh}, where the same variations were made 
at different center-of-mass energies $Q$.  We find that varying $a$ ($Q$) down (up) from high (low) values 
leads to an enhanced sensitivity to the relative effects of $\alpha_{s}$ and $\overline\Omega_{1}$ variation.  
This suggests that extractions from angularity data
at a single centre-of-mass energy $Q$, but different values of $a$, will be able to discriminate between 
$\alpha_{s}$ and $\overline\Omega_{1}$ in a 
similar way to varying $Q$.  In fact, angularities for $-2 \le a \leq 0.5 $ exhibit a factor of six 
variance in the overall non-perturbative shift, which represents an essentially equivalent sensitivity 
to measurements made between $Q = 35$ GeV and  
$Q= 207$ GeV, as analyzed for thrust in~\cite{Abbate:2010xh}.   Unfortunately, angularity data is  
currently only available down to $a = -1.0$ and only at two values of $Q$.  
Thus a reanalysis of additional LEP data would be highly desirable. Of course, a high-precision 
determination of $\as$ and $\Omega_1$ should ideally also include further theory improvements beyond 
those we have already implemented here, such as hadron mass effects or higher-precision numerical 
(or even analytical) determinations of $c_{\tilde J}^2$ and $r^2(\tau_a)$.
New methods and calculations for $\cO(\as^3)$ jet and soft functions at $a\neq 0$ would be needed to go to N$^3$LL$(')$ accuracy, while $\cO(\as^3)$ fixed-order matching to QCD could be obtained from a program like \texttt{EERAD3} \cite{Ridder:2014wza}.
In conclusion, we are optimistic that the $a$-dependence of the angularities may ultimately 
be helpful in lifting degeneracies between $\alpha_{s}$ and ${\Omega}_{1}$ in the two-parameter 
event-shape fits.  We leave it to future work to bring such an analysis to full culmination.

%%%%%%%%%%%%%%%%%%%%%%%%%%%%%%%%%%%%%%%%%%%%%%%%%%%%%%%%%%%%%%%

\acknowledgments

We thank the organizers of the 2015 SCET Workshop where this project began and the 2018 SCET Workshop, 
which helped draw it to its conclusion. We would like to thank Grigory Ovanesyan for collaboration on early 
stages of this project, and Jonathan Walsh who, in collaboration on related past 
work, wrote some of the codes we used to analyze the \event\ data. We also thank Massimilano Procura
for discussions. JT thanks the T-2 Group at LANL, and CL in particular, 
for hospitality and support while much of this work was completed.  JT  also acknowledges research and travel 
support from DESY and the Senior Scholarship Trust of Hertford College, Oxford.  
The work of AH and CL was supported by the U.S. Department of Energy, Office of Science, Office of Nuclear 
Physics, under Contract DE-AC52-06NA25396 and an Early Career Research Award, and by the LANL/LDRD Program.

Finally, we would like to dedicate this work to the memory of Andrew Hornig (1982--2018) who passed away suddenly before its final publication.

%%%%%%%%%%%%%%%%%%%%%%%%%%%%%%%%%%%%%%%%%%%%%%%%%%%%%%%%%%%%%%%

\appendix

\section{Laplace transforms}
\label{app:Laplace}

In this appendix we collect results for the Laplace transforms and inverse Laplace transforms
between the logarithms
\be
L\equiv \ln \frac{1}{\tau} \,,\qquad \tilde L \equiv \ln(\nu e^{\gamma_E})\,.
\ee
The Laplace transforms, defined by
\be
\wt F(\nu) \equiv \LP \{ F\} (\nu) = \int_0^\infty d\tau \,e^{-\nu\tau} F(\tau)
\ee
are given by
\begin{align}
\label{eq:LPs}
\LP\bigl\{ 1\bigr\} &= \frac{1}{\nu}\,, 
& \LP\bigl\{ L^3\bigr\}  &= \frac{1}{\nu} \biggl\{ \tilde L^3 + \frac{\pi^2}{2} \tilde L + 2\zeta_3 \biggr\}\,, \\
\LP\bigl\{ L\bigr\} &= \frac{1}{\nu} \, \tilde L\,,
 & \LP\bigl\{ L^4\bigr\} &= \frac{1}{\nu} \biggl\{ \tilde L^4 + \pi^2 \tilde L^2 + 8\zeta_3 \tilde L  + \frac{3\pi^4}{20} \biggr\}  \nn \,, \\
\LP\bigl\{ L^2\bigr\}&= \frac{1}{\nu} \biggl\{ \tilde L^2 + \frac{\pi^2}{6} \biggr\} \nn\,,
\end{align}
and so on. 
The inverse Laplace transforms, defined by
\be
\iLP \{ \wt F\}(\tau) = \int_{\gamma-i\infty}^{\gamma+i\infty}\,
\frac{d\nu}{2\pi i} \,e^{\nu\tau} \wt F(\nu)\,,
\ee
where $\gamma$ lies to the right of all the poles of $\wt F$ in the complex plane,
are given by
\begin{align}
\label{eq:iLPs}
\iLP\Bigl\{\frac{1}{\nu} \Bigr\} &= 1\,, 
& \iLP\Bigl\{\frac{1}{\nu}\tilde L^3\Bigr\}  &= L^3 - \frac{\pi^2}{2} L - 2\zeta_3\,, \\
\iLP\Bigl\{\frac{1}{\nu}\tilde L\Bigr\} &= L\,, 
& \iLP\Bigl\{\frac{1}{\nu}\tilde L^4\Bigr\}  &= L^4 - \pi^2 L^2 - 8\zeta_3 L  + \frac{\pi^4}{60} \nn \,, \\
\iLP\Bigl\{\frac{1}{\nu}\tilde L^2\Bigr\} &= L^2 - \frac{\pi^2}{6}\,,  & & \nn
\end{align}
and so on. The results explicitly tabulated in \eqs{LPs}{iLPs} are needed to transform 
logarithms in the fixed-order expansions of event-shape distributions in QCD up to 
$\cO(\as^2)$.
\section{Anomalous dimensions}%
\label{app:anomalous}

The coefficients of the beta function up to four-loop order in the $\overline{\mathrm{MS}}$  scheme are given by~\cite{Tarasov:1980au, Larin:1993tp,vanRitbergen:1997va}
%%%
\begin{align} \label{eq:betacoeffs}
\beta_0 &= \frac{11}{3}\,C_A -\frac{4}{3}\,T_F\,n_f
\,,\\
\beta_1 &= \frac{34}{3}\,C_A^2  - \Bigl(\frac{20}{3}\,C_A\, + 4 C_F\Bigr)\, T_F\,n_f
\,, \nn\\
\beta_2 &=
\frac{2857}{54}\,C_A^3 + \Bigl(C_F^2 - \frac{205}{18}\,C_F C_A
 - \frac{1415}{54}\,C_A^2 \Bigr)\, 2T_F\,n_f
 + \Bigl(\frac{11}{9}\, C_F + \frac{79}{54}\, C_A \Bigr)\, 4T_F^2\,n_f^2
\,,\nn \\
\beta_3 &= 4826.16\quad (N_C=3,n_f = 5)\,.
\end{align}
The running coupling $\as(\mu)$ up to four-loop accuracy is given by the formula,
\begin{align}
\label{eq:runningalpha}
& \frac{\as(m_Z)}{\as(\mu)} = X + \frac{\as(m_Z)}{4\pi}\frac{\beta_1}{\beta_0}\ln X + \Bigl(\frac{\as(m_Z)}{4\pi}\Bigr)^2 \biggl[\frac{\beta_2}{\beta_0}\Bigl( 1 \minus \frac{1}{X}\Bigr) + \frac{\beta_1^2}{\beta_0^2}\Bigl(\frac{\ln X}{X} \plus \frac{1}{X} \minus 1\Bigr)\biggr] \\
&+ \Bigl(\frac{\as(m_Z)}{4\pi}\Bigr)^3 \frac{1}{X^2} \biggl[ \frac{\beta_3}{2\beta_0}( X^2 \minus 1) + \frac{\beta_1\beta_2}{\beta_0^2} (X \plus  \ln X \minus X^2) + \frac{\beta_1^3}{2\beta_0^3}\bigl((1\minus X)^2 \minus \ln^2 X\bigr)\biggr] \,, \nn
\end{align} 
where
\be
X\equiv 1+ \frac{\as(m_Z)\beta_0}{2\pi}\ln\frac{\mu}{m_Z}\,.
\ee
The four-loop term in \eq{runningalpha} agrees with the numerical form given in \cite{Abbate:2010xh}, and is rederived below in \appx{invariance}.

The cusp anomalous dimension coefficients are given up to 3-loop order ~\cite{Korchemsky:1987wg, Moch:2004pa}
\begin{align}\label{eq:Gacuspexp}
\Gamma^q_0 &= 4C_F
\,,\\
\Gamma^q_1 &= 4C_F \Bigl[\Bigl( \frac{67}{9} -\frac{\pi^2}{3} \Bigr)\,C_A  -
   \frac{20}{9}\,T_F\, n_f \Bigr]
\,,\nn\\
\Gamma^q_2 &= 4C_F \Bigl[
\Bigl(\frac{245}{6} -\frac{134 \pi^2}{27} + \frac{11 \pi ^4}{45}
  + \frac{22 \zeta_3}{3}\Bigr)C_A^2 
  + \Bigl(- \frac{418}{27} + \frac{40 \pi^2}{27}  - \frac{56 \zeta_3}{3} \Bigr)C_A\, T_F\,n_f
\nn\\* & \hspace{8ex}
  + \Bigl(- \frac{55}{3} + 16 \zeta_3 \Bigr) C_F\, T_F\,n_f
  - \frac{16}{27}\,T_F^2\, n_f^2 \Bigr] \,. \nn 
\end{align}
%%%

The $\overline{\mathrm{MS}}$ non-cusp anomalous dimension 
for the hard function can be obtained~\cite{Idilbi:2006dg, Becher:2006mr} from the infrared divergences of the on-shell massless quark form factor, 
which are known to three loops~\cite{Moch:2005id},
%%%
\begin{align} \label{eq:gaHexp}
\gamma_H^0 &= -12 C_F
\,,\\
\gamma_H^1
&= - 2C_F 
\Bigl[
  \Bigl(\frac{82}{9} - 52 \zeta_3\Bigr) C_A
+ (3 - 4 \pi^2 + 48 \zeta_3) C_F
+ \Bigl(\frac{65}{9} + \pi^2 \Bigr) \beta_0 \Bigr]
\,,\nn\\
\gamma_H^2
&= -4C_F \Bigl[
  \Bigl(\frac{66167}{324} - \frac{686 \pi^2}{81} - \frac{302 \pi^4}{135} - \frac{782 \zeta_3}{9} + \frac{44\pi^2 \zeta_3}{9} + 136 \zeta_5\Bigr) C_A^2
\nn\\ & 
\quad + \Bigl(\frac{151}{4} - \frac{205 \pi^2}{9} - \frac{247 \pi^4}{135} + \frac{844 \zeta_3}{3} + \frac{8 \pi^2 \zeta_3}{3} + 120 \zeta_5\Bigr) C_F C_A
\nn\\ & 
\quad + \Bigl(\frac{29}{2} + 3 \pi^2 + \frac{8\pi^4}{5} + 68 \zeta_3 - \frac{16\pi^2 \zeta_3}{3} - 240 \zeta_5\Bigr) C_F^2 \nn \\
&\quad + \Bigl(-\frac{10781}{108} + \frac{446 \pi^2}{81} + \frac{449 \pi^4}{270} - \frac{1166 \zeta_3}{9} \Bigr) C_A \beta_0
\nn\\ & 
\quad + \Bigl(\frac{2953}{108} - \frac{13 \pi^2}{18} - \frac{7 \pi^4 }{27} + \frac{128 \zeta_3}{9}\Bigr)\beta_1
+ \Bigl(-\frac{2417}{324} + \frac{5 \pi^2}{6} + \frac{2 \zeta_3}{3}\Bigr)\beta_0^2
\Bigr] \nn
\,.\end{align}
In the text we found it convenient to split up the Abelian and non-Abelian pieces of $\gamma_H^1$. We define
\be
\label{eq:gammaHsplit}
\begin{split}
\gamma_H^1\bigr\rvert_{C_F} &\equiv -2C_F^2 (3-4\pi^2+48\zeta_3)\,, \\
\gamma_H^1\bigr\rvert_{\text{n.A.}} &\equiv -2C_F \biggl[ \biggl( \frac{82}{9} - 52\zeta_3\biggr) C_A + \biggl( \frac{65}{9} + \pi^2\biggr)\beta_0\biggr]\,.
\end{split}
\ee

%%%
The non-cusp anomalous dimensions for the $a=0$ soft function are given explicitly by
\begin{align}
\label{eq:gammaS2}
\gamma_{S}^0  &= 0\,, \\
\gamma_{S}^1 &= -2C_F \Bigl[ \Big(\frac{64}{9} - 28\zeta_3\Bigr) C_A + \Bigl(\frac{56}{9} - \frac{\pi^2}{3}\Bigr)\beta_0 \Bigr ]\,, \nn \\
\gamma_{S}^2
&= - 2C_F\bigg[ C_A^2 \bigg( \frac{37871}{162} - \frac{310 \pi^2}{81}-\frac{8 \pi^4}{5} -\frac{2548 \zeta_3}{9}+ \frac{88\pi^2 \zeta_3}{9} +192 \zeta_5\bigg) \nn \\
&\qquad \quad + C_A \beta_0 \bigg( -\frac{4697}{54}-\frac{242\pi^2}{81}+\frac{56\pi^4}{45}-\frac{220 \zeta_3}{9}\bigg) \nn
\\ & \qquad\quad
+ \beta_1 \bigg( \frac{1711}{54}-\frac{\pi^2}{3}-\frac{4\pi^4}{45}-\frac{152 \zeta_3}{9}\bigg)
+ \beta_0^2  \bigg(-\frac{520}{81}-\frac{5\pi^2}{9}+\frac{28 \zeta_3}{3} \bigg)\bigg] \,. \nn
\end{align}
For $a\neq 0$, we have $\gamma_S^0(a)=0$ \cite{Hornig:2009vb}
and the non-cusp two-loop soft anomalous dimension was given in \eq{gammaS1a}. 
The non-cusp jet anomalous dimension is then given by the consistency relation
$2\gamma_J = -\gamma_H - \gamma_S$. 

\section{RG invariance of total evolution kernels}
\label{app:invariance}

%%%%%%%%%%%%%%%%%%%%%%%%%%%%%%%%%%%%%%%%%%%%%%%%%%%%%%%
\subsection{Residual scale dependence due to standard change of variables}

Renormalization group (RG) invariance of the physical cross section guarantees that the combination of all evolution kernels on the right-hand side of \eq{differential} is independent of the scale $\mu$. This is in fact true not only in the limit of infinite accuracy in perturbation theory in $\as$, but at each \emph{finite} order of resummed and fixed-order accuracy, as a consequence of the anomalous dimensions satisfying the consistency relation $\gamma_H(\mu) + 2\gamma_J(\mu) + \gamma_S(\mu) = 0$ at each finite order.\footnote{The dependence of the factorized cross sections on the individual scales $\mu_{H,J,S}$, by contrast, inherits the usual residual dependence on the running of $\alpha_s$ at any truncated order, canceled only by higher-order terms.} The evaluation of $K_\Gamma$ defined in \eq{Keta}, after the change of integration variables \eq{dmu},
\be
K_\Gamma(\mu,\mu_F) = \int_{\as(\mu_F)}^{\as(\mu)} \frac{d\alpha}{\beta[\alpha]} \Gcusp[\alpha] \int_{\as(\mu_F)}^\alpha\frac{d\alpha'}{\beta[\alpha']}\,,
\ee
order-by-order in logarithmic accuracy gives the well-known results (e.g. \cite{Ligeti:2008ac,Abbate:2010xh}):
\begin{subequations}
\label{eq:Kclosedform}
\begin{align}
K_\Gamma^{\text{LL}}(\mu,\mu_F) &=  \frac{\Gamma_0}{4\beta_0^2} \frac{4\pi}{\alpha_s(\mu_F)} \biggl\{  \ln r + \frac{1}{r} - 1  \biggr\}\,, \\
K_\Gamma^{\text{NLL}}(\mu,\mu_F)  &=  \frac{\Gamma_0}{4\beta_0^2} \biggl\{ \left(\frac{\Gamma_1}{\Gamma_0} \minus \frac{\beta_1}{\beta_0}\right) ( r \minus 1\minus \ln r) \minus \frac{\beta_1}{2\beta_0} \ln^2 r  \biggr\}\,,\\
K_\Gamma^{\text{NNLL}}(\mu,\mu_F)  &=  \frac{\Gamma_0}{4\beta_0^2} \frac{\as(\mu_F)}{4\pi} \biggl\{  B_2\left( \frac{r^2 \minus 1}{2} - \ln r\right) + \left(\frac{\beta_1 \Gamma_1}{\beta_0\Gamma_0} \minus  \frac{\beta_1^2}{\beta_0^2}\right) ( r  \minus 1\minus r\ln r)  \\
& \qquad\qquad\qquad + \left(\frac{\Gamma_2}{\Gamma_0} - \frac{\beta_1\Gamma_1}{\beta_0\Gamma_0}\right) \frac{(1-r)^2}{2}  \biggr\}\,, \nn \\
K_\Gamma^{\text{N$^3$LL}}(\mu,\mu_F) &= \frac{\Gamma_0}{4\beta_0^2} \Bigl(\frac{\as(\mu_F)}{4\pi}\Bigr)^2 \biggl\{ \Bigl( \frac{\Gamma_3}{\Gamma_0} - \frac{\Gamma_2\beta_1}{\Gamma_0\beta_0} + \frac{\Gamma_1}{\Gamma_0}B_2 + B_3\Bigr) \frac{r^3-1}{3} - \frac{B_3}{2}\ln r \\
&\qquad \qquad \qquad - B_2 \Bigl(\frac{\Gamma_1}{\Gamma_0} - \frac{\beta_1}{\beta_0}\Bigr)(r-1) - \frac{\beta_1}{2\beta_0} \Bigl( \frac{\Gamma_2}{\Gamma_0} - \frac{\Gamma_1\beta_1}{\Gamma_0\beta_0} + B_2\Bigr) r^2 \ln r \nn \\
&\qquad \qquad \qquad + \Bigl(-\frac{2\Gamma_3}{\Gamma_0} + \frac{3\Gamma_2\beta_1}{\Gamma_0\beta_0} - \frac{\Gamma_1\beta_1^2}{\Gamma_0\beta_0^2} + \frac{\beta_3}{\beta_0} - \frac{\beta_1\beta_2}{\beta_0^2}\Bigr) \frac{r^2-1}{4} \biggr\} \,, \nn
\end{align}
\end{subequations}
where $r$, $B_2$ and $B_3$ were defined in \eq{rB} and the corresponding 
expressions for the kernel $\eta_\Gamma$ were given in \eq{etaclosedform}.  From these formulae we can explicitly check the property of $\mu$-invariance.

In their general form \eq{KOmega}, we can check the $\mu$-independence that $K$ and $\Omega$ \emph{should} satisfy. For instance, for $\Omega$:
\begin{align}
\label{eq:Omegaproof}
\Omega &= 2\omega_J(\mu,\mu_J) + \omega_S(\mu,\mu_S) \\
&= -2\kappa_J \int_{\mu_J}^\mu d\ln\mu' \, \Gcusp[\as(\mu')] - \kappa_S \int_{\mu_S}^\mu d\ln\mu' \, \Gcusp[\as(\mu')] \nn \\
&= -\kappa_S \int_{\mu_S}^{\mu_J} d\ln\mu' \,\Gcusp[\as(\mu')]\,, \nn
\end{align}
using that $2\kappa_J = -\kappa_S$ from \eq{jFkF}, and thus $\Omega$ is exactly $\mu$-independent. The same is then clearly true of the sum of the non-cusp parts of $K$. Focusing, then, on the cusp terms in $K$, they form a $\mu$-independent combination with all the $\omega$-dependent factors on the first line of \eq{differential}:
\begin{subequations}
\label{eq:Kproof}
\begin{align}
\label{eq:Kproof1}
&e^{-\kappa_H K_\Gamma(\mu,\mu_H)}  \biggl (\frac{\mu_H}{Q}\biggr)^{\omega_H(\mu,\mu_H)} 
e^{-2j_J\kappa_J K_\Gamma(\mu,\mu_J)} \biggl(\frac{\mu_J}{Q\tau_a^{1/j_J}}\biggr)^{2j_J\omega_J(\mu,\mu_J)}
e^{-\kappa_S K_\Gamma(\mu,\mu_S)} \biggl(\frac{\mu_S}{Q\tau_a}\biggr)^{\omega_S(\mu,\mu_S)} \nn  \\
&\quad = -\kappa_H \int_{\mu_H}^\mu d\ln\mu' \, \Gcusp[\as(\mu')] \ln\frac{\mu'}{\mu_H} - \kappa_H \int_{\mu_H}^\mu d\ln\mu' \, \Gcusp[\as(\mu')] \ln\frac{\mu_H}{Q}   \\
&\qquad - 2j_J\kappa_J  \int_{\mu_J}^\mu d\ln\mu' \, \Gcusp[\as(\mu')] \ln\frac{\mu'}{\mu_J} - 2j_J\kappa_J \int_{\mu_J}^\mu d\ln\mu' \, \Gcusp[\as(\mu')] \ln\frac{\mu_J}{Q\tau_a^{1/j_J}} \nn \\
&\qquad - \kappa_S \int_{\mu_S}^\mu d\ln\mu'\, \Gcusp[\as(\mu')]\ln\frac{\mu'}{\mu_S} - \kappa_S\int_{\mu_S}^\mu d\ln\mu' \, \Gcusp[\as(\mu')]\ln\frac{\mu_S}{Q\tau_a} \,, \nn
\end{align}
wherein we group together the terms depending on $Q$ and on $\tau_a$:
\begin{align}
&\quad =  -\kappa_H \int_{\mu_H}^\mu d\ln\mu' \, \Gcusp[\as(\mu')] \ln\frac{\mu'}{Q} - 2j_J\kappa_J \int_{\mu_J}^\mu d\ln\mu'\, \Gcusp[\as(\mu')] \ln\frac{\mu'}{Q} \nn \\
&\qquad - \kappa_S \int_{\mu_S}^\mu d\ln\mu' \, \Gcusp[\as(\mu')] \ln\frac{\mu'}{Q} \nn \\
&\qquad + \ln\tau_a\biggl\{  2\kappa_J\int_{\mu_J}^\mu d\ln\mu' \, \Gcusp[\as(\mu')]  + \kappa_S \int_{\mu_S}^\mu d\ln\mu' \, \Gcusp[\as(\mu')]  \biggr\} \nn\,,
\end{align}
which, using the relations $\kappa_H + 2j_J\kappa_J + \kappa_S = 0$ and $2\kappa_J + \kappa_S = 0$ from \eq{jFkF}, simplifies to
\begin{align}
\label{eq:Kproof2}
&= - \kappa_H \int_{\mu_H}^{\mu_J} d\ln\mu' \, \Gcusp[\as(\mu')] \ln \frac{\mu'}{Q} - \kappa_S \int_{\mu_S}^{\mu_J} d\ln\mu' \, \Gcusp[\as(\mu')]\ln\frac{\mu'}{Q\tau_a} \,,
\end{align}
\end{subequations}
which is explicitly $\mu$-independent. The properties \eqs{Omegaproof}{Kproof} \emph{should} hold not only at infinite order in $\as$, but at every truncated order, since the anomalous dimensions satisfy consistency at every order. Thus the expression in \eq{differential} for the resummed cross section should be exactly $\mu$-independent at any truncated order of perturbation theory.

If, however, one uses the closed-form expressions in \eqs{Kclosedform}{etaclosedform} for $K_\Gamma,\eta_\Gamma$ to evaluate the individual kernels $K_F(\mu,\mu_F),\omega_F(\mu,\mu_F)$ at a truncated resummed accuracy, one will find a small residual dependence on $\mu$ left over in \eq{differential} in practice (see \tab{muvariation} below). This residual dependence is subleading to the order at which the pieces of \eq{differential} are truncated, but the proof in \eq{Kproof} should have guaranteed the dependence to be exactly vanishing. What happened?

Examining the steps in \eq{Kproof}, we observe that the cancellation of $\mu$ dependence between the different pieces of the total evolution kernel relies, among other things, on the 
scales $\mu_{H,J,S}$ in the logarithms inside each integral canceling between the logarithms of $\mu/\mu_F$ in the $K$ part of the kernels and the logarithms of $\mu_F/Q_F$ in the $\omega_F$ parts of the kernels. In obtaining the closed forms of $K_\Gamma$ in \eq{Kclosedform}, however, we made the substitution (see \eq{Ketaalpha}):
\be
\label{eq:lnmu}
\ln\frac{\mu}{\mu_F} = \int_{\as(\mu_F)}^{\as(\mu)}\frac{d\alpha}{\beta[\alpha]} \,,
\ee
using the definition of the beta function in \eq{beta}, $d\mu/\mu = d\alpha_s/\beta[\alpha_s]$. It turns out that this innocent-looking relation is only approximate at a given order of truncated accuracy (wherein $\beta$ and $\as$ are themselves truncated), with nonzero corrections that are subleading. The exceptions are at LL accuracy and N$^\infty$LL accuracy, where it is exact. If \eq{lnmu} is violated, then the $\mu$ dependence in the combination of RG evolution kernels evaluated in \eq{Kproof} will no longer be exactly zero. While numerically this is not a serious issue, since corrections are subleading, our observation 
may help to explain any perplexing residual $\mu$-dependence of a cross section evaluated using an equation like \eq{differential}. To obtain a purely $\mu$-independent expression for the cross section, one commonly cancels it in the generic forms for the evolution kernels as in \eq{Kproof2}, and then uses the closed-form expressions in \eqs{Kclosedform}{etaclosedform} to evaluate them, which amounts just to picking, e.g., $\mu=\mu_J$, as is the case in \eq{Kproof2}. The resulting expression, then, still implicitly depends on this hidden choice. Can we remove the dependence on such a choice of $\mu$ \emph{exactly}?

Let us proceed to evaluate \eq{lnmu} explicitly at finite orders of accuracy up to N$^3$LL. The expansion of the beta function to the needed accuracy is:
\be
\beta[\as] = -2\as\biggl[ \beta_0 \frac{\as}{4\pi} + \blue{ \beta_1 \Bigl(\frac{\as}{4\pi}\Bigr)^2} + \red{\beta_2 \Bigl(\frac{\as}{4\pi}\Bigr)^3}  + \green{\beta_3 \Bigl(\frac{\as}{4\pi}\Bigr)^4} \biggr]\,,
\ee
where we will keep track of terms of LL, \blue{NLL}, \red{NNLL}, and \green{N$^3$LL} accuracy by color coding. Plugged into \eq{lnmu}, and Taylor expanding the denominator, we obtain the relation
\begin{align}
\label{eq:betaintegral}
\ln\frac{\mu}{\mu_F} &= -\frac{2\pi}{\beta_0} \int_{\as(\mu_F)}^{\as(\mu)} \frac{d\alpha}{\alpha^2} \frac{1}{ 1 + \blue{\frac{\alpha}{4\pi}\frac{\beta_1}{\beta_0}} + \red{\bigl(\frac{\alpha}{4\pi}\bigr)^2 \frac{\beta_2}{\beta_0}} + \green{\bigl(\frac{ \alpha}{4\pi}\bigr)^3 \frac{\beta_3}{\beta_0} }}\\
&= -\frac{2\pi}{\beta_0} \int_{\as(\mu_F)}^{\as(\mu)} \frac{d\alpha}{\alpha^2}\biggl[ 1 - \blue{ \frac{\alpha}{4\pi}\frac{\beta_1}{\beta_0}} + \red{\Bigl(\frac{\alpha}{4\pi}\Bigr)^2 \Bigl( \frac{\beta_1^2}{\beta_0^2} - \frac{\beta_2}{\beta_0}\Bigr)} - \green{\Bigl(\frac{\alpha}{4\pi}\Bigr)^3 \Bigl( \frac{\beta_1^3}{\beta_0^3} - \frac{2\beta_1\beta_2}{\beta_0^2} + \frac{\beta_3}{\beta_0}\Bigr)} \biggr] \nn \\
&= \frac{2\pi}{\beta_0}\biggl\{ \frac{1}{\as(\mu)} - \frac{1}{\as(\mu_F)} + \blue{\frac{\beta_1}{4\pi\beta_0} \ln \frac{\as(\mu)}{\as(\mu_F)}} + \red{\frac{1}{16\pi^2}\Bigl(\frac{\beta_2}{\beta_0} - \frac{\beta_1^2}{\beta_0^2} \Bigr) [ \as(\mu) - \as(\mu_F)] } \nn \\
&\qquad \qquad + \green{\frac{1}{128\pi^3} \Bigl(\frac{\beta_3}{\beta_0} - \frac{2\beta_1\beta_2}{\beta_0^2} + \frac{\beta_1^3}{\beta_0^3}\Bigr) [ \as^2(\mu) - \as^2(\mu_F)]  }\biggr\} \,. \nn
\end{align}
Now, the running coupling $\as(\mu)$ must itself have a perturbative expansion in $\as(\mu_F)$ such that this relation is satisfied order-by-order. At LL, we have from \eq{runningalpha} 
\be
\frac{1}{\as(\mu)} - \frac{1}{\as(\mu_F)} = \frac{1}{\as(\mu_F)} (X-1) = \frac{\beta_0}{2\pi}\ln\frac{\mu}{\mu_F}\,,
\ee
so at LL, \eq{betaintegral} is satisfied exactly (all colored terms are truncated away at this order). Beyond LL, let us pretend the coefficients in \eq{runningalpha} are unknown and use \eq{betaintegral} to solve for them:
\be
\label{eq:Xn}
\frac{\as(\mu_F)}{\as(\mu)} = X + \blue{\frac{\as(\mu_F)}{4\pi} X_1 } + \red{\Bigl(\frac{\as(\mu_F)}{4\pi}\Bigr)^2 X_2 } + \green{\Bigl(\frac{\as(\mu_F)}{4\pi}\Bigr)^3 X_3 }+ \cdots\,,
\ee
which plugged into \eq{betaintegral} gives the condition
\begin{align}
\label{eq:zerocondition}
0 &= \frac{1}{\as(\mu_F)} \biggl[ \blue{\frac{\as(\mu_F)}{4\pi} X_1} + \red{\Bigl(\frac{\as(\mu_F)}{4\pi}\Bigr)^2 X_2 } + \green{\Bigl(\frac{\as(\mu_F)}{4\pi}\Bigr)^3 X_3 + \cdots}  \biggr] \\
&\quad - \blue{ \frac{\beta_1}{4\pi\beta_0} \ln\biggl[ X} + \red{\frac{\as(\mu_F)}{4\pi} X_1} + \green{\Bigl(\frac{\as(\mu)}{4\pi}\Bigr)^2 X_2} + \cdots \biggr] \nn \\
&\quad + \red{ \frac{\as(\mu_F)}{16\pi^2} \Bigl(\frac{\beta_2}{\beta_0} - \frac{\beta_1^2}{\beta_0^2} \Bigr)  \biggl[ \biggl( X} + \green{\frac{\as(\mu_F)}{4\pi} X_1} + \cdots\biggr)^{-1} - \red{1} \biggr] \nn \\
&\quad + \green{ \frac{\as^2(\mu_F) }{128\pi^3} \Bigl(\frac{\beta_3}{\beta_0} - \frac{2\beta_1\beta_2}{\beta_0^2} + \frac{\beta_1^3}{\beta_0^3}\Bigr)  \Bigl( \frac{1}{X^2}  - 1\Bigr) } + \cdots \,, \nn
\end{align}
where the $\cdots$ indicate terms beyond N$^3$LL accuracy. Expanding the logarithm and reciprocal functions on the second and third lines in Taylor series and solving for $X_n$ iteratively order-by-order, we obtain precisely the coefficients of the running coupling in \eq{runningalpha},
\begin{align}
&\blue{X_1 = \frac{\beta_1}{\beta_0} \ln X}\,, \\
&\red{X_2 = \frac{\beta_2}{\beta_0} \Bigl( 1 - \frac{1}{X}\Bigr) + \frac{\beta_1^2}{\beta_0^2} \Bigl( \frac{\ln X}{X} + \frac{1}{X} - 1\Bigr)}\,, \nn \\
&\green{X_3 = \frac{1}{X^2} \biggl[ \frac{\beta_3}{2\beta_0}( X^2 \minus 1) + \frac{\beta_1\beta_2}{\beta_0^2} (X \plus  \ln X \minus X^2) + \frac{\beta_1^3}{2\beta_0^3}\bigl((1\minus X)^2 \minus \ln^2 X\bigr)\biggr]} \,. \nn
\end{align}
However, imagine truncating \eq{zerocondition}, and thus \eq{betaintegral}, at a finite N$^k$LL accuracy.  At NLL, we keep the black and blue terms of \eq{betaintegral} and truncate red and above. Plugging in up to the blue $X_1$ term of \eq{Xn} for $\as(\mu)$, we retain the blue terms on the first two lines of \eq{zerocondition} but as a byproduct also the red term of the second line:
\be
\label{eq:zeroNLL}
\blue {0 = X_1 - \frac{\beta_1}{\beta_0} \ln X } - \red{\frac{\beta_1}{\beta_0} \ln\biggl[ 1 + \frac{\as(\mu_F)}{4\pi}\frac{X_1}{X}\biggr]}\,,
\ee
which is satisfied up to the \blue{blue (NLL)} terms, but violated at \red{NNLL (red)}. We would have needed to keep the red NNLL term on the first line of \eq{zerocondition} to cancel this last term, but it was truncated at NLL. Thus at NLL the relation \eq{lnmu} is only numerically correct up to terms of NLL accuracy, and violated at subleading order beginning with the red term in \eq{zeroNLL}. In effect, the relation \eq{lnmu} at NLL accuracy is making the replacement
\be
\label{eq:mismatch}
\ln\frac{\mu}{\mu_F} \rightarrow \ln\frac{\mu}{\mu_F} \red{ -  \frac{\beta_1}{2\beta_0^2} \ln\biggl[1 + \frac{\as(\mu_F)}{4\pi}\frac{X_1}{X}\biggr]} \,,
\ee
and the presence of the red term causes violations of $\mu$-independence when plugged into \eq{Kproof}. Namely, this violation adds to the RHS of \eq{Kproof} the terms:
\begin{align}
\red{\frac{\beta_1}{2\beta_0^2}}\biggl\{ & \kappa_H \int_{\mu_H}^\mu \red{ \ln \biggl[ 1 + \frac{\as(\mu_H)\beta_1}{4\pi\beta_0}\frac{\ln X_H}{X_H}\biggr]} \\
&+ 2j_J\kappa_J \int_{\mu_J}^\mu \red{\ln \biggl[ 1 + \frac{\as(\mu_J)\beta_1}{4\pi\beta_0}\frac{\ln X_J}{X_J}\biggr]} \nn \\
& + \kappa_S\int_{\mu_S}^\mu \red{\ln \biggl[ 1 + \frac{\as(\mu_S)\beta_1}{4\pi\beta_0}\frac{\ln X_S}{X_S}\biggr] } \biggr\}\Gcusp[\as(\mu')]d\ln\mu' \,,\nn
\end{align}
where $X_F \equiv 1+ \as(\mu_F)\beta_0/(2\pi) \ln \mu' / \mu_F$, with nothing to cancel the $\mu$ dependence of these terms.
Similarly, at NNLL, \eq{zerocondition} is satisfied up to the red terms on the first three lines, but violated by the green terms on the second and third lines, which would have needed the truncated $X_3$ term on the first line. This pattern continues to higher orders.  The exact relation \eq{lnmu} is not restored beyond LL until infinite resummed accuracy.

In principle this numerical mismatch is not a problem, as the offending terms are always of subleading order to the accuracy one is working. But it is useful to be aware of, in the case one expected the cross section \eq{differential} to be exactly $\mu$-independent based on the properties shown in \eq{Kproof}, but finds it is not when using the closed-form expressions \eqs{Kclosedform}{etaclosedform} for the evolution kernels. These compact forms are extremely useful in capturing all logarithms at a given order summed into closed expressions, but the inexactness at subleading order of the relation \eq{lnmu} is the small price to pay for them.

%%%%%%%%%%%%%%%%%%%%%%%%%%%%%%%%%%%%%%%%%%%%%%%%%%%%%%%
\subsection{Alternative organization restoring RG invariance}
\label{sec:reorganize}

The observations in the previous subsection suggest a way to restore exact RG invariance to the cross section \eq{differential} at every order in resummed perturbation theory, while still using closed-form expressions as \eq{Kclosedform} for the cusp evolution kernel. Going back to the general demonstration of exact RG invariance in \eq{Kproof} (before changing integration variables), note that the only properties of the integrands that we needed for this proof were that, after combining the $K$ and $\omega_F$ parts of the RG evolution kernels,
\begin{enumerate}
\item The three integrands in the last equality of \eq{Kproof1} dependent on $\ln\mu'/Q$ are all the same.
\item The integrals multiplying $\ln\tau_a$ in the last line of \eq{Kproof1} do not contain any extra $\ln\mu'$ in the integrands, and they are proportional to $\Omega = 2\omega_J + \omega_S$, which is exactly $\mu$-invariant even after the change of variables and closed-form evaluation in \eq{etaclosedform}.
\end{enumerate}
Property 1 means that we could still apply the change of variables \eq{lnmu} to $\ln\mu'/Q$ at the end of \eq{Kproof1} and preserve exact $\mu$ invariance. Property 2 means that we do not need to apply \eq{lnmu} to the $\ln\tau_a$ factor at the end of \eq{Kproof1}---in fact, we should not.\footnote{One might be led to think that we should just apply \eq{lnmu} to the entire factor $\ln\mu'/Q\tau^{1/j_F}$ in evaluating \eq{Kproof1}, so the canceling logs are treated symmetrically. But it is straightforward to show that this actually leads to a final expression for the cross section still plagued by  mismatches like \eq{mismatch} and consequently a (subleading) violation of $\mu$-invariance: the problem is that $\ln(1/\tau_a^{1/j_F})$ is still not exactly equal to $\int_{\as(Q\tau_a^{1/j_F})}^{\as(Q)}\frac{d\alpha}{\beta[\alpha]}$ in general.}

Instead, Properties 1 and 2 lead us to write the resummed cross section \eq{differential} in terms of the following reshuffled combinations of $K_\Gamma,\eta_\Gamma$ in \eq{Keta}:
\be
\begin{split}
\wt K_\Gamma(\mu,\mu_F;Q) &\equiv K_\Gamma(\mu,\mu_F) + \eta_\Gamma(\mu,\mu_F) \ln\frac{\mu_F}{Q} \\
&= \int_{\mu_F}^\mu \frac{d\mu'}{\mu'} \Gcusp[\as(\mu')] \ln\frac{\mu'}{Q} \,,
\end{split}
\ee
to which we can freely apply the change of variables \eq{lnmu} and express it as:
\be
\wt K_\Gamma(\mu,\mu_F;Q) = \int_{\as(\mu_F)}^{\as(\mu)} \frac{d\alpha}{\beta[\alpha]} \Gcusp[\alpha] \int_{\as(Q)}^\alpha\frac{d\alpha'}{\beta[\alpha']}\,,
\ee
similar to \eq{Ketaalpha}. It is now easy to show that the combination
\be
\label{eq:Ktilde}
\wt K(\mu_H,\mu_J,\mu_S;Q) \equiv -\kappa_H \wt K_\Gamma(\mu,\mu_H;Q) - 2j_J\kappa_J \wt K_\Gamma(\mu,\mu_J;Q) - \kappa_S \wt K_\Gamma(\mu,\mu_S;Q)
\ee
is by itself exactly $\mu$-invariant, at any order of resummed accuracy, not just LL. This leads us to reorganize the pieces of the cross section \eq{differential},
\begin{align}
\label{eq:sigmainvariant}
\frac{\sigma_\text{sing}(\tau_a)}{\sigma_0} &= e^{\wt K(\mu_H,\mu_J,\mu_S;Q) + K_\gamma(\mu_H,\mu_J,\mu_S)}  \biggl(\frac{1}{\tau_a}\biggr)^{\Omega(\mu_J,\mu_S)} H(Q^2,\mu_H)  \\
&\quad\times \wt J\Bigl(\partial_\Omega + \ln\frac{\mu_J^{2-a}}{Q^{2-a}\tau_a},\mu_J\Bigr)^2 \;\wt S\Bigl(\partial_\Omega + \ln\frac{\mu_S}{Q\tau_a},\mu_S\Bigr)
\times
\begin{cases}
\frac{1}{\tau_a}\mathcal{F}(\Omega) & \sigma = \frac{d\sigma}{d\tau_a} \\
\mathcal{G}(\Omega) & \sigma = \sigma_c
\end{cases}\,, \nn 
\end{align}
where $K_\gamma$ is the sum of just the non-cusp evolution kernels in \eq{Keta}:
\be
K_\gamma(\mu_H,\mu_J,\mu_S) \equiv K_{\gamma_H}(\mu,\mu_H) + 2K_{\gamma_J}(\mu,\mu_J) + K_{\gamma_S}(\mu,\mu_S)\,,
\ee
which is exactly $\mu$-invariant by itself at any order of accuracy, and where we will evaluate $\wt K,\Omega$ using closed-form expressions like \eqs{Kclosedform}{etaclosedform}, replacing \eq{Kclosedform} for $K_\Gamma$ with the expansion for $\wt K_\Gamma$ we have given in \eq{Ktildeclosedform}.

Note the expressions in \eq{Ktildeclosedform} for $\wt K_\Gamma$ reduce to those for 
$K_\Gamma$ in \eq{Kclosedform} for $Q=\mu_F$ or $r_Q=1$, as they must. The departure of $r_Q$ from 1 is an $\cO(\as)$ effect, so the difference between $K_\Gamma$ and $\wt K_\Gamma$ at each order is always subleading, consistent with our observations above. However, the $r_Q$ terms precisely compensate for the subleading terms that violate the $\mu$ invariance arising from using \eq{lnmu}, restoring exact $\mu$ invariance of $\wt K$ in \eq{Ktilde}.

%----------------------------------------------------------------------------
\begin{table}[t]
\centering
\begin{tabular}{|c||c|c|}
\hline 
$\frac{1}{\sigma_\text{tot}}\sigma_{c,\text{sing}}(\tau_a = 0.1)$ & \eq{sigmainvariant} &  \eq{differential}
\\ \hline \hline
NLL & 0.8154 &  $\{0.8164,0.8167,0.8169\} $
\\ \hline
NLL$'+\cO(\as)$ & 0.8451 & $\{0.8462,0.8465,0.8467\}$
\\ \hline
NNLL$+\cO(\as)$ &  0.85769 & $\{0.85775,0.85778,0.85781\}$
\\ \hline
NNLL$'+\cO(\as^2)$ &  0.87638 & $\{0.87643,0.87647,0.87650\}$
\\ \hline
\end{tabular}
\caption{Prediction of $\mu$-invariant form of  resummed perturbative integrated distribution in \eq{sigmainvariant} versus standard form \eq{differential}, which has residual $\mu$-dependence at truncated orders of resummed accuracy. These values are for $a=0$, $\tau_a = 0.1$, $Q=m_Z$, $\as(m_Z) =0.11$, and canonical scales $\mu_{H,J,S}$. The three values for \eq{differential} at each order are for the scale choices $\mu=\{\mu_J/2,\mu_J,2\mu_J\}$.}
\label{tab:muvariation}
\end{table}
%-----------------------------------------------------------------------------
To see the small numerical effect of using the $\mu$-invariant \eq{sigmainvariant} in place of the original form \eq{differential}, we show in \tab{muvariation} the predictions of these two formulas for the purely perturbative resummed integrated distributions $\sigma_{c,\text{sing}}/\sigma_0$ given by \eqs{differential}{sigmainvariant}  for $a=0$, $Q=m_Z$, $\as(m_Z)=0.11$, at $\tau_a = 0.1$, and canonical scales $\mu_H = Q, \mu_J = Q\tau_a^{1/(2-a)}, \mu_S = Q\tau_a$, and with $\mu$ varied amongst $\mu=\{\mu_J/2,\mu_J,2\mu_J\}$. We see the small effect both of reorganizing terms in \eq{differential} to get \eq{sigmainvariant} and also of varying $\mu$ in \eq{differential}, which retains subleading $\mu$-dependence. 
In the succession of results at different orders, we observe that the differences between \eqs{sigmainvariant}{differential} as well as the $\mu$-dependence of \eq{differential} itself, decrease at higher orders, as they should. This level of numerical variation is negligible in any practical application, and the use of the $\mu$-invariant reorganized \eq{sigmainvariant} is, again, mostly for purely mathematical aesthetics, though it does remove one pesky parameter to worry about.

\section{Angularity of symmetric four-particle final state}
\label{app:tetra}

In this appendix we determine the upper kinematic endpoint $\tau_a^{2,\text{max}}$ of the 
$\cO(\as^2)$ angularity distributions, which enters the remainder functions through \eq{REMAINDERfunc}.

\begin{figure}[t]
\vspace{-5mm}
\begin{center}
\includegraphics[width=.45\columnwidth]{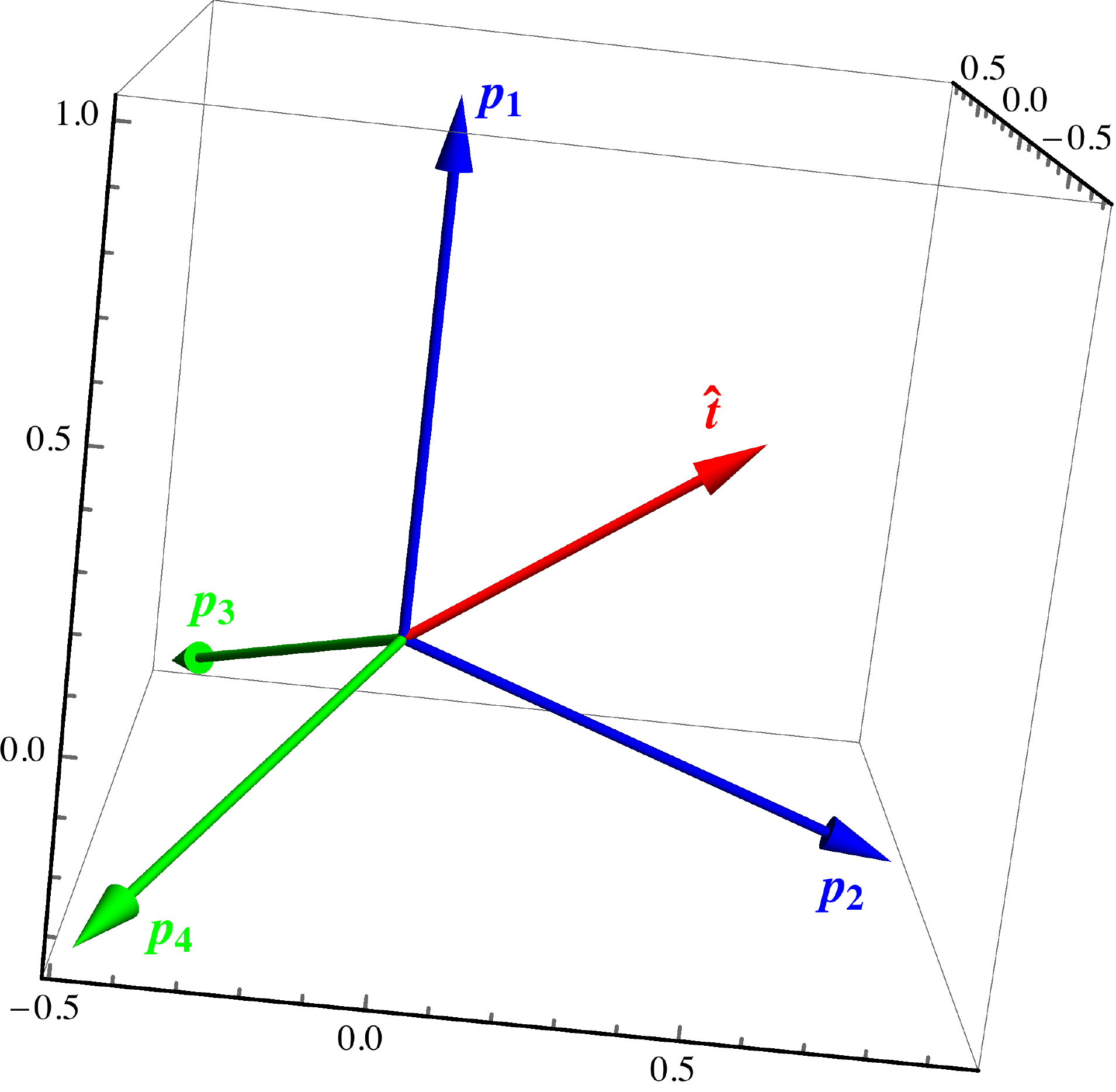}
\end{center}
\vspace{-5mm}
\caption{Maximally symmetric four-particle configuration. The blue and green pairs of particles are in back-to-back hemispheres, and the thrust axis $\vect{\hat t}$ is along the red vector, which is aligned with $\vect{p}_1+\vect{p}_2$ or $-(\vect{p}_3+\vect{p}_4)$.}
\vspace{10mm}
\label{fig:tetra}
\end{figure}

At $\cO(\as^2)$ there are up to four particles in the final state. To compute any $\tau_a$, we first need to determine the thrust axis. The maximum thrust $\tau_0$ occurs for the maximally symmetric four-particle configuration shown in \fig{tetra}. In the frame where the momentum of particle 1 is aligned with the $z$-axis and the momentum of particle 2 lies in the $xz$-plane, the four particles have three-momenta $\vect{p} = (p_x,p_y,p_z)$:
\be
\begin{split}
\vect{p}_1 &= \frac{Q}{4}(0,0,1)\,, \\
\vect{p}_2 &= \frac{Q}{4}(\sin\theta,0,\cos\theta) = \frac{Q}{4}\Bigl(\frac{2\sqrt{2}}{3},0,-\frac{1}{3}\Bigr)\,, \\
\vect{p}_3 &= \frac{Q}{4}(\sin\theta\cos\phi,\sin\theta\sin\phi,\cos\theta) = \frac{Q}{4}\Bigl(-\frac{\sqrt{2}}{3},\frac{\sqrt{6}}{3},-\frac{1}{3}\Bigr)\,, \\
\vect{p}_4 &= \frac{Q}{4}(\sin\theta\cos(2\phi),\sin\theta\sin(2\phi),\cos\theta) = \frac{Q}{4}\Bigl(-\frac{\sqrt{2}}{3},-\frac{\sqrt{6}}{3},-\frac{1}{3}\Bigr)\,, \\
\theta &= \cos^{-1}\Bigl(-\frac{1}{3}\Bigr)\approx 109.5^\circ\,,\quad \phi = \frac{2\pi}{3}\,,
\end{split}
\ee
where $\theta$ is the angle between any two particles and $\phi$ is the azimuthal angle between the planes of particles 1-2 and 3-4.
The thrust axis then lies along $\vect{p}_1+\vect{p}_2 = -(\vect{p}_3+\vect{p}_4)$.
In these coordinates,
\be
\vect{\hat t} = \Bigl(\frac{\sqrt{6}}{3},0,\frac{\sqrt{3}}{3}\Bigr)\,,
\ee
and the thrust of this configuration is 
\be
\tau_0^{2,\text{max}} = 1 -\frac{\sqrt{3}}{3}  \approx 0.4227\,.
\ee
In a coordinate system in which the thrust axis is aligned along the $z$-axis,
$\vect{\hat t} = (0,0,1)$, the four particles have three-momenta:
\be
\begin{split}
\vect{p}_1 &= \frac{Q}{4}(-\sin\vartheta,0,\cos\vartheta) = \frac{Q}{4}\Bigl(-\sqrt{\frac{2}{3}},0,\sqrt{\frac{1}{3}}\Bigr)\,,  \\
\vect{p}_2 &= \frac{Q}{4}(\sin\vartheta,0,\cos\vartheta) = \frac{Q}{4}\Bigl(\sqrt{\frac{2}{3}},0,\sqrt{\frac{1}{3}}\Bigr)\,,  \\
\vect{p}_3 &= \frac{Q}{4}(\sin\vartheta\cos\phi,\sin\vartheta\sin\phi,-\cos\vartheta) = \frac{Q}{4}\Bigl(\sqrt{\frac{1}{6}},\sqrt{\frac{1}{2}},-\sqrt{\frac{1}{3}}\Bigr)\,, \\
\vect{p}_4 &= \frac{Q}{4}(\sin\vartheta\cos(2\phi),\sin\vartheta\sin(2\phi),-\cos\vartheta) = \frac{Q}{4}\Bigl(-\sqrt{\frac{1}{6}},-\sqrt{\frac{1}{2}},-\sqrt{\frac{1}{3}}\Bigr)\,, \\
\vartheta &= \cos^{-1}\Bigl(-\frac{\sqrt{3}}{3}\Bigr)\approx 54.7^\circ\,,\quad \phi = \frac{2\pi}{3}\,,
\end{split}
\ee
where $\vartheta$ is now the angle between the $z$-axis and any particle.

%---------------------------------------------------------------------------------------
%---------------------------------------------------------------------------------------

\begin{figure}[t]
\vspace{-5mm}
\begin{center}
\raisebox{-7em}{\includegraphics[width=.5\columnwidth]{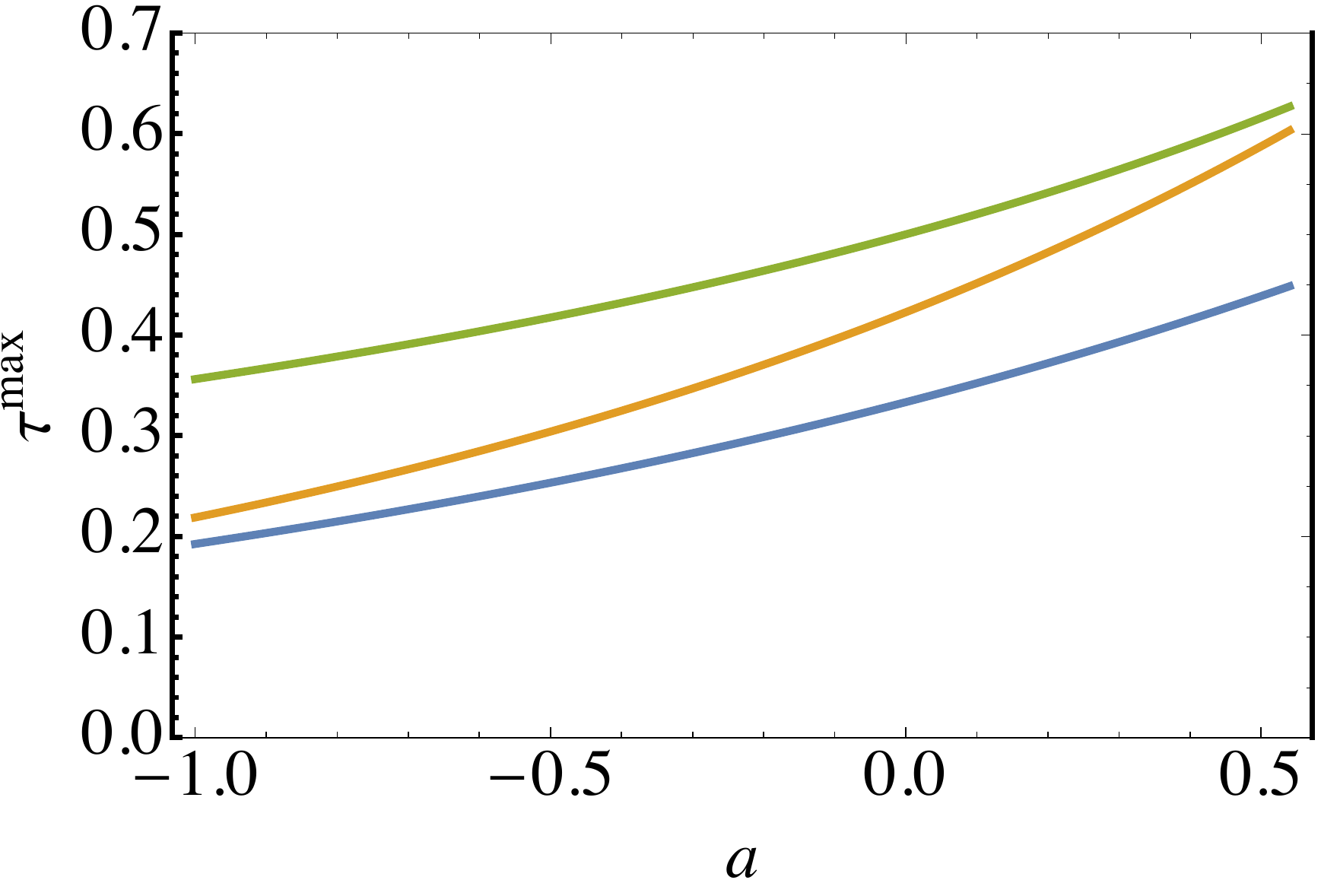}} \qquad 
\begin{tabular}{|c||c|c|c|}
\hline 
$a$ & $\tau_a^{1,\text{max}}$ & $\tau_a^{2,\text{max}}$ & $\tau_a^\text{sph.}$ 
\\ \hline \hline
$-1$ & 0.192&  0.219 & 0.356 \\
$-0.75$ & 0.221 &  0.258 & 0.385 \\
$-0.5$ & 0.253 & 0.304 & 0.417\\
$-0.25$ & 0.291 &  0.358 & 0.455\\
0 & 0.333 & 0.423 & 0.5 \\
0.25 & 0.382 &  0.498 & 0.553 \\
0.5 & 0.439 & 0.587 & 0.616\\
\hline
\end{tabular}
\end{center}
\vspace{-5mm}
\caption{The maximum values of $\tau_a$ for the three-particle (blue), four-particle (orange) and spherical distribution of particles (green) as a function of $a$.}
%\vspace{10mm}
\label{fig:taumax}
\end{figure}

The angularity of this configuration can easily be computed using
the representation
\be
\tau_a 
= \frac{1}{Q} \, \sum_{i} \, E_{i} \,
 (1-|\cos\theta_{i}|)^{1-a}\; |\sin\theta_{i}|^{a} \,,
\ee
where $E_i$ is the energy and $\theta_{i}$ the angle of the i'th particle with respect 
to the thrust axis. This yields
\be
\tau_a^{2,\text{max}} = 4\times \frac{1}{Q}\frac{Q}{4} (1-\cos\vartheta)^{1-a}\sin^a\vartheta = \Bigl(1 - \frac{\sqrt{3}}{3}\Bigr)^{1-a} \Bigl(\frac{2}{3}\Bigr)^{a/2} \,,
\ee
or in terms of the three-particle maximum $\tau_a^{1,\text{max}} = 1/3^{1-a/2}$,
\be
\tau_a^{2,\text{max}} = 
\tau_a^{1,\text{max}}\times 2^{a/2}(3-\sqrt{3})^{1-a}\,.
\ee
For some extreme values of $a$, it is possible that the maximum $\tau_a$ configuration is not the symmetric one in \fig{tetra}, as occurs for the three-particle configuration for $a\lesssim -2.6$ \cite{Hornig:2009vb}. We assume that this does not occur in the four-particle configuration for the values of $a$ we consider, $-1\leq a\leq 0.5$.  We in fact did not notice any anomalous deviation in the endpoints of our $\cO(\as^2)$ distributions away from this formula.
The maximum three- and four-particle values of $\tau_a$ are illustrated in \fig{taumax}.

%%%%%%%%%%%%%%%%%%%%%%%%%%%%%%%%%%%%%%%%%%%%%%%%%%%%%%%%%%%%%%%

\bibliography{angularitiesv3}

\providecommand{\href}[2]{#2}\begingroup\raggedright\begin{thebibliography}{10}

\bibitem{Dasgupta:2003iq}
M.~Dasgupta and G.~P. Salam, {\it {Event shapes in $e^+ e^-$ annihilation and
  deep inelastic scattering}},  {\em J.Phys.G} {\bf G30} (2004) R143,
  [\href{http://arxiv.org/abs/hep-ph/0312283}{{\tt hep-ph/0312283}}].

\bibitem{GehrmannDeRidder:2007hr}
A.~Gehrmann-De~Ridder, T.~Gehrmann, E.~W.~N. Glover, and G.~Heinrich, {\it
  {NNLO corrections to event shapes in $e^+ e^-$ annihilation}},  {\em JHEP}
  {\bf 12} (2007) 094, [\href{http://arxiv.org/abs/0711.4711}{{\tt
  arXiv:0711.4711}}].

\bibitem{Weinzierl:2009ms}
S.~Weinzierl, {\it {Event shapes and jet rates in electron-positron
  annihilation at NNLO}},  {\em JHEP} {\bf 06} (2009) 041,
  [\href{http://arxiv.org/abs/0904.1077}{{\tt arXiv:0904.1077}}].

\bibitem{Ridder:2014wza}
A.~Gehrmann-De~Ridder, T.~Gehrmann, E.~W.~N. Glover, and G.~Heinrich, {\it
  {EERAD3: Event shapes and jet rates in electron-positron annihilation at
  order $\alpha_s^3$}},  {\em Comput. Phys. Commun.} {\bf 185} (2014) 3331,
  [\href{http://arxiv.org/abs/1402.4140}{{\tt arXiv:1402.4140}}].

\bibitem{DelDuca:2016ily}
V.~Del~Duca, C.~Duhr, A.~Kardos, G.~Somogyi, Z.~Sz\~or, Z.~Tr\'ocs\'anyi, and
  Z.~Tulip\'ant, {\it {Jet production in the CoLoRFulNNLO method: event shapes
  in electron-positron collisions}},  {\em Phys. Rev.} {\bf D94} (2016), no.~7
  074019, [\href{http://arxiv.org/abs/1606.03453}{{\tt arXiv:1606.03453}}].

\bibitem{Catani:1992ua}
S.~Catani, L.~Trentadue, G.~Turnock, and B.~R. Webber, {\it Resummation of
  large logarithms in {$e^+ e^-$} event shape distributions},  {\em Nucl.
  Phys.} {\bf B407} (1993) 3--42.

\bibitem{Bauer:2000ew}
C.~W. Bauer, S.~Fleming, and M.~E. Luke, {\it Summing {S}udakov logarithms in
  {$ B\to X_s \gamma$} in effective field theory},  {\em Phys. Rev.} {\bf D63}
  (2000) 014006, [\href{http://arxiv.org/abs/hep-ph/0005275}{{\tt
  hep-ph/0005275}}].

\bibitem{Bauer:2000yr}
C.~W. Bauer, S.~Fleming, D.~Pirjol, and I.~W. Stewart, {\it An effective field
  theory for collinear and soft gluons: Heavy to light decays},  {\em Phys.
  Rev.} {\bf D63} (2001) 114020,
  [\href{http://arxiv.org/abs/hep-ph/0011336}{{\tt hep-ph/0011336}}].

\bibitem{Bauer:2001yt}
C.~W. Bauer, D.~Pirjol, and I.~W. Stewart, {\it Soft-collinear factorization in
  effective field theory},  {\em Phys. Rev.} {\bf D65} (2002) 054022,
  [\href{http://arxiv.org/abs/hep-ph/0109045}{{\tt hep-ph/0109045}}].

\bibitem{Beneke:2002ph}
M.~Beneke, A.~Chapovsky, M.~Diehl, and T.~Feldmann, {\it {Soft collinear
  effective theory and heavy to light currents beyond leading power}},  {\em
  Nucl.Phys.} {\bf B643} (2002) 431--476,
  [\href{http://arxiv.org/abs/hep-ph/0206152}{{\tt hep-ph/0206152}}].

\bibitem{deFlorian:2004mp}
D.~de~Florian and M.~Grazzini, {\it The back-to-back region in {$e^+ e^-$}
  energy energy correlation},  {\em Nucl. Phys.} {\bf B704} (2005) 387--403,
  [\href{http://arxiv.org/abs/hep-ph/0407241}{{\tt hep-ph/0407241}}].

\bibitem{Becher:2012qc}
T.~Becher and G.~Bell, {\it {NNLL Resummation for Jet Broadening}},  {\em JHEP}
  {\bf 1211} (2012) 126, [\href{http://arxiv.org/abs/1210.0580}{{\tt
  arXiv:1210.0580}}].

\bibitem{Banfi:2014sua}
A.~Banfi, H.~McAslan, P.~F. Monni, and G.~Zanderighi, {\it {A general method
  for the resummation of event-shape distributions in $e^+ e^-$ annihilation}},
   {\em JHEP} {\bf 05} (2015) 102, [\href{http://arxiv.org/abs/1412.2126}{{\tt
  arXiv:1412.2126}}].

\bibitem{Becher:2008cf}
T.~Becher and M.~D. Schwartz, {\it A precise determination of {$\alpha_s$} from
  {LEP} thrust data using effective field theory},  {\em JHEP} {\bf 07} (2008)
  034, [\href{http://arxiv.org/abs/0803.0342}{{\tt arXiv:0803.0342}}].

\bibitem{Chien:2010kc}
Y.-T. Chien and M.~D. Schwartz, {\it {Resummation of heavy jet mass and
  comparison to LEP data}},  {\em JHEP} {\bf 1008} (2010) 058,
  [\href{http://arxiv.org/abs/1005.1644}{{\tt arXiv:1005.1644}}].

\bibitem{Hoang:2014wka}
A.~H. Hoang, D.~W. Kolodrubetz, V.~Mateu, and I.~W. Stewart, {\it
  {$C$-parameter distribution at $N^3LL'$ including power corrections}},  {\em
  Phys. Rev.} {\bf D91} (2015), no.~9 094017,
  [\href{http://arxiv.org/abs/1411.6633}{{\tt arXiv:1411.6633}}].

\bibitem{Davison:2008vx}
R.~A. Davison and B.~R. Webber, {\it Non-perturbative contribution to the
  thrust distribution in {$e^+ e^-$} annihilation},  {\em Eur. Phys. J.} {\bf
  C59} (2009) 13--25, [\href{http://arxiv.org/abs/0809.3326}{{\tt
  arXiv:0809.3326}}].

\bibitem{Abbate:2010xh}
R.~Abbate, M.~Fickinger, A.~H. Hoang, V.~Mateu, and I.~W. Stewart, {\it {Thrust
  at N$^3$LL with Power Corrections and a Precision Global Fit for
  $\alpha_s(m_Z)$}},  {\em Phys. Rev.} {\bf D83} (2011) 074021,
  [\href{http://arxiv.org/abs/1006.3080}{{\tt arXiv:1006.3080}}].

\bibitem{Abbate:2012jh}
R.~Abbate, M.~Fickinger, A.~H. Hoang, V.~Mateu, and I.~W. Stewart, {\it
  {Precision Thrust Cumulant Moments at $N^3$LL}},  {\em Phys. Rev.} {\bf D86}
  (2012) 094002, [\href{http://arxiv.org/abs/1204.5746}{{\tt
  arXiv:1204.5746}}].

\bibitem{Gehrmann:2012sc}
T.~Gehrmann, G.~Luisoni, and P.~F. Monni, {\it {Power corrections in the
  dispersive model for a determination of the strong coupling constant from the
  thrust distribution}},  {\em Eur. Phys. J.} {\bf C73} (2013), no.~1 2265,
  [\href{http://arxiv.org/abs/1210.6945}{{\tt arXiv:1210.6945}}].

\bibitem{Hoang:2015hka}
A.~H. Hoang, D.~W. Kolodrubetz, V.~Mateu, and I.~W. Stewart, {\it {Precise
  determination of $\alpha_s$ from the $C$-parameter distribution}},  {\em
  Phys. Rev.} {\bf D91} (2015), no.~9 094018,
  [\href{http://arxiv.org/abs/1501.04111}{{\tt arXiv:1501.04111}}].

\bibitem{Lee:2006fn}
C.~Lee and G.~F. Sterman, {\it {Universality of nonperturbative effects in
  event shapes}},  {\em eConf} {\bf C0601121} (2006) A001,
  [\href{http://arxiv.org/abs/hep-ph/0603066}{{\tt hep-ph/0603066}}].

\bibitem{Lee:2006nr}
C.~Lee and G.~Sterman, {\it Momentum flow correlations from event shapes:
  Factorized soft gluons and {Soft-Collinear Effective Theory}},  {\em Phys.
  Rev.} {\bf D75} (2007) 014022,
  [\href{http://arxiv.org/abs/hep-ph/0611061}{{\tt hep-ph/0611061}}].

\bibitem{Becher:2013iya}
T.~Becher and G.~Bell, {\it {Enhanced nonperturbative effects through the
  collinear anomaly}},  {\em Phys. Rev. Lett.} {\bf 112} (2014), no.~18 182002,
  [\href{http://arxiv.org/abs/1312.5327}{{\tt arXiv:1312.5327}}].

\bibitem{Salam:2001bd}
G.~Salam and D.~Wicke, {\it {Hadron masses and power corrections to event
  shapes}},  {\em JHEP} {\bf 0105} (2001) 061,
  [\href{http://arxiv.org/abs/hep-ph/0102343}{{\tt hep-ph/0102343}}].

\bibitem{Mateu:2012nk}
V.~Mateu, I.~W. Stewart, and J.~Thaler, {\it {Power Corrections to Event Shapes
  with Mass-Dependent Operators}},  {\em Phys.Rev.} {\bf D87} (2013) 014025,
  [\href{http://arxiv.org/abs/1209.3781}{{\tt arXiv:1209.3781}}].

\bibitem{PhysRevD.98.030001}
{\bf Particle Data Group} Collaboration, M.~Tanabashi et~al., {\it Review of
  particle physics},  {\em Phys. Rev. D} {\bf 98} (Aug, 2018) 030001.

\bibitem{Bethke:2011tr}
S.~Bethke et~al., {\it {Workshop on Precision Measurements of alphas}},
  \href{http://arxiv.org/abs/1110.0016}{{\tt arXiv:1110.0016}}.

\bibitem{Berger:2003iw}
C.~F. Berger, T.~Kucs, and G.~Sterman, {\it Event shape / energy flow
  correlations},  {\em Phys. Rev.} {\bf D68} (2003) 014012,
  [\href{http://arxiv.org/abs/hep-ph/0303051}{{\tt hep-ph/0303051}}].

\bibitem{Dokshitzer:1998kz}
Y.~L. Dokshitzer, A.~Lucenti, G.~Marchesini, and G.~P. Salam, {\it On the {QCD}
  analysis of jet broadening},  {\em JHEP} {\bf 01} (1998) 011,
  [\href{http://arxiv.org/abs/hep-ph/9801324}{{\tt hep-ph/9801324}}].

\bibitem{Larkoski:2014uqa}
A.~J. Larkoski, D.~Neill, and J.~Thaler, {\it {Jet Shapes with the Broadening
  Axis}},  {\em JHEP} {\bf 1404} (2014) 017,
  [\href{http://arxiv.org/abs/1401.2158}{{\tt arXiv:1401.2158}}].

\bibitem{Procura:2018zpn}
M.~Procura, W.~J. Waalewijn, and L.~Zeune, {\it {Joint resummation of two
  angularities at next-to-next-to-leading logarithmic order}},  {\em JHEP} {\bf
  10} (2018) 098, [\href{http://arxiv.org/abs/1806.10622}{{\tt
  arXiv:1806.10622}}].

\bibitem{Bauer:2008dt}
C.~W. Bauer, S.~Fleming, C.~Lee, and G.~Sterman, {\it Factorization of
  {$e^+e^-$} event shape distributions with hadronic final states in {Soft
  Collinear Effective Theory}},  {\em Phys. Rev.} {\bf D78} (2008) 034027,
  [\href{http://arxiv.org/abs/0801.4569}{{\tt arXiv:0801.4569}}].

\bibitem{Hornig:2009vb}
A.~Hornig, C.~Lee, and G.~Ovanesyan, {\it Effective predictions of event
  shapes: Factorized, resummed, and gapped angularity distributions},  {\em
  JHEP} {\bf 05} (2009) 122, [\href{http://arxiv.org/abs/0901.3780}{{\tt
  arXiv:0901.3780}}].

\bibitem{Almeida:2014uva}
L.~G. Almeida, S.~D. Ellis, C.~Lee, G.~Sterman, I.~Sung, and J.~R. Walsh, {\it
  {Comparing and counting logs in direct and effective methods of QCD
  resummation}},  {\em JHEP} {\bf 1404} (2014) 174,
  [\href{http://arxiv.org/abs/1401.4460}{{\tt arXiv:1401.4460}}].

\bibitem{Becher:2011pf}
T.~Becher, G.~Bell, and M.~Neubert, {\it {Factorization and Resummation for Jet
  Broadening}},  {\em Phys.Lett.} {\bf B704} (2011) 276--283,
  [\href{http://arxiv.org/abs/1104.4108}{{\tt arXiv:1104.4108}}].

\bibitem{Chiu:2012ir}
J.-Y. Chiu, A.~Jain, D.~Neill, and I.~Z. Rothstein, {\it {A Formalism for the
  Systematic Treatment of Rapidity Logarithms in Quantum Field Theory}},  {\em
  JHEP} {\bf 1205} (2012) 084, [\href{http://arxiv.org/abs/1202.0814}{{\tt
  arXiv:1202.0814}}].

\bibitem{Korchemsky:1998ev}
G.~P. Korchemsky, {\it {Shape functions and power corrections to the event
  shapes}},  in {\em {Continuous advances in QCD. Proceedings, 3rd Workshop,
  QCD'98, Minneapolis, USA, April 16-19, 1998}}, pp.~489--498, 1998.
\newblock \href{http://arxiv.org/abs/hep-ph/9806537}{{\tt hep-ph/9806537}}.

\bibitem{Korchemsky:1999kt}
G.~P. Korchemsky and G.~Sterman, {\it Power corrections to event shapes and
  factorization},  {\em Nucl. Phys.} {\bf B555} (1999) 335--351,
  [\href{http://arxiv.org/abs/hep-ph/9902341}{{\tt hep-ph/9902341}}].

\bibitem{Berger:2003pk}
C.~F. Berger and G.~Sterman, {\it Scaling rule for nonperturbative radiation in
  a class of event shapes},  {\em JHEP} {\bf 09} (2003) 058,
  [\href{http://arxiv.org/abs/hep-ph/0307394}{{\tt hep-ph/0307394}}].

\bibitem{Berger:2004xf}
C.~F. Berger and L.~Magnea, {\it Scaling of power corrections for angularities
  from dressed gluon exponentiation},  {\em Phys. Rev.} {\bf D70} (2004)
  094010, [\href{http://arxiv.org/abs/hep-ph/0407024}{{\tt hep-ph/0407024}}].

\bibitem{Gardi:2001di}
E.~Gardi, {\it Dressed gluon exponentiation},  {\em Nucl. Phys.} {\bf B622}
  (2002) 365--392, [\href{http://arxiv.org/abs/hep-ph/0108222}{{\tt
  hep-ph/0108222}}].

\bibitem{Dokshitzer:1995qm}
Y.~L. Dokshitzer, G.~Marchesini, and B.~R. Webber, {\it Dispersive approach to
  power-behaved contributions in {QCD} hard processes},  {\em Nucl. Phys.} {\bf
  B469} (1996) 93--142, [\href{http://arxiv.org/abs/hep-ph/9512336}{{\tt
  hep-ph/9512336}}].

\bibitem{Dokshitzer:1995zt}
Y.~L. Dokshitzer and B.~R. Webber, {\it Calculation of power corrections to
  hadronic event shapes},  {\em Phys. Lett.} {\bf B352} (1995) 451--455,
  [\href{http://arxiv.org/abs/hep-ph/9504219}{{\tt hep-ph/9504219}}].

\bibitem{Dokshitzer:1998pt}
Y.~L. Dokshitzer, A.~Lucenti, G.~Marchesini, and G.~P. Salam, {\it {On the
  universality of the Milan factor for 1 / Q power corrections to jet shapes}},
   {\em JHEP} {\bf 05} (1998) 003,
  [\href{http://arxiv.org/abs/hep-ph/9802381}{{\tt hep-ph/9802381}}].

\bibitem{Bell:2015lsf}
G.~Bell, R.~Rahn, and J.~Talbert, {\it {Automated Calculation of Dijet Soft
  Functions in Soft-Collinear Effective Theory}},  {\em PoS} {\bf RADCOR2015}
  (2016) 052, [\href{http://arxiv.org/abs/1512.06100}{{\tt arXiv:1512.06100}}].

\bibitem{Bell:2018vaa}
G.~Bell, R.~Rahn, and J.~Talbert, {\it {Two-loop anomalous dimensions of
  generic dijet soft functions}},  {\em Nucl. Phys.} {\bf B936} (2018)
  520--541, [\href{http://arxiv.org/abs/1805.12414}{{\tt arXiv:1805.12414}}].

\bibitem{Bell:2018oqa}
G.~Bell, R.~Rahn, and J.~Talbert, {\it {Generic dijet soft functions at
  two-loop order: correlated emissions}},
  \href{http://arxiv.org/abs/1812.08690}{{\tt arXiv:1812.08690}}.

\bibitem{Banfi:2018mcq}
A.~Banfi, B.~K. El-Menoufi, and P.~F. Monni, {\it {The Sudakov radiator for jet
  observables and the soft physical coupling}},
  \href{http://arxiv.org/abs/1807.11487}{{\tt arXiv:1807.11487}}.

\bibitem{Bell:2017wvi}
G.~Bell, A.~Hornig, C.~Lee, and J.~Talbert, {\it {Angularities from LEP to
  FCC-ee}},  in {\em {Proceedings, Parton Radiation and Fragmentation from LHC
  to FCC-ee: CERN, Geneva, Switzerland, November 22-23, 2016}}, pp.~90--96,
  2017.

\bibitem{Hoang:2007vb}
A.~H. Hoang and I.~W. Stewart, {\it Designing gapped soft functions for jet
  production},  {\em Phys. Lett.} {\bf B660} (2008) 483--493,
  [\href{http://arxiv.org/abs/0709.3519}{{\tt arXiv:0709.3519}}].

\bibitem{Achard:2011zz}
P.~Achard, O.~Adriani, M.~Aguilar-Benitez, J.~Alcaraz, G.~Alemanni, et~al.,
  {\it {Generalized event shape and energy flow studies in $e^+ e^-$
  annihilation at $\sqrt{s} = $91.2--208.0 GeV}},  {\em JHEP} {\bf 1110} (2011)
  143.

\bibitem{Berger:2003zh}
C.~F. Berger, {\em {Soft gluon exponentiation and resummation}}.
\newblock PhD thesis, SUNY, Stony Brook, 2003.
\newblock \href{http://arxiv.org/abs/hep-ph/0305076}{{\tt hep-ph/0305076}}.

\bibitem{Fleming:2007xt}
S.~Fleming, A.~H. Hoang, S.~Mantry, and I.~W. Stewart, {\it Top jets in the
  peak region: Factorization analysis with {NLL} resummation},  {\em Phys.
  Rev.} {\bf D77} (2008) 114003, [\href{http://arxiv.org/abs/0711.2079}{{\tt
  arXiv:0711.2079}}].

\bibitem{Becher:2006mr}
T.~Becher, M.~Neubert, and B.~D. Pecjak, {\it Factorization and momentum-space
  resummation in deep-inelastic scattering},  {\em JHEP} {\bf 01} (2007) 076,
  [\href{http://arxiv.org/abs/hep-ph/0607228}{{\tt hep-ph/0607228}}].

\bibitem{Becher:2006nr}
T.~Becher and M.~Neubert, {\it Threshold resummation in momentum space from
  effective field theory},  {\em Phys. Rev. Lett.} {\bf 97} (2006) 082001,
  [\href{http://arxiv.org/abs/hep-ph/0605050}{{\tt hep-ph/0605050}}].

\bibitem{Ligeti:2008ac}
Z.~Ligeti, I.~W. Stewart, and F.~J. Tackmann, {\it Treating the {$b$} quark
  distribution function with reliable uncertainties},  {\em Phys. Rev.} {\bf
  D78} (2008) 114014, [\href{http://arxiv.org/abs/0807.1926}{{\tt
  arXiv:0807.1926}}].

\bibitem{Manohar:2003vb}
A.~V. Manohar, {\it Deep inelastic scattering as {$x \to 1$} using
  {Soft-Collinear Effective Theory}},  {\em Phys. Rev.} {\bf D68} (2003)
  114019, [\href{http://arxiv.org/abs/hep-ph/0309176}{{\tt hep-ph/0309176}}].

\bibitem{Bauer:2003di}
C.~W. Bauer, C.~Lee, A.~V. Manohar, and M.~B. Wise, {\it Enhanced
  nonperturbative effects in {Z} decays to hadrons},  {\em Phys. Rev.} {\bf
  D70} (2004) 034014, [\href{http://arxiv.org/abs/hep-ph/0309278}{{\tt
  hep-ph/0309278}}].

\bibitem{Idilbi:2006dg}
A.~Idilbi, X.-d. Ji, and F.~Yuan, {\it {Resummation of threshold logarithms in
  effective field theory for DIS, Drell-Yan and Higgs production}},  {\em
  Nucl.Phys.} {\bf B753} (2006) 42--68,
  [\href{http://arxiv.org/abs/hep-ph/0605068}{{\tt hep-ph/0605068}}].

\bibitem{Lee:2010cga}
R.~N. Lee, A.~V. Smirnov, and V.~A. Smirnov, {\it {Analytic Results for
  Massless Three-Loop Form Factors}},  {\em JHEP} {\bf 04} (2010) 020,
  [\href{http://arxiv.org/abs/1001.2887}{{\tt arXiv:1001.2887}}].

\bibitem{Gehrmann:2005pd}
T.~Gehrmann, T.~Huber, and D.~Maitre, {\it {Two-loop quark and gluon
  form-factors in dimensional regularisation}},  {\em Phys. Lett.} {\bf B622}
  (2005) 295--302, [\href{http://arxiv.org/abs/hep-ph/0507061}{{\tt
  hep-ph/0507061}}].

\bibitem{Matsuura:1987wt}
T.~Matsuura and W.~L. van Neerven, {\it {Second Order Logarithmic Corrections
  to the {Drell-Yan} Cross-section}},  {\em Z. Phys.} {\bf C38} (1988) 623.

\bibitem{Matsuura:1988sm}
T.~Matsuura, S.~C. van~der Marck, and W.~L. van Neerven, {\it {The Calculation
  of the Second Order Soft and Virtual Contributions to the Drell-Yan
  Cross-Section}},  {\em Nucl. Phys.} {\bf B319} (1989) 570.

\bibitem{Moch:2005id}
S.~Moch, J.~Vermaseren, and A.~Vogt, {\it {The Quark form-factor at higher
  orders}},  {\em JHEP} {\bf 0508} (2005) 049,
  [\href{http://arxiv.org/abs/hep-ph/0507039}{{\tt hep-ph/0507039}}].

\bibitem{Kelley:2011ng}
R.~Kelley, M.~D. Schwartz, R.~M. Schabinger, and H.~X. Zhu, {\it {The two-loop
  hemisphere soft function}},  {\em Phys.Rev.} {\bf D84} (2011) 045022,
  [\href{http://arxiv.org/abs/1105.3676}{{\tt arXiv:1105.3676}}].

\bibitem{Monni:2011gb}
P.~F. Monni, T.~Gehrmann, and G.~Luisoni, {\it {Two-Loop Soft Corrections and
  Resummation of the Thrust Distribution in the Dijet Region}},  {\em JHEP}
  {\bf 1108} (2011) 010, [\href{http://arxiv.org/abs/1105.4560}{{\tt
  arXiv:1105.4560}}].

\bibitem{Bauer:2003pi}
C.~W. Bauer and A.~V. Manohar, {\it Shape function effects in {$B \to X_s
  \gamma$} and {$B \to X_u l \bar\nu$} decays},  {\em Phys. Rev.} {\bf D70}
  (2004) 034024, [\href{http://arxiv.org/abs/hep-ph/0312109}{{\tt
  hep-ph/0312109}}].

\bibitem{Becher:2009th}
T.~Becher and M.~D. Schwartz, {\it {Direct photon production with effective
  field theory}},  {\em JHEP} {\bf 02} (2010) 040,
  [\href{http://arxiv.org/abs/0911.0681}{{\tt arXiv:0911.0681}}].

\bibitem{Becher:2006qw}
T.~Becher and M.~Neubert, {\it {Toward a NNLO calculation of the
  anti-B$\to$X(s) gamma decay rate with a cut on photon energy. II. Two-loop
  result for the jet function}},  {\em Phys.Lett.} {\bf B637} (2006) 251--259,
  [\href{http://arxiv.org/abs/hep-ph/0603140}{{\tt hep-ph/0603140}}].

\bibitem{Becher:2010pd}
T.~Becher and G.~Bell, {\it {The gluon jet function at two-loop order}},  {\em
  Phys. Lett.} {\bf B695} (2011) 252--258,
  [\href{http://arxiv.org/abs/1008.1936}{{\tt arXiv:1008.1936}}].

\bibitem{Bruser:2018rad}
R.~Br{\"u}ser, Z.~L. Liu, and M.~Stahlhofen, {\it {The Three-loop Quark Jet
  Function}},  {\em Phys. Rev. Lett.} {\bf 121} (2018) 072003,
  [\href{http://arxiv.org/abs/1804.09722}{{\tt arXiv:1804.09722}}].

\bibitem{Banerjee:2018ozf}
P.~Banerjee, P.~K. Dhani, and V.~Ravindran, {\it {Gluon jet function at three
  loops in QCD}},  {\em Phys. Rev.} {\bf D98} (2018), no.~9 094016,
  [\href{http://arxiv.org/abs/1805.02637}{{\tt arXiv:1805.02637}}].

\bibitem{Catani:1996jh}
S.~Catani and M.~H. Seymour, {\it The dipole formalism for the calculation of
  {QCD} jet cross sections at next-to-leading order},  {\em Phys. Lett.} {\bf
  B378} (1996) 287--301, [\href{http://arxiv.org/abs/hep-ph/9602277}{{\tt
  hep-ph/9602277}}].

\bibitem{Catani:1996vz}
S.~Catani and M.~H. Seymour, {\it A general algorithm for calculating jet cross
  sections in {NLO QCD}},  {\em Nucl. Phys.} {\bf B485} (1997) 291--419,
  [\href{http://arxiv.org/abs/hep-ph/9605323}{{\tt hep-ph/9605323}}].

\bibitem{Hoang:2008fs}
A.~H. Hoang and S.~Kluth, {\it Hemisphere soft function at
  {$\mathcal{O}(\alpha_s^2)$} for dijet production in {$e^+e^-$} annihilation},
   \href{http://arxiv.org/abs/0806.3852}{{\tt arXiv:0806.3852}}.

\bibitem{Chetyrkin:1979bj}
K.~G. Chetyrkin, A.~L. Kataev, and F.~V. Tkachov, {\it {Higher Order
  Corrections to $\sigma_{\rm tot}(e^+ e^- \rightarrow$ Hadrons) in Quantum
  Chromodynamics}},  {\em Phys. Lett.} {\bf B85} (1979) 277.

\bibitem{Dine:1979qh}
M.~Dine and J.~R. Sapirstein, {\it {Higher Order QCD Corrections in $e^+ e^-$
  Annihilation}},  {\em Phys. Rev. Lett.} {\bf 43} (1979) 668.

\bibitem{Celmaster:1979xr}
W.~Celmaster and R.~J. Gonsalves, {\it {An Analytic Calculation of Higher Order
  Quantum Chromodynamic Corrections in $e^+ e^-$ Annihilation}},  {\em Phys.
  Rev. Lett.} {\bf 44} (1980) 560.

\bibitem{Chien:2015cka}
Y.-T. Chien, A.~Hornig, and C.~Lee, {\it {Soft-collinear mode for jet cross
  sections in soft collinear effective theory}},  {\em Phys. Rev.} {\bf D93}
  (2016), no.~1 014033, [\href{http://arxiv.org/abs/1509.04287}{{\tt
  arXiv:1509.04287}}].

\bibitem{Kang:2013nha}
D.~Kang, C.~Lee, and I.~W. Stewart, {\it {Using 1-Jettiness to Measure 2 Jets
  in DIS 3 Ways}},  {\em Phys.Rev.} {\bf D88} (2013) 054004,
  [\href{http://arxiv.org/abs/1303.6952}{{\tt arXiv:1303.6952}}].

\bibitem{Korchemsky:1994is}
G.~P. Korchemsky and G.~Sterman, {\it Nonperturbative corrections in resummed
  cross-sections},  {\em Nucl. Phys.} {\bf B437} (1995) 415--432,
  [\href{http://arxiv.org/abs/hep-ph/9411211}{{\tt hep-ph/9411211}}].

\bibitem{Beneke:1994sw}
M.~Beneke and V.~M. Braun, {\it {Heavy quark effective theory beyond
  perturbation theory: Renormalons, the pole mass and the residual mass term}},
   {\em Nucl. Phys.} {\bf B426} (1994) 301--343,
  [\href{http://arxiv.org/abs/hep-ph/9402364}{{\tt hep-ph/9402364}}].

\bibitem{Jain:2008gb}
A.~Jain, I.~Scimemi, and I.~W. Stewart, {\it {Two-loop Jet-Function and
  Jet-Mass for Top Quarks}},  {\em Phys. Rev.} {\bf D77} (2008) 094008,
  [\href{http://arxiv.org/abs/0801.0743}{{\tt arXiv:0801.0743}}].

\bibitem{Hoang:2008yj}
A.~H. Hoang, A.~Jain, I.~Scimemi, and I.~W. Stewart, {\it Infrared
  renormalization group flow for heavy quark masses},  {\em Phys. Rev. Lett.}
  {\bf 101} (2008) 151602, [\href{http://arxiv.org/abs/0803.4214}{{\tt
  arXiv:0803.4214}}].

\bibitem{VM}
V.~Mateu. Private communication.

\bibitem{Kang:2014qba}
D.~Kang, C.~Lee, and I.~W. Stewart, {\it {Analytic calculation of 1-jettiness
  in DIS at $ \mathcal{O}\left({\alpha}_s\right) $}},  {\em JHEP} {\bf 1411}
  (2014) 132, [\href{http://arxiv.org/abs/1407.6706}{{\tt arXiv:1407.6706}}].

\bibitem{Hornig:2016ahz}
A.~Hornig, Y.~Makris, and T.~Mehen, {\it {Jet Shapes in Dijet Events at the LHC
  in SCET}},  {\em JHEP} {\bf 04} (2016) 097,
  [\href{http://arxiv.org/abs/1601.01319}{{\tt arXiv:1601.01319}}].

\bibitem{Freedman:2013vya}
S.~M. Freedman, {\it {Subleading Corrections To Thrust Using Effective Field
  Theory}},  \href{http://arxiv.org/abs/1303.1558}{{\tt arXiv:1303.1558}}.

\bibitem{Moult:2018jjd}
I.~Moult, I.~W. Stewart, G.~Vita, and H.~X. Zhu, {\it {First Subleading Power
  Resummation for Event Shapes}},  {\em JHEP} {\bf 08} (2018) 013,
  [\href{http://arxiv.org/abs/1804.04665}{{\tt arXiv:1804.04665}}].

\bibitem{Bertolini:2017eui}
D.~Bertolini, M.~P. Solon, and J.~R. Walsh, {\it {Integrated and Differential
  Accuracy in Resummed Cross Sections}},  {\em Phys. Rev.} {\bf D95} (2017),
  no.~5 054024, [\href{http://arxiv.org/abs/1701.07919}{{\tt
  arXiv:1701.07919}}].

\bibitem{Tarasov:1980au}
O.~Tarasov, A.~Vladimirov, and A.~Y. Zharkov, {\it {The Gell-Mann-Low Function
  of QCD in the Three Loop Approximation}},  {\em Phys.Lett.} {\bf B93} (1980)
  429--432.

\bibitem{Larin:1993tp}
S.~Larin and J.~Vermaseren, {\it {The Three loop QCD Beta function and
  anomalous dimensions}},  {\em Phys.Lett.} {\bf B303} (1993) 334--336,
  [\href{http://arxiv.org/abs/hep-ph/9302208}{{\tt hep-ph/9302208}}].

\bibitem{vanRitbergen:1997va}
T.~van Ritbergen, J.~A.~M. Vermaseren, and S.~A. Larin, {\it {The Four loop
  beta function in quantum chromodynamics}},  {\em Phys. Lett.} {\bf B400}
  (1997) 379--384, [\href{http://arxiv.org/abs/hep-ph/9701390}{{\tt
  hep-ph/9701390}}].

\bibitem{Korchemsky:1987wg}
G.~P. Korchemsky and A.~V. Radyushkin, {\it {Renormalization of the Wilson
  Loops Beyond the Leading Order}},  {\em Nucl. Phys.} {\bf B283} (1987)
  342--364.

\bibitem{Moch:2004pa}
S.~Moch, J.~Vermaseren, and A.~Vogt, {\it {The Three loop splitting functions
  in QCD: The Nonsinglet case}},  {\em Nucl.Phys.} {\bf B688} (2004) 101--134,
  [\href{http://arxiv.org/abs/hep-ph/0403192}{{\tt hep-ph/0403192}}].

\end{thebibliography}\endgroup

\end{document}